\documentclass[epj]{svjour}
\usepackage{graphicx}
\usepackage{epstopdf}
\usepackage{subfigure}
\usepackage{bm}
\usepackage{xspace}
\usepackage{multirow}
\usepackage{tikz}
\usepackage{amsmath,amssymb}
\usepackage[amsmath,amssymb,textstyle]{SIunits}
\usepackage{relsize}
\usepackage{hyperref}

\makeatletter
\def\contribution#1#2{\@startsection{subsection}{2}{\z@}%
    {-21dd plus-4pt minus-4pt}{10.5dd plus 4pt minus4pt}
    {\normalsize\sffamily\bfseries}{#1\\\mdseries #2}}
\makeatother

\begin{document}
\title{Multi-Parton Interactions at the LHC}
\author{P. Bartalini \inst{1}
\and E. L. Berger \inst{2}
\and B. Blok \inst{3} 
\and G. Calucci \inst{4}
\and R. Corke \inst{5} 
\and M. Diehl \inst{6}
\and Yu. Dokshitzer \inst{7}
\and L. Fan\`o \inst{8}
\and L. Frankfurt \inst{9}
\and J. R. Gaunt \inst{10}
\and S. Gieseke \inst{11}
\and G. Gustafson \inst{5}
\and D. Kar \inst{12}
\and C.-H. Kom \inst{17}
\and A. Kulesza \inst{13}
\and E. Maina \inst{14}
\and Z. Nagy \inst{6}
\and Ch. R\"ohr \inst{11}
\and A. Si\'odmok \inst{11}
\and M. Schmelling \inst{15}
\and W. J. Stirling \inst{10}
\and M. Strikman \inst{16}
\and D. Treleani \inst{4}\\[1em]
edited by A. Kulesza and Z. Nagy\\[1em]
}                     
%
%
\institute{
National Taiwan University, Taipei,Taiwan
\and High Energy Physics Division, Argonne National Laboratory, Argonne, IL 60439, USA
\and Department of Physics, Technion---Israel Institute of Technology, 32000 Haifa, Israel
\and Dipartimento di
Fisica dell'Universit\`a di Trieste and INFN, Sezione di
Trieste, Strada Costiera 11, Miramare-Grignano, I-34151 Trieste,
Italy
\and Theoretical High Energy Physics, Department of Astronomy and Theoretical Physics,Lund University, S\"olvegatan 14A, S-223 62 Lund, Sweden
\and Deutsches Elektronen-Synchroton DESY, 22603 Hamburg, Germany
\and Laboratory of Theoretical High Energy Physics (LPTHE), University Paris 6, Paris, France;
on leave of absence: PNPI, St.\ Petersburg, Russia
\and INFN and Universit\`a degli Studi di Perugia, Italy
\and School of Physics and Astronomy, Raymond and Beverly
Sackler Faculty of Exact Sciences, Tel Aviv University, 69978 Tel
Aviv, Israel
\and Cavendish Laboratory, J.J. Thomson Avenue, University of Cambridge, Cambridge CB3 0HE, UK
\and  Karlsruhe Institute of Technology, 76128 Karlsruhe, Germany
\and IKTP, TU Dresden, Germany
\and Institute for Theoretical Particle Physics and
  Cosmology, RWTH Aachen University D-52056 Aachen, Germany
\and INFN, Sezione di Torino, and Dipartimento di Fisica Teorica, Universit\`a di Torino, Italy
\and MPI for Nuclear Physics, Saupfercheckweg 1, D-69117 Heidelberg, Germany
\and Physics Department, Penn State University, University Park, PA, USA
\and Division of Theoretical Physics, Department of Mathematical Sciences, University of Liverpool, Liverpool L69 7ZL, UK 
}
\date{Received: date / Revised version: date}
%
\abstract{We review the recent progress in the theoretical description and
 experimental observation of multiple parton interactions.  Subjects
 covered include experimental measurements of minimum bias interactions
 and of the underlying event, models of soft physics implemented in Monte
 Carlo generators, developments in the theoretical description of multiple
 parton interactions and phenomenological studies of double parton
 scattering. This article stems from contributions presented at the
 Helmholtz Alliance workshop on "Multi-Parton Interactions at the LHC",
 DESY Hamburg, 13-15 September 2010.
} 
\maketitle
\onecolumn

\section{\label{sec:intro}Introduction}

The Large Hadron Collider (LHC) began operation in 2008, opening a new chapter in particle physics. The basis for understanding hadronic collisions at high energy is provided by the QCD improved parton model. In this framework each hadron is described as a collection of essentially free
elementary constituents. The interactions between constituents belonging to different colliding hadrons
are the seeds of the complicated process which eventually leads to the particles observed in the detector.
Due to the composite nature of hadrons, it is possible to have
multiple parton hard-scatterings, i.e. events in which two or more
distinct hard parton interactions occur simultaneously in a single
hadron-hadron collision. At fixed final state invariant masses, such
cross sections tend to increase with collision energy because partons
with successively lower momentum fraction $x$, hence rapidly
increasing fluxes, are being probed.  As a result, events with
relatively low invariant masses could receive enhanced contributions
from multiple hard scatterings. This class of events is known as
Multiple Parton Interactions (MPI), while those in which only a single pair of partons produce a hard scattering are referred as Single Parton Scattering (SPS).

The MPI can manifest themselves in various ways in high energy
hadronic collisions. It is natural to expect a relation between the
multiplicity of simultaneous partonic scatterings and their typical
scale. In particular, large hadronic activity is observed in the soft
regime, characterized by small transverse momenta ($p_T$) of the
produced particles. For relatively large $p_T$ values, the observation
of MPI will mostly focus on two simultaneous
scatterings, i.e. on Double Parton Scattering (DPS).
Unfortunately, also depending on the scale of the partonic subprocess
our ability to describe MPI in the pQCD framework is challenging.  Whereas it is
most legitimate to use pQCD methods for the description of MPI at large $p_T$, 
it is necessary to supplement the pQCD picture in the soft regime with models of soft physics .

The evidence for MPI comes from high $p_T$ events observed in hadron
collisions at the ISR at CERN \cite{Akesson:1986iv} and later at the Fermilab Tevatron collider\cite{Abe:1997bp,Abe:1997xk,Abazov:2009gc}.  At lower $p_T$, underlying event (UE) observables have been measured in $p\bar{p}$ collisions in dijet and Drell-Yan
events at CDF in Run~I~\cite{Acosta:2004wqa} and
Run~II~\cite{Aaltonen:2010rm} at center-of-mass energies of
$\sqrt{s}=1.8$~TeV and $1.96$~TeV respectively, and in $pp$ collisions at
$\sqrt{s}=900$~GeV in a detector-specific study by CMS~\cite{Khachatryan:2010pv}.

At small transverse momentum MPI have been shown to be necessary for the 
successful description of the UE in Monte Carlo generators such as {\sc Pythia}
\cite{Sjostrand:1987su,Sjostrand:2004pf,Sjostrand:2004ef} or {\sc Herwig}
\cite{Butterworth:1996zw,Bahr:2008dy}. Additionally, MPI are currently invoked to account for observations at hadron colliders that would not be explained otherwise: the cross sections of multi-jet production, the survival probability of large rapidity gaps in hard diffraction, etc. \cite{Bartalini:2010su}.
The wide range of phenomena in which MPI
are involved highlights the urgency of a more thorough understanding of these
reactions both experimentally and from a theoretical point of view.

The last few years have proven to be a renaissance for research work on MPI. The renewed interest in the field follows from the expected abundance of MPI phenomena at the LHC and thus their importance for the full picture of hadronic collisions, as well as opportunities provided by the LHC to measure multiple parton hard-scatterings. In particular, given the inability to
describe the very soft regime with perturbative methods, the relevance of experimental measurements by the LHC collaborations of observables containing information on MPI, sensitive to the underlying event or minimum bias events, cannot be overstated.

Ultimately, one would strive for a uniform and coherent description of MPI in both soft and hard regimes. At present we are still far away from this goal, as so far essentially separate research efforts focus on specific aspects of MPI.
In the present article we attempt to bridge this gap by reviewing the current status of the field. The article is a result of the Helmholtz Alliance workshop ``Multi-Parton Interactions at the LHC'' which took place at DESY in September 2010. Its main goal was to bring the experimental and theoretical MPI community together, providing a forum for a discussion and scientific exchange. 

This article is organized as follows. We begin with a review of the experimental measurements (available at the time of the workshop) in Chapter~\ref{sec:exp}.   Progress in the implementation of MPI in the Monte Carlo event generators is described in Chapter~\ref{sec:mc}. We then focus on the theoretical aspects of MPI in Chapter~\ref{sec:theory} and discuss the phenomenology of selected DPS processes at the LHC in Chapter~\ref{sec:pheno}.


\section{Experimental situation}
\label{sec:exp}

A complete description of hadronic activity in high energy collisions requires understanding of the UE, as it constitutes the unavoidable background to most observables. From an experimental point of view, the UE gathers all the activity accompanying the actual hard scattering one is interested in measuring. In this sense the UE consists of MPI and the interactions between consituents of beam remnants, left behind after the scattering partons have been pulled out.

Even more inclusive measurements probe the so-called Minimum Bias (MB) events. These are events which are collected with a relatively non-restrictive trigger which accepts large fraction of events, of both the elastic and inelastic nature, inelastic involving the diffractive as well as soft and hard ``hard-core'' events. 

Since it is impossible to uniquely separate the UE from the
hard scattering process on an event-by-event basis, the topological
structure of the outcome of hadronic collisions is focused on
instead. Typically, studies of UE rely on measurements of the
properties of charged particle production, while the charged
multiplicity and $p_T$ spectrum are basic MB observables. Observing
charged particles allows one to investigate the region of very low $p_T$, crucial for exploring soft and semi-hard physics. At higher values of $p_T$, it is possible to directly observe hard MPI in the form of double parton scattering (DPS). In this chapter, we review the experimental studies of UE, MB, forward measurements, as well as discuss the prospects for observing hard DPS at the LHC.

\newcommand{\dNchgdetadphi}{\ensuremath {\langle \mathrm{d}^2N_\text{ch}/\mathrm{d}\eta\,\mathrm{d}\phi\rangle} \xspace}
\newcommand{\dpTsumdetadphi}{\ensuremath{\langle \mathrm{d}^2\!\sum\!\pt/\mathrm{d}\eta\,\mathrm{d}\phi \rangle}\xspace}
\newcommand{\Pythia}{{\sc Pythia}\xspace}
\newcommand{\Phojet}{{\sc Phojet}\xspace}
\newcommand{\Herwig}{{\sc Herwig}\xspace}
\newcommand{\Jimmy}{{\sc Jimmy}\xspace}
\newcommand{\HerwigJimmy}{{\sc Herwig+Jimmy}\xspace}
\newcommand{\Nchg}{\ensuremath{N_\text{ch}}\xspace}
\newcommand{\pt}{\ensuremath{p_\text{T}}\xspace}
\newcommand{\etamod}{\ensuremath{|\eta\mspace{0.2mu}|}\xspace}
\newcommand{\ptlead}{\ensuremath{\pt^\text{lead}}\xspace}
\newcommand{\ptmean}{\ensuremath{\langle\pt\rangle}\xspace}
\newcommand{\ptsum}{\ensuremath{\sum\mspace{-0.8mu}\pt}\xspace}

\graphicspath{{dkar/figures/}}

\contribution{\label{exp:atlas}Underlying Event Measurements at ATLAS}
{Contributing author: D. Kar (on behalf of the ATLAS Collaboration)}

\label{dkar}




This section reports on the measurement of UE observables, performed with the ATLAS
detector
at the LHC using proton--proton collisions at
center-of-mass energies of $900$~GeV and $7$~TeV~\cite{Aad:2010fh}.

At the detector level, charged particles are observed as tracks in the inner tracking system.
The direction of the track with the largest \pt in the event -- referred to as
the ``leading'' track -- is used to define regions of the $\eta$--$\phi$ plane
which have different sensitivities to the UE. 
The axis given by the leading track is well-defined for all events, and is highly
correlated with the axis of the hard scattering in high-\pt events.
A single track is used as opposed to a jet or the decay products of a massive gauge
boson, as it allows significant results to be derived with limited luminosity
and avoids the systematic measurement complexities of alignment with more
complex objects.

\begin{figure}[tbp]
  \begin{center}
\resizebox{0.45\textwidth}{!}{%
    \begin{tikzpicture}[>=stealth, very thick, scale=1.1]
      \footnotesize\smaller

      \draw[color=blue!80!black] (0, 0) circle (3.0);
      \draw[rotate= 30, color=gray] (-3.0, 0) -- (3.0, 0);
      \draw[rotate=-30, color=gray] (-3.0, 0) -- (3.0, 0);

      \draw[->, color=black, rotate=-2] (0, 3.5) arc (90:47:3.5) node[right] {$\Delta{\phi}$};
      \draw[->, color=black, rotate=2] (0, 3.5) arc (90:133:3.5) node[left] {$-\Delta{\phi}$};

      \draw[->, color=red, ultra thick] (0, 2) -- (0, 4) node[above] {\textcolor{black}{leading track}};
      \draw[->, color=green!70!black, ultra thick] (0, -2) -- ( 0.0, -4);
      \draw[->, color=green!70!black, ultra thick, rotate around={-15:(0,-0.3)}] (0, -2) -- ( 0.0, -4);
      \draw[->, color=green!70!black, ultra thick, rotate around={ 15:(0,-0.3)}] (0, -2) -- ( 0.0, -4);

      \draw (0,  1.55) node[text width=2cm] {\begin{center} toward \end{center}};
      \draw (0,  1.3) node[text width=2cm] {\begin{center} $|\Delta\phi| < 60^\circ$ \end{center}};
      \draw (0, -0.85) node[text width=2cm] {\begin{center} away \end{center}};
      \draw (0, -1.2) node[text width=2cm] {\begin{center} $|\Delta\phi| > 120^\circ$ \end{center}};
      \draw ( 1.7, 0.4) node[text width=3cm] {\begin{center} transverse \end{center}};
      \draw ( 1.7, 0.05) node[text width=3cm] {\begin{center} $60^\circ < |\Delta\phi| < 120^\circ$ \end{center}};
      \draw (-1.7, 0.4) node[text width=3cm] {\begin{center} transverse \end{center}};
      \draw (-1.7, 0.05) node[text width=3cm] {\begin{center} $60^\circ < |\Delta\phi| < 120^\circ$ \end{center}};
    \end{tikzpicture}
    }
    \caption{Definition of regions in the azimuthal angle with respect to the leading track.}
    \label{fig:ueregions}
  \end{center}
\end{figure}
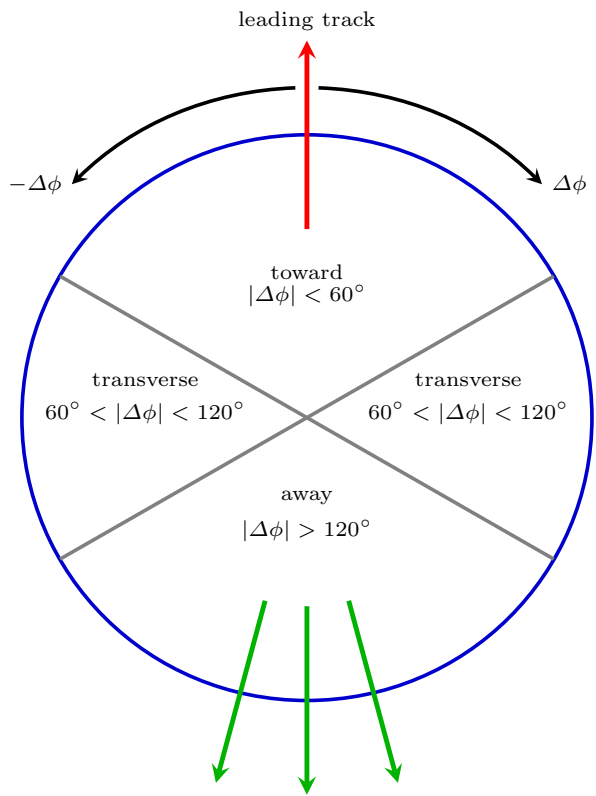

As illustrated in Fig.~\ref{fig:ueregions}, the azimuthal angular difference between charged tracks and the leading track,
$|\Delta\phi|=|\phi-\phi_\text{leading~track}|$, is used to define the following three azimuthal regions~\cite{Acosta:2004wqa}:
\begin{itemize}
\item $|\Delta\phi| < 60^{\circ}$, the ``toward region'';
\item $60^{\circ} < |\Delta\phi| < 120^{\circ}$, the ``transverse region''; and
\item $|\Delta\phi| > 120^{\circ}$, the ``away region''.
\end{itemize}

\noindent
The transverse regions are most sensitive to the underlying event, since they
are generally perpendicular to the axis of hardest scattering.

\subsubsection{Analysis Details}

All data used in this paper were taken during the LHC running periods with
stable beams and defined beam-spot values, between 6th--15th~December~2009 for the
analysis at $\sqrt{s} = 900$~GeV, and from 30th~March to 27th~April~2010
for the $7$~TeV analysis.
Only events with leading track \pt $>1$~GeV  within the inner detector, $\etamod < 2.5$, 
were considered, 
in order to reject events where the leading track selection can potentially 
introduce large systematic effects, and also to
reduce the contribution from diffractive hard scattering processes.
All the other tracks were required to have \pt $>500$~MeV and the same $\eta$ range.

The $900$~GeV and $7$~TeV data respectively correspond to integrated luminosities of
$7~\micro b^{-1}$ and $168~\micro b^{-1}$, respectively, and the effects of pileup was negligible.
For the MC models considered here, the contribution of diffractive events
to the underlying event observables was less than~$1\%$.

The data were corrected back to charged primary particle spectra satisfying the
event-level requirement of at least one primary charged particle within $\pt >
1$~GeV and $\etamod < 2.5$.  A two step correction process was used,
where first the event and track efficiency corrections were applied, then an
additional bin-by-bin unfolding was performed to account for possible bin
migrations and any remaining detector effects.

\subsubsection{Results}

In this section, corrected distributions of underlying event
observables are compared to model predictions tuned to a wide range of measurements.
The transverse, toward and away regions each have
an area of $\Delta\phi \, \Delta\eta = 10\, \pi/3$ in $\eta$--$\phi$ space, so
the density of particles \dNchgdetadphi and transverse momentum sum \dpTsumdetadphi
are constructed by dividing the mean values by the corresponding area. The leading charged
particle is included in the toward region distributions, unless otherwise stated.

The data, corrected back to particle level in the transverse, toward and away regions
are compared with predictions by \Pythia~\cite{Sjostrand:2006za} with the ATLAS~MC09~\cite{ATL-PHYS-PUB-2010-002}, DW~\cite{CDFtuneA}, and Perugia0~\cite{Skands:2009zm} tunes, by
\HerwigJimmy~\cite{Corcella:2002jc,Butterworth:1996zw} with the ATLAS~MC09 tune, and by \Phojet~\cite{Engel:1994vs}. The ratios of the MC
predictions to the data are shown at the bottom of these plots.
The error bars show the statistical
uncertainty while the shaded area shows the combined statistical and systematic
uncertainties.

\begin{figure}[pbt]
 \begin{center}
   \includegraphics[width=.41\textwidth]{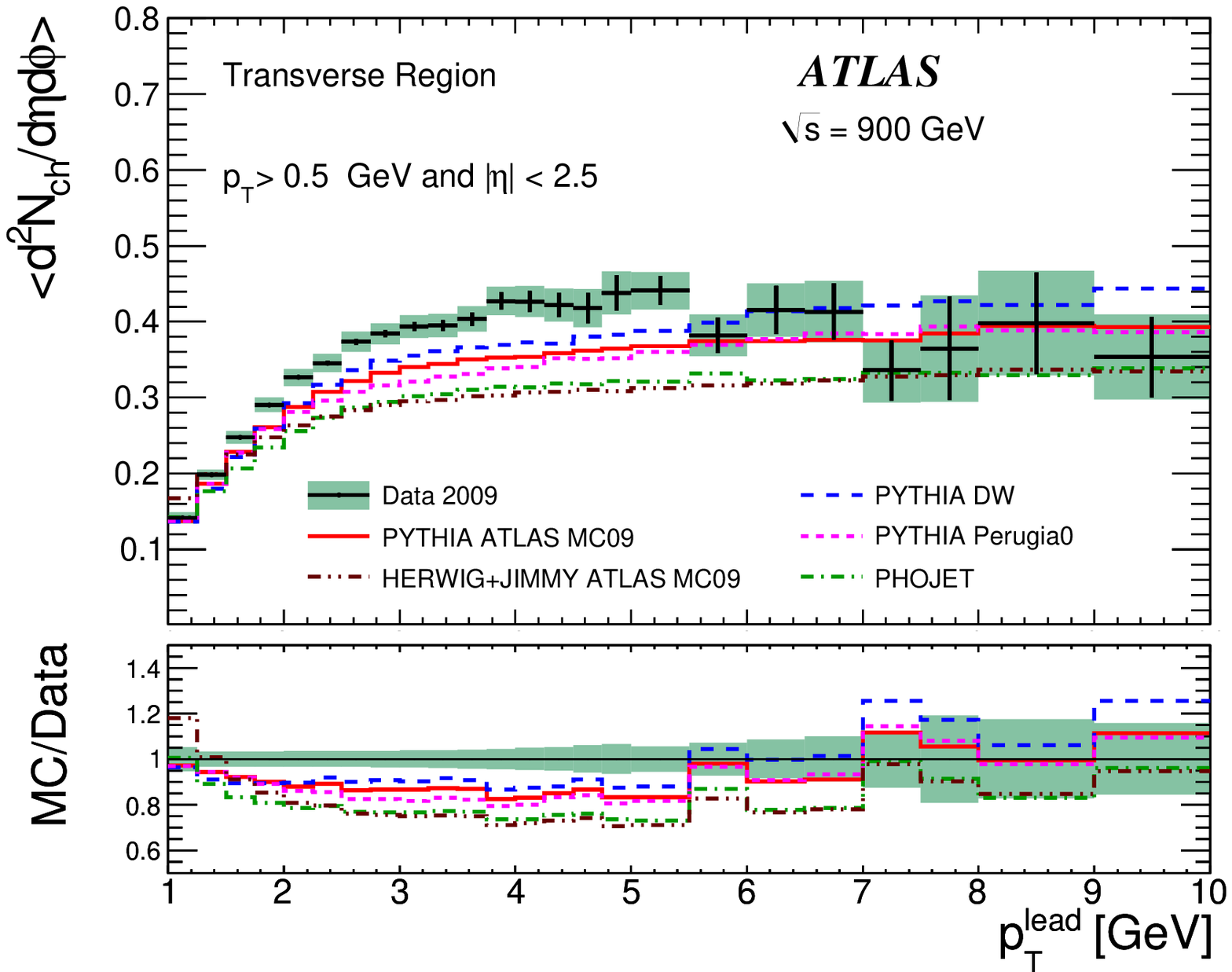}\hfill
   \includegraphics[width=.41\textwidth]{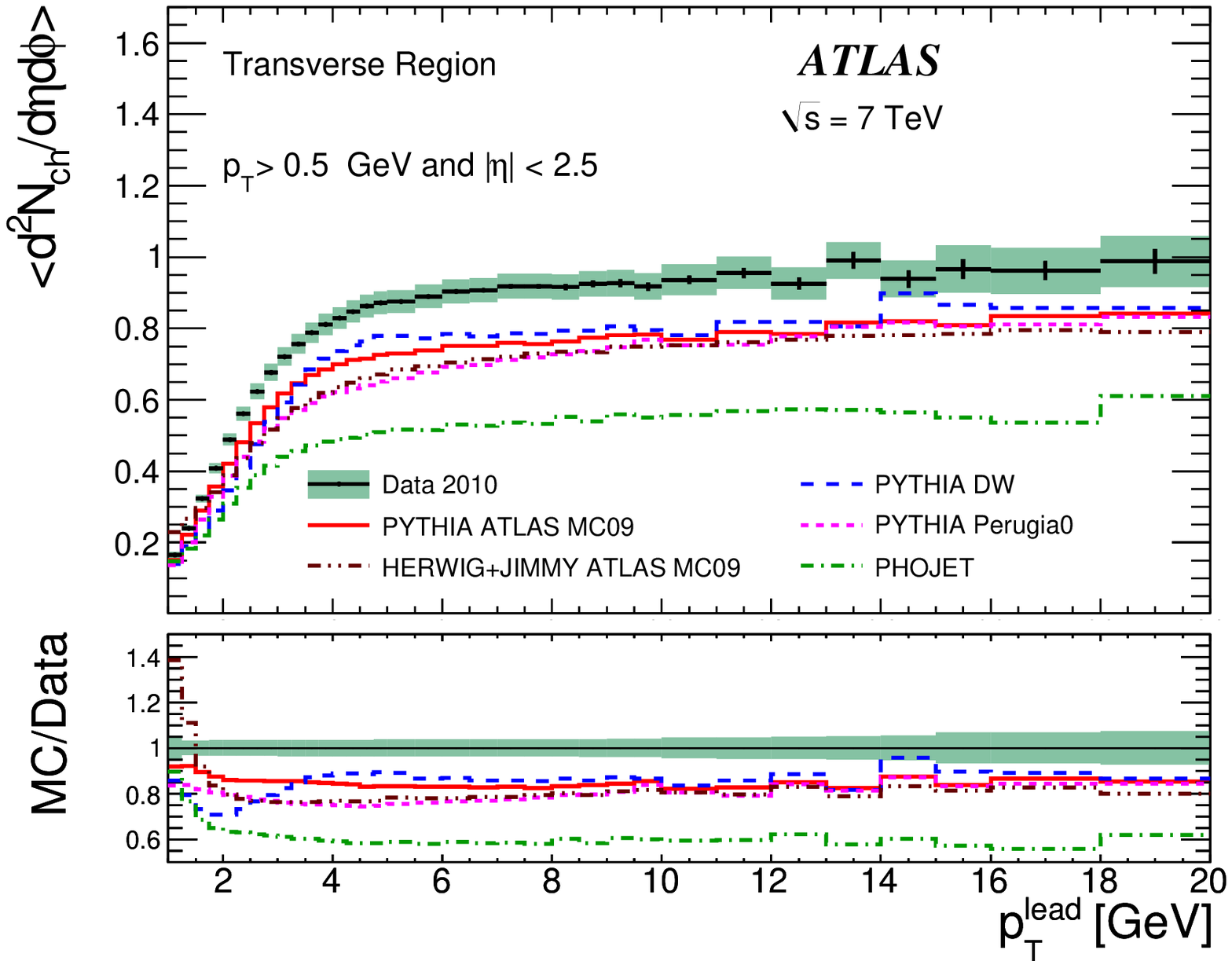}\hfill
   \includegraphics[width=.39\textwidth]{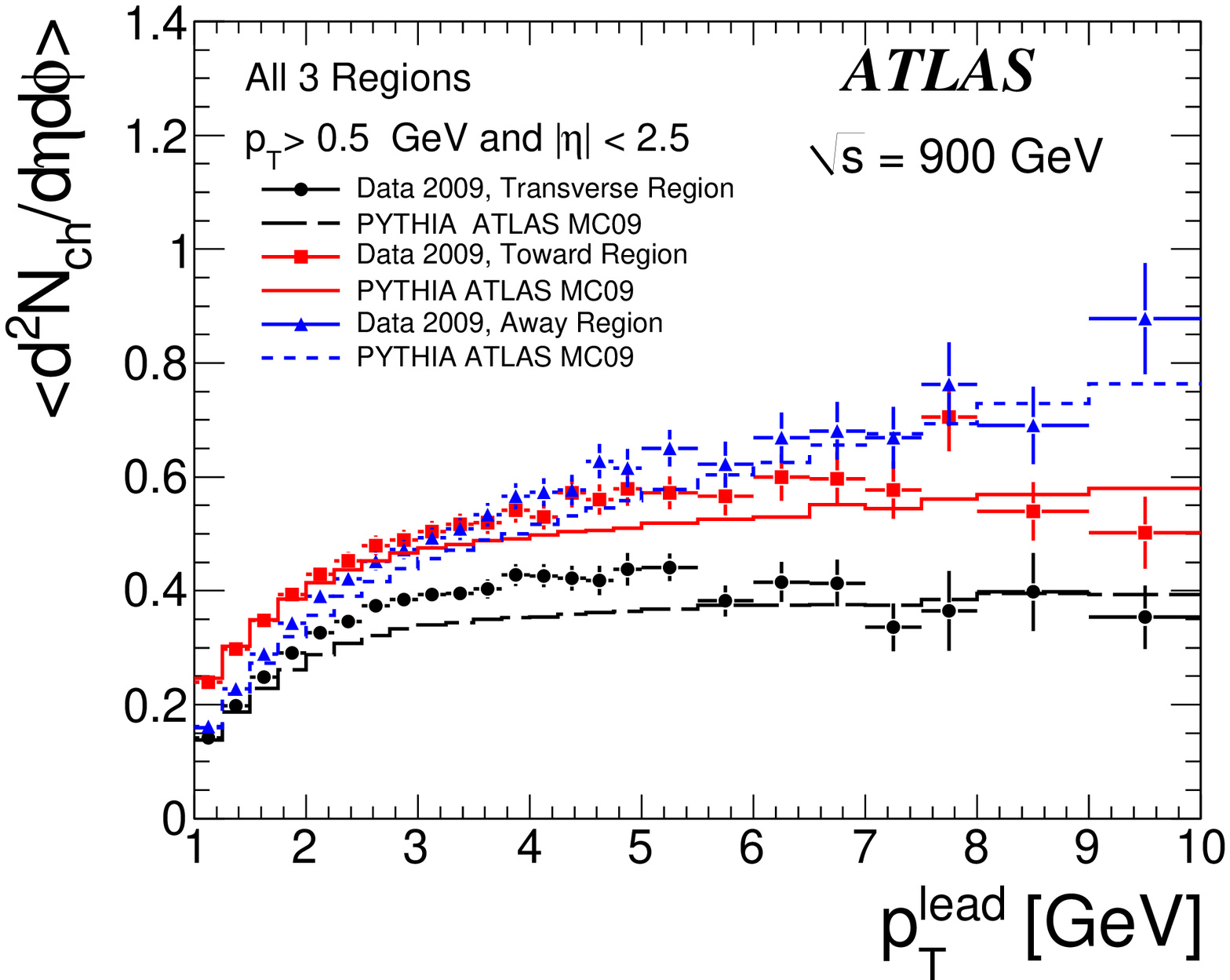}\hfill
   \includegraphics[width=.39\textwidth]{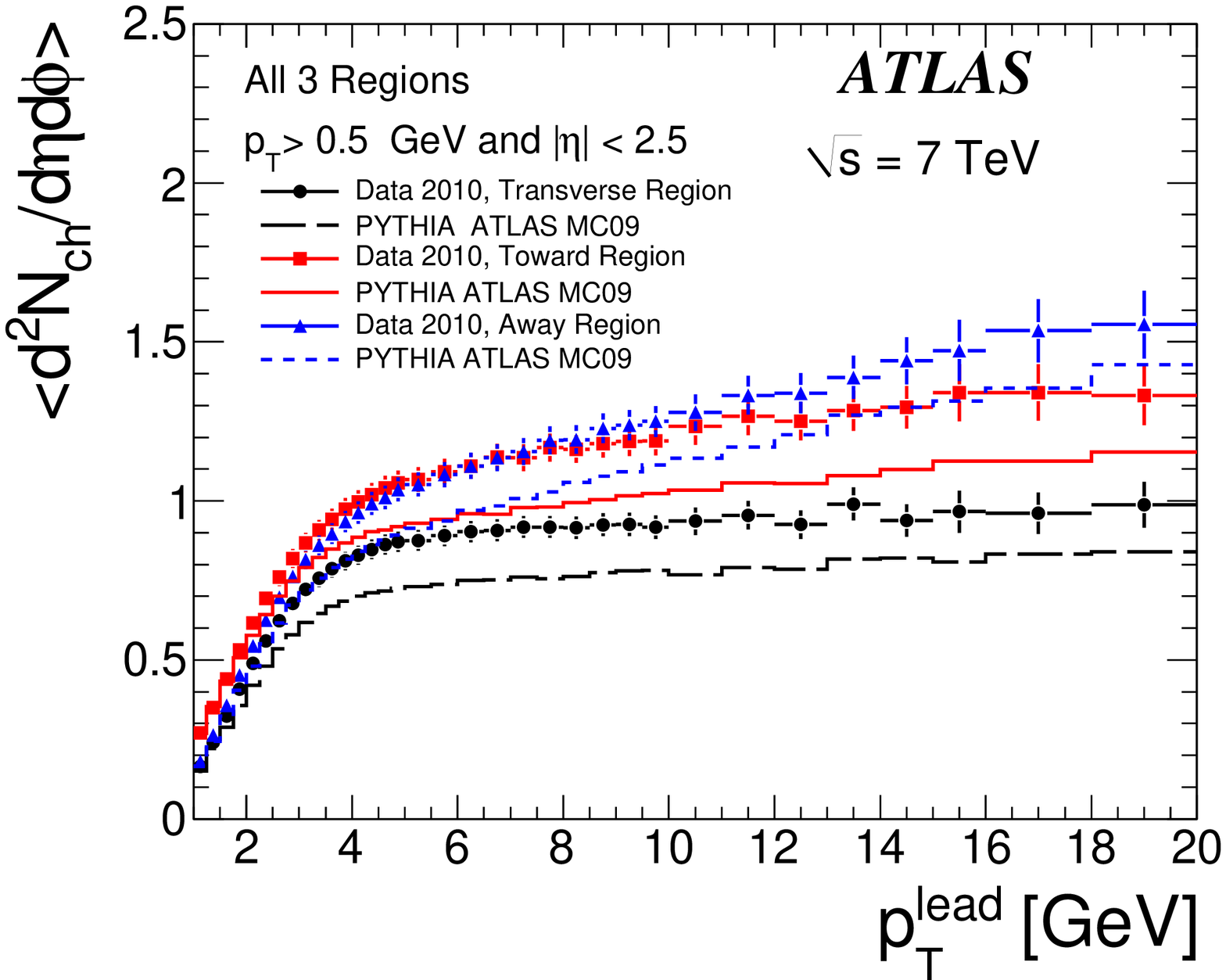}
 \end{center}

 \caption[]{ATLAS data at $\sqrt{s}=$ 900 GeV (left) and 7 TeV (right),
   showing the density of the charged particles in the transverse region (top row), and in all three regions (bottom row)}
 \label{fig:nchg}
\end{figure}

The charged particle multiplicity density,
is shown in Fig.~\ref{fig:nchg}.

The average number of charged particles in the transverse region increases with leading \pt, until it reaches an approximately constant ``plateau''. 
All the pre-LHC MC tunes considered show at least 10--15\% lower activity than the data in the transverse region plateau.
The \Pythia~DW tune is seen to be the closest model to data for the transverse region.
The toward and away regions are dominated by jet-like activity, yielding
gradually rising number densities. 
The $900$~GeV and $7$~TeV ATLAS data show the same trend. 
The underlying event activity is seen to increase by a factor of approximately
two between the $900$~GeV and $7$~TeV data. This is roughly
consistent with the rate of increase predicted by MC models tuned to Tevatron
data.

\begin{figure}[pbt]
 \begin{center}
   \includegraphics[width=.41\textwidth]{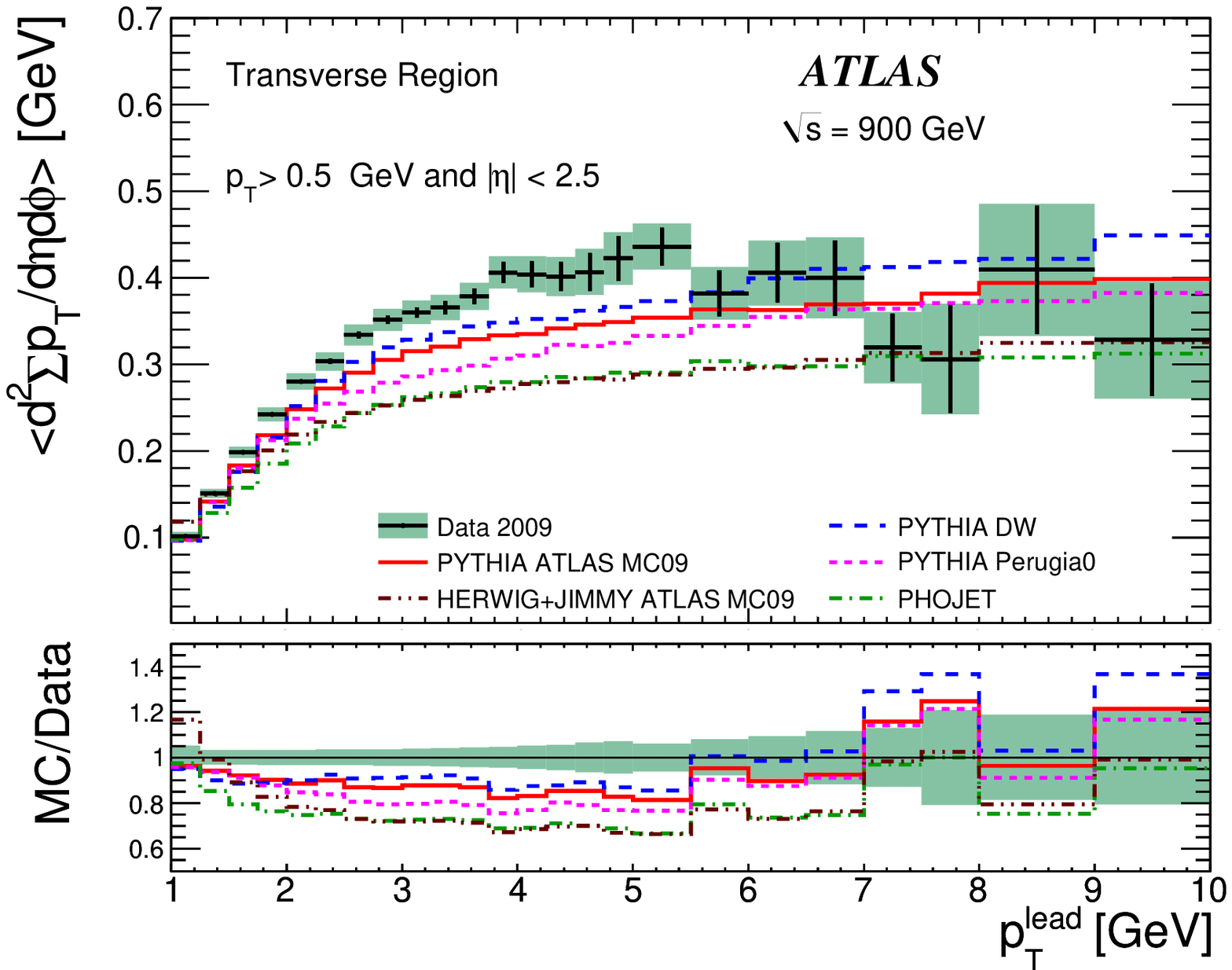}\hfill
   \includegraphics[width=.41\textwidth]{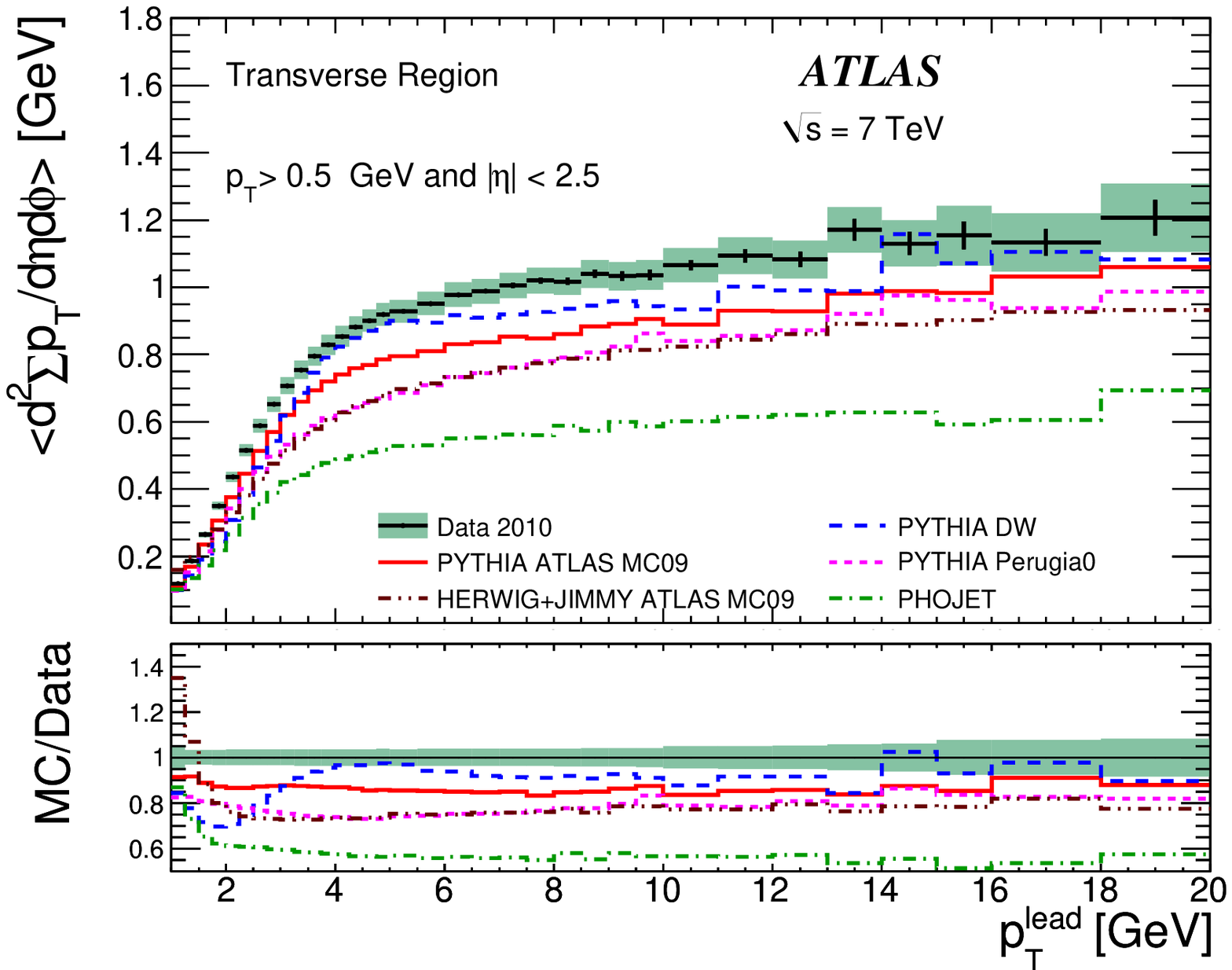}\hfill
   \includegraphics[width=.38\textwidth]{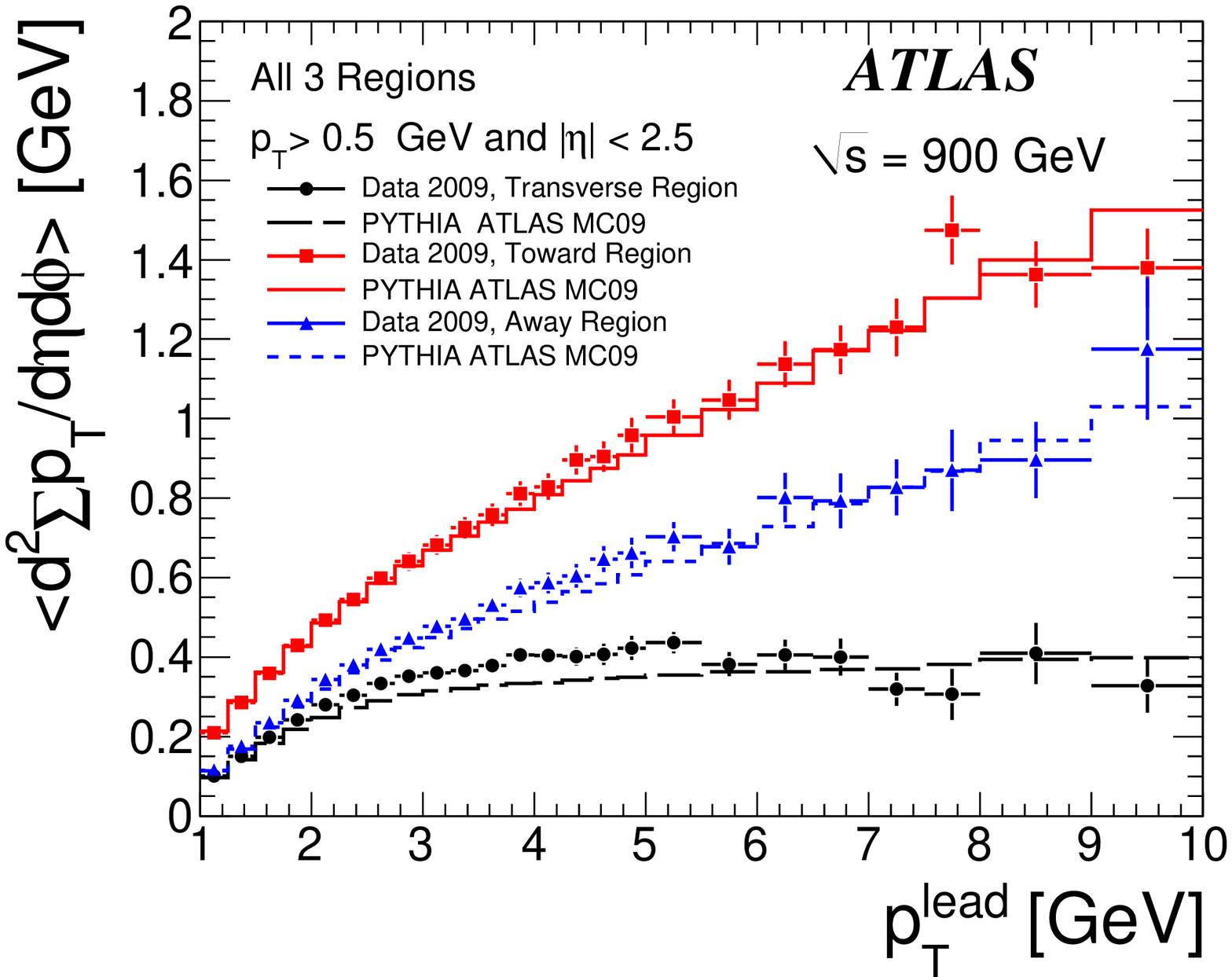}\hfill
   \includegraphics[width=.38\textwidth]{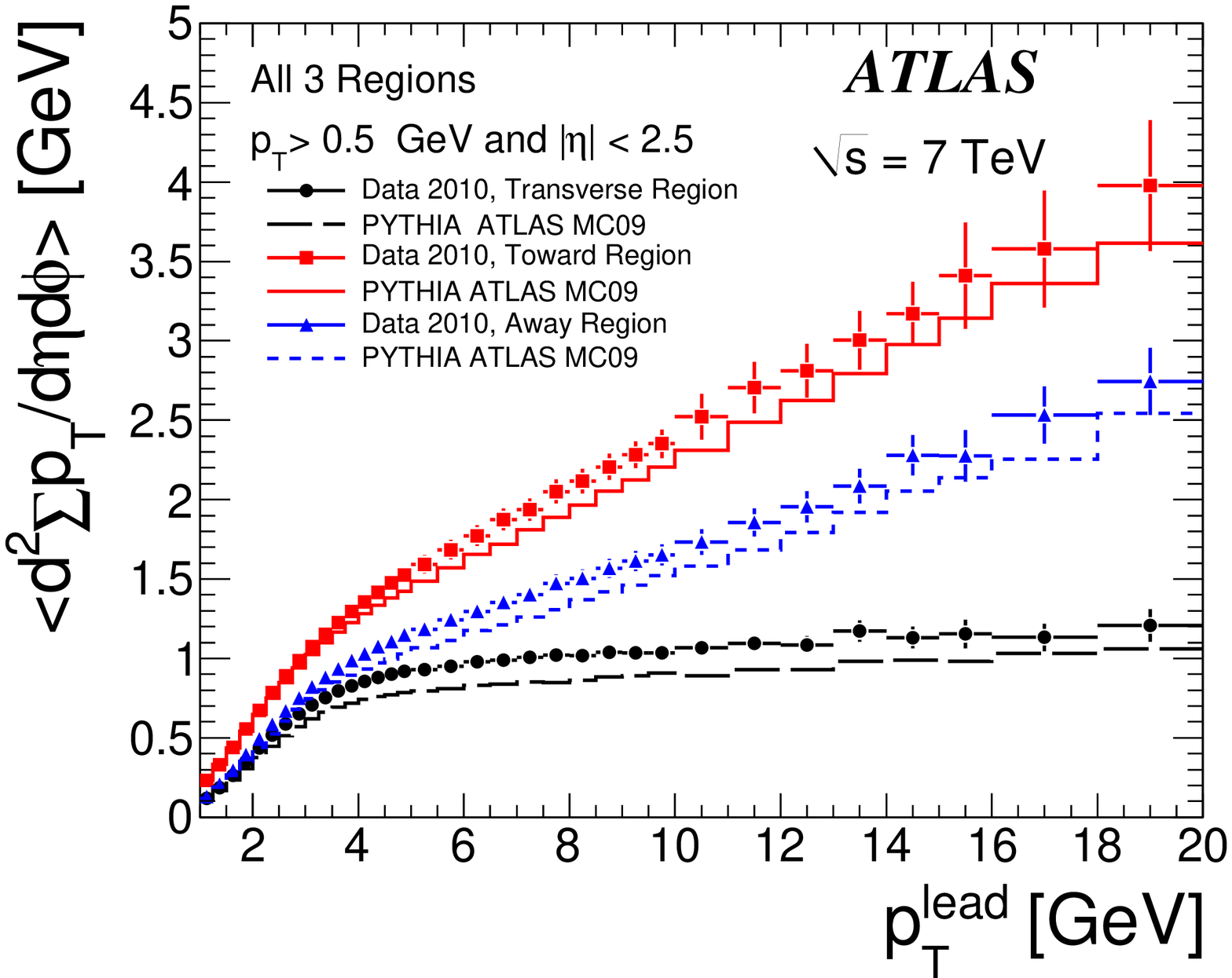}
 \end{center}
 \caption[]{ATLAS data at $\sqrt{s}=$ 900 GeV (left) and 7 TeV (right),
   showing the scalar $\ptsum$ density of the charged particles in the transverse region (top row), and in all three regions (bottom row)}
 \label{fig:ptsum}
\end{figure}

The charged particle scalar \pt sum, \ptsum density,
is shown in Fig.~\ref{fig:ptsum}.
The summed charged particle \pt in the plateau characterizes the mean
contribution of the underlying event to jet energies. Again, we can see  
that pre-LHC tunes model CDF data better than ATLAS data.
The higher number density implies a higher \pt density as well.  
In the toward and away regions, jet-like rising profiles are observed,
in contrast to the plateau in the transverse region. The toward region includes
the leading charged particle, and has a higher \ptsum than the away region as
there is higher probability of high-\pt particles being produced in association
with the leading \pt charged particle. In the toward region the highest fraction
of energy has been allocated to a single charged particle. This implicitly
reduces the number of additional charged particles in that region, since there
is less remaining energy to be partitioned. As a result the multiplicity of
charged particles is slightly lower in the toward region by comparison to the
away region for high \ptlead.  The increase of the \pt densities in the toward
and away regions indicates the extent of the variation in the charged fraction
of the total energy in each region.

\begin{figure}[pbt]
 \begin{center}
   \includegraphics[width=.41\textwidth]{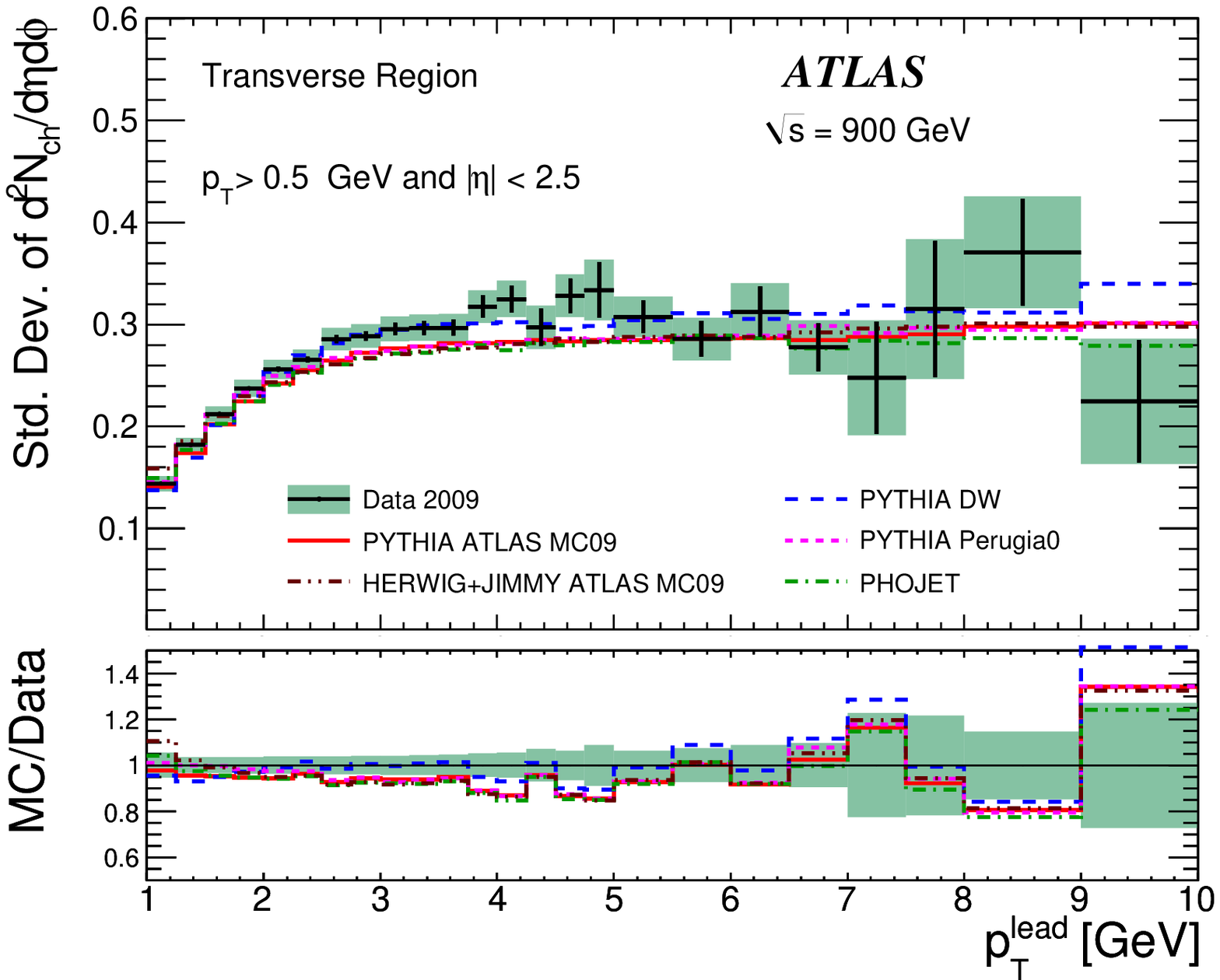}\hfill
   \includegraphics[width=.41\textwidth]{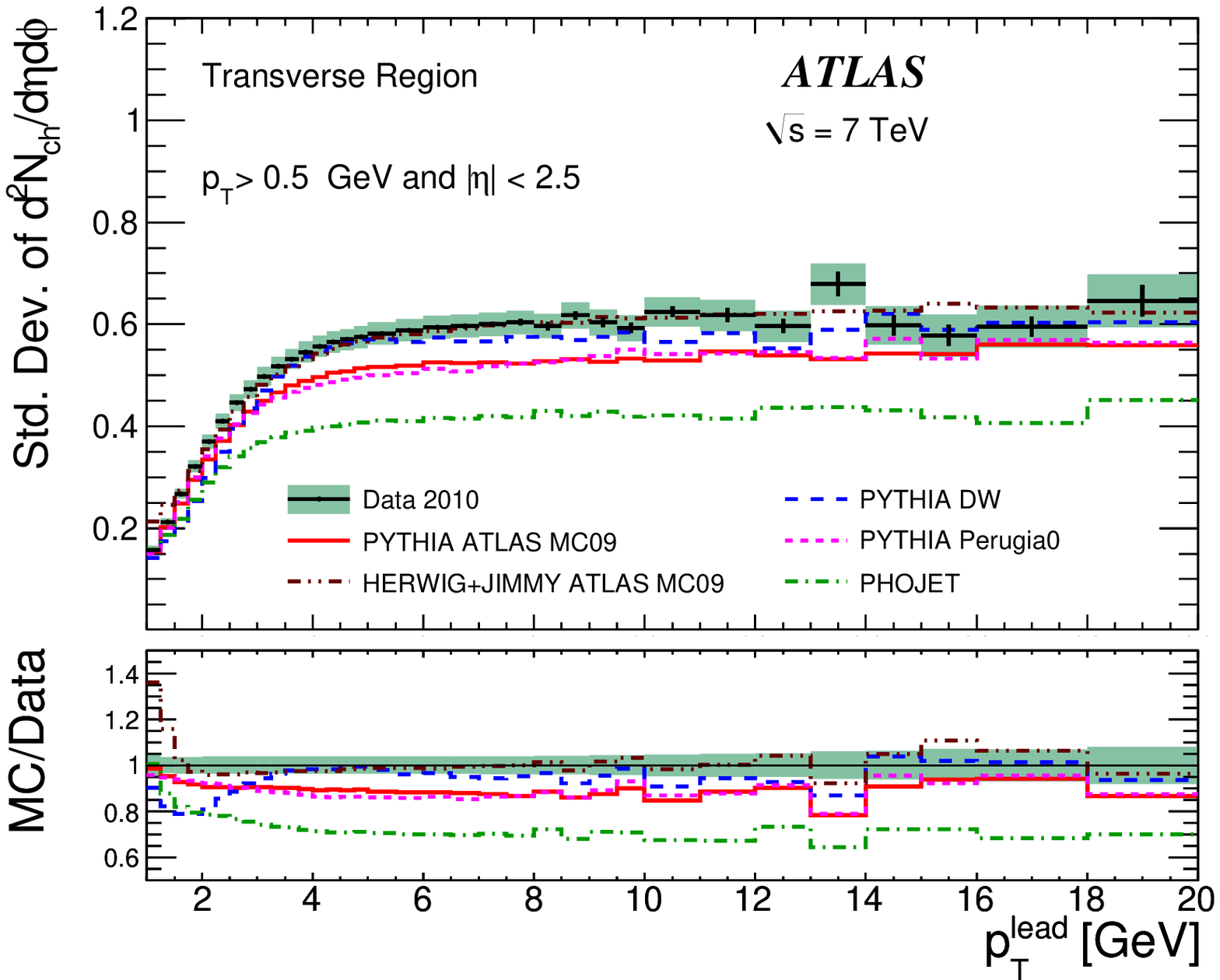}\hfill
   \includegraphics[width=.41\textwidth]{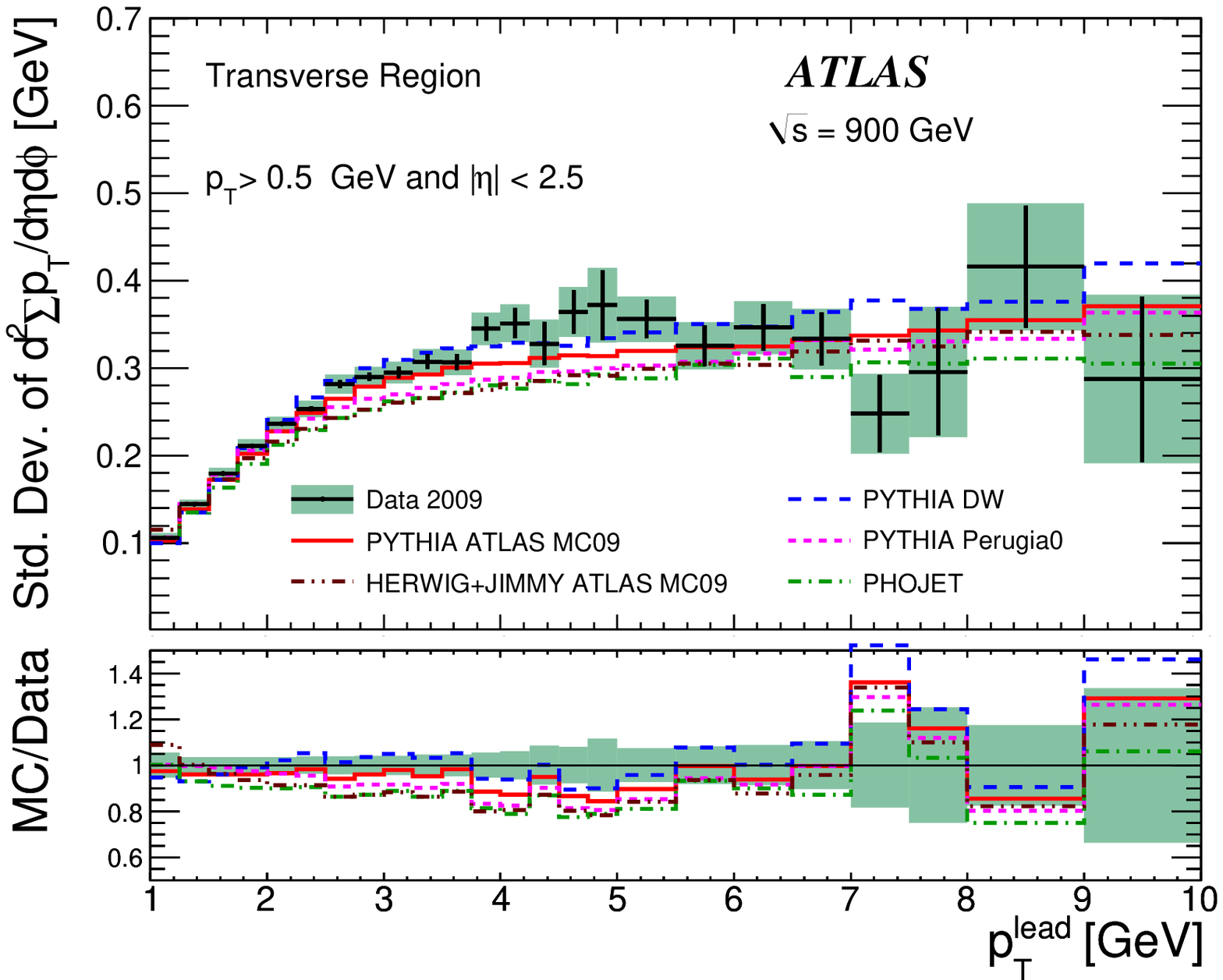} \hfill
   \includegraphics[width=.41\textwidth]{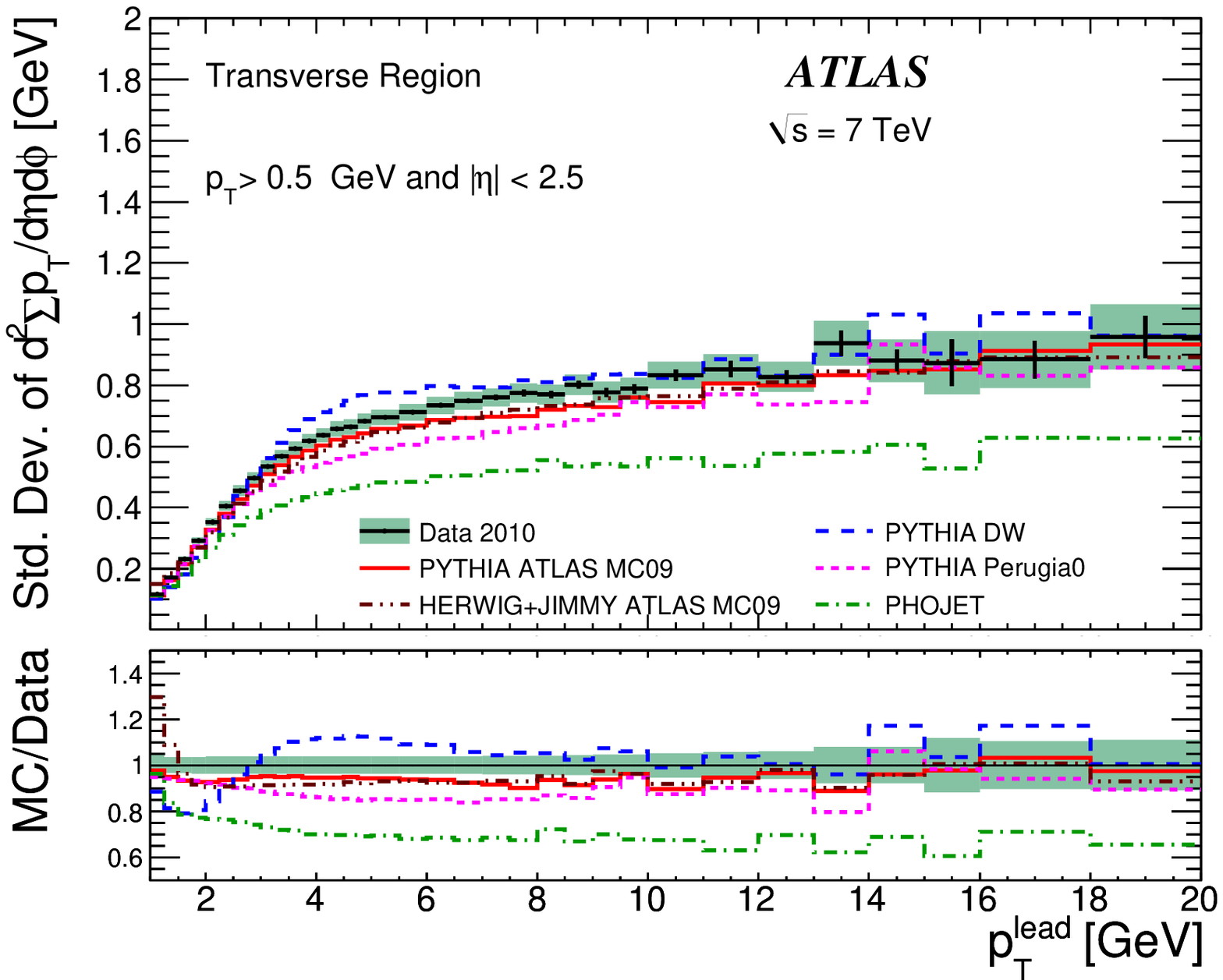}
 \end{center}
 \caption[]{TLAS data at $\sqrt{s}=$ 900 GeV (left) and 7 TeV (right), showing the standard deviation of the
   density of the charged particles (top row) and the standard
   deviation of the scalar $\ptsum$ density of charged particles
   (bottom row) in the transverse region.}
 \label{fig:SD}
\end{figure}

In Fig.~\ref{fig:SD}, the standard deviation 
of the charged particle multiplicity and charged particle scalar \ptsum densities,
are shown. 
The mean and standard deviation of the \pt
density in the transverse region characterize a range of additional energy that
jets might acquire if the underlying event were uniformly distributed.
The confirmation that the magnitude of the standard deviations of the
distributions are comparable to the magnitudes of the mean values indicates
that a subtraction of the underlying event from jets should be done on an
event by event basis, rather than by the subtraction of an invariant average value.

The correlation between the mean \pt of charged particles and the charged
particle multiplicity in that region is sensitive to the amount of hard (perturbative QCD)
versus soft (non-perturbative QCD) processes contributing to the underlying
event. 

\begin{figure}[pbt]
 \begin{center}
   \includegraphics[width=.45\textwidth]{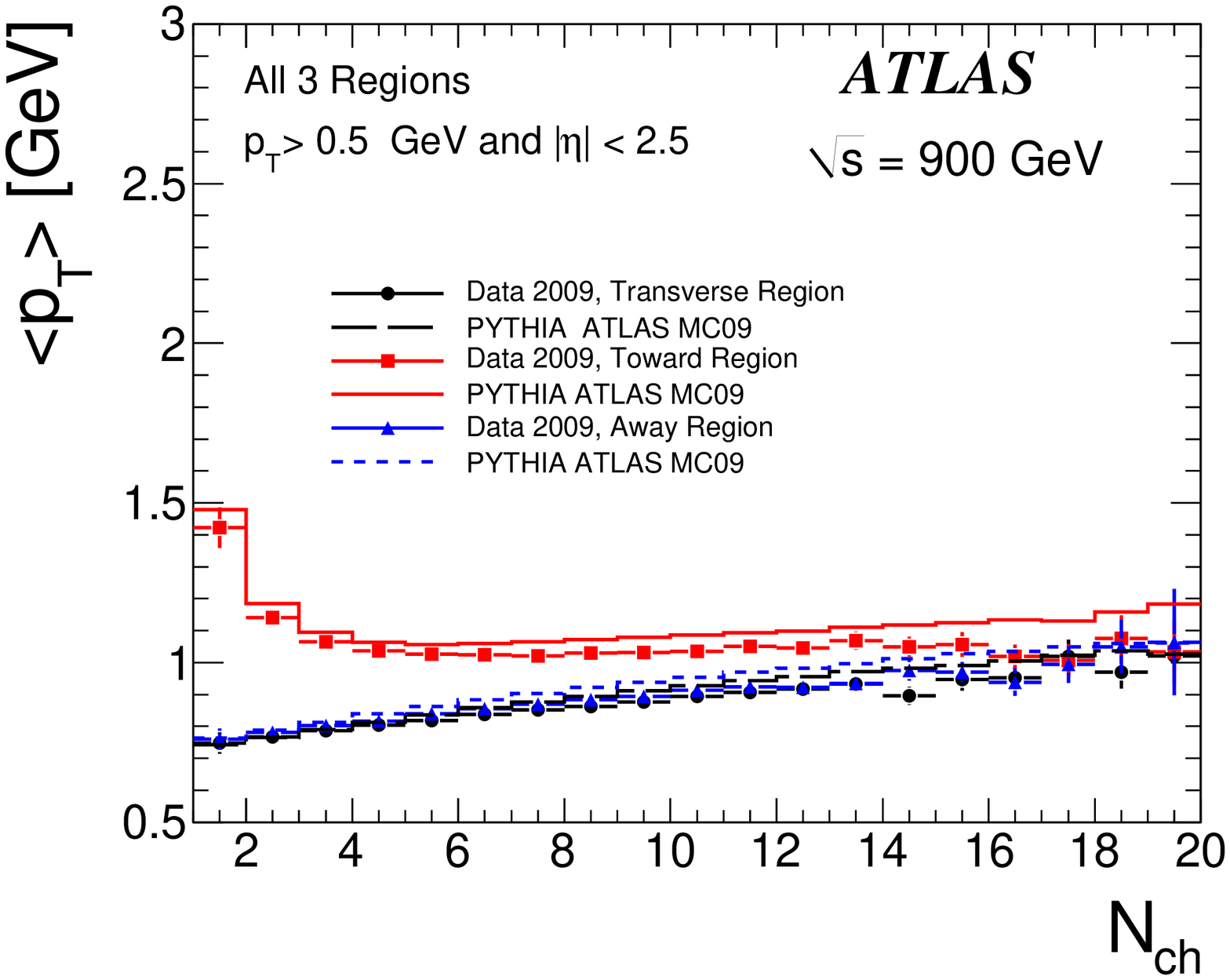}\hfill
   \includegraphics[width=.45\textwidth]{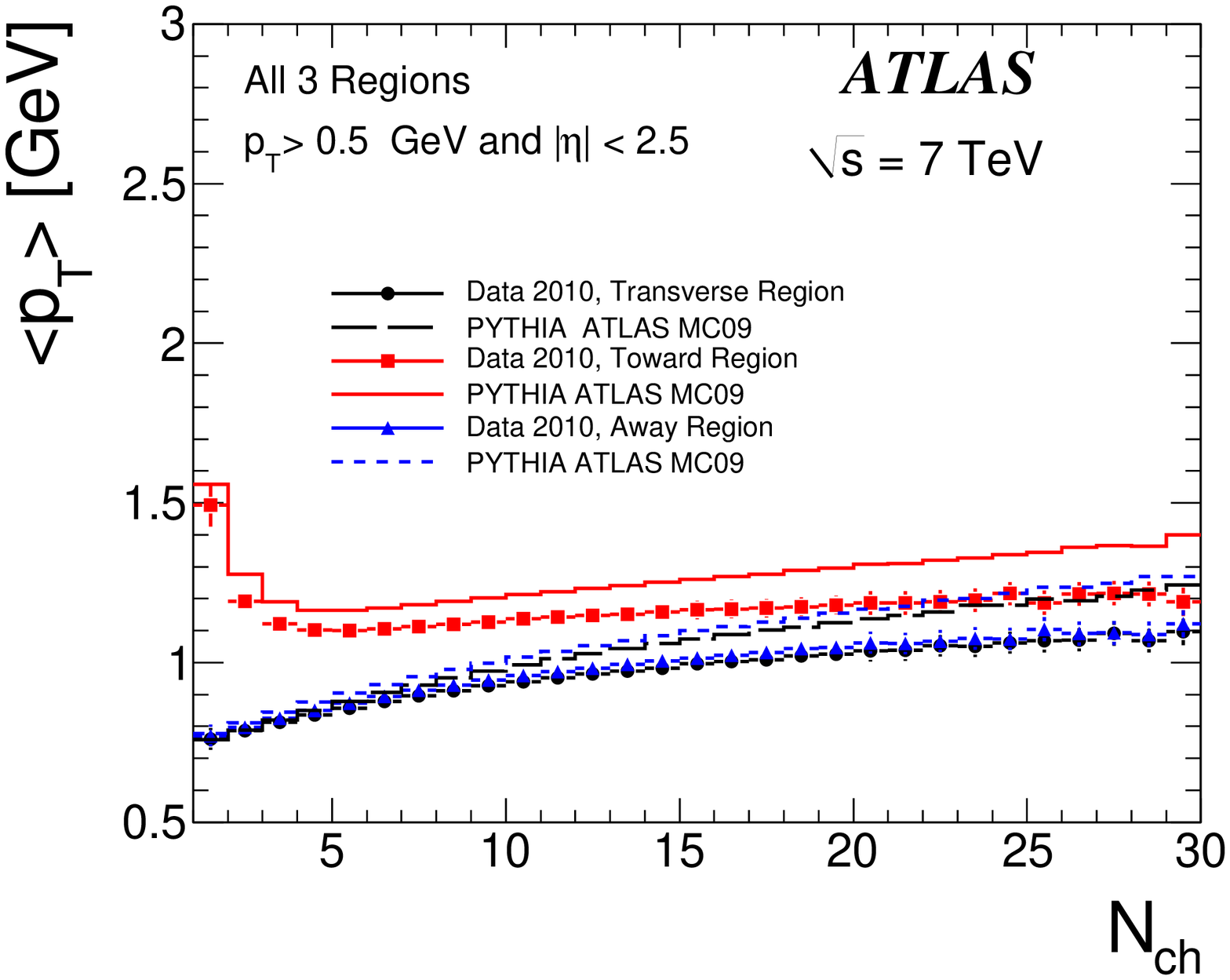}
    \end{center}
 \caption[]{ATLAS data at $\sqrt{s}=$ 900 GeV (left) and 7 TeV (right), showing the mean \pt of the charged
   particles against the charged multiplicity in all three regions.}
 \label{fig:corr}
\end{figure}

In Fig.~\ref{fig:corr}, the ATLAS profiles in the transverse and away regions are very similar, showing a
monotonic increase of \ptmean with \Nchg. The profile of the toward region is
different, as it is essentially determined by the requirement of a
track with $\pt > 1$~GeV. For $N_{ch} =1$, it
contains only the leading charged particle and as the $N_{ch}$ is increased by
inclusion of soft charged particles the average is reduced. However, for $N_{ch} > 5$
jet-like structure begins to form, and the weak rise of the mean \pt is observed.
Comparing the  $900$~GeV and $7$~TeV  data, it is seen that the mean
charged particle \pt vs. \Nchg profiles are largely independent of the energy scale of
the collisions.

\begin{figure}[pbt]
 \begin{center}
   \includegraphics[width=.4\textwidth]{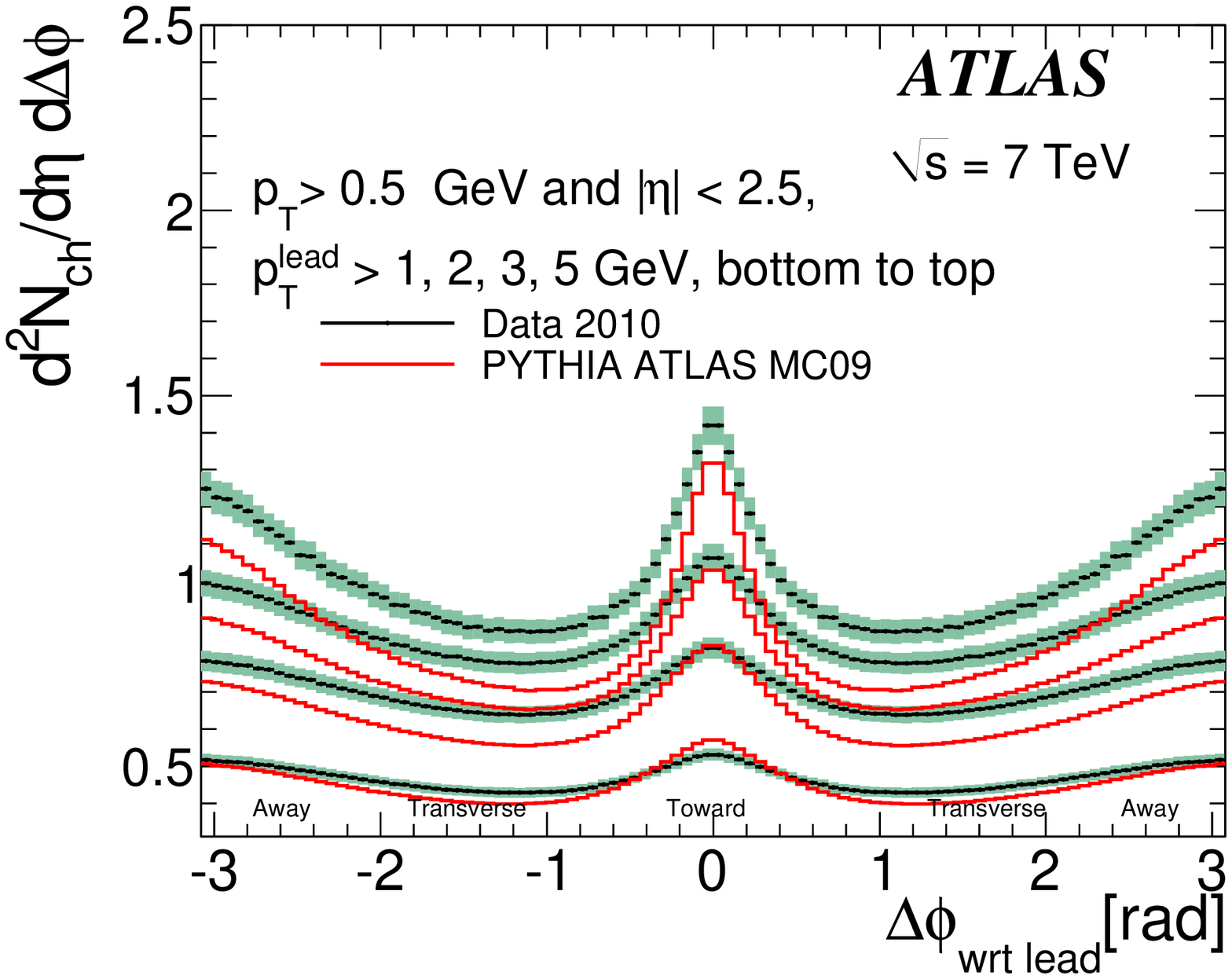}
   \includegraphics[width=.4\textwidth]{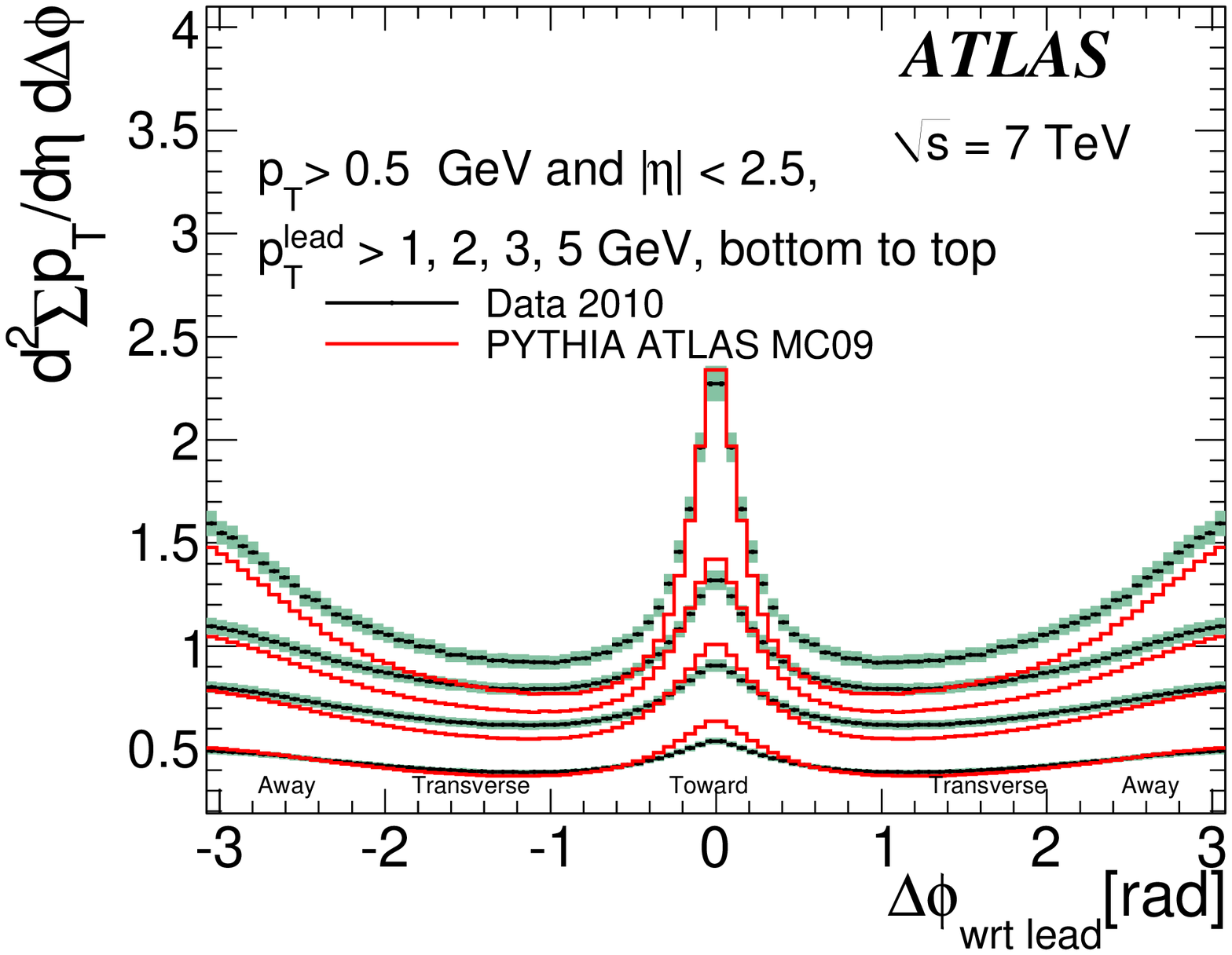}
 \end{center}
 \caption[]{ATLAS data showing the $\phi$ distribution of
   charged particle multiplicity (left) and scalar \ptsum density (right), 
   with respect to the leading charged particle rotated to
   $\phi_\text{leading} = 0$, excluding the leading charged particle and compared
   to different MC model predictions. The distributions obtained by restricting the
   minimum leading charged particle \pt to different values are shown separately.
   The plots were symmetrized by reflecting them about $\Delta\phi=0$}
 \label{fig:deltaphi}
\end{figure}

The angular distributions with respect to the leading charged particle of the charged
particle number and \ptsum densities at the center-of-mass energy of 
$7$~TeV at ATLAS, are plotted in Fig~\ref{fig:deltaphi}. 
The leading charged particle taken to be at $\Delta\phi = 0$ has
been excluded from the distributions. The data are shown for four different
lower cut values in leading charged particle \pt. These distributions are constructed by
reflecting $|\Delta\phi|$ about zero, i.e. the region $-\pi \le \Delta\phi < 0$ is an exact
mirror image of the measured $|\Delta\phi|$ region shown in $0 \le \Delta\phi
\le \pi$.

These distributions show a significant difference in shape between data and MC
predictions.  With the increase of the leading charged particle \pt, the development of
jet-like structure can be observed, and the corresponding sharper rise in
transverse regions compared to the MC. MC models essentially predict a stronger correlation than is seen in the data, and
this discrepancy in toward region associated particle density was also observed
at CDF~\cite{:RDFlead}.

\begin{figure}[htbp]
\centering
\subfigure[]{
\includegraphics[scale=0.2]{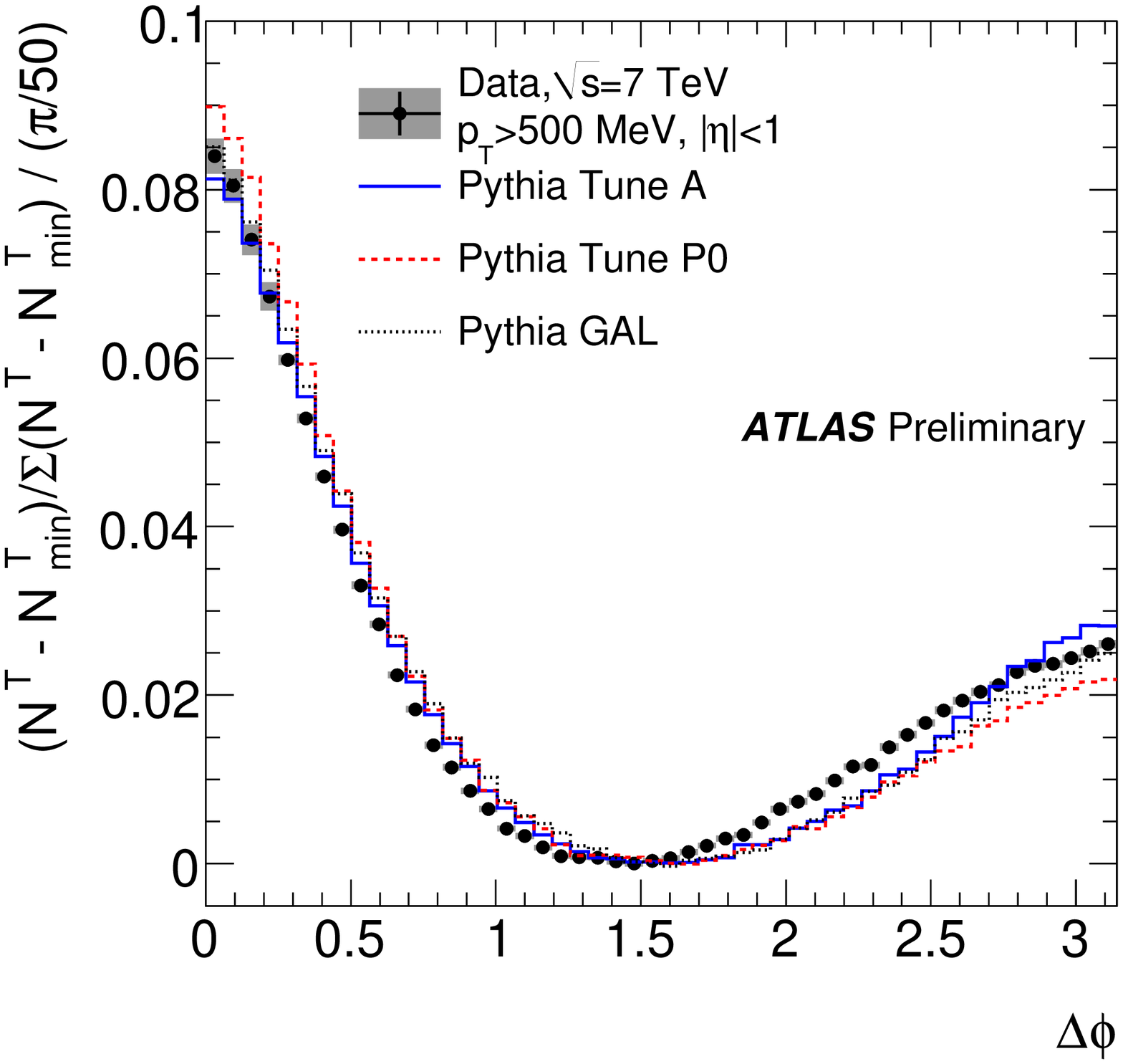}
}
\subfigure[]{
\includegraphics[scale=0.2]{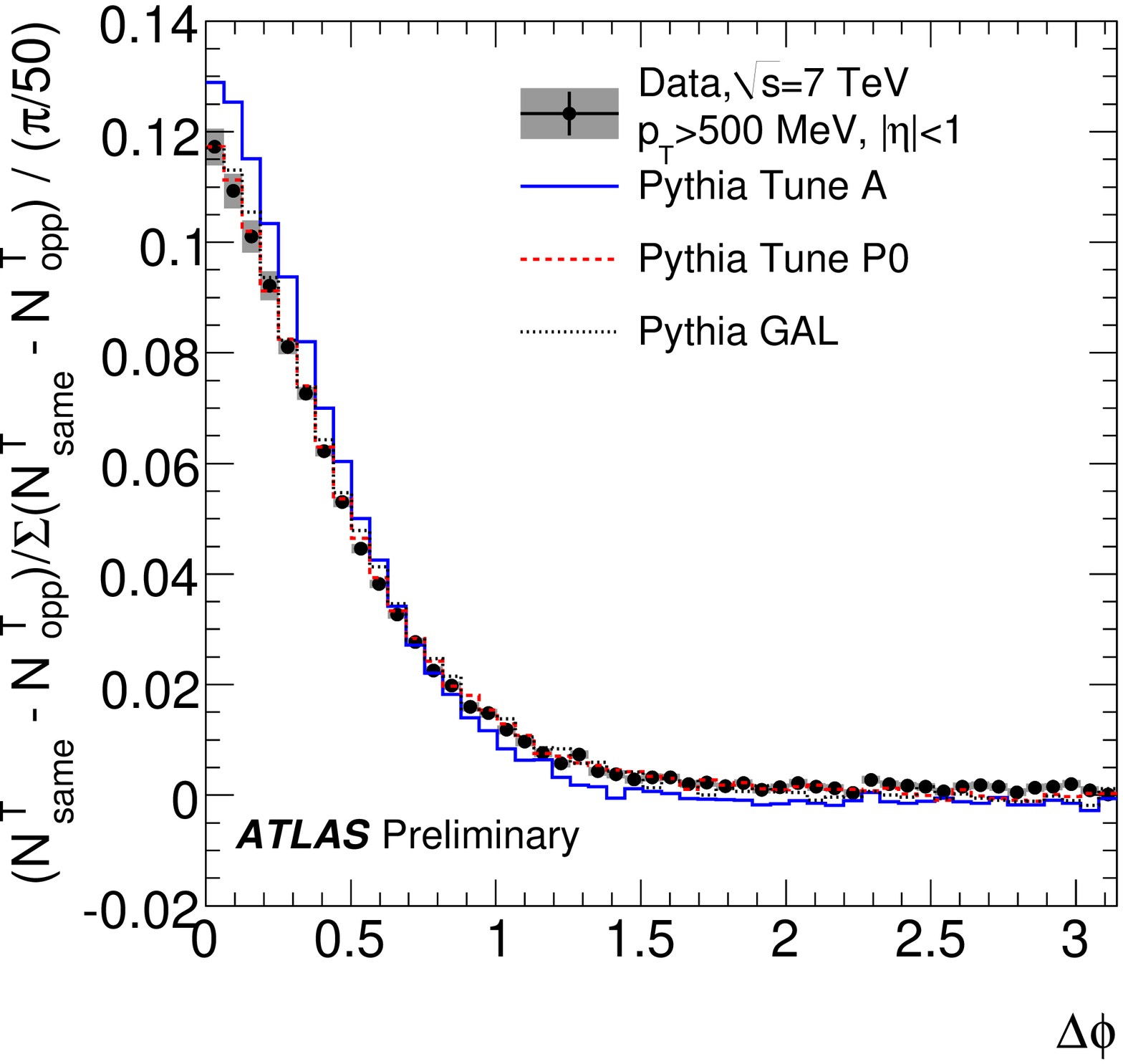}
}
\subfigure[]{
\includegraphics[scale=0.2]{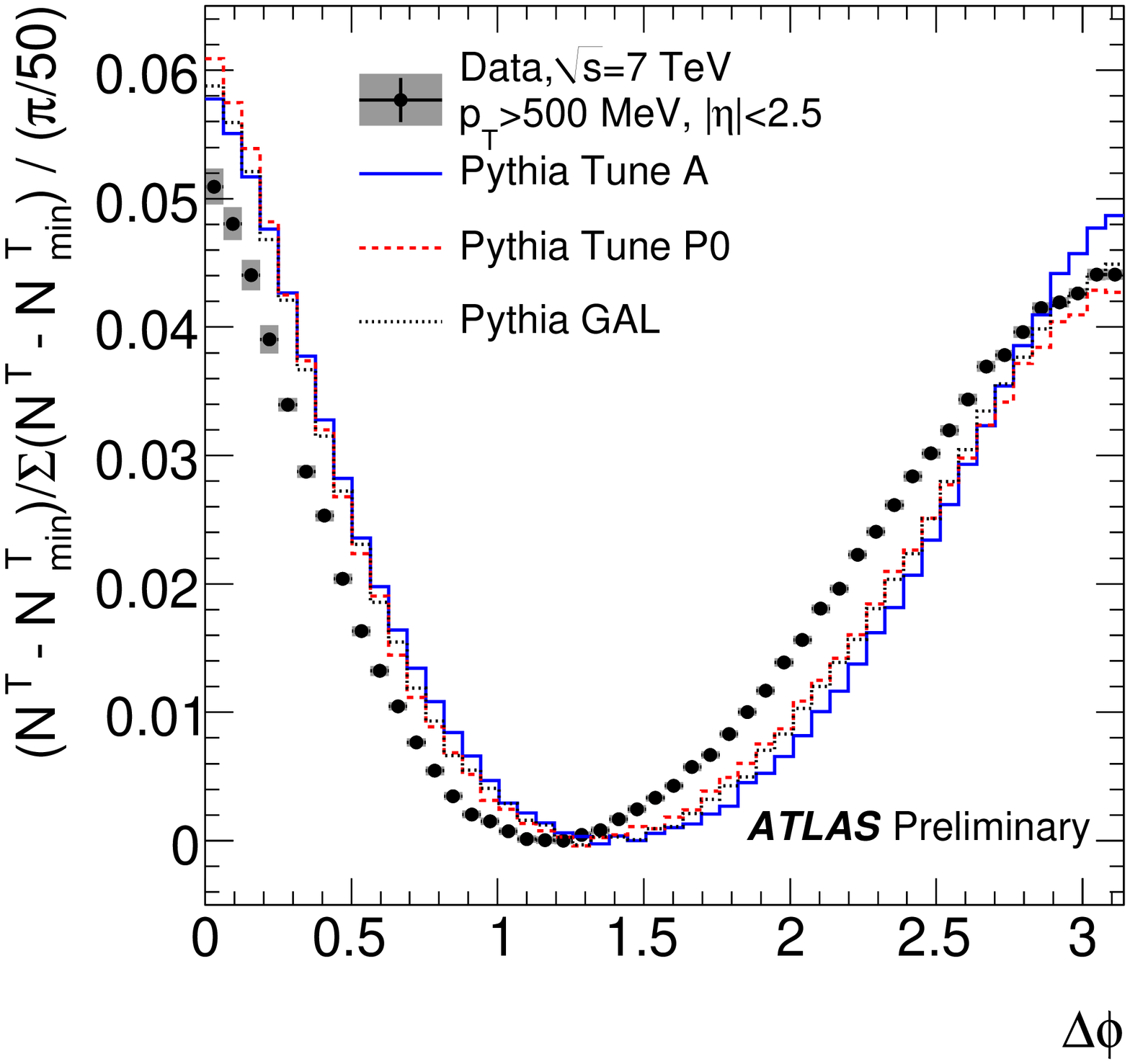}
}
\subfigure[]{
\includegraphics[scale=0.2]{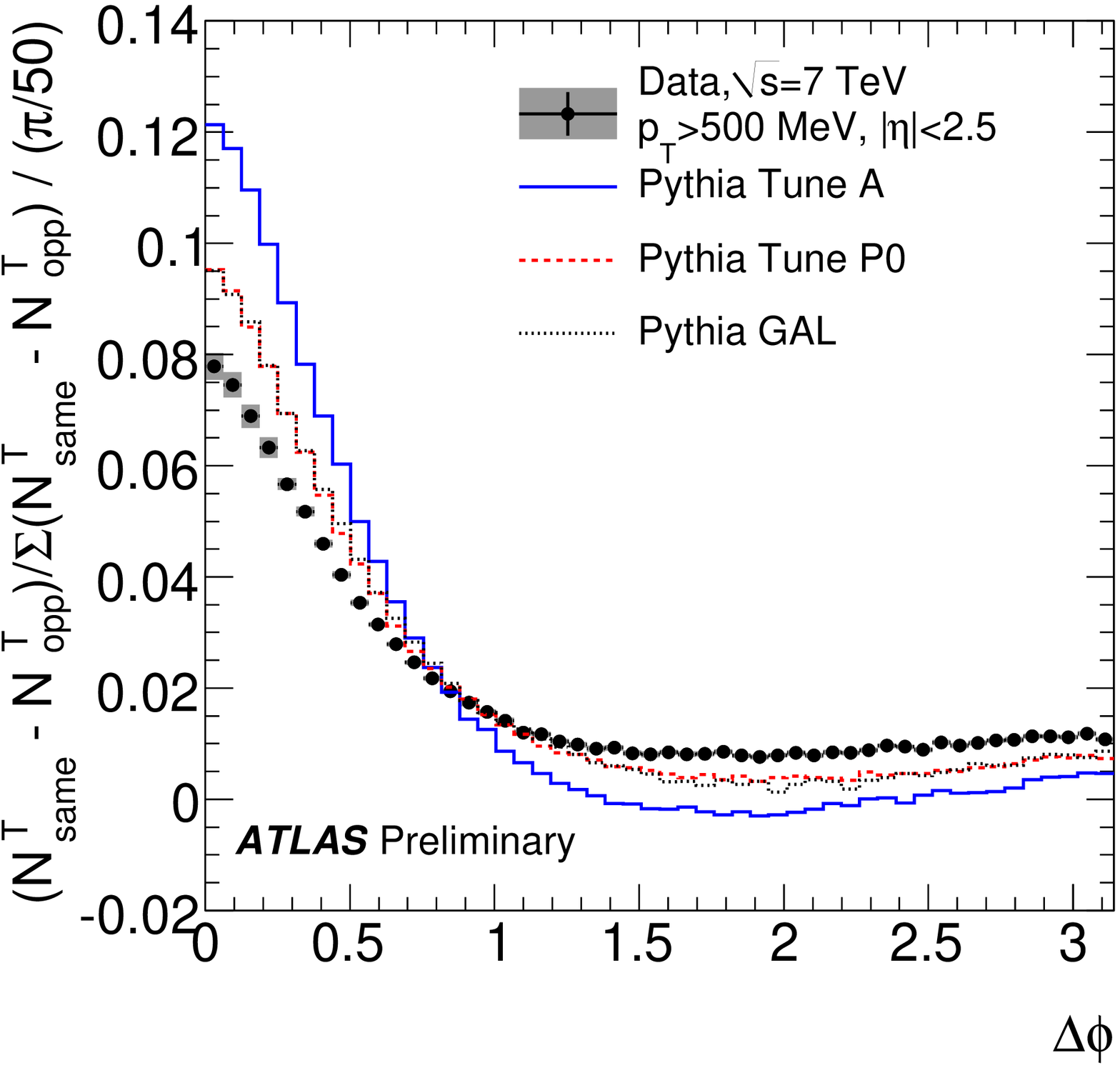}
}
\caption[]{The $\Delta \phi$ crest shape obtained by subtracting the minimum are shown in (a) and (c), while those obtained by the subtracting the `opposite from same' are shown in (b) and (d). The left two plots are for $|\eta| < 1.0$ and the right two plots are for $|\eta| < 2.5$.}
\label{fig:dphi-dpak}
\end{figure}

A complementary way~\cite{ATLAS-CONF-2010-082} to look at the angular correlation is by either subtracting the minimum of the distribution (determined by a second-order polynomial fit), or by subtracting the opposite side distribution (defined according to if pseudorapidity has the same or the opposite sign as the leading track) from the same side distribution and normalizing to unity. In Fig.~\ref{fig:dphi-dpak}, it is seen that the models are better at lower $\eta$ than at higher.

\subsubsection{Summary and Conclusions}

One of the goals of these analyses is to provide data that can be used to test
and improve MC models for current and future physics studies at the
LHC. The underlying event observables presented here are particularly important
for constraining the energy evolution of multiple partonic interaction models,
since the plateau heights of the underlying event profiles are highly correlated
to multiple parton interaction activity. The data at $7$~TeV are crucial for MC tuning, since measurements are
needed with at least two energies to constrain the energy evolution of MPI
activity. 

There is no current
standard MC tune which adequately describes all the early ATLAS data.  However,
using diffraction-limited minimum bias distributions and the plateau of the
underlying event distributions presented here, ATLAS has developed a new \Pythia
tune AMBT1 (ATLAS Minimum Bias Tune 1) and a new \HerwigJimmy tune AUET1
(ATLAS Underlying Event Tune 1) which model the \pt and charged multiplicity
spectra significantly better than the pre-LHC tunes of those
generators~\cite{ATLAS-CONF-2010-031,ATLAS:1303025}. It is critical to have
sensible underlying event models containing our best physical knowledge and intuition, tuned
to all relevant available data.

\graphicspath{{bartalini/}}

\contribution{\label{exp:cms}Multiple Parton Interactions Studies at CMS}
{Contributing authors: P. Bartalini and Livio Fan\`o (on behalf of the CMS Collaboration)}

\label{cms}

This section summarizes the early Underlying Event and forward measurements of the CMS collaboration using pp collision data up to highest energies of $\sqrt{s}=7$ TeV. It also reports along the feasibility study for the direct measurement of double parton scattering phenomena focusing on the $3jet + \gamma$ channel. 

A detailed description of the CMS detector is available in Ref.~\cite{:2008zzk}. Generator level Monte Carlo (MC) predictions are compared to the data corrected with a bayesian unfolding technique taking into account the detector effects~\cite{agostini}. 

The predictions for inelastic events are provided here by several tunes of the  {\sc Pythia} program, versions 6.420~\cite{Sjostrand:1986ep,Sjostrand:2006za} and 8.135\footnote{{\sc Pythia} version 8.108 is used in the feasibility studies reported in section {\ref{feas}}.}~\cite{Sjostrand:2008vc,Corke:2009pm}. {\sc Phojet} \cite{Bopp:1998rc} is also used in the forward measurements:

The pre-LHC tune D6T~\cite{Field:2008zz,Bartalini:2010su} of {\sc Pythia 6}, which describes the
lower energy UA5 and Tevatron data, is a widely used reference that will
also be used for most of the presented analyses.
The tunes DW~\cite{Bartalini:2010su} and CW~\cite{Khachatryan:2010pv}, which were found to
describe best the UE CMS data at $0.9$ TeV whereas D6T predictions were
too low~\cite{Khachatryan:2010pv}, will also be discussed for the $7$ TeV data.
The pre-LHC tune Perugia-0 \cite{Skands:2010ak} and the new tune, Z1~\cite{Field:2010kx}, adopt $p_T$ ordering of parton showers
and the new {\sc Pythia} MPI model~\cite{Skands:2007zg}. It includes the results of the Professor tunes~\cite{Buckley:2009bj} considering LEP fragmentation and the color reconnection parameters of the AMBT1 tune \cite{:2010ir}, while with the first CMS UE results \cite{Khachatryan:2010pv,QCD-10-010} have been used to tune the parameters governing the value and the $\sqrt{s}$ dependence of the cut-off transverse momentum that in {\sc Pythia} regularizes the divergence of  the leading order scattering amplitude 
as the final state parton transverse momentum $\hat{p}_T$ approaches $0$.
The tune Z2 is similar to Z1, except for the transverse momentum cut-off  at the nominal energy of  $\sqrt{s_0} = 1.8$ TeV which is decreased from $1.932$ GeV/c to $1.832$ GeV/c.
{\sc Pythia 8} also uses the new {\sc Pythia} MPI model, which is interleaved with parton 
showering. The default Tune 1 is adopted here.
{\sc Pythia 8} includes soft and hard diffraction~\cite{Navin:2010fk}, 
whereas only soft diffraction is included in {\sc Pythia 6}; the diffractive 
contributions are, however, heavily suppressed by the trigger and event 
selection requirements, especially for large $p_T$ values of the leading 
track-jet.
The parton distribution functions used to describe the interacting protons are
the CTEQ6LL set for D6T and Z2 and the CTEQ5L set for the other 
simulations~\cite{Lai:1999wy,Pumplin:2002vw}.


\subsubsection{The Early Underlying Event Measurements}

In the presence of a hard process, characterized by particles or clusters of particles with a large transverse  momentum $p_T$ with respect to the beam direction, the final state of hadron-hadron interactions can be described  as the superposition of several contributions:
products of the partonic hard scattering with the highest $p_T$, including initial and final state radiation;
hadrons produced in additional MPI; ``beam-beam remnants" (BBR) resulting from the hadronization of the partonic constituents that did not participate in other scatterings.
Products of MPI and BBR form the UE, which cannot be separated from initial and final state radiation.

The early CMS UE measurements focus on the understanding of the UE dynamics studying charged particle production with two different approaches.
The first (traditional) approach \cite{Khachatryan:2010pv,QCD-10-010} concentrates on the study of the transverse region, which is defined considering the azimuthal distance of the reconstructed tracks  with respect to the leading track or leading track-jet of the event: 
$60^\circ <  |\Delta\phi| < 120^\circ$. The jet reconstruction algorithm used in these studies is SisCone~\cite{Salam:2007xv}. 
On top of the traditional approach, a new methodology using anti-$k_T$ jets~\cite{Cacciari:2008gp}  and relying on the measurement of their area~\cite{Cacciari:2009dp} is adopted for the first time by CMS using charged particles in pp collision data collected at $\sqrt{s} = 0.9$ TeV \cite{QCD-10-005}. The new set of UE observables consider the whole pseudorapidity-azimuth plane instead of the transverse region and inherently take into account the leading jets of an event. 
\begin{figure}[t]
\begin{center}
\includegraphics[angle=0, width=0.4 \textwidth]{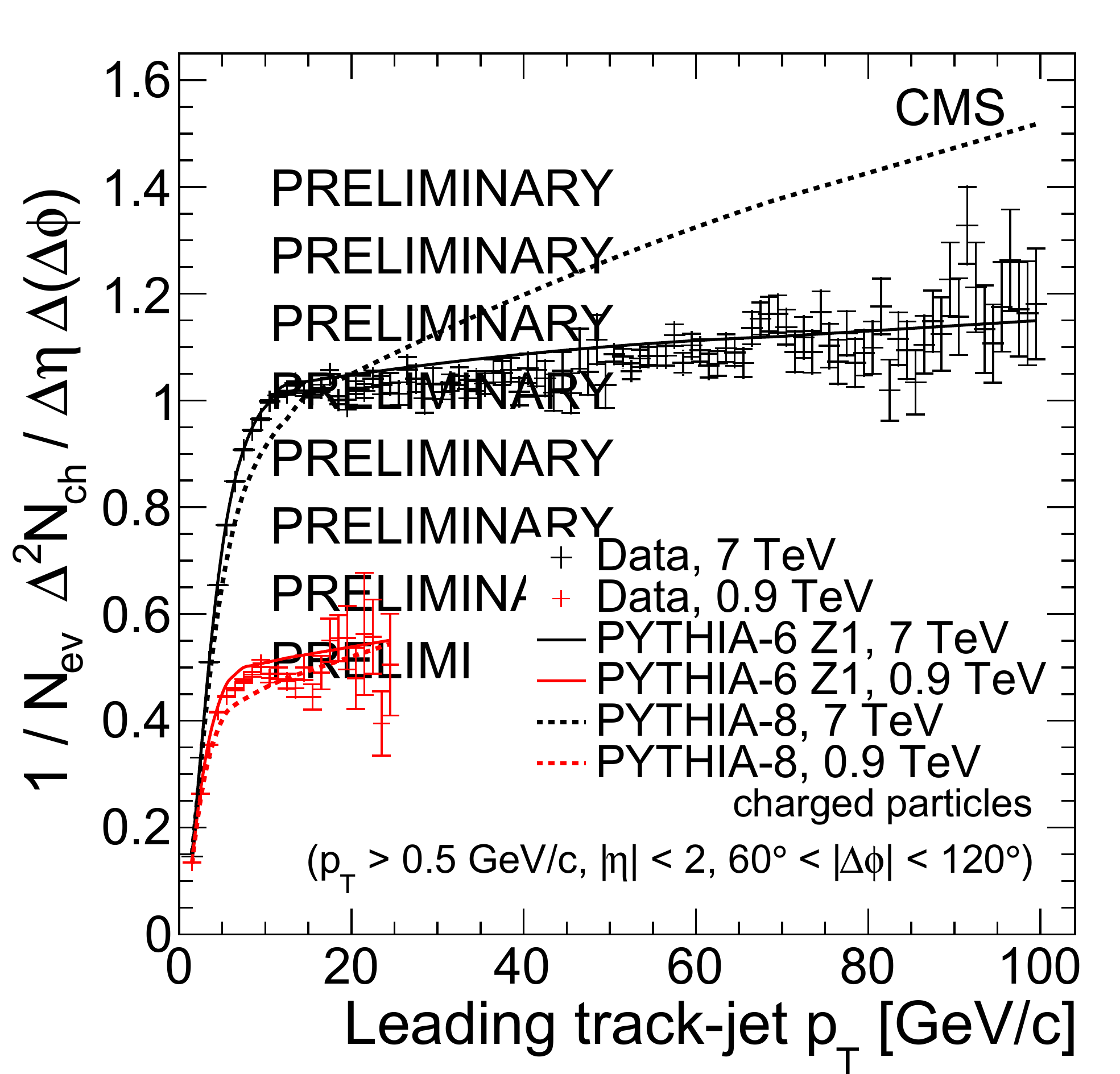}
\includegraphics[angle=0, width=0.4 \textwidth]{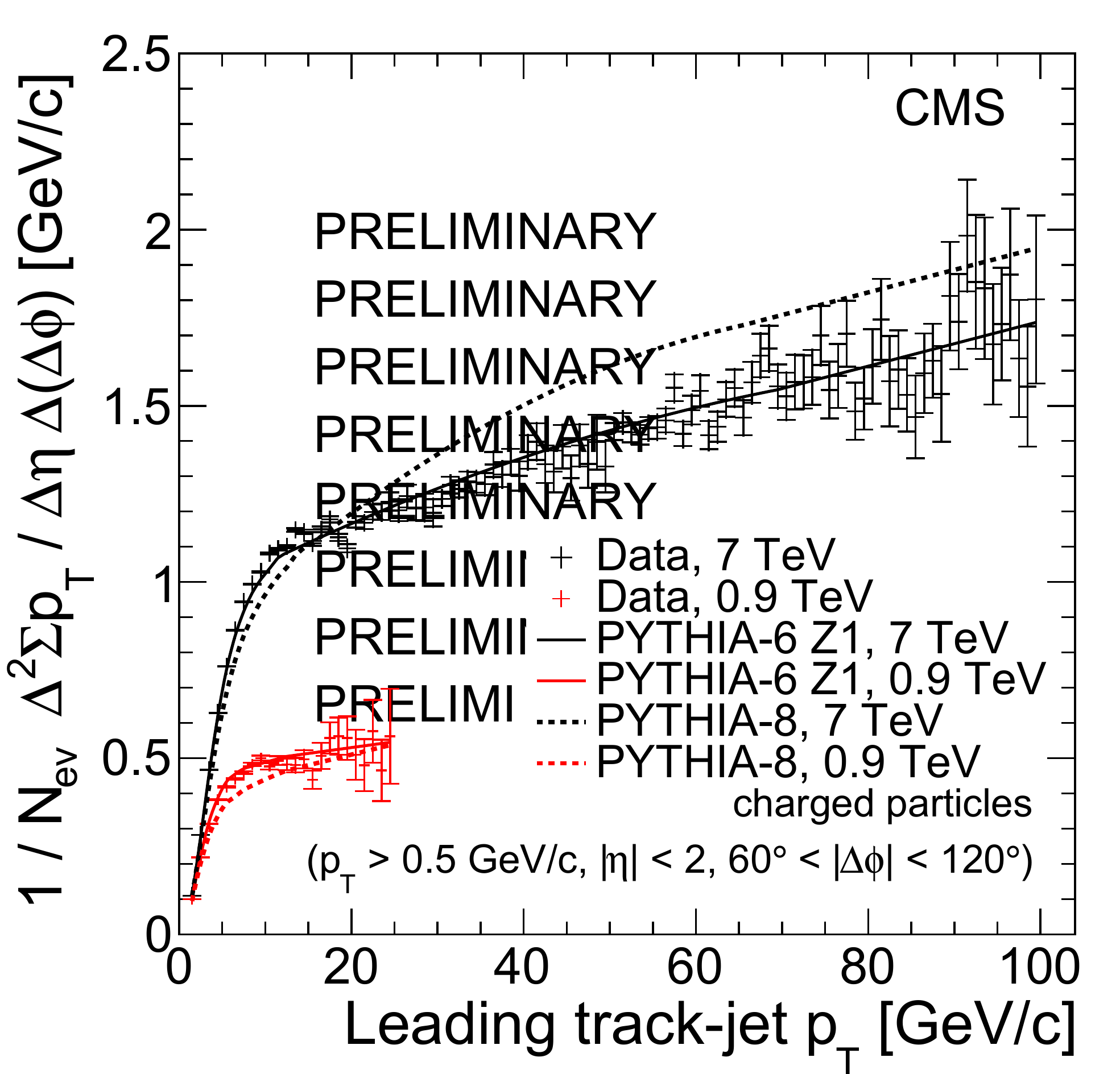} 
\includegraphics[angle=0, width=0.4 \textwidth]{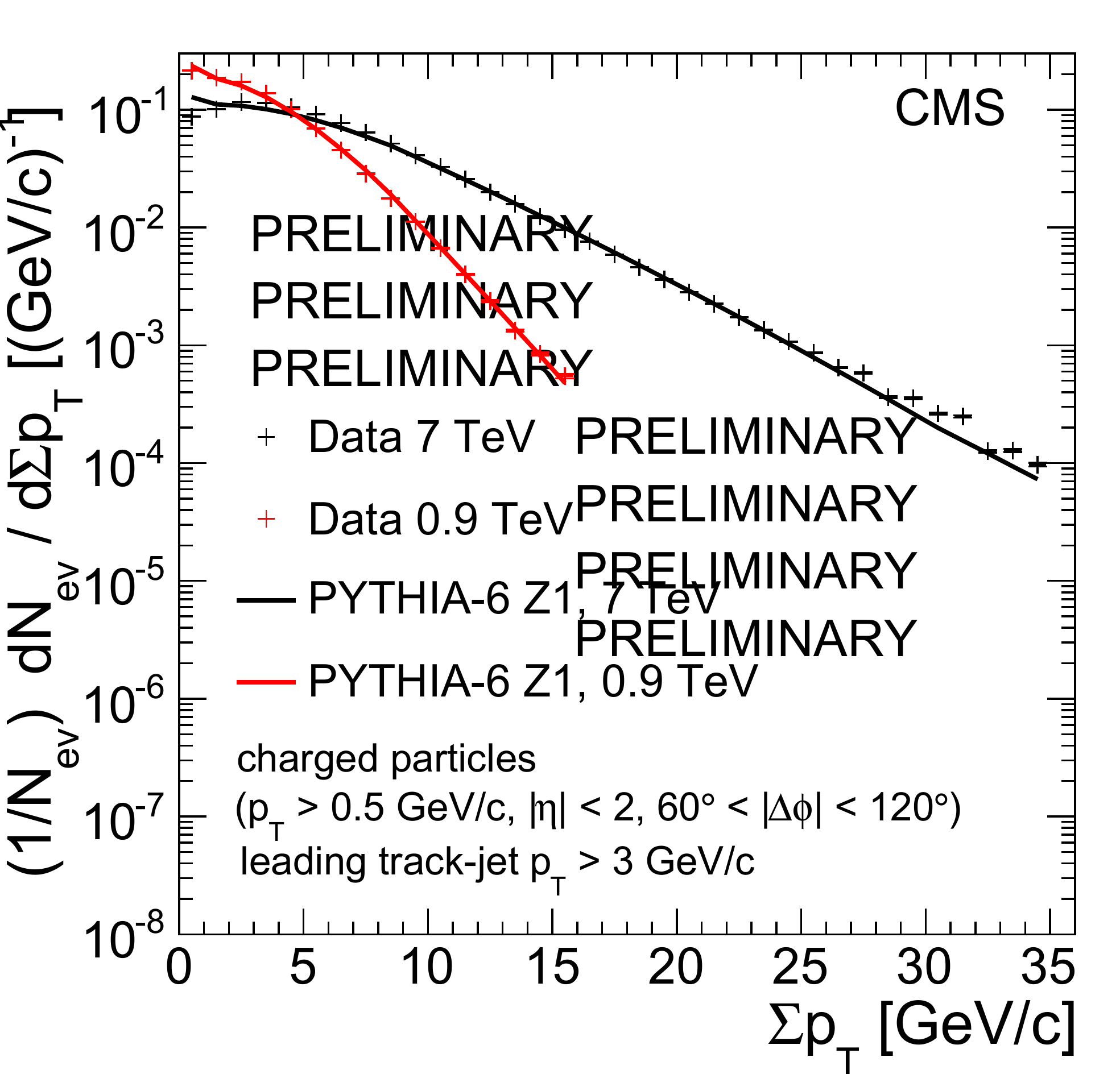} 
\includegraphics[angle=0, width=0.4 \textwidth]{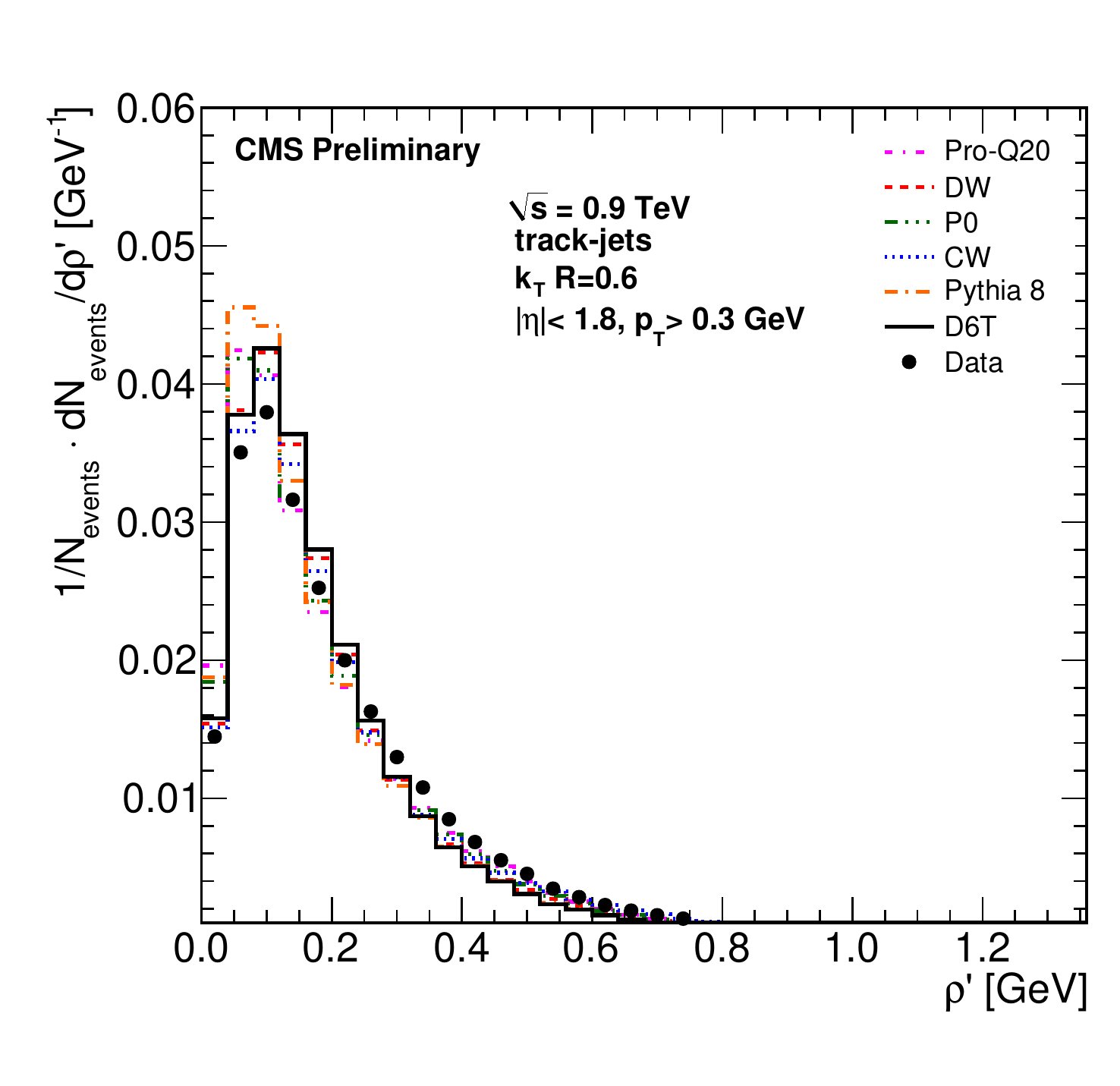} 
\end{center}
\caption{ 
(Upper plots) average multiplicity and average scalar $\sum p_T$ in the transverse region as a function of the leading track-jet $p_T$, for
data at $\sqrt{s} = 0.9$ TeV and $\sqrt{s} =7$ TeV.
(Bottom left plot)~normalized scalar $\sum p_T$ distribution in the transverse region for
data at $\sqrt{s} = 0.9$ TeV and $\sqrt{s} =7$ TeV; the leading track-jet is required to have $p_T > 3$ GeV/c.
Predictions from  {\sc Pythia 6} tune Z1 and {\sc Pythia 8.135} Tune 1 are compared to the corrected data. 
The inner error bars indicate the statistical uncertainties affecting the measurements, the outer error bars thus represent the statistical uncertainties on the measurements and the systematic uncertainties affecting the MC predictions added in quadrature.
(Bottom right plot)~normalized median of $p_T$ over jet area for track-jets reconstructed from
collision data at $\sqrt{s} = 0.9$ TeV~(black circles).
Predictions from several {\sc Pythia 6} tunes and {\sc Pythia 8} Tune 1 are compared to data. 
\label{fig:three-prime}}
\end{figure}

The centre-of-mass energy dependence of the hadronic activity in the transverse 
region is presented on the two top Figures~\ref{fig:three-prime} as a function of the $p_T$ of the leading track-jet.
The data points represent the average multiplicity and average scalar track-$p_T$ sum dependence, for $\sqrt{s} = 0.9$ TeV and $\sqrt{s} = 7$ TeV using tracks with a pseudorapidity $|\eta| <  2.0$ and $p_T > 0.5$ GeV/c.
A significant growth of the average multiplicity and of the average scalar $p_T$ 
sum of charged particles transverse to that of the leading track-jet is observed with increasing 
scale provided by the leading track-jet $p_T$, followed by saturation at large 
values of the scale (more evident for multiplicity profile than average scalar $p_T$ sum). 
A significant growth of the activity in the transverse region is also observed,
for the same value of the leading track-jet $p_T$, from $\sqrt{s} = 0.9$ TeV
to $\sqrt{s} = 7$ TeV.
These observations are consistent with the ones obtained at Tevatron \cite {Affolder:2001xt}.
The evolution with the hard scale of the ratio of the UE activity at 7 TeV and 0.9 TeV is remarkably well described by the Z1 MC. The trend is also reproduced by {\sc Pythia 8}. 
The Z2 predictions at $\sqrt{s}= 0.9$ TeV (not shown here) agree with Z1 in shape, but the normalization is 5-10\% higher for both the observables; this trend is opposite with respect to the one observed at 7 TeV and indicates that a less pronounced $\sqrt{s} $ dependence of the transverse momentum cut-off should be adopted for tunes using the CTEQ6LL PDF set than for the tunes optimized for the CTEQ5L set.

The strong growth of UE activity with	 charged particles is also striking in the comparison of the normalized distributions of charged particle multiplicity (not shown here) and of scalar $p_T$ sum which is presented in bottom-left plot of Figure~\ref{fig:three-prime} for events at	$\sqrt{s} = 0.9$ TeV and $\sqrt{s} = 7$ TeV with leading track-jet $p_T > 3$ GeV/c. The particle $p_T$ spectrum (not shown) extends up to $p_T > 10$ GeV/c, indicating the presence of 
a hard component in particle production in the transverse region.
The distributions for track-jet $p_T > 3$ GeV/c, which extend up to quite large values of the selected observables in the transverse region 
are quite well described by the various MC models, over several orders of 
magnitude. This observation gives support to the implementation of MPI in {\sc Pythia}.

The novel technique to quote the UE activity 
relies on the introduction of ``ghosts'', virtual deposits of very low energy filling the overall phase space which are taken into account by the jet clustering algorithm. 
The estimator of the overall soft background activity in an event can be derived as the median of the ratio between the transverse momentum and the area of the jets. One of the advantages of using the median compared to the mean is that it turns out to be less sensitive to the influence of outliers, i.e.\ in particular the leading jets in an event.
In order to cope with the low occupancy observed at $\sqrt{s} = $ 0.9 TeV, CMS redefines such observable restricting the median only to those jets which have physical deposits on top of ghosts:
\begin{equation}
  \rho'=
  {\mathrm{median}}\left[\left\{{\frac{p_{\textrm{T}j}}{A_j}}\right\}\right] \cdot C
\end{equation}
where $C$ is the occupancy of the event, which is the summed area $\sum_{j}{A_{j}}$ covered these jets divided by the considered detector region $A_{\rm tot}$. 
In the CMS analysis at $\sqrt{s} = 0.9$ TeV, jets are reconstructed with the anti-$K_T$ algorithm~\cite{Cacciari:2008gp} using tracks with $|\eta| <  2.0$ and $p_T > 0.3$ GeV/c.  In the right bottom plot of Figure~\ref{fig:three-prime} the $\rho'$ observable is presented for minimum bias events. The general pattern of deviations from data with respect to the considered {\sc Pythia} tunes looks rather similar to the one observed with the traditional UE measurement. 

\subsubsection{Study of the Activity in the Forward Region}

CMS reports a measurement of the energy flow in the forward region ($ 3.15<|\eta|< 4.9$, where $\eta$ denotes the pseudorapidity) \cite{FWD-10-002} for minimum bias and dijet events in pp interactions with centre-of-mass energies $\sqrt{s}$ of 0.9 TeV, 2.36 TeV and 7 TeV. This measurement is connected to the ones reported in the previous sections as the basic philosophy is the same: it concentrates on the complementary activity of a pp interaction for different energy scales of the reconstructed leading objects.

\begin{figure}[htbp]
\begin{center} 
\includegraphics[angle=90, width=0.435 \textwidth]{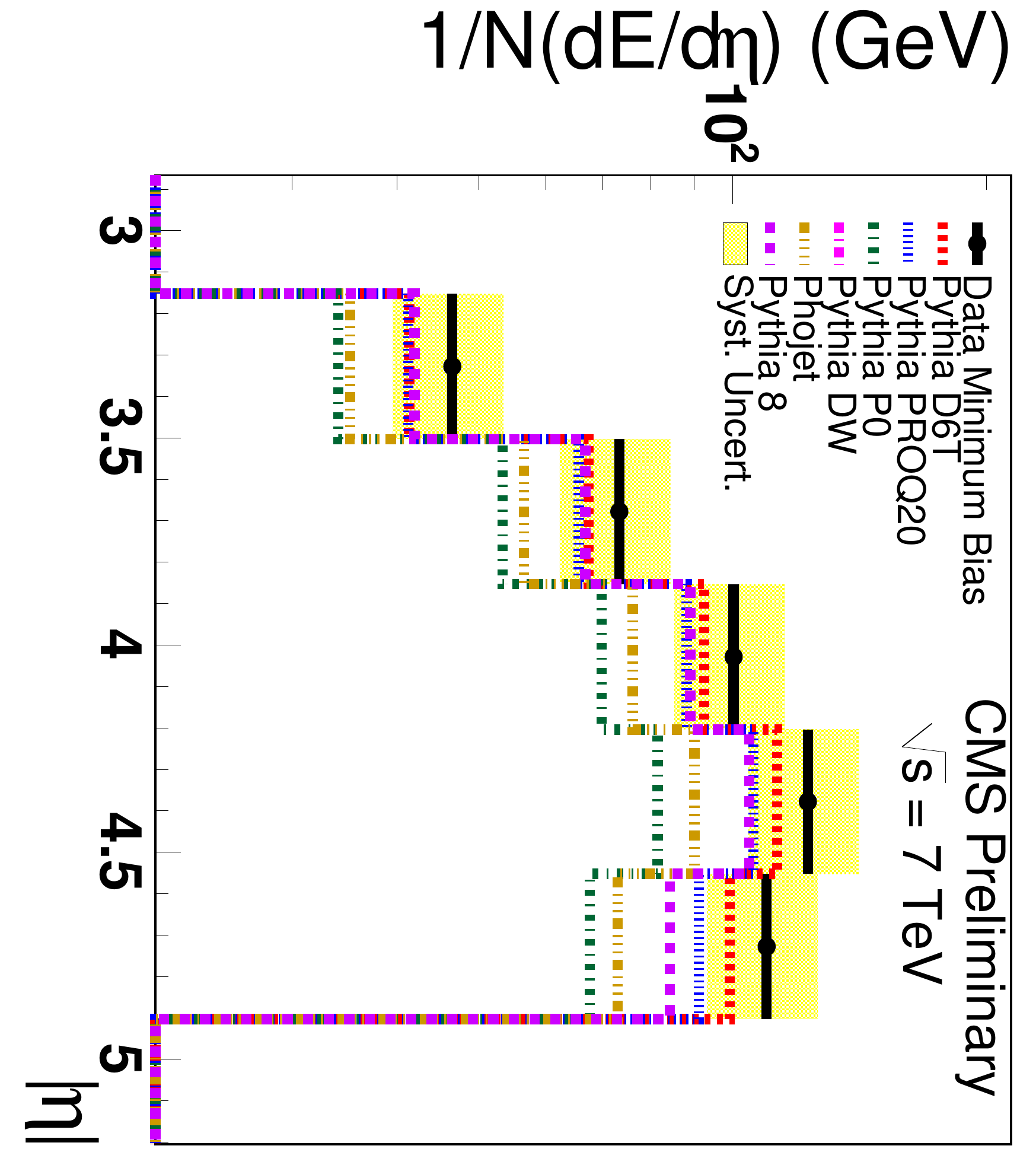}
\includegraphics[angle=90, width=0.4 \textwidth]{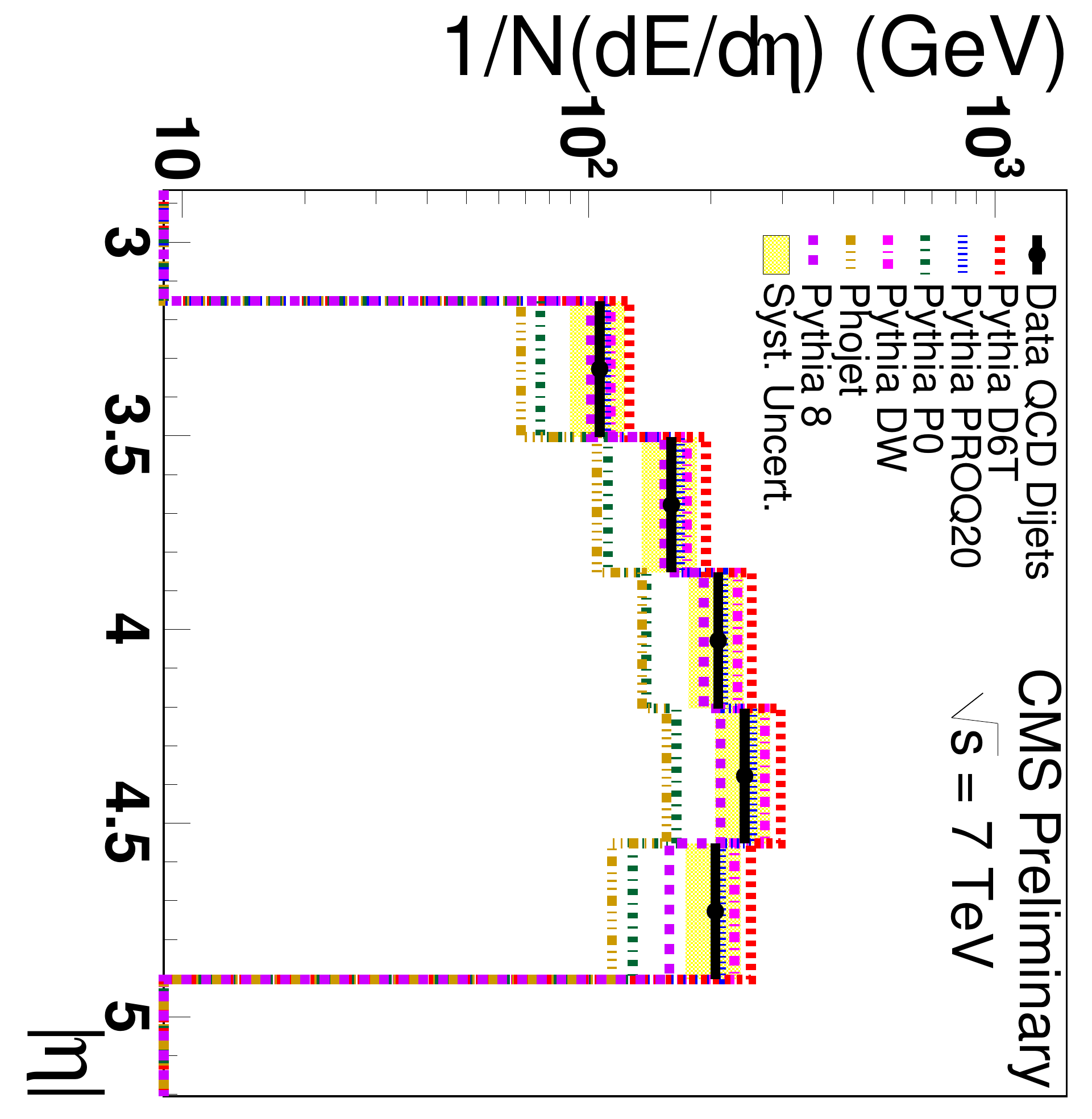}
\caption{Energy flow in the minimum bias (left) and di-jet (right) samples as a function of $\eta_{obs}$ at $\sqrt{s}=7$ TeV. Uncorrected data are shown as points, the histograms correspond to the MC predictions. Error bars corresponds to statistical errors. The shaded bands in these plots represent the systematic uncertainties of the measurements, which are largely correlated point-to-point.}
\label{resultsMBMCE}
\end{center}
\end{figure}

The energy flow in the region of the Hadron Forward detector is measured in two different event classes: in minimum bias events and in events with a  hard scale provided by a dijet system at central pseudorapidities ($|\eta| < 2.5$) and with transverse energy $E_{T,jet}>8$~GeV for $\sqrt{s} = $ 0.9~TeV and 2.36~TeV; the dijet threshold is increased to $20$~GeV for $\sqrt{s} =$ 7~TeV.
The results are qualitatively similar at all the investigated centre of mass energies.
Fig.~\ref{resultsMBMCE} shows the results of the forward energy flow at $\sqrt{s} =$ 7~TeV for the two event classes compared to predictions from Monte Carlo event generators.
The measured forward energy flow is found to be significantly different between the two event classes, with a sensitive increase and a more central activity seen in dijet events. 

\subsubsection{Multiple Parton Interactions in High-$p_T$ Phenomenology \label{feas}}

Quantifying the MPI cross sections basically deals with the measurement of $\sigma_{eff}$, the scale factor which characterizes the inclusive rate of the interactions~\cite{Paver:1982yp,Calucci:1997ii}, cf. Chapters \ref{sec:theory} and \ref{sec:pheno}. From a phenomenological point of view $\sigma_{eff}$ is a non perturbative quantity related to the transverse size of the hadrons and has the dimensions of a cross section. The measurements performed by the AFS, CDF and D0 collaborations~\cite{Akesson:1986iv,Abe:1997xk,Alitti:1991rd,Abe:1993rv} favor smaller values of $\sigma_{eff}$ with respect to the naive expectations. The consequently increased rates of multiple parton interactions can be interpreted as an effect of the hadron structure in transverse plane~\cite{Calucci:2008jw}. 
Extending such measurements at the LHC and studying the possible scale dependency of $\sigma_{eff}$ is definitely of great interest and may have a deep impact on the data driven estimations of the MPI backgrounds to new physics.

The production of four high-$p_T$ jets is the most prominent process to search for multiple high $p_T$ scatterings: two independent scatters (i.e. DPS) in the same $pp$ or $p\bar{p}$ collision, each producing two jets. Such a signature has been searched for by the AFS experiment at the CERN ISR, by the UA2 experiment at the CERN S$\bar{p}p$S and by the CDF and D0 experiment at the Fermilab Tevatron. 

\begin{figure}[htpb]
\begin{center}
\includegraphics[width=0.28\columnwidth]{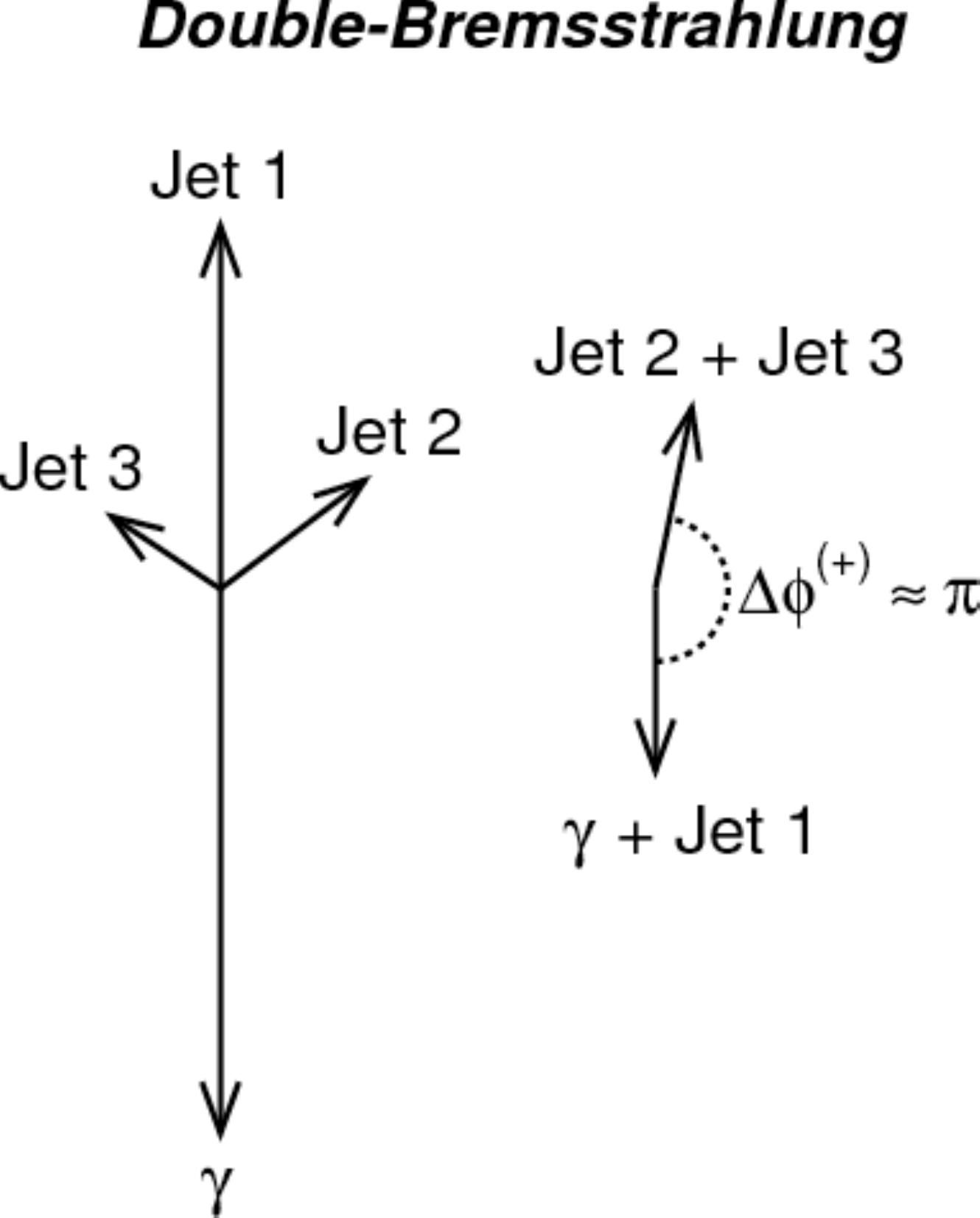} \ \ \ \ \ %
\includegraphics[width=0.28\columnwidth]{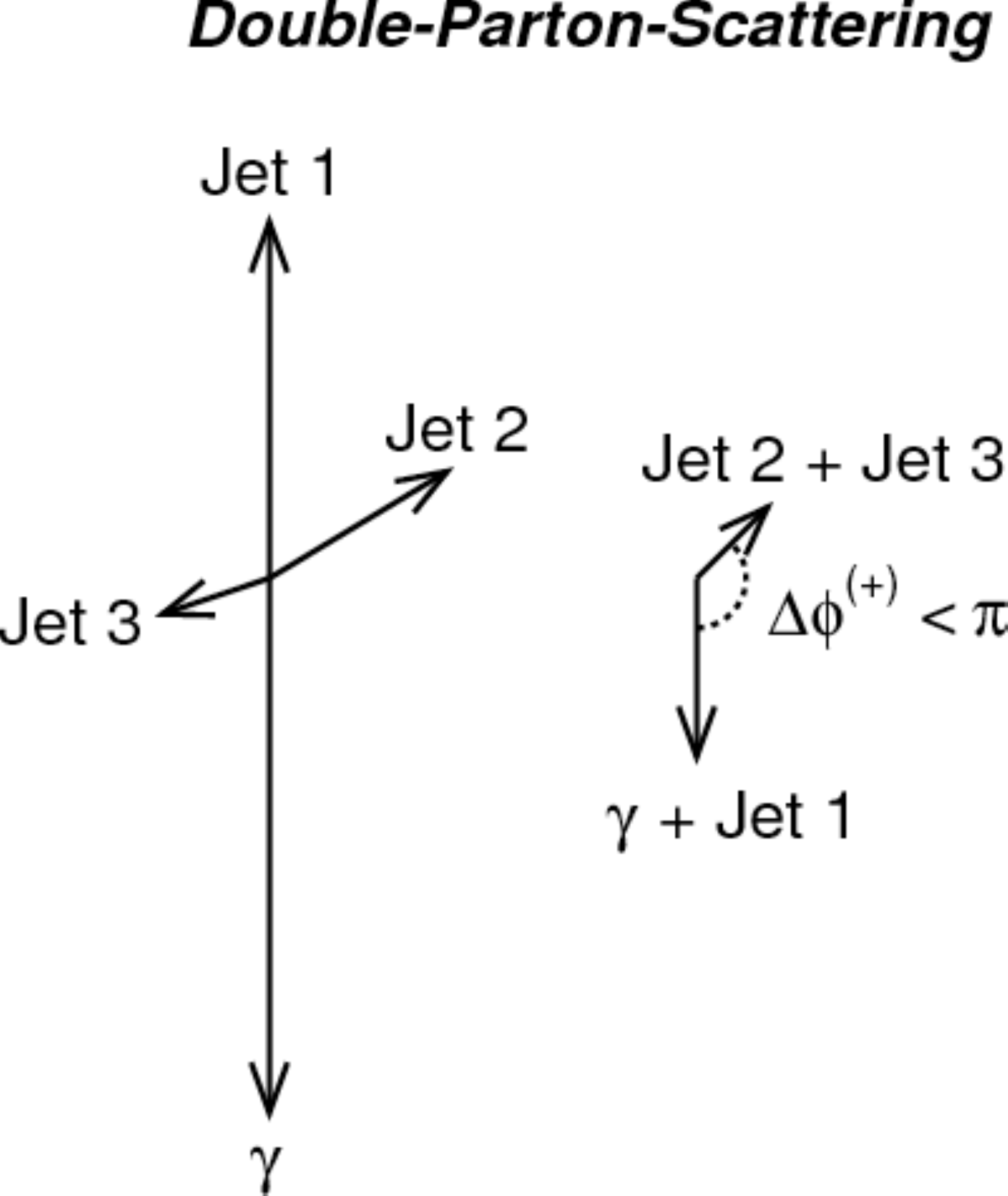}
\caption{Definition of azimuthal angle between pairs, together with typical configurations of double-bremsstrahlung (left) and double-parton scattering events (right).}
\label{fig:deltaS-sketch}
\end{center}
\end{figure}

However, searches for double-parton scattering in four-jet events at hadron colliders may face significant backgrounds from other sources of jet production, in particular from QCD brems\-strahlung (Fig.~\ref{fig:deltaS-sketch}-left). Typical thresholds employed in jet triggers bias the event sample towards hard scatterings. However, a high-$p_T$ jet parton is more likely to radiate additional partons, thus producing further jets. Thus, the relative fraction of jets from final-state showers above a given threshold is enlarged in jet trigger streams which is an unwanted bias. On the other hand, looking for four jets in a minimum-bias stream will yield little statistics. 

\begin{figure}[htpb]
  \includegraphics[width=0.49\columnwidth]{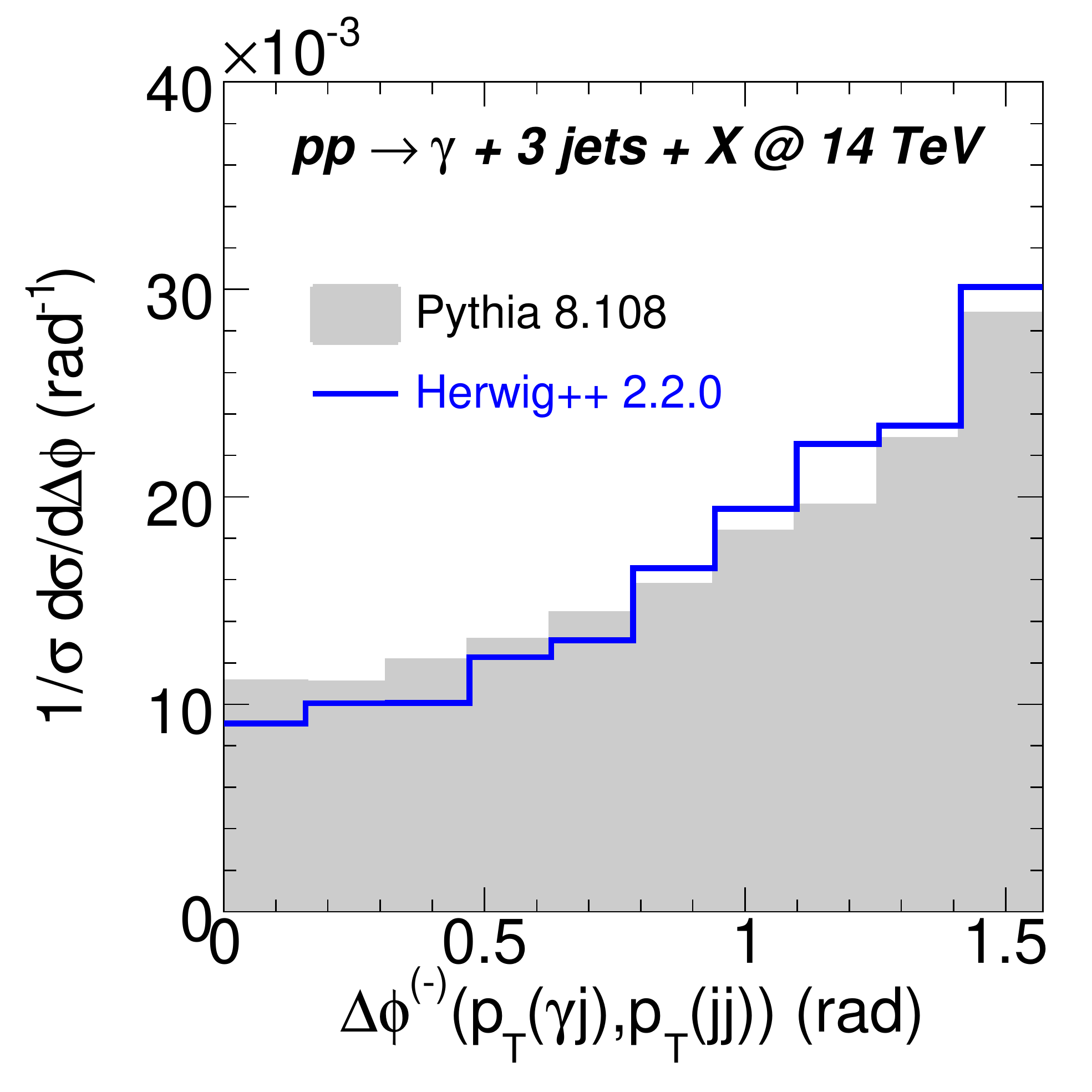}\hfill
  \includegraphics[width=0.49\columnwidth]{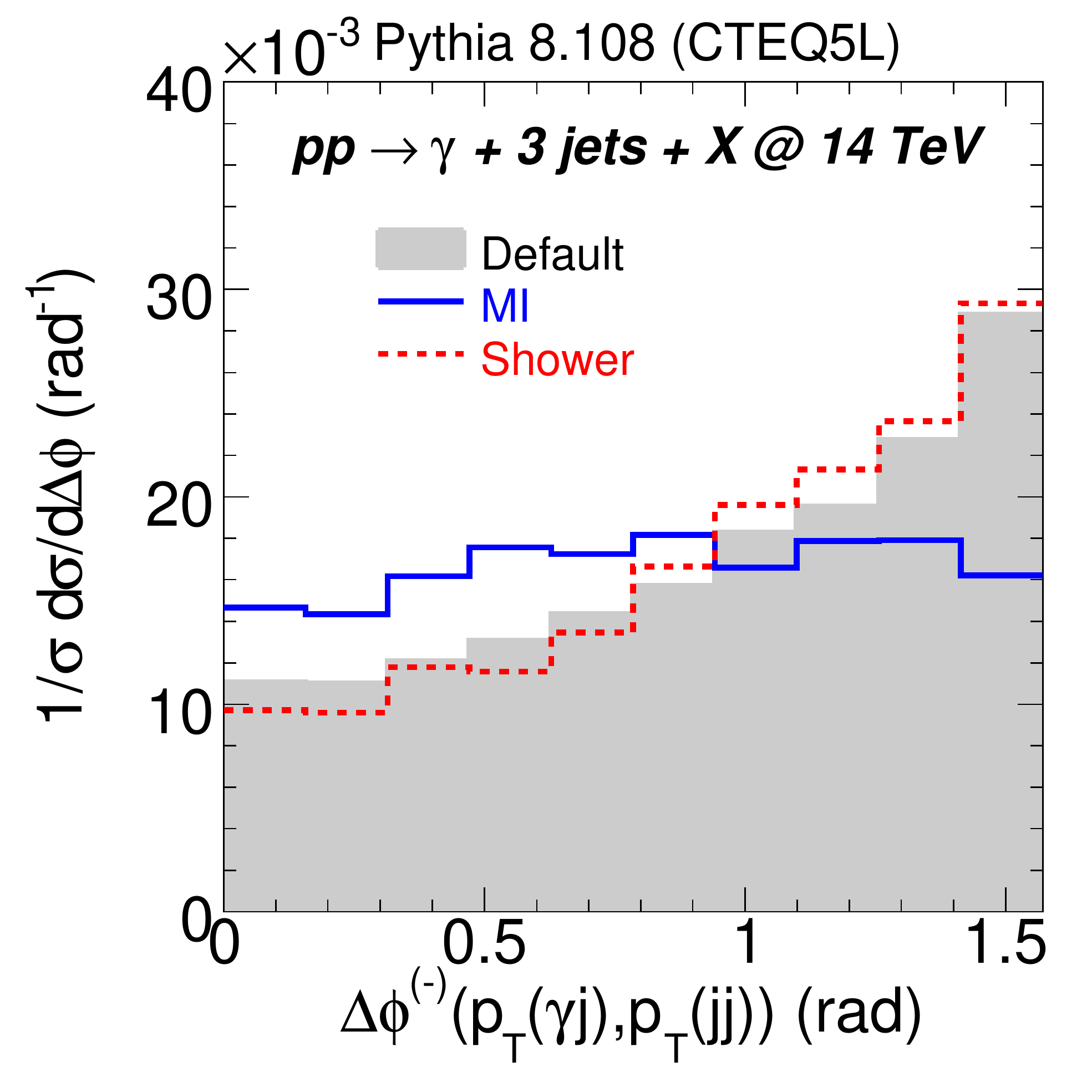} \hfill
  \includegraphics[width=0.49\columnwidth]{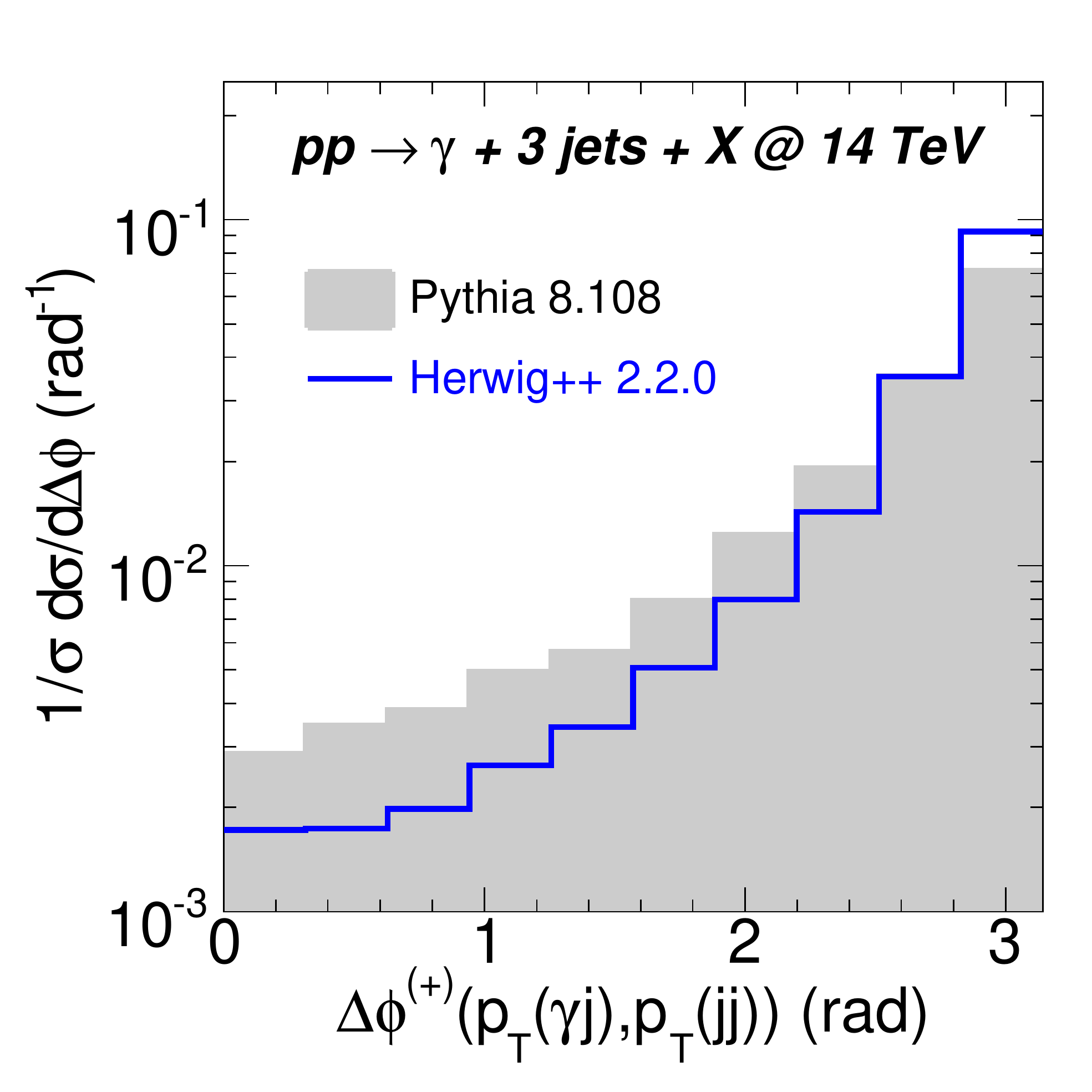}\hfill
  \includegraphics[width=0.49\columnwidth]{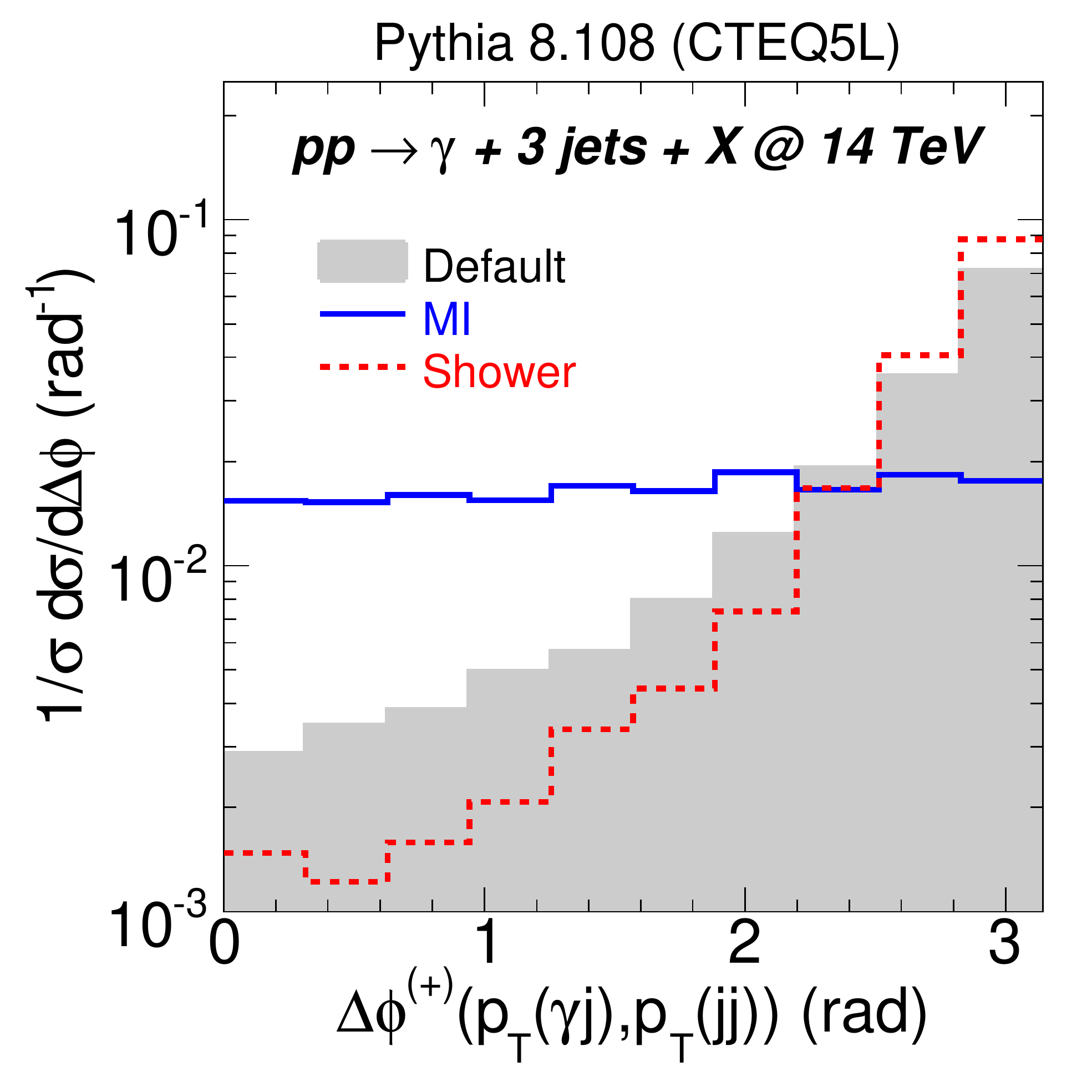}\hfill
  \caption%
  {Differential cross section shape as a function of $\Delta \phi^{(-)}$ (upper plots) and $\Delta \phi^{(+)}$ (bottom plots) variables. Predictions from {\sc Pythia} 8.108 (\emph{Default} scenario) and {\sc Herwig} 2.2.0 (left panel) and from three different {\sc {\sc Pythia}} settings (right panel) shown.}
  \label{fig:dphi}
\end{figure}
Therefore the strategy to directly measure the MPI rate in high-$p_T$ regime at hadron colliders also includes the study of multi-jets or jets+photon final states. Indeed the CDF and D0 collaborations studied final states with one photon and three jets looking for pairwise balanced photon-jet and dijet combinations. The data sample was selected with the experiment's inclusive photon trigger, thereby avoiding a bias on the jet energy. The better energy resolution of photons compared to jets purifies the identification of $E_T$ balanced pairs. Tevatron found an excess in pairs that are uncorrelated in azimuth with respect to the predictions from models without additional hard parton scatters per proton-proton scatter. CDF interpreted this result as an observation of double-parton-scatters.

Analyses trying to identify two hard scatters in multi-jet events typically rely on methodologies which overcome combinatorics. There are three possible ways to group four objects into two pairs: combinations are commonly selected pairing objects which are balanced in azimuth and energy. The flavor or other specific features of the jets may be used to decrease the combinatorics and to make looser the constraints on the balancing. One example of such a final state is constituted by events with two $b$ jets and two additional light jets. 

In order to discriminate double-parton scatters against double-brems\-strah\-lung events, CMS studies 
the observables $\Delta \phi^{(-)}$, employed by AFS, and $\Delta \phi^{(+)}$, employed by CDF, probing the azimuthal angle between pairs (Fig.~\ref{fig:deltaS-sketch}).
Expectations for the above described variables are therefore $\Delta \phi^{(-)} \approx \pi / 2$ and $\Delta \phi^{(+)} \approx \pi$ if additional jets come from double-bremsstrahlung. Otherwise, i.~e.~if additional jets come from multiple interactions, both variables should be distributed uniformly.

Differential cross section shape predictions for the $\Delta \phi$ observables in pp interactions at $14$ TeV are shown in Fig.~\ref{fig:dphi}. {\sc Herwig 2.2.0} \cite{Corcella:2002jc,Bahr:2008pv} and {\sc Pythia 8.108} with default settings which include multiple interactions and showering predict similar shapes (Fig.~\ref{fig:dphi}-left). The discrimination power of the selected observables to Multiple Parton Interaction patterns is clearly shown in Fig.~\ref{fig:dphi}-right, where events with MPI switched off (\emph{Shower} scenario) are compared to events with parton shower switched off (\emph{MI} scenario). The differences are particularly pronounced when selecting the $\Delta \phi^{(+)}$ observable. 


\subsubsection{Conclusions}

A strong growth of the UE activity is observed with increasing leading track-jet $p_T$ for both  $\sqrt{s} = 7$ TeV and $\sqrt{s} = 0.9$ TeV.
At 7 TeV this fast rise is followed above $\sim$8 GeV/c by a saturation region with nearly constant multiplicity and small $p_T$ increase. The same pattern is observed at 0.9 TeV, with the saturation region starting at $\sim$4 GeV/c.
A strong growth of the activity is also observed with increasing centre-of-mass energy.
The large increase of activity in the transverse region is also observed in the $\sum p_T$ distribution, indicating the presence of a hard component in the transverse region. Very good post-LHC MC tunes are available for the description of the UE in the central region.

A measurement of the underlying event using the jet-area/median approach is also reported, demonstrating its sensitivity to different underlying event scenarios. 

Complementary underlying event measurements in the forward region are also presented. The energy flow in the forward direction is measured for minimum bias and central di-jets events. A more global UE description including both the central and the forward regions is certainly one of the next MC tuning challenges, with deep impact on the understanding of the MPI dynamics.


The Multiple Parton Interactions measurement strategy in the high-$p_T$ regime is also briefly discussed focusing on the $3jet + \gamma$ topology. The very good performances of the LHC machine should allow to have soon the integrated luminosity conditions adequate to perform these measurements over a wide range of energy scales, with deep impact on the data driven estimation of the MPI backgrounds to searches.

\graphicspath{{mschmelling/}}

\contribution{Minimum Bias Physics at LHCb}
{Contributing author: M. Schmelling (on behalf of the LHCb Collaboration)}

\label{mschmelling}

At the startup in 2009 the LHC provided proton-proton collisions at a
center-of-mass energy of $\sqrt{s}=0.9$\,TeV. Although higher
collision energies have previously been reached at proton-antiproton
colliders, it was for the first time that the TeV-scale was studied in
proton-proton collisions. Using a data sample with an integrated
luminosity of only $6.8\pm1.0\,\mu$b$^{-1}$\/ recorded by the LHCb 
detector, a first measurement of the production cross-section of neutral
K-mesons was performed in a kinematic range not accessible to the
other LHC experiments. In 2010 the collision energy was moved up to
$7$\,TeV and the performance of the machine improved exponentially.
Until September LHCb collected an integrated luminosity of over
$3$\,pb$^{-1}$\/ with a data acquisition efficiency larger than 90\%.
At the same time the detector calibration approached its
design values. In the following, after a brief description of the 
detector the first measurements on strangeness 
production and studies of baryon number transport and baryon 
suppression in the fragmentation will be presented, before finally 
discussing prospects for doing diffractive physics with LHCb.

\subsubsection{The LHCb Experiment}
\begin{figure}[h]
\centering
\includegraphics[width=0.8\textwidth]{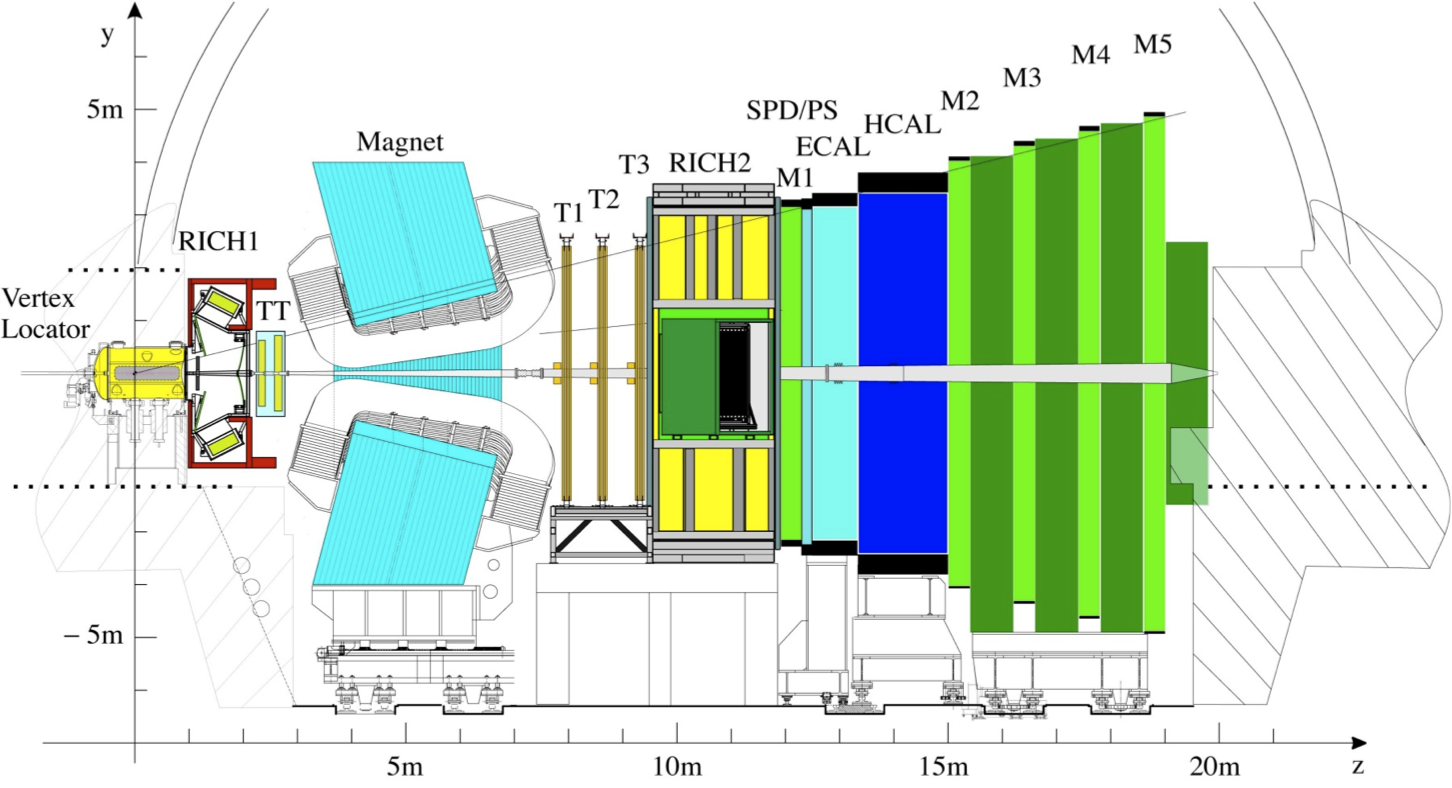}
\caption{
Schematic view of the LHCb single arm forward spectrometer. 
The interaction region is located on the left inside the Vertex 
Locator. The tracking system and the RICH detectors for particle 
identification are installed both before and after the dipole 
magnet, calorimetry and the muon system are located downstream
of the magnet.}
\label{fig:lhcb}
\end{figure}

The LHCb detector \cite{Alves:2008zz} is a forward
spectrometer, covering the angular range of $15 < \theta < 300$\,mrad
with respect to the beam axis. A schematic view of the experiment is
shown in Fig.\,\ref{fig:lhcb}. The detector offers tracking, 
calorimetry and particle identification over most of its forward 
acceptance. Momenta of charged particles are determined from the 
deflection by a dipole magnet with a field integral of 4\,Tm. The
interaction region is surrounded by the Vertex Locator (VeLo).
Going downstream, a first RICH detector and the so-called TT tracking
station are still located in front of the magnet. Immediately behind
the magnet follows the second part of the tracking system, consisting
of a high granularity Inner Tracker (IT) in the region of large
particle densities close to the beam pipe and the Outer Tracker system
at larger transverse distances. VeLo, TT and IT are silicon strip
detectors, the OT consists of straw tubes. Following the tracking
system is a second RICH detector, a pre-shower and scintillating pad
detector (SPD/PS), electromagnetic calorimeter (ECAL), hadron
calorimeter(HCAL) and muon system for the identification of electrons
and photons, neutral hadrons and muons, respectively. The RICH
detectors allow pion, kaon, proton separation in the momentum range
between $2 < p < 100$\,GeV/$c$.

\begin{figure}[t]
\centering
\includegraphics[width=0.7\textwidth]{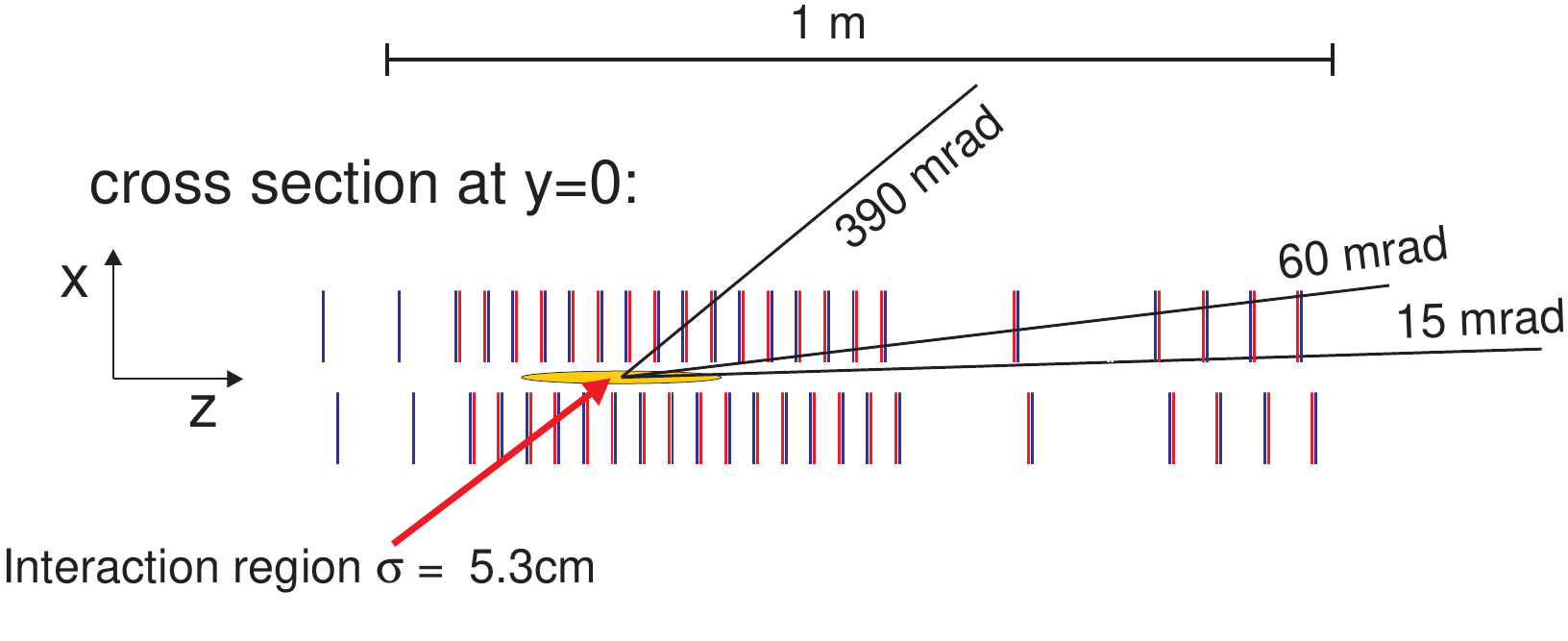}
\caption{
Layout of the LHCb Vertex Locator (VeLo) in the horizontal 
$(x,z)$-plane, with the $z$-axis along the direction of the 
proton beams. 21 sensor planes measure space points, 
the two most backward ($-z$) layers provide only radial 
coordinates of charged particle tracks.}
\label{fig:velo}
\end{figure}

The VeLo has 21 double-layer sensor planes for measuring space points
and two single-layer planes providing only radial track coordinates.
Its layout is shown in Fig.\,\ref{fig:velo}. The angular acceptance is
larger than for the rest of the tracking system and covers also part
of the backward hemisphere. However, being located outside of the
magnetic field, VeLo track segments do not have momentum information.
Furthermore, since at least three planes are required to reconstruct a
track segment, the VeLo is blind in the central region. Charged
particle tracks are reconstructed in the rapidity ranges
$-4<\eta<-1.5$\/ and $1.5<\eta<5$.

With the data recorded until summer 2010 a first precise calibration
of the various subdetector systems was performed. With the
tracking system, for example, impact parameter resolutions for
high-$p_T$ particles around $16\,\mu$m were reached. Tracking
efficiencies above 95\% for charged particles with transverse momenta
above $p_T=200$\,MeV/$c$ were determined by a tag and probe approach
using daughter particles from $K^0_S$\/ and $J/\psi$\/ decays. The
measurements were found to be in good agreement with Monte Carlo
simulations, allowing to set a limit on the systematic uncertainty 
of the tracking efficiency to 3\% for high momentum tracks and 4\% for 
soft particles.

\subsubsection{Measurements of Strangeness Production}
The study of strangeness production is of particular interest for
early measurements since, compared to heavy flavours, the cross
section is large, and $V^0$-decays ($K^0_s,\Lambda,\bar{\Lambda}$) 
have a very clean experimental signature which allows to identify them
unambiguously using only kinematical information even when only a 
coarse calibration of the detector exists. Furthermore, since 
strange quarks are not present as valence quarks in the initial 
state and have a mass in an intermediate range where QCD 
predictions have large uncertainties, they directly probe the 
mechanism of multi-particle production in high energy collisions
where the theory is least well understood.

\begin{figure}[t]
\centering
\includegraphics[width=0.8\textwidth]{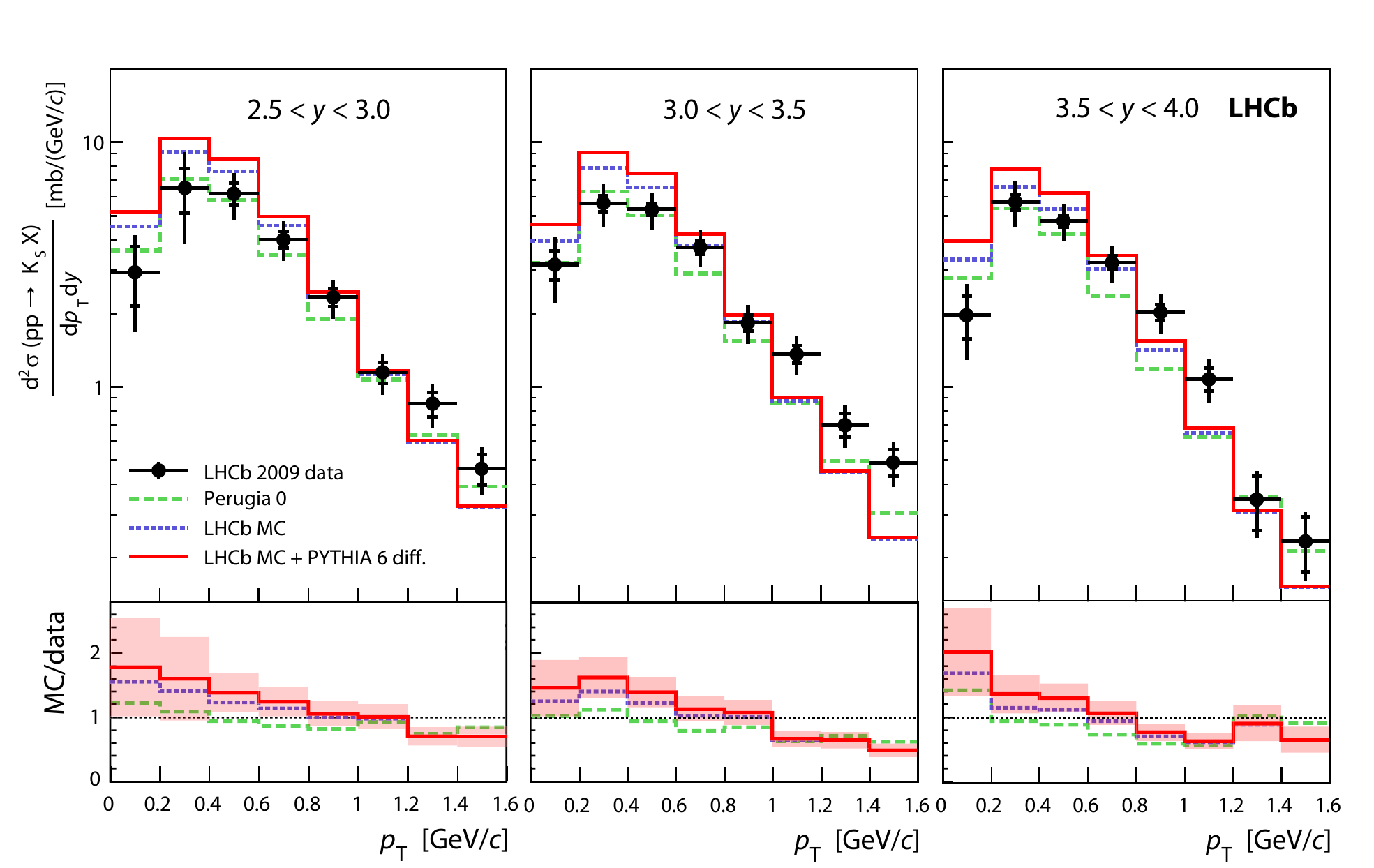}
\caption{
Differential cross-section for $K^0_s$-production 
in $pp$ collisions at $\sqrt{s}=0.9$~TeV as a function 
of transverse momentum $p_{\rm T}$ and rapidity $y$.  
The vertical bars on the data points show the total 
uncertainties, the purely statistical errors are indicated
by the tick marks. The histograms are predictions 
from different settings of the {\sc Pythia} model.  
The lower plots show the MC/data ratios, with the shaded 
bands the experimental uncertainty.}
\label{fig:ksxsec}
\end{figure}

Results from the first cross-section measurements of prompt
$K^0_s$-production in proton-proton collisions at a center-of-mass
energy $\sqrt{s}=0.9$\,TeV \cite{Aaij:2010nx} are shown in
Fig.\,\ref{fig:ksxsec}. Uncorrelated errors, mainly due to 
finite statistics and the modeling of the shape of
the differential cross section in the determination of the correction
factors, range between 6 and 28\,\%. The correlated errors are between
16 and 23\,\% and are dominated by uncertainties in the track finding
efficiencies and the luminosity measurement. Comparing experimental
results and Monte Carlo predictions, one clearly sees that the data
favour a harder $p_T$-spectrum than the models that were considered.
The best description is obtained with the {\tt Perugia0}-tune of
{\sc Pythia}, which does not include diffractive processes.

\subsubsection{Baryon Number Transport and Baryon Suppression}
\begin{figure}[t]
\centering
\centerline{%
\includegraphics[width=0.32\textwidth]{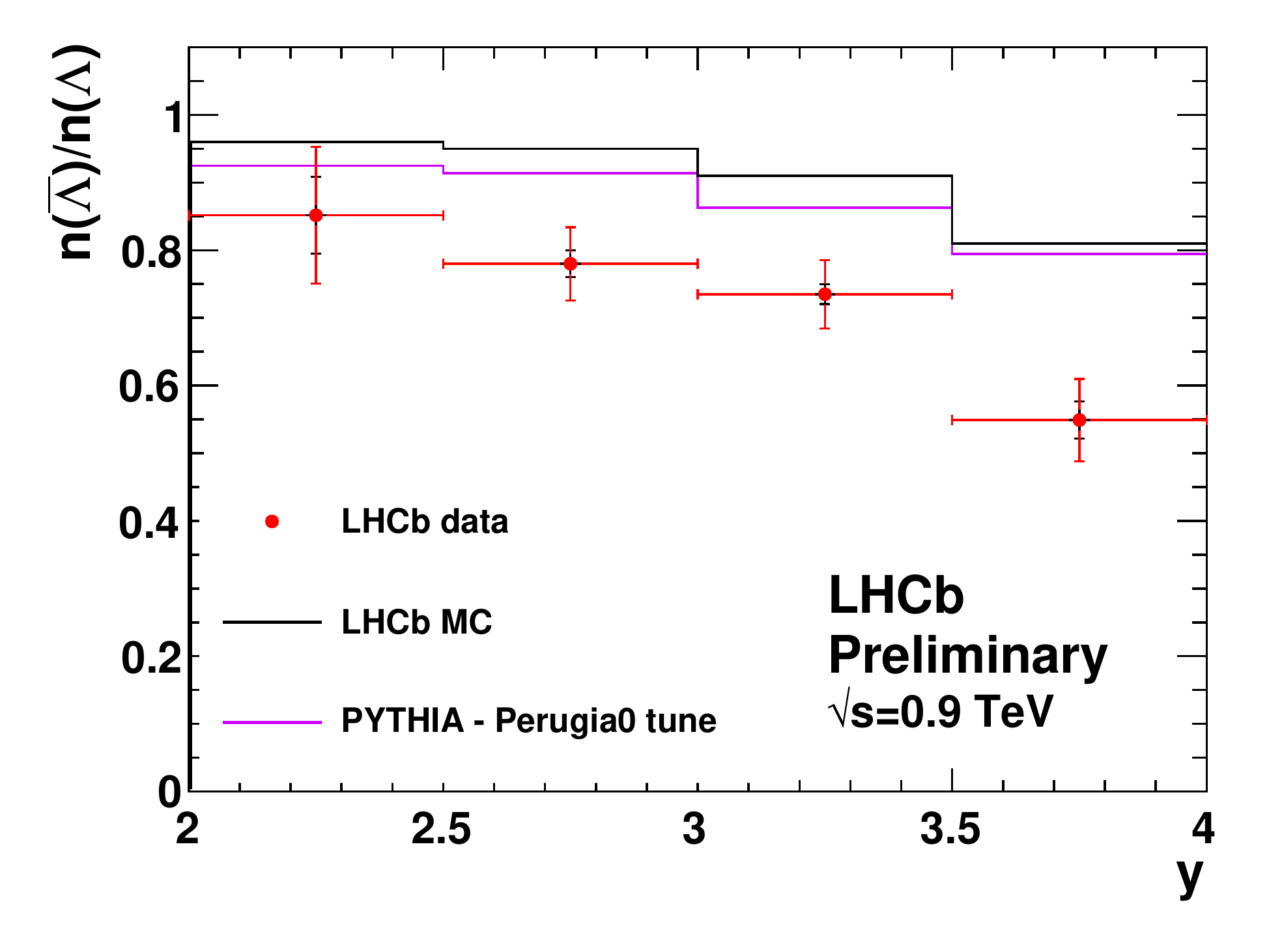}\hfill%
\includegraphics[width=0.32\textwidth]{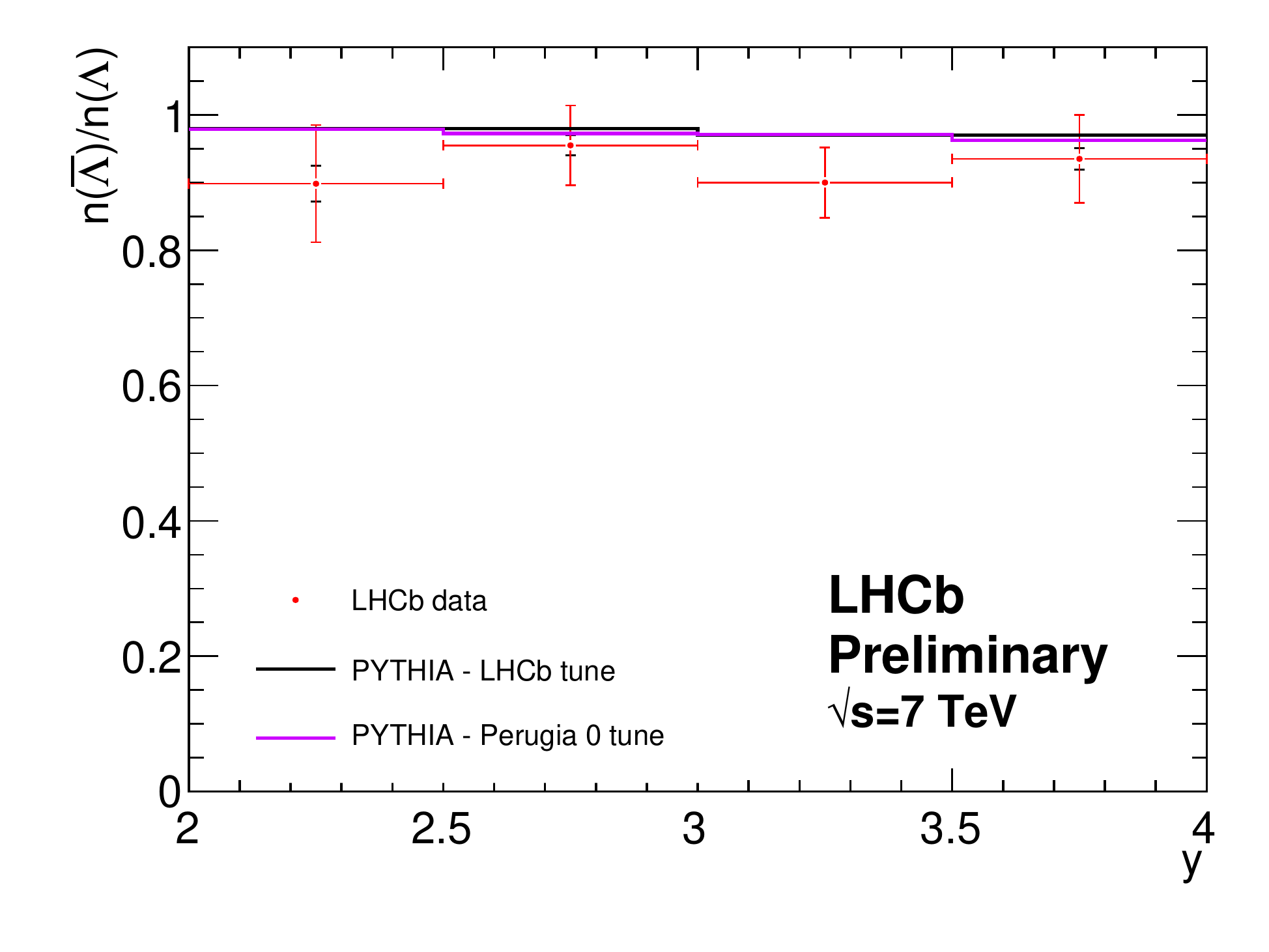}\hfill%
\includegraphics[width=0.33\textwidth]{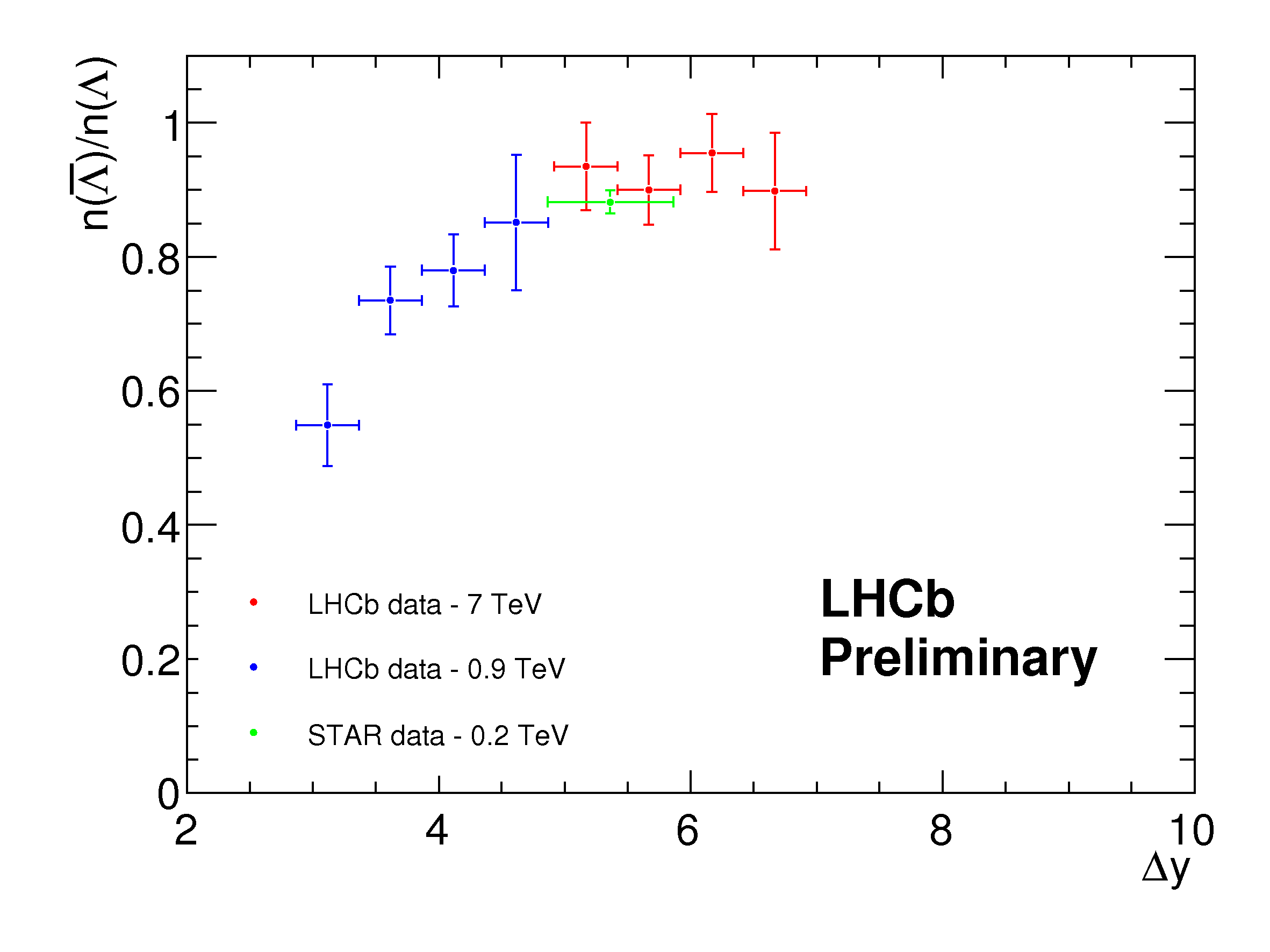}}
  \caption{$\bar{\Lambda}/\Lambda$\/ cross-section ratio as a 
           function of rapidity for center-of-mass energies 
           $\sqrt{s}=0.9$\,TeV and $\sqrt{s}=7$\,TeV compared 
           with expectations from Monte Carlo models (left), and
           as a function of the rapidity distance to the beam 
           particle (right).}
\label{fig:alaratio}
\end{figure}

Other observables probing the dynamics of particle production in high
energy hadron collisions are cross-section ratios, where luminosity
and many systematic uncertainties cancel. Results from the study of
the $\bar{\Lambda}/\Lambda$\/ cross-section ratio for center-of-mass
energies $\sqrt{s}=0.9$\,TeV and $\sqrt{s}=7$\,TeV are shown in
Fig.\ref{fig:alaratio}. While $\Lambda$-baryons have two of their
three valence quarks in common with the proton, all three antiquarks
of the $\bar{\Lambda}$\/ have to be produced in the collision. The
ratio of the production cross-sections thus measures the baryon-number
transport from the beam particles to the final state. In general one
observes that the measured ratio is lower, i.e. larger baryon
number transport, than the expectation from the Monte Carlo models.
The effect becomes stronger in the forward region and when going to
lower center-of-mass energies.

A natural variable to study baryon number transport is the rapidity
difference to the beam. While for $\sqrt{s}=0.9$\,TeV the beam rapidity
is at $y_{b}=6.6$\/ it rises to $y_{b}=8.3$\/ at $\sqrt{s}=7$\,TeV. At
the lower energy the LHCb acceptance thus reaches much closer to the
beam than at $\sqrt{s}=7$\,TeV. Plotting the $\bar{\Lambda}/\Lambda$\/
cross-section ratio, as a function of $\Delta y = y_{b} -y$, also 
shown in Fig.\,\ref{fig:alaratio}, scaling is observed. The LHCb results 
are in good agreement with measurements by the STAR collaboration.
 
\begin{figure}[t]
\centering
\centerline{%
\includegraphics[width=0.4\textwidth]{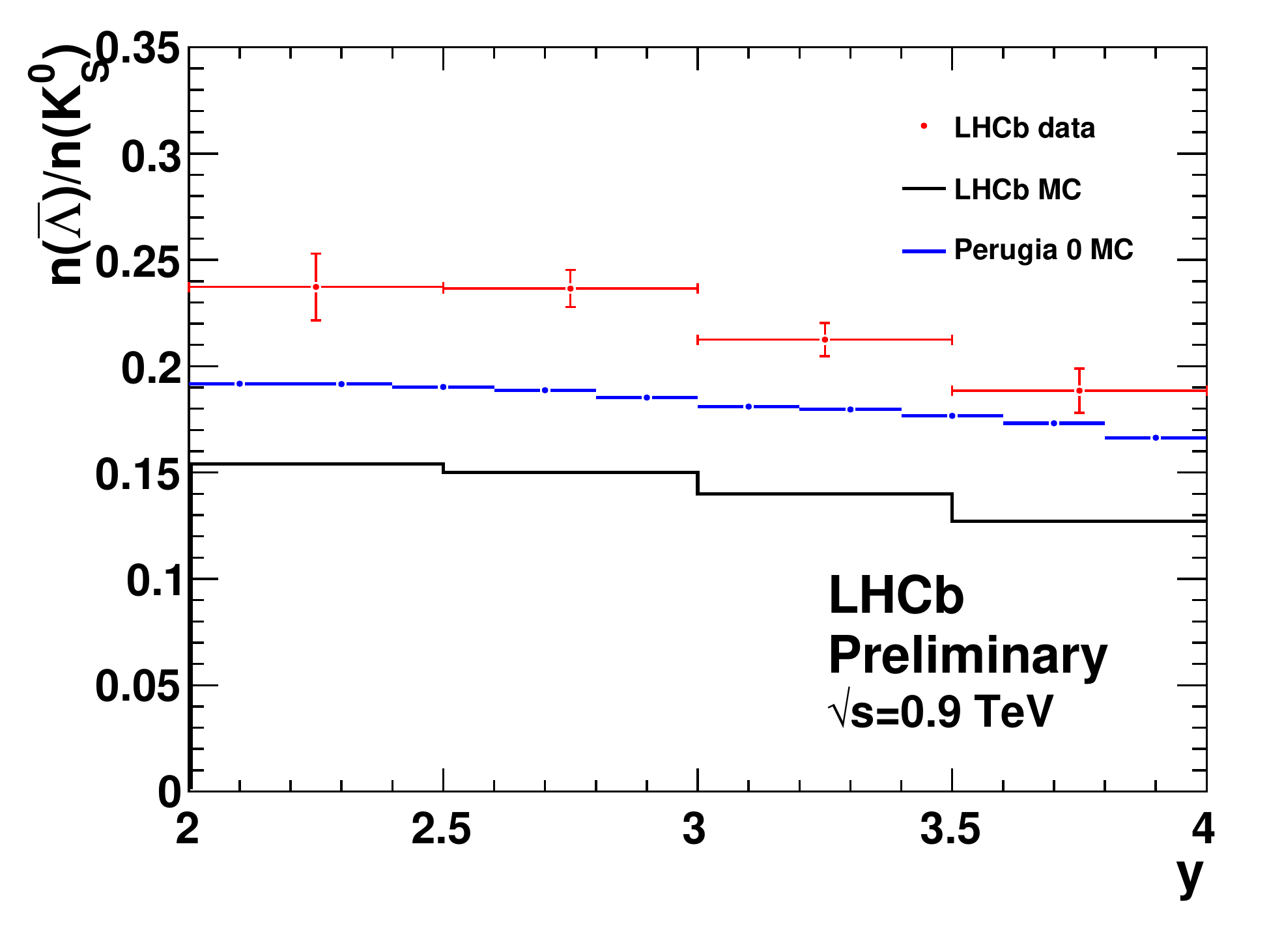}\hspace{10mm}%
\includegraphics[width=0.4\textwidth]{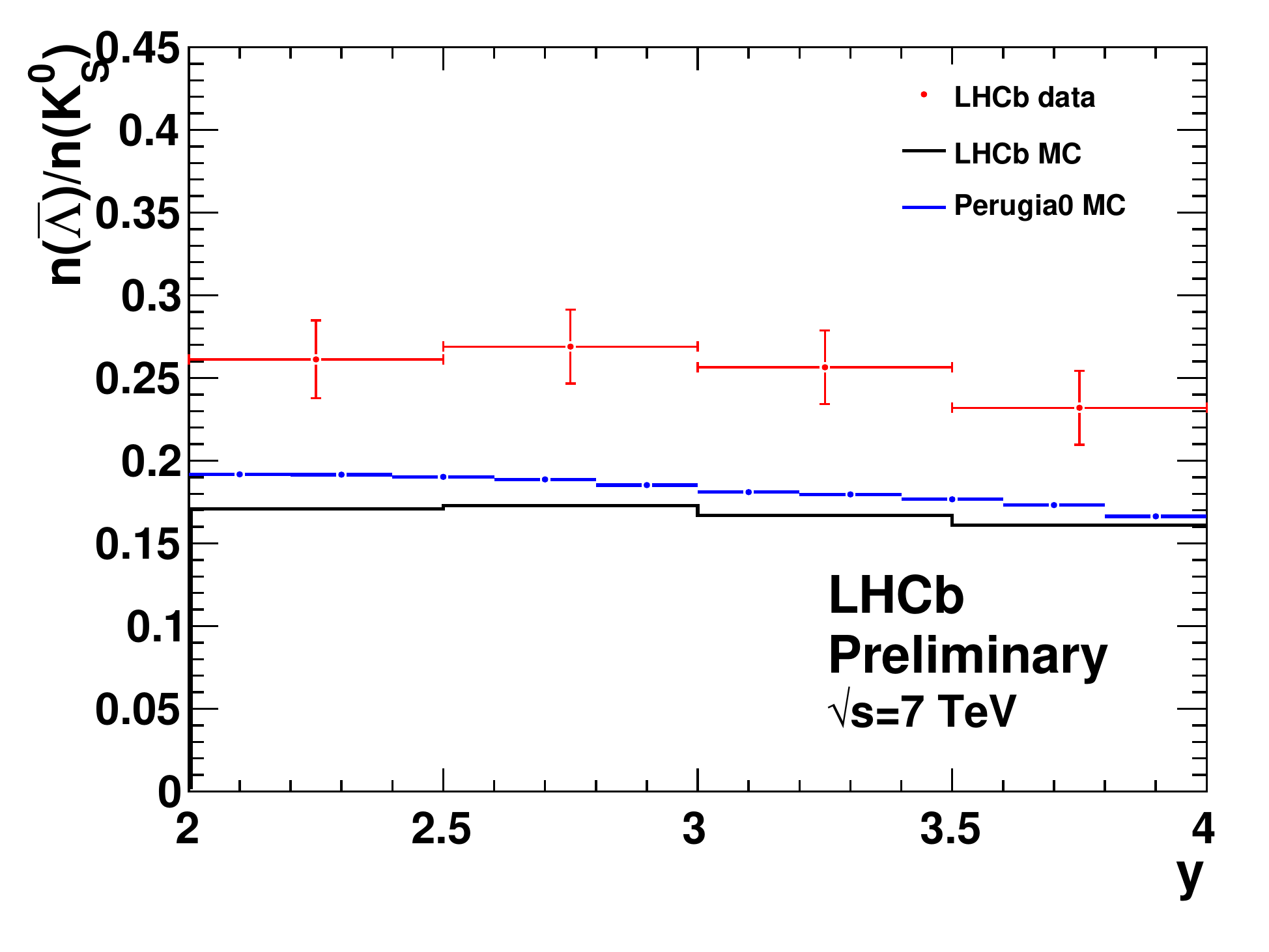}}
  \caption{$\bar{\Lambda}/K^0_S$\/ cross-section ratio as a 
           function of rapidity for center-of-mass energies
           $\sqrt{s}=0.9$\,TeV and $\sqrt{s}=7$\,TeV. The 
           measurements are compared with expectations from 
           Monte Carlo models. (Note the slightly different 
           vertical scales in the two plots.)}
\label{fig:bsupp}
\end{figure}

Another ratio to look at is the $\bar{\Lambda}/K^0_S$\/ cross-section
ratio. Since the $\bar{\Lambda}$\/ has no valence quarks in common
with the initial state protons, this ratio measures the suppression of
baryon- relative to meson-production. Experimental results compared to
Monte Carlo predictions are shown in Fig.\ref{fig:bsupp}. Both data
and Monte Carlo show a slight increase in the ratio when going from
$\sqrt{s}=0.9$\,TeV to $\sqrt{s}=7$\,TeV. This is plausible, since
particle masses and kinematic factors in general should become less
important at higher energies. It is, however, striking that in both
cases the ratio is significantly underestimated by the Monte Carlo
models. Since both the $\bar{\Lambda}$\/ and the $K^0_S$\/ contain a
single (anti)strange valence quark, the discrepancy cannot be
explained by a mismatch in strangeness suppression between data and
Monte Carlo. Instead it probes directly the understanding baryon
formation in the fragmentation process.

\subsubsection{Prospects for the Study of Diffractive Interactions}
\begin{figure}[t]
\centering
\includegraphics[width=0.8\textwidth]{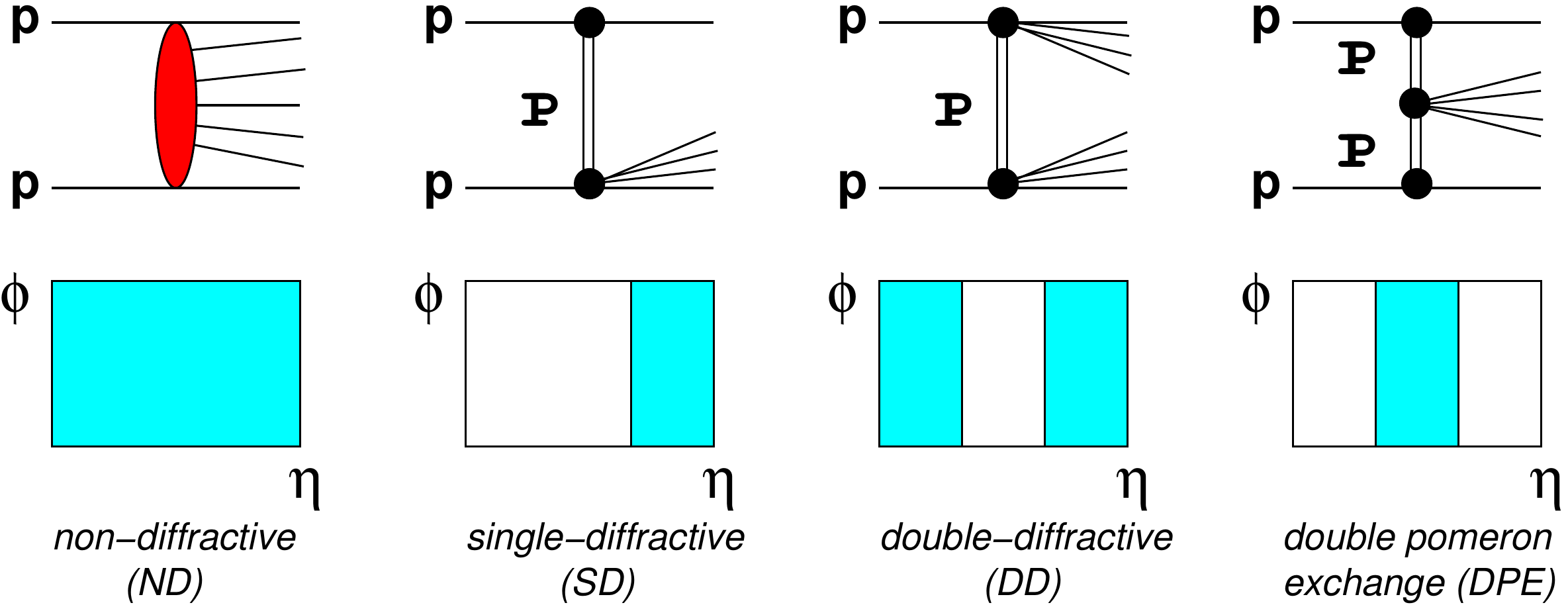}
\caption{
Schematic classification of inelastic proton-proton collisions.
The upper row shows some born-level type diagrams for different 
classes of interactions, the lower row illustrates the angular
range into which particles produced in the collision are emitted.
Note that in the upper row rapidity runs from top to bottom,
while it goes from left to right in the lower row.}
\label{fig:ppint}
\end{figure}

A schematic view of different types of inelastic $pp$-collisions 
contributing to minimum bias interactions is presented in 
Fig.\,\ref{fig:ppint}. Here the basic distinction is colour-octet
(gluon) and colour-singlet (pomeron) exchange between the 
colliding protons. In the language of QCD the pomeron is understood
as a color-singlet two-gluon state. Colour exchange implies that the structure of 
both protons is resolved with the consequence that the colour fields 
stretched between the partons lead to particle production in the full 
rapidity range. In contrast, colour-singlet exchange is 
phenomenologically described by pomerons coupling to the protons as 
a whole. No colour is transferred and the protons can either scatter 
elastically or be excited into a high mass state which then decays 
to produce multi-particle final states. Depending on whether only 
one or both protons are excited these processes are referred to as 
single- or double-diffractive scattering. An example for a higher 
order process involving pomerons is double pomeron exchange, where 
both protons stay intact and the two pomerons interact to form 
a massive central system.

The above classification of $pp$-interactions is most adequate at
small momentum transfers. With increasing momentum transfer,
however, the simple picture breaks down. The two protons are resolved
into an increasing number of partons and the interaction is described
by ladder-diagrams of all possible topologies. Diffractive and
non-diffractive scattering is no longer an unambiguous classification,
and even the notion of colour singlet exchange becomes frame dependent
\ref{sec:gosta}.

Another conceptual problem with Fig.\,\ref{fig:ppint} is that the
upper row represents Born-level type diagrams, i.e. amplitudes, while
the lower one is a pictorial representation of the cross-section. In
many Monte Carlo models, such as e.g. {\sc Pythia}
\cite{Sjostrand:2006za}, the different contributions to the total
cross-section are generated independently and interference terms are
ignored. Furthermore, parameter tuning generally allows to trade e.g.
a larger diffractive component against a smaller non-diffractive
part. It follows that event classification into diffractive and
non-diffractive parts based on a Monte Carlo implementation is to 
some extent arbitrary and always model dependent.
A better approach would be to avoid any such artificial
classifications, and instead perform measurements subject to
experimental cuts which enhance or suppress certain contributions to
the cross section.

\begin{figure}[t]
\centering
\includegraphics[width=0.75\textwidth]{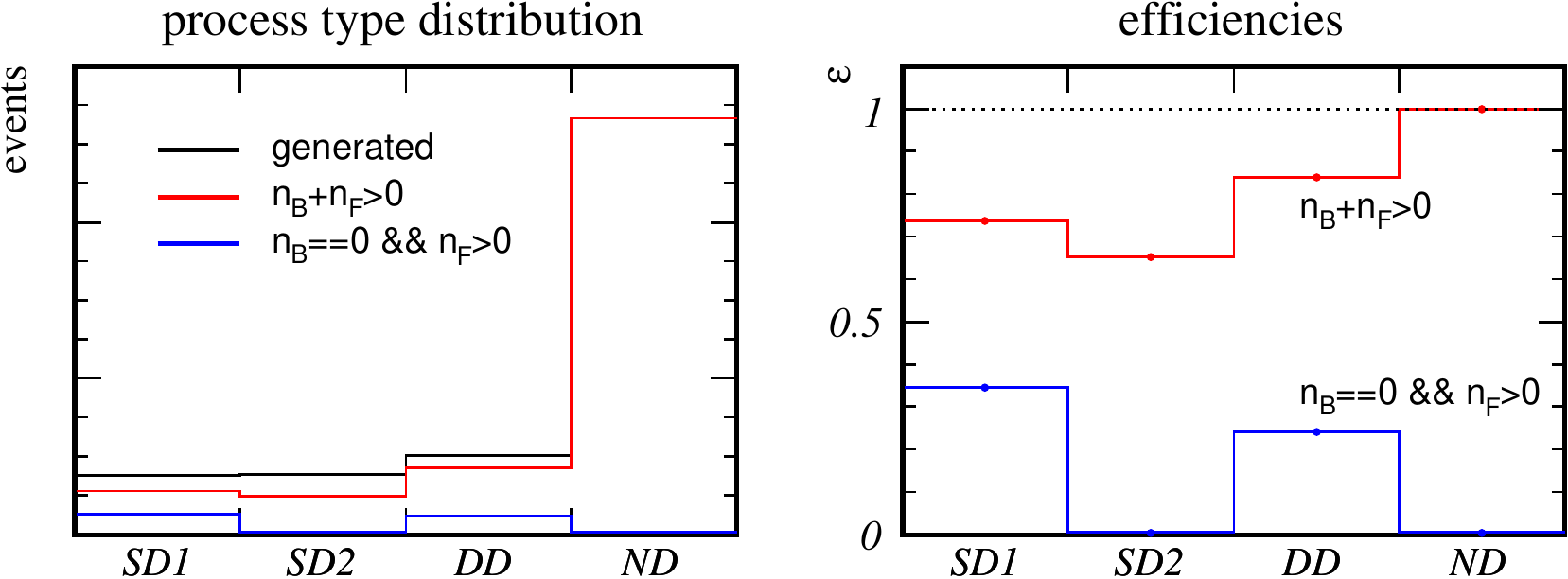}
\caption{
Generator level MC study: Mix of single diffractive, double diffractive
and non-diffractive events generated by {\sc Pythia 8.135}. Here {\tt SD1}
refers to single diffractive scattering where the excited proton travels 
in the direction of the LHCb detector, in {\tt SD2} the decaying heavy 
mass moves into the opposite direction. The left hand plot shows the mix 
of events generated (black), passing the VeLo track-segment trigger(red) and 
the diffraction selection (blue). The  right hand plot shows the selection 
efficiencies for the two cases.}
\label{fig:compeff}
\end{figure} 

\begin{figure}[t]
\centering
\includegraphics[width=0.8\textwidth]{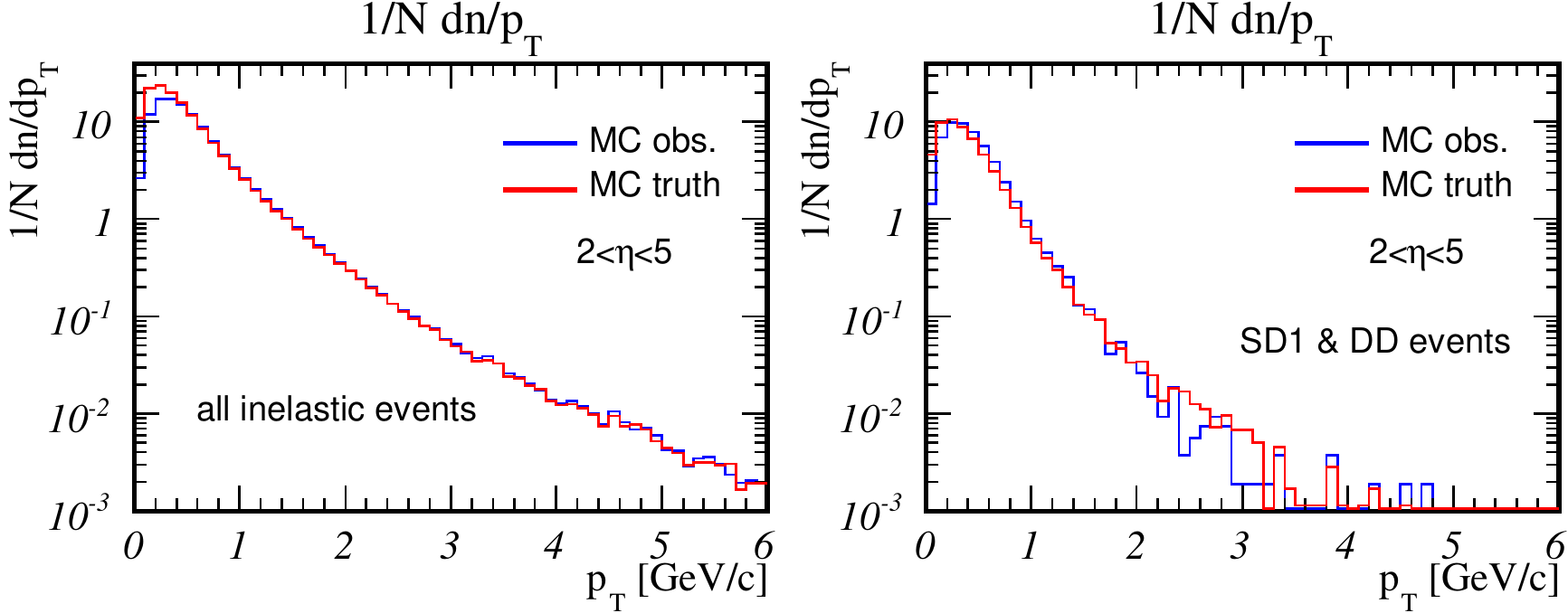}
\caption{
Generator level MC study: comparison of generated and observable
transverse momentum spectra of charged particles in the 
pseudo rapidity range $2<\eta<5$\/ when asking for a single VeLo 
track (left) or, in addition, a rapidity gap (right). The spectra are 
normalized to the number of accepted events. 
The losses at small transverse momenta are due to 
incomplete geometric coverage at low $p_T$\/ and low $\eta$.}
\label{fig:genrec}
\end{figure} 

To study the prospects for measuring the properties of events with
dominantly diffractive contributions, a simple generator level
analysis, ignoring finite resolution and imperfect efficiencies,
of charged particle transverse momentum spectra has been
performed. The study is based on {\sc Pythia 8.135} available from
\cite{Sjostrand:2010zz}. Single proton-proton collisions with a
center-of-mass energy $\sqrt{s}=7$\,TeV were generated with process
selection {\tt pythia.readString("SoftQCD:all=on")}. The VeLo was
simulated with its nominal geometry, and a track was assumed to be
measured if at least three stations were hit. 

The event selection was based on the VeLo track segments only.
Denoting by $n_B$\/ and $n_F$\/ the number of VeLo track segments in
the backward ($\eta<0$) and forward ($\eta>0$) hemispheres, two
selection criteria were studied: (a) $n_B+n_F>0$\/ and (b) $n_B==0
\;\&\&\; n_F>0$. Both criteria ask for at least one track segment in
the VeLo, the second corresponds to the additional requirement of a
rapidity gap $\Delta\eta=2.5$\/ for charged tracks and thus is
expected to enhance the fraction of diffractive events in the sample.
To obtain a quantitative measure for the level of enrichment which can
be achieved, the {\sc Pythia} process type was analyzed for all
events. While evidently giving model dependent estimates for the
fractions of different events, the qualitative picture is expected to
be generic.

For accepted events the transverse momentum spectrum then was
determined using all tracks with a VeLo-segment and within the
acceptance of the tracking system behind the magnet. The latter was
approximated by the requirement $p>2$\,GeV/$c$\/ the pseudo-rapidity
range $2<\eta<5$. Results are shown in Figs.\,\ref{fig:compeff} and
\ref{fig:genrec}. The left hand plot of Fig.\,\ref{fig:compeff} shows
the mix of single diffractive, double diffractive and non-diffractive
events generated by {\sc Pythia 8.135}. One clearly sees that the
requirement of a rapidity gap in the backwards region almost
completely suppresses non-diffractive events while keeping between
$20\%$\/ and $30\%$\/ of diffractive interactions. The comparison of
the transverse momentum spectra in Fig.\,\ref{fig:genrec} shows very
good agreement between generated and observed distribution, i.e. a
robust measurement comparing the fully inclusive transverse momentum
spectra and the spectra in events dominated by diffraction seems
feasible. Measurements of production cross-sections for identified
particles and particle ratios are a natural extension of these
studies.

\subsubsection{Summary and Outlook}
First results from the study of minimum bias events by the LHCb
experiment at center-of-mass energies $\sqrt{s}=0.9$\,TeV and
$\sqrt{s}=7$\,TeV have been presented. Production cross-sections for
$K^0_S$-mesons at $\sqrt{s}=0.9$\,TeV were found to have harder 
transverse momentum spectra than expected from Monte Carlo models. 
The baryon number transport from the
beam particles to the final state was found to scale with rapidity
difference to the beam particles and to be more pronounced that
expected from the currently used models. The models also feature a
stronger baryon suppression in the fragmentation than is observed in
the experiment. Finally, a model independent approach towards the
study of diffractive particle production in minimum bias events has
been presented. Generator level Monte Carlo studies suggest that
asking for a rapidity gap of $\Delta\eta=2.5$\/ in the backward region
of the VeLO allows to select event samples dominated by diffractive
processes which, making use of the excellent tracking and particle
identification capabilities of the LHCb detector, then can be studied
in detail in the pseudo-rapidity range $2<\eta<5$.

\section{Multi-parton Interactions in Event Generators}
\label{sec:mc}
 The description of low $p_T$ hadronic activity used in experimental
 analyses relies on models implemented in Monte Carlo (MC) event
 generators. These generators combine a perturbative description in terms
 of multiple scatterings with phenomenological models for soft,
 non-perturbative physics. Recently, the implementation of MPI effects in
 Monte Carlo models has quickly progressed through an increasing level of
 sophistication and complexity that has deep general implications for the
 LHC physics. In this chapter recent developments within the {\sc Pythia} and
 {\sc Herwig} frameworks are reviewed.

\graphicspath{{corke/}}

\newlength{\captivewidth}
\setlength{\captivewidth}{\textwidth}
\addtolength{\captivewidth}{-10mm}
\newcommand{\captive}[1]{\rule{5mm}{0mm}%
\begin{minipage}{\captivewidth}%
\caption[small]{#1}\end{minipage}}

\newcommand{\mrm}[1]{\mathrm{#1}}
\newcommand{\mbf}[1]{\mathbf{#1}}
\newcommand{\mtt}[1]{\mathtt{#1}}
\newcommand{\tsc}[1]{\textsc{#1}}
\newcommand{\tbf}[1]{\textbf{#1}}
\newcommand{\ttt}[1]{\texttt{#1}}
\newcommand{\br}[1]{\overline{#1}}
\newlength{\tmplen}
\newcommand{\clab}[1]{\tiny\settowidth{\tmplen}{\scriptsize#1}%
\colorbox{white}{\textcolor{white}{#1}}\hspace*{-1.27\tmplen}\scriptsize#1}

\def\lsim{\mathrel{\rlap{\lower4pt\hbox{\hskip1pt$\sim$}}
    \raise1pt\hbox{$<$}}}                
\def\gsim{\mathrel{\rlap{\lower4pt\hbox{\hskip1pt$\sim$}}
    \raise1pt\hbox{$>$}}}                
\newcommand{\half}{$\frac{1}{2}$}        
 
\newcommand{\alphas}{\alpha_{\mathrm{s}}}
\newcommand{\alphaem}{\alpha_{\mathrm{em}}}
\newcommand{\pT}{\ensuremath{p_{\perp}}}
\newcommand{\pTi}{\ensuremath{p_{\perp i}}}
\newcommand{\pTs}{\ensuremath{p^2_{\perp}}}
\newcommand{\kT}{\ensuremath{k_{\perp}}}
\newcommand{\pTmin}{p_{\perp\mathrm{min}}}
\newcommand{\pTmax}{p_{\perp\mathrm{max}}} 
\newcommand{\pTp}{{p'}_{\perp}} 
\newcommand{\pTo}{p_{\perp 0}}
\newcommand{\pTevol}{p_{\perp\mathrm{evol}}}
\newcommand{\pTPOW}{p_{\perp\mathrm{POWHEG}}}
\newcommand{\ECM}{E_{\mathrm{CM}}}
\newcommand{\mmin}{\mathrm{min}}
\newcommand{\mmax}{\mathrm{max}}
\newcommand{\MeV}{\ensuremath{\!\ \mathrm{MeV}}}
\newcommand{\GeV}{\ensuremath{\!\ \mathrm{GeV}}}
\newcommand{\TeV}{\ensuremath{\!\ \mathrm{TeV}}}
\newcommand{\mb}{\ensuremath{\!\ \mathrm{mb}}}
\renewcommand{\rem}{\ensuremath{\mathrm{rem}}}
 
\renewcommand{\b}{\mathrm{b}}
\renewcommand{\c}{\mathrm{c}}
\renewcommand{\d}{\mathrm{d}}
\newcommand{\e}{\mathrm{e}}
\newcommand{\f}{\mathrm{f}}
\newcommand{\g}{\mathrm{g}}
\renewcommand{\j}{\mathrm{j}}
\newcommand{\J}{\mathrm{J}}
\newcommand{\hrm}{\mathrm{h}}
\newcommand{\n}{\mathrm{n}}
\newcommand{\p}{\mathrm{p}}
\newcommand{\q}{\mathrm{q}}
\newcommand{\s}{\mathrm{s}}
\renewcommand{\t}{\mathrm{t}}
\renewcommand{\u}{\mathrm{u}}
\newcommand{\A}{\mathrm{A}}
\newcommand{\D}{\mathrm{D}}
\renewcommand{\H}{\mathrm{H}}
\newcommand{\K}{\mathrm{K}}
\newcommand{\Q}{\mathrm{Q}}
\newcommand{\W}{\mathrm{W}}
\newcommand{\Z}{\mathrm{Z}}
\renewcommand{\bbar}{\overline{\mathrm{b}}}
\newcommand{\cbar}{\overline{\mathrm{c}}}
\newcommand{\dbar}{\overline{\mathrm{d}}}
\newcommand{\fbar}{\overline{\mathrm{f}}}
\newcommand{\nbar}{\overline{\mathrm{n}}}
\newcommand{\pbar}{\overline{\mathrm{p}}}
\newcommand{\qbar}{\overline{\mathrm{q}}}
\newcommand{\sbar}{\overline{\mathrm{s}}}
\newcommand{\tbar}{\overline{\mathrm{t}}}
\newcommand{\ubar}{\overline{\mathrm{u}}}
\newcommand{\Bbar}{\overline{\mathrm{B}}}
\newcommand{\Dbar}{\overline{\mathrm{D}}}
\newcommand{\Qbar}{\overline{\mathrm{Q}}}
\newcommand{\qval}{\ensuremath{\q_{\mrm{v}}}}
\newcommand{\qsea}{\ensuremath{\q_{\mrm{s}}}}
\newcommand{\qcmp}{\ensuremath{\q_{\mrm{c}}}}
\newcommand{\val}{\ensuremath{{\mrm{v}}}}
\newcommand{\sea}{\ensuremath{{\mrm{s}}}}
\newcommand{\cmp}{\ensuremath{{\mrm{c}}}}
 
\newcommand{\sg}{\tilde{\mathrm{g}}}
\newcommand{\sq}{\tilde{\mathrm{q}}}
\newcommand{\sqd}{\tilde{\mathrm{d}}}
\newcommand{\squ}{\tilde{\mathrm{u}}}
\newcommand{\sqc}{\tilde{\mathrm{c}}}
\newcommand{\sqs}{\tilde{\mathrm{s}}}
\newcommand{\st}{\tilde{\mathrm{t}}}
\newcommand{\schi}{\tilde{\chi}}

\newenvironment{Itemize}{\begin{list}{$\bullet$}%
{\setlength{\topsep}{0.2mm}\setlength{\partopsep}{0.2mm}%
\setlength{\itemsep}{0.2mm}\setlength{\parsep}{0.2mm}}}%
{\end{list}}
\newcounter{enumct}
\newenvironment{Enumerate}{\begin{list}{\arabic{enumct}.}%
{\usecounter{enumct}\setlength{\topsep}{0.2mm}%
\setlength{\partopsep}{0.2mm}\setlength{\itemsep}{0.2mm}%
\setlength{\parsep}{0.2mm}}}{\end{list}}

\contribution{Multiparton Interactions in {Pythia 8}}
{Contributing author: R. Corke}
\label{pythia}




\subsubsection{MPI in \tsc{Pythia 8}}
\label{sec:MPI}
The original MPI model, first introduced in previous versions of
\tsc{Pythia}, featured $\pT$ ordering, perturbative
QCD cross sections dampened in the $\pT \to 0$ limit, and a variable impact
parameter formalism \cite{Sjostrand:1987su}. These features remain in the
MPI framework of \tsc{Pythia 8} \cite{Sjostrand:2007gs} and have been
extended to give a wide range of possible underlying-event processes,
including all $2 \to 2$ QCD processes, prompt photon production and others
\cite{Sjostrand:2004pf}. This newer model was developed after the
introduction of transverse-momentum-ordered parton showers, opening the way
to have a common $\pT$ evolution scale for initial-state radiation (ISR),
final-state radiation (FSR) and MPI \cite{Sjostrand:2004ef}. This common
evolution is most important for ISR and MPI, which both directly compete
for momentum from the beams.

With such an interleaving, the probability for the $i^{th}$ interaction or
shower branching to take place at $\pT = \pTi$ is given by the combined
evolution equation
\begin{eqnarray}
\frac{\d \mathcal{P}}{\d \pT} &=& 
\left( \frac{\vphantom{\left(\right)} \d\mathcal{P}_{\mrm{MPI}}}{\d \pT}  + 
\sum   \frac{\vphantom{\left(\right)} \d\mathcal{P}_{\mrm{ISR}}}{\d \pT}  +
\sum   \frac{\vphantom{\left(\right)} \d\mathcal{P}_{\mrm{FSR}}}{\d \pT}
\right) \nonumber \\ 
 & \times & \exp \left( - \int_{\pT}^{p_{\perp i - 1}}
\left( \frac{\vphantom{\left(\right)} \d\mathcal{P}_{\mrm{MPI}}}{\d \pT'}  + 
\sum   \frac{\vphantom{\left(\right)} \d\mathcal{P}_{\mrm{ISR}}}{\d \pT'}  +
\sum   \frac{\vphantom{\left(\right)} \d\mathcal{P}_{\mrm{FSR}}}{\d \pT'} 
\right) \d \pT' \right) ~,
\label{eq:combinedevol}
\end{eqnarray}
where contributions from MPI, ISR and FSR are unitarised by a Sudakov-like
factor. The sums for ISR and FSR run over all initiator and final-state
partons respectively, including those for MPI subsystems, giving full
showers from these interactions.

Focusing on just the MPI contribution, the probability for an interaction
is given by
\begin{equation}
 \frac{\d \mathcal{P}_{\mrm{MPI}}}{\d \pT} =
 \frac{1}{\sigma_{\mrm{ND}}} \frac{\d \sigma}{\d \pT} \;
 \exp \left( - \int_{\pT}^{p_{\perp i-1}}
  \frac{1}{\sigma_{\mrm{ND}}} \frac{\d \sigma}{\d \pT'} \d \pT' \right)
 ~,
 \label{eqn:MPIevol}
\end{equation}
where $\d \sigma / \d \pT$ is given by the convolution of PDF factors with
the partonic QCD $2 \to 2$ cross section. This cross section is dominated
by $t$-channel gluon exchange, and diverges roughly as $\d \pT^2 / \pT^4$.
To avoid this divergence, the idea of colour screening is introduced into
the model.  The concept of a perturbative cross section is based on the
assumption of free incoming states, which is not the case when partons are
confined in colour-singlet hadrons. One therefore expects a colour charge
to be screened by the presence of nearby anti-charges; that is, if the
typical charge separation is $d$, gluons with a transverse wavelength $\sim
1 / \pT > d$ are no longer able to resolve charges individually, leading to
a reduced effective coupling. This is introduced by regularising the
interaction cross section according to
\begin{equation}
 \frac{\d \hat{\sigma}}{\d \pT^2} \propto
 \frac{\alpha_S^2(\pT^2)}{\pT^4} \rightarrow
 \frac{\alpha_S^2({\pT^2}_0 + \pT^2)}{({\pT^2}_0 + \pT^2)^2} ~,
 \label{eqn:pt0}
\end{equation}
where $\pTo$ (related to $1 / d$ above) is now a free parameter in the
model. This parameter is expected to scale with energy, and the ansatz
is that it does so in a manner similar to the total cross section, an
effective power related to the Pomeron intercept. The form of the scaling
is given by
\begin{equation}
 \pTo(E_{\mathrm{CM}}) = p_{\perp0}^{\mathrm{ref}} \times \left(
 \frac{E_{\mathrm{CM}}}{E_{\mathrm{CM}}^{\mathrm{ref}}}\right)%
 ^{E_{\mathrm{CM}}^{\mathrm{pow}}} ~,
\end{equation}
where a reference $\pTo$ is given at some reference energy, and scaled
according to $E_{\mathrm{CM}}^{\mathrm{pow}}$.

Up to this point, all parton-parton interactions have been assumed to be
independent, such that the probability to have $n$ interactions in an event,
$\mathcal{P}_n$, is given by Poissonian statistics. This picture is now
changed, first by requiring that there is at least one interaction, such
that there is a physical event, and second by including an impact parameter,
$b$. In general, the amount of MPI activity in an event will be directly
related to the time-integrated overlap of the incoming hadrons during
collision, given by
\begin{equation}
 \mathcal{O}(b) = \int \d t \int \d^3 x \;
 \rho(x, y, z) \; \rho(x + b, y, z + t) ~,
\end{equation}
after a suitable scale transformation to compensate for the boosted nature
of the incoming hadrons. Different matter distributions are available,
including a single Gaussian, double Gaussian and an intermediate overlap
function. While requiring at least one interaction results in
$\mathcal{P}_n$ being narrower than Poissonian, when the impact parameter
dependence is added, the overall effect is that $\mathcal{P}_n$ is broader
than Poissonian.

When the $\pT$ evolution has come to an end, colour reconnection is
performed. It has been noted, especially by Rick Field \cite{Field:2002vt},
that a large amount of colour reconnection is necessary to correctly
describe data, such as the mean $\pT$ as a function of charged multiplicity
in minimum-bias events. In \tsc{Pythia 8}, this is performed by giving each
MPI subsystem a probability to reconnect with a harder system
\begin{equation}
 \mathcal{P} = \frac{{\pT}_{\mrm{rec}}^2}{({\pT}_{\mrm{rec}}^2 + \pT^2)},
 ~~~~~~~~~~
 {\pT}_{\mrm{rec}} = R * \pTo,
 \label{eqn:crec}
\end{equation}
where $R$ is a user-tunable parameter and ${\pT}_{0}$
is the same parameter as in eq.~(\ref{eqn:pt0}). The idea of colour
reconnection can be motivated by noting that MPI leads to many colour strings
that will overlap in physical space. Moving from the limit of
$N_C \rightarrow \infty$ to $N_C = 3$, it is perhaps not unreasonable to
consider these strings to be connected differently due to a coincidence of
colour, so as to reduce the total string length and thereby the potential
energy. With the above probability for reconnection, it is easier to
reconnect low $\pT$ systems, which can be viewed as them having a larger
spatial extent such that they are more likely to overlap with other colour
strings. Currently, however, given the lack of a firm theoretical basis,
the need for colour reconnection has only been established within the
context of specific models.

\subsubsection{Rescattering}
\label{sec:rescattering}
\begin{figure}
 \begin{center}
\includegraphics[angle=0, width=0.35 \textwidth]{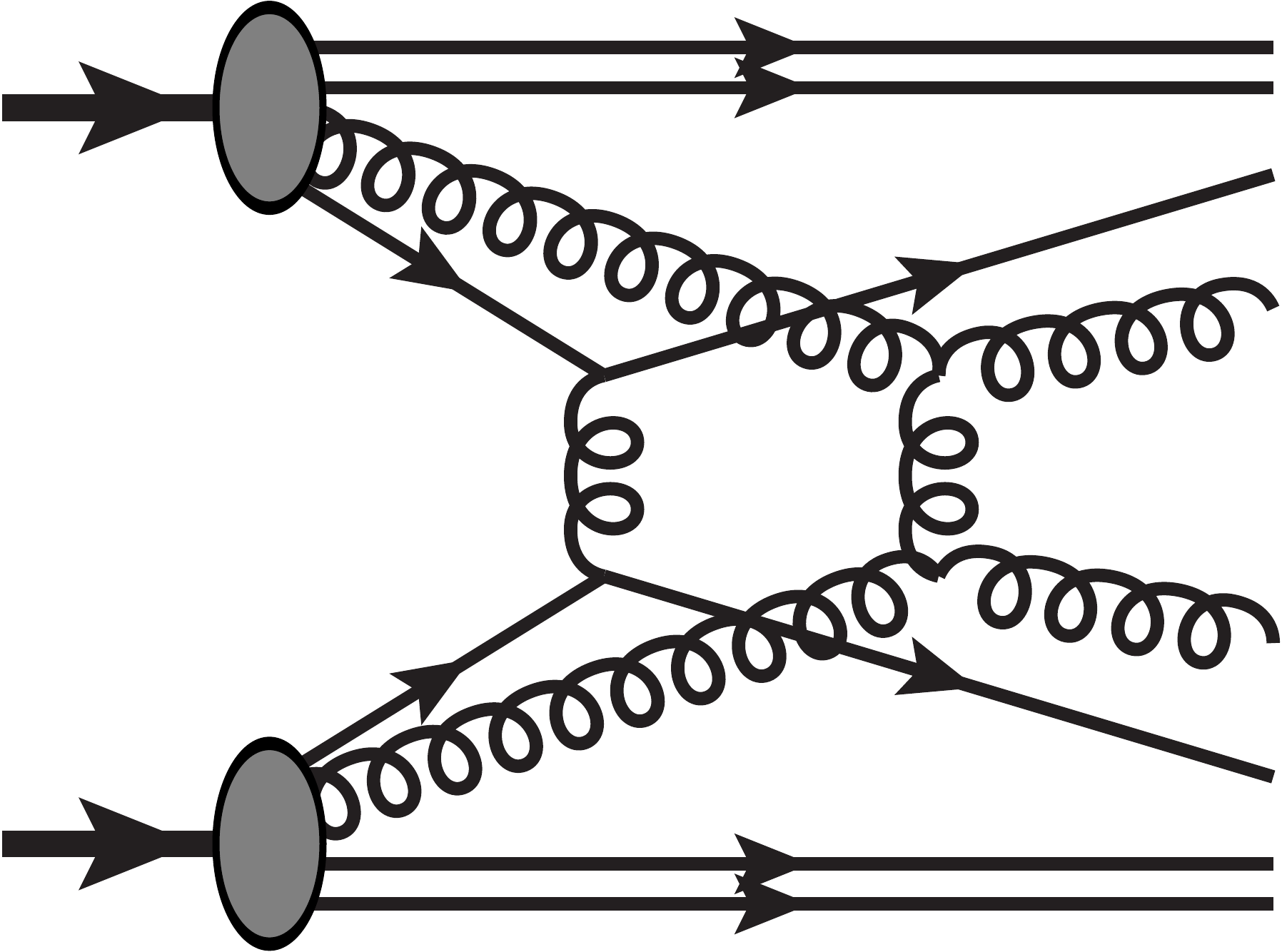}\hspace*{10mm}
\includegraphics[angle=0, width=0.35 \textwidth]{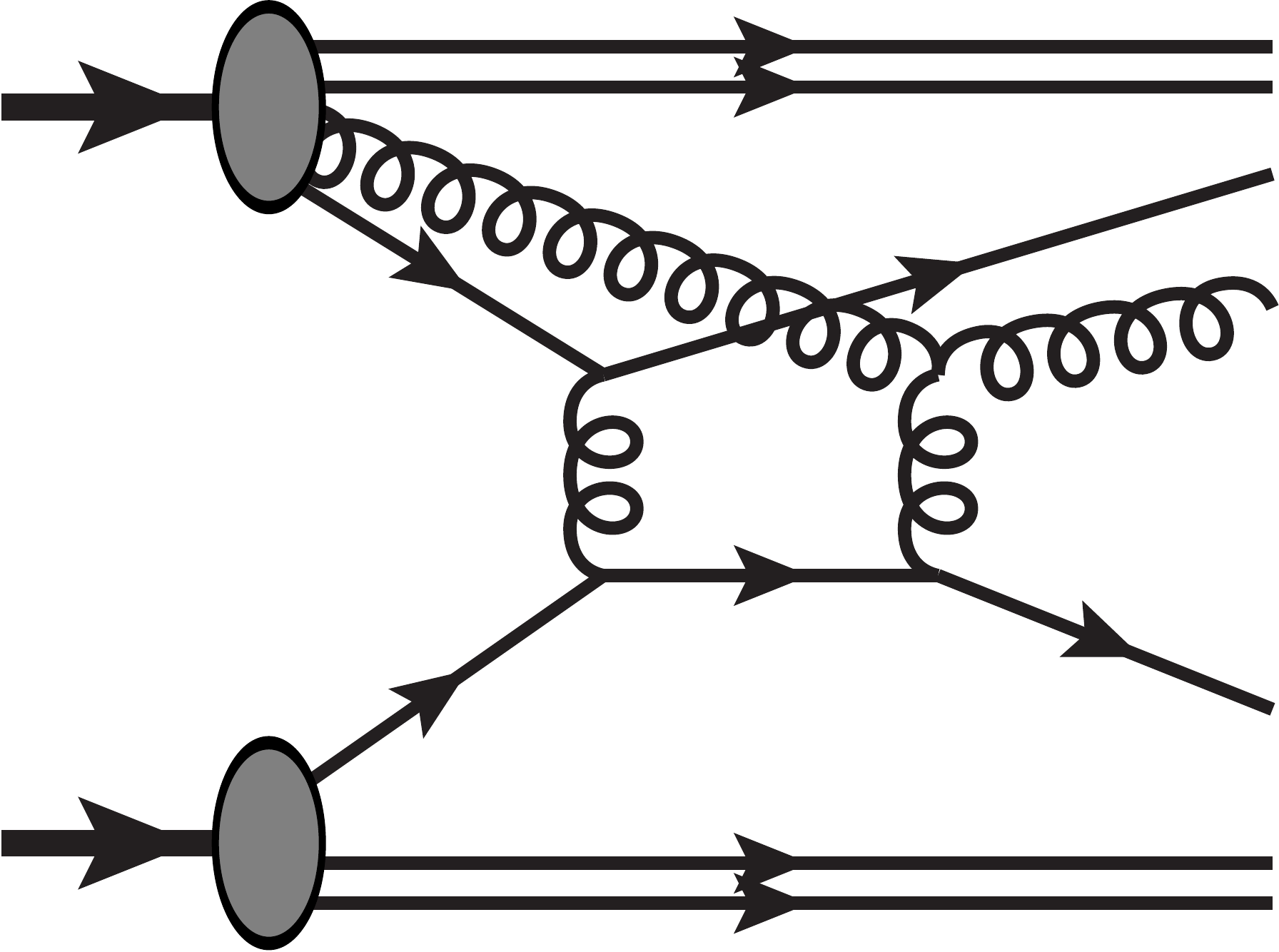}
\end{center}
\hspace{50mm}(a)
\hspace{60mm}(b)
\caption{(a) Two $2 \rightarrow 2$ scatterings and (b) a $2 \rightarrow 2$
scattering followed by a rescattering}
\label{fig:res}
\end{figure}

A process with a rescattering occurs when an outgoing state from one
scattering is allowed to become the incoming state of another.
This is illustrated schematically in Fig.~\ref{fig:res}, where (a) shows
two independent $2 \to 2$ processes while (b) shows a rescattering
process. An estimate for the size of such rescattering effects is given by
Paver and Treleani \cite{Paver:1983hi}. Although rescattering is expected
to be a small effect when compared to independent $2 \to 2$ processes, it
should be allowed to take place. It would show up in the collective
effects of MPI, manifesting itself as changes to multiplicity, $\pT$
and other distributions, although after a retuning of $\pTo$ and other
model parameters, it is likely that its impact is significantly reduced.
A full implementation of rescattering has been made, allowing the
generation of fully hadronic final states \cite{Corke:2009tk}, and an
outline is given below.

The starting point for the implementation of rescattering is the typical
case of small-angle $t$-channel gluon scattering. In this case, the
combination of a scattered parton and one of the hadron remnants will
closely match one of the original incoming hadrons, and the PDF can then be
written as
\begin{equation}
 f(x,Q^2) \rightarrow
 f_{rescaled}(x,Q^2) + \sum_n \delta(x - x_n) ~.
\end{equation}
Here, each time a scattering occurs, one parton becomes fixed at a specific
$x_n$ value, while the remainder is still a continuous probability
distribution, although rescaled to take into account the momentum
taken from it. In this way, the original disjoint $2 \to 2$ MPI are
supplemented by single rescatterings, where one parton is taken from the
rescaled PDF and the other is a delta function, and double rescatterings,
where both partons are delta functions.

Of course, in general, it is not possible to uniquely identify a scattered
parton with one of the hadron remnants, so an approximate prescription must
be used. In particular, we have studied a rapidity based scheme,
where at one extreme, partons with rapidity $y > 0$ belong to beam A and
those with $y < 0$ to beam B (``step'') and at the other, all partons
belong to both beams simultaneously (``sim''). A natural suppression in the
amount of single rescattering for the simultaneous case means that results
do not differ greatly compared to the step prescription. The amount of
rescattering is also dependent on the amount of underlying activity per
event, as the more branchings and scatterings in an event, the more partons
that are available to rescatter. To study this effect, two different tunes
have been studied. These are labelled ``old'' and ``new'', where the
primary difference is a reduced amount of MPI activity in the new tune.

\begin{figure}
\begin{center}
\includegraphics[clip=true,trim=20 0 0 0,angle=270,scale=0.7]%
                {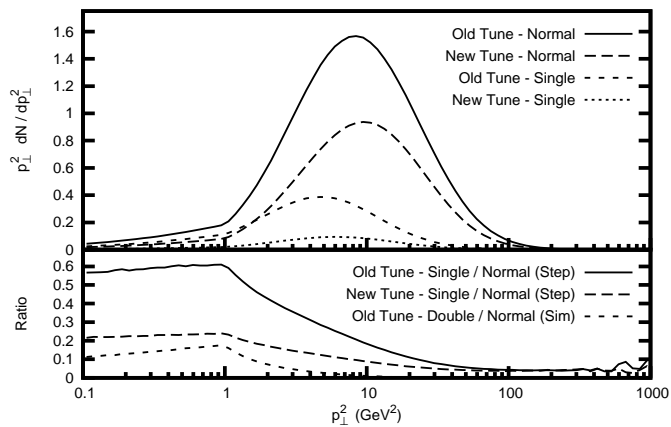}
\end{center}
\caption{$\pT$ distribution of normal MPI and single rescatterings ($\p\p$,
$\sqrt{s} = 14\TeV$, old and new tunes). Double rescattering with the
simultaneous prescription and the old tune is also shown in the ratio plot
\label{fig:pTres}}
\end{figure}

In Fig.~\ref{fig:pTres}, the $\pT$ distribution of single rescatterings is
shown compared to normal MPI ($\p\p$, $\sqrt{s} = 14\TeV$) for the two
tunes. In the ratio plot, double rescattering with the simultaneous
prescription and the old tune is also shown (with these settings, the
effect is largest). As the $\pT$ evolution progresses downwards, more and
more partons become available to rescatter and the rate grows. As expected,
however, the rate of single rescattering is small compared to normal MPI.
As an indicator of the effect of energy on the growth of rescattering,
Tab.~\ref{tab:noRes} shows the average number of scatterings and
rescatterings for different types of event at Tevatron and LHC energies
(step option only, old and new tunes). Double rescattering is always a very
small effect and was neglected in the subsequent hadron-level studies.

\renewcommand{\arraystretch}{1.15}
\begin{table}
\begin{center}
\begin{tabular}{c|l|c|c|c|c|}
\cline{3-6}
\multicolumn{2}{c|}{}
& \multicolumn{2}{|c|}{\textbf{Tevatron}}
& \multicolumn{2}{|c|}{\textbf{LHC}} \\
\cline{3-6}
\multicolumn{2}{c|}{}
& \textbf{Min Bias} & \textbf{QCD Jets} & \textbf{Min Bias}
& \textbf{QCD Jets} \\
\cline{2-6}
\multirow{3}{*}{\textbf{Old}\quad}
& \textbf{Scatterings}          & 2.81  & 5.09  & 5.19  & 12.19\phantom{0} \\
& \textbf{Single rescatterings} & 0.41  & 1.32  & 1.03  & 4.10             \\
& \textbf{Double rescatterings} & 0.01  & 0.04  & 0.03  & 0.15             \\
\cline{2-6}
\multirow{3}{*}{\textbf{New}\quad}
& \textbf{Scatterings}          & 2.50  & 3.79  & 3.40  & 5.68  \\
& \textbf{Single rescatterings} & 0.24  & 0.60  & 0.25  & 0.66  \\
& \textbf{Double rescatterings} & 0.00  & 0.01  & 0.00  & 0.01  \\
\cline{2-6}
\end{tabular}
\end{center}
\caption{Average number of scatterings, single rescatterings and double
rescatterings in minimum bias and QCD jet events at
Tevatron ($\p\pbar$, $\sqrt{s} = 1.96 \TeV$,
QCD jet $\hat{p}_{\perp \mrm{min}} = 20 \GeV$)
and LHC ($\p\p$, $\sqrt{s} = 14.0 \TeV$,
QCD jet $\hat{p}_{\perp \mrm{min}} = 50 \GeV$)
energies for both the old and new tunes
\label{tab:noRes}
}
\end{table}
\renewcommand{\arraystretch}{1}

With the full framework implemented, a range of different observables were
studied to look for definitive signatures of rescattering, including
exclusive jet rates, effects on the amount of colour reconnection required
to match data, enhancement in the $\pT$ spectra of final-state hadrons and
$\Delta R$/$\Delta \phi$ distributions. Unfortunately we were unable to
find any ``smoking-gun'' signatures. One possible approach to further
study where rescattering plays a role would be to tune the generator, both
with and without rescattering, and to examine differences in the overall
fit.

\subsubsection{Tuning prospects}
\label{sec:tuning}
In this section, tunes of the generator are briefly presented, first
for Tevatron data only, and then for an early set of LHC data
\cite{Corke:2010yf}. Parameters relating to final-state
showers and hadronisation have been left unchanged, having been previously
tuned to LEP data. It should be stressed that the parameter space used to
make these tunes is somewhat limited, and the tunes themselves have been
made ``by hand'' following the principles outlined in \cite{Corke:2010yf}.

\subsubsection*{Tevatron data}
Two tunes to Tevatron data have been produced, Tune 2C using the CTEQ6L1
PDF set and Tune 2M using MRST LO** \cite{Pumplin:2002vw,Sherstnev:2007nd}. The MRST LO** PDF set has a relaxed momentum sum rule
such that it contains more momentum that CTEQ6L1, leading to lower
$\alphas$ and higher $\pTo$ parameters. For both tunes, the matter
distribution uses an overlap function, intermediate between the single
Gaussian and the default double Gaussian settings. The balance between ISR
and MPI activity is primarily tuned by comparisons to $\pT(\Z^0)$ and
jet-jet azimuthal angle distributions, both of which are driven strongly by
ISR.

\begin{figure}
\begin{center}
\includegraphics[scale=0.6]{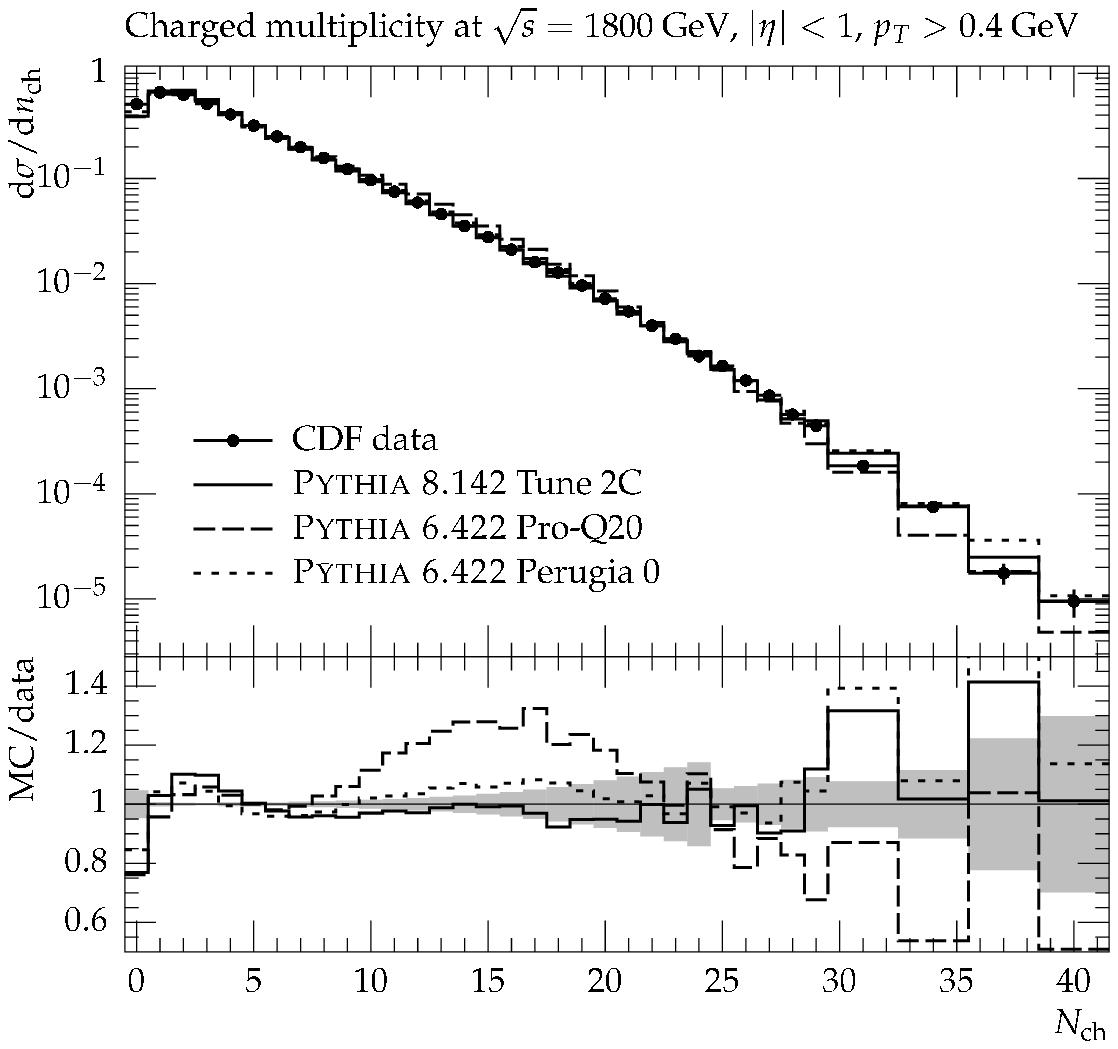}\hspace{3ex}
\includegraphics[scale=0.6]{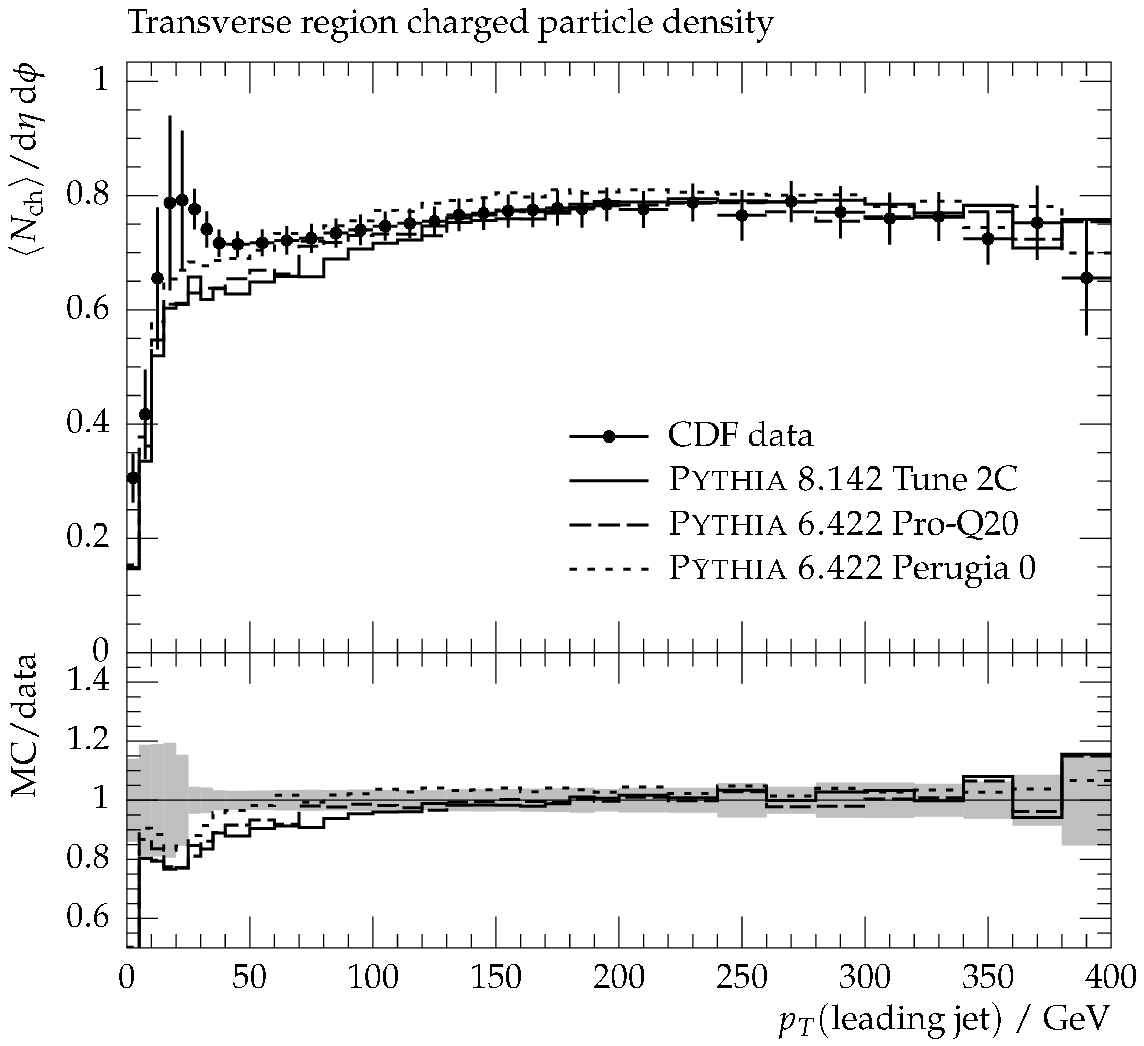}
\end{center}
\vspace{-4mm}\hspace{44mm}(a)\hspace{68mm}(b)
\caption{Results from Tune 2C for (a) the charged multiplicity at $\sqrt{s}
= 1.8\TeV$ and (b) the transverse region charged particle density in Rick
Field's leading jet analysis
\label{fig:2C}}
\end{figure}

In Fig.~\ref{fig:2C}, results from Tune 2C are shown using the Rivet
analysis framework for (a) the charged multiplicity at $\sqrt{s} =
1.8\TeV$ and (b) the transverse region charged particle density in Rick
Field's leading jet analysis \cite{Buckley:2010ar,Field:2002vt,Acosta:2001rm}. Comparisons are made both to data, and to the Pro-Q20 and
Perugia 0 tunes of \tsc{Pythia 6} \cite{Buckley:2009bj,Skands:2010ak}.
In both cases, Tune 2C is in good agreement.

\subsubsection*{LHC data}
Tune 4C is a modification of Tune 2C, where MPI and colour reconnection
parameters have been changed to give a good match to LHC data (with ISR
settings left unchanged). Additionally, the diffractive cross section has
been slightly damped, to better match an early ATLAS study \cite{AtlasDiff}.
One notable aspect of this tune is a reduced amount of colour reconnection
required to match $\langle \pT \rangle (N_{\mrm{CH}})$ data in
minimum-bias events. This reduction also affects the balance between the
charged particle density and the scalar $\pT$ sum density in the transverse
region of the underlying event, leading to a slightly worse description of
this data.

\begin{figure}
\begin{center}
\includegraphics[scale=0.6,angle=270]{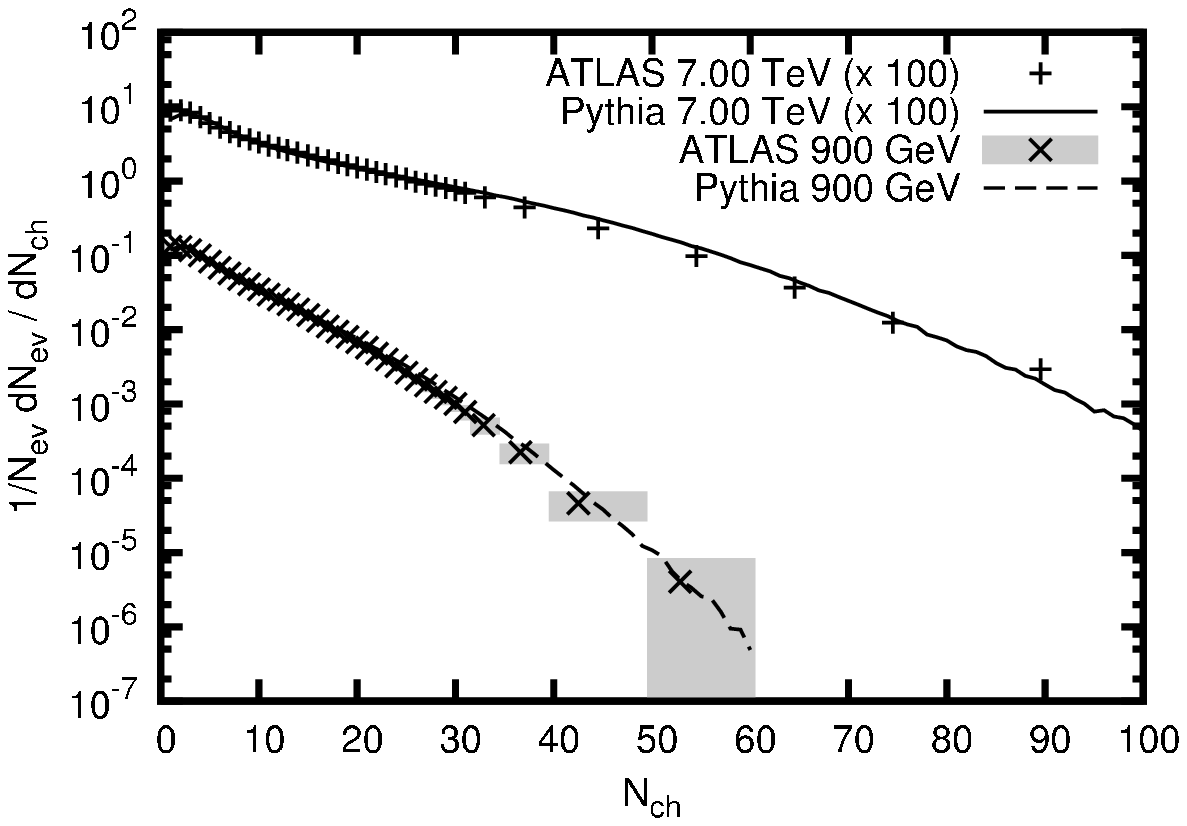}
\includegraphics[scale=0.6,angle=270]{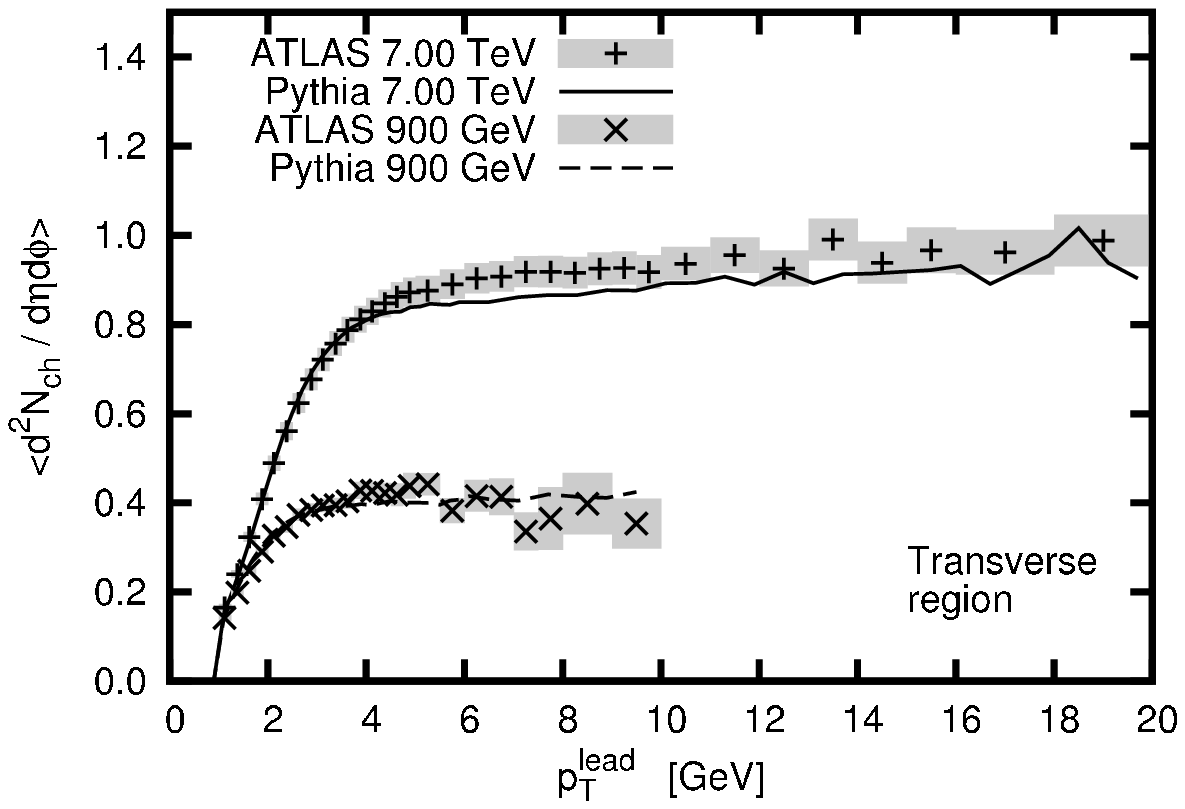}
\end{center}
\vspace{-4mm}\hspace{44mm}(a)\hspace{68mm}(b)
\caption{Results from Tune 4C compared to ATLAS data at $\sqrt{s} = 0.9$
and $7\TeV$ showing (a) the charged multiplicity and (b) the transverse
region charged particle density
\label{fig:4C}}
\end{figure}

In Fig.~\ref{fig:4C}, results from Tune 4C are shown against ATLAS data at
$\sqrt{s} = 0.9$ and 7~TeV showing (a) the charged multiplicity and (b)
the transverse region charged particle density
\cite{Aad:2010rd,AtlasMB7,AtlasUE}. The broad features of the
data are reproduced, and it is hoped that a more complete tuning will
increase agreement further. It is noted that Tunes 2C and 2M give too
little activity when compared to LHC data and Tune 4C gives too much
activity when compared to Tevatron data.

\subsubsection{Conclusions}
\label{sec:conclusions}

The MPI model in \tsc{Pythia 8} is an evolution of the original framework
introduced in earlier versions of \tsc{Pythia} and has been well proven
in comparisons to experimental data. Of course, it is not the final word in
the modelling of soft MPI and undoubtedly there are further physics aspects
that can be included. One possible extension we have implemented is
rescattering, outlined above. Although we were unable to find a distinct
signature for these processes, there is still future scope for the
study of effects on an overall generator tune.

Long awaited LHC data are now being published and will hopefully help to
constrain different models and model parameters. One current concern is the
need for different tunes to describe Tevatron and LHC data,
and it remains to be seen if there is some region in parameter space where
this is possible. So far it is not possible to rule out differences due to
experimental effects or deficiencies in the model, e.g. related to the
energy dependence of $\pTo$. We look forward to future data that may help
to resolve these issues.

\newcommand{\ptmin}{p_t^{\rm min}}
\newcommand{\refline}[1]{{\hfill\tiny\textcolor{NavyBlue}{[#1]}}}
\newcommand{\Sinc}{\sigma^{\rm inc}}
\newcommand{\diff}[2]{\frac{\mathrm{d}#1}{\mathrm{d}#2}}
\newcommand{\dd}{\mathrm{d}}
\graphicspath{{gieseke/figs/}}

\contribution{Multiple Partonic Interactions in {\sc Herwig++}}
{Contributing authors: S. Gieseke, Ch. R\"ohr, A. Si\'odmok}
\label{herwig}

The modelling of underlying events in {\sc Herwig++} is based on the fact that
at high enough energies the hard inclusive cross section will eventually
exceed the total cross section \cite{Bahr:2008pv}.  Generally speaking, we write
the hard inclusive cross section for partonic $2\to 2$ scatters as
\begin{equation}
  \Sinc(s;\ptmin) = \sum_{i,j} \int_{{\ptmin}^2} {\dd p_t^2}
  f_{i/h_1}(x_1, \mu^2) \otimes \diff{\hat\sigma_{i,j}}{p_t^2} \otimes
  f_{j/h_2}(x_2, \mu^2) \, ,
  \label{eq:sigmainc}
\end{equation}
which is the usual collinear factorization ansatz.  With recent parton
distribution functions and lower limits for the transverse momentum,
$\ptmin$, in the perturbative regime of a few GeV, Eq.~\ref{eq:sigmainc}
results in values for $\Sinc(s; \ptmin)$ which exceed the
Donnachie--Landshoff (DL) parametrization \cite{Donnachie:1992ny} of the
total cross section $\sigma_{\rm tot}$. 
The simplest way out is the observation that
the proton is a spatially extended object, allowing for independent
multiple hard interactions, which are strictly all taken into account in
the calculation of the inclusive cross section.  Therefore, we calculate
the average number of hard interactions from an eikonal ansatz as
\begin{equation}
  \bar n(\vec b, s) = A(\vec b; \mu^2) \Sinc (s; \ptmin)\ .
\end{equation}
The overlap function $A(\vec b; \mu^2)$ describes the spatial overlap of
the two colliding hadrons (protons) as a function of the impact
parameter $\vec b$.  The parameter $\mu^2$ characterizes the size of the
proton and is proportional to the squared inverse radius.  For
simplicity, we assume a spatial distribution following the functional
form deduced from the electromagnetic elastic form factor.  But we do
allow for a different width parameter $\mu$ of the distribution, as the
colour might be distributed differently from the electric charges.  The
basic ideas for this multiple interaction model follow the model in
Ref.~\cite{Butterworth:1996zw}, which in turn introduces a model similar to
the one discussed in Ref.~\cite{Sjostrand:1987su}.

The extension to soft scatterings, similar to the model of
Ref.~\cite{Borozan:2002fk}, is
kept as simple as possible. First, the transverse momentum of scattered
particles is extended to transverse momenta below $\ptmin$.  The
additional soft contribution to the inclusive cross section is also
eikonalized, such that we can as well calculate an average number of
soft scatters from the resulting $\Sinc_{\rm soft}$ and an overlap
function $A_{\rm soft}(\vec b)$ for the soft scattering centers.  The
functional form $A_{\rm soft}(\vec b)$ is assumed to be the same as for
the hard scatters, but we allow for a different inverse radius,
$\mu_{\rm soft}^2$.  The phase space of the soft scatters is determined
with a simple Gaussian ansatz for the transverse momentum, of which the
parameter is determined by the value of $\Sinc_{\rm soft}$.

The consistency of this model with unitarity is given by fixing the two
additional parameters $\Sinc_{\rm soft}$ and $\mu_{\rm soft}^2$ from two
additional constraints.  First, we can calculate the total cross section
from the eikonal model and fix it to be consistent with the DL
parametrization.  In addition, using the optical theorem, we can
calculate the $t$--slope parameter from the eikonal model as well and
fix it to a reasonable parametrization.

After in a first step only the model for hard multiple partonic
interactions has been introduced \cite{Bahr:2008dy}, we also studied its
implications from Tevatron data and total cross section data in a
simplified version \cite{Bahr:2008wk}.  Finally, the extension of the
model to include soft scatters has been implemented in {\sc Herwig++} and is
the default underlying event model since version 2.3.  The consistency
of the model predictions with current Tevatron data has been studied in
detail in Ref.~\cite{Bahr:2009ek}.

\subsubsection{{\sc Herwig++} against first LHC data}

Equipped with the good description of the Tevatron data, we can now take
a first look at the ATLAS measurements made at the 900\,GeV and 7\,TeV
runs at the LHC \cite{Aad:2010rd,:2010ir}.  We anticipate the
possibility that the assumptions made in order to extend the model into
the soft region may be far too simple.  Nevertheless, we have been able
to accommodate the detailed underlying event analyses carried out at the
Tevatron.  There we have come up with regions in the two--dimensional
parameter plane of $\ptmin$ and $\mu^2$, where we obtain a similarly
good overall $\chi^2$ for the underlying event data and still are
consistent with our constraints from the total cross section and the
elastic slope parameter.  This region roughly follows a line.  We now
had a first look at Minimum Bias data, particularly the relatively
simple distribution of charged particles in pseudorapidity.

As a first step, we have varied our model parameters and compared the
results against the 900\,GeV data.  We find that the shape of the
pseudorapidity distribution in {\sc Herwig++} is by far too much peaked in the
forward directions.  In addition, there is not enough freedom in our
parameter space to describe $\langle p_\perp\rangle (N_{\rm ch})$.

\begin{figure}
  \begin{center}
    \resizebox{0.5\textwidth}{!}{\includegraphics{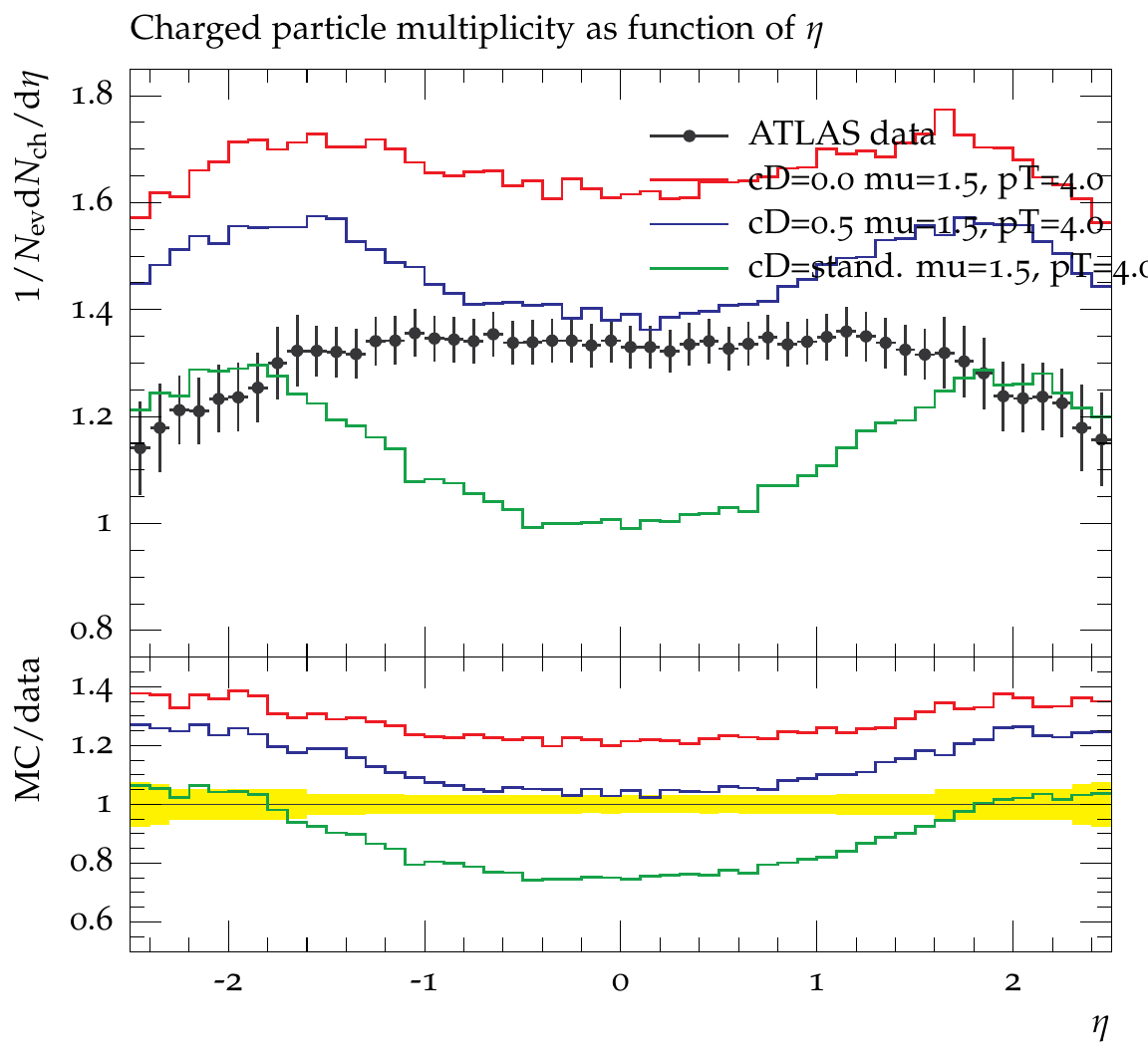}}
    \caption{Pseudorapidity distribution of charged particles in Minimum
      Bias events at 900\,GeV compared to ATLAS data.  The most sensitive
      model parameter was the colour disruption probability for soft
      events.\label{Fig:disrupt}}
  \end{center}
\end{figure}

A first hint towards the possible improvement of our description of data
was found when we varied the probability that any of the additional soft
scatters gets disconnected in colour space from the rest of the event
and the beam remnants in particular.  The value $\texttt{cD} = 1$ was
used as a default, saying that the soft scatters have always been
disconnected.  Physically this means that there are no colour strings
built up between the beam remnants and the soft particles produced in
the soft underlying event.  When they are build up more and more, as we
see when we vary the parameter towards the other extreme value 0 (always
connected), we find that we produce many additional soft particles,
resulting in an evenly filled plateau in rapidity.  Having checked also
other parameters, such as parton distribution functions and their
behaviour at small $x$ values, we found that the effect of the colour
disruption parameter was most important.  Fig.~\ref{Fig:disrupt} shows
the sensitivity to the colour disruption parameter.

A second hint is given by the unability to describe $\langle
p_\perp\rangle (N_{\rm ch})$ which is considered to be very sensitive to
non-perturbative colour reconnections.  So, as final
additional modification we have considered a newly implemented model for
soft colour reconnections in {\sc Herwig++}.  We find that only with the two
latter modifications we can give a sensible description of minimum bias
events.

A colour reconnection model is a very significant modification of the final
state as the production of charged particles is affected in its
multiplicity as well as in its phase space distribution, once a multiple
partonic interaction model is used.  Hence, before moving on to LHC data,
we have checked the new model against data that we previously described
quite well.  First we considered LEP final states and found no sensitivity
whatsoever.  This was expected as the colour structure of the event is
well-defined by the perturbative parton shower evolution.  The Tevatron
underlying event analysis, that we were always able to reproduce, was
slightly improved, as expected.  In particular, the tension between
$p_t^{\rm sum}$ and $N_{\rm ch}$ against the leading-jet transverse
momentum has been reduced.  

\begin{figure}
  \begin{center}
    \resizebox{0.45\textwidth}{!}{\includegraphics{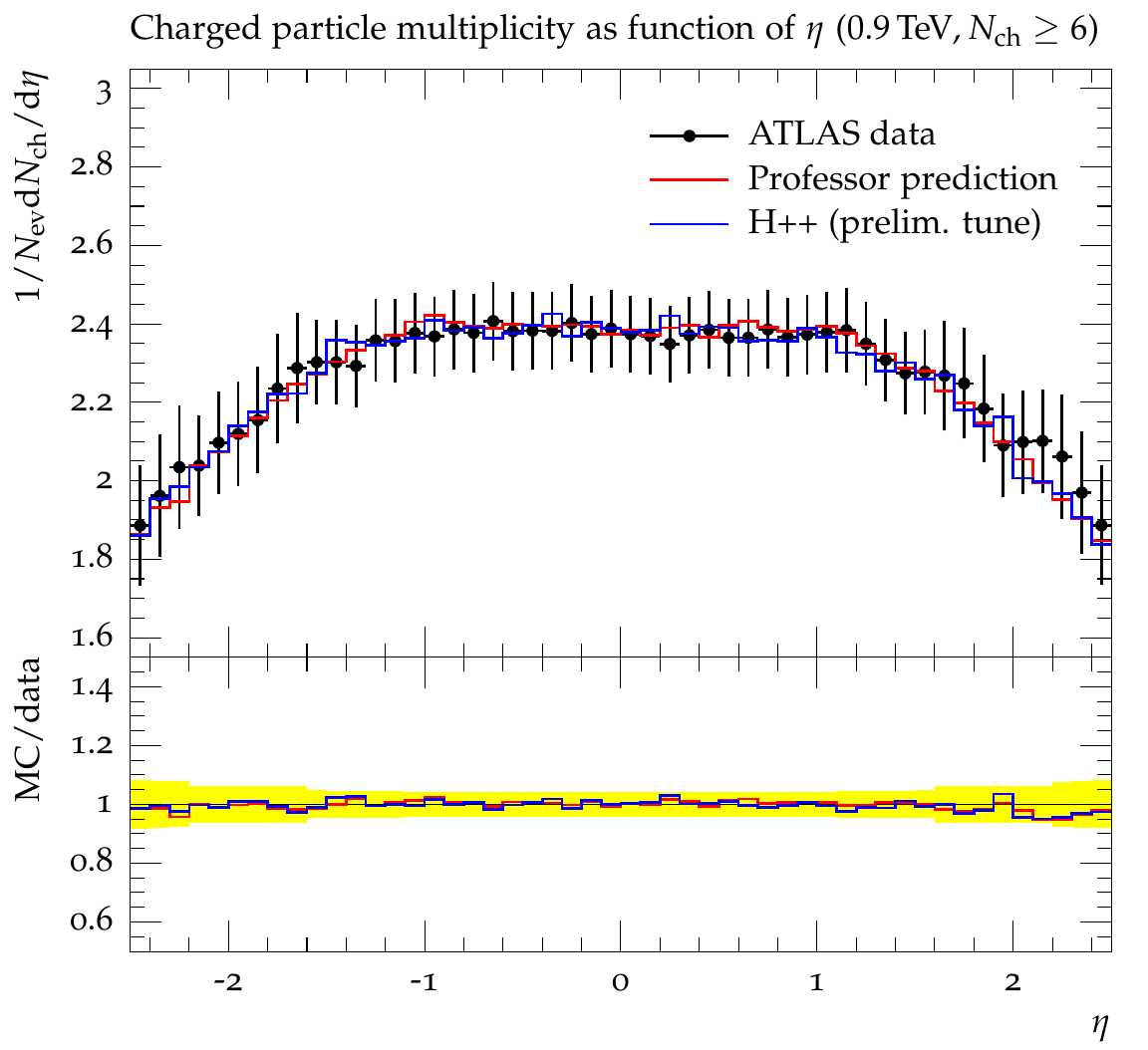}}
    \resizebox{0.45\textwidth}{!}{\includegraphics{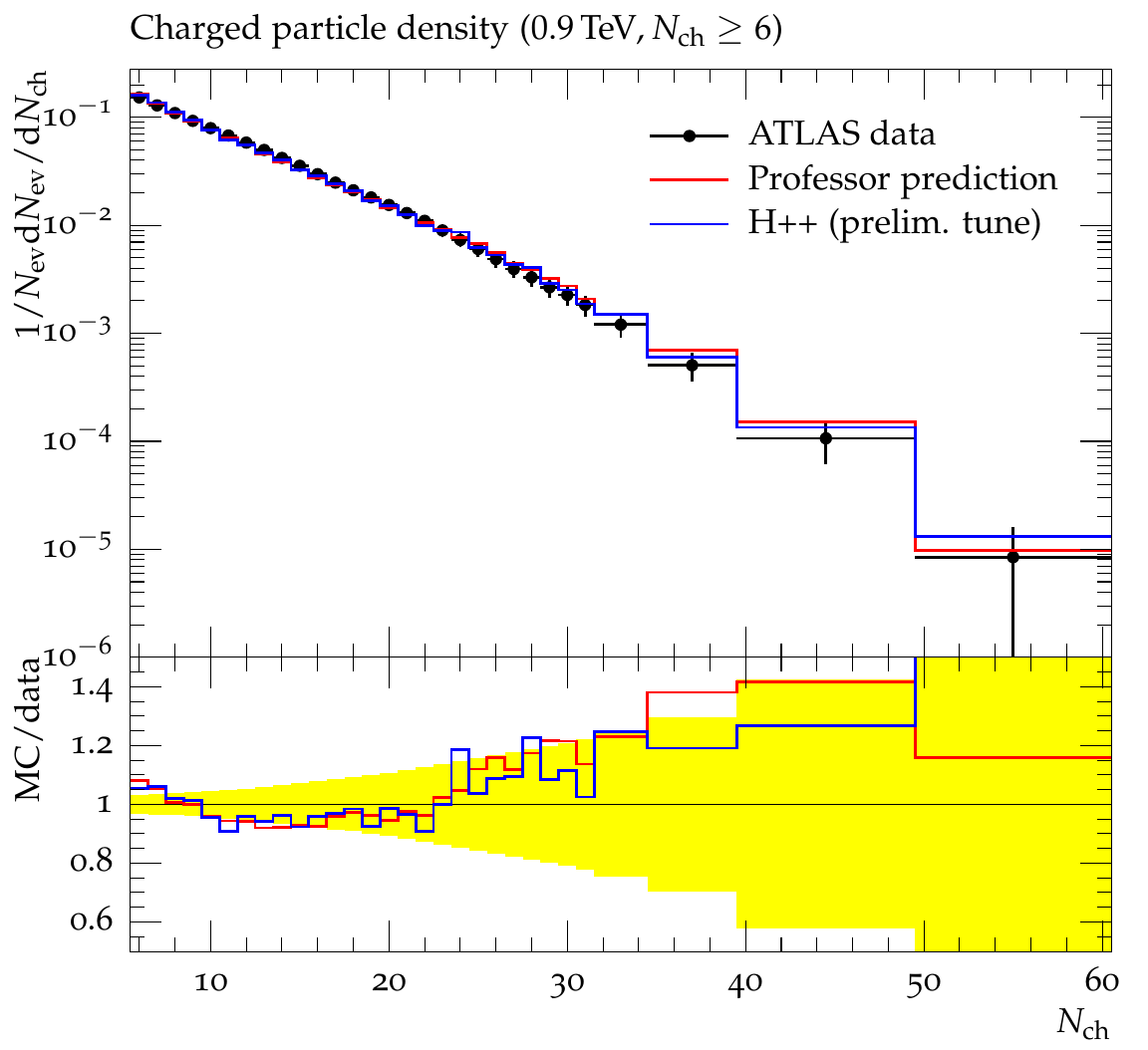}}
  \end{center}
  \caption{Pseudorapidity and charged particle distribution from the
    ATLAS $N_{\rm ch}\geq 6$ analysis at 900\,GeV compared to
    {\sc Herwig++}. The data points are read off preliminary, but publicly
    available, ATLAS figures. \label{Fig:900gev}}
\end{figure}

\begin{figure}[hb]
  \begin{center}
    \resizebox{0.45\textwidth}{!}{\includegraphics{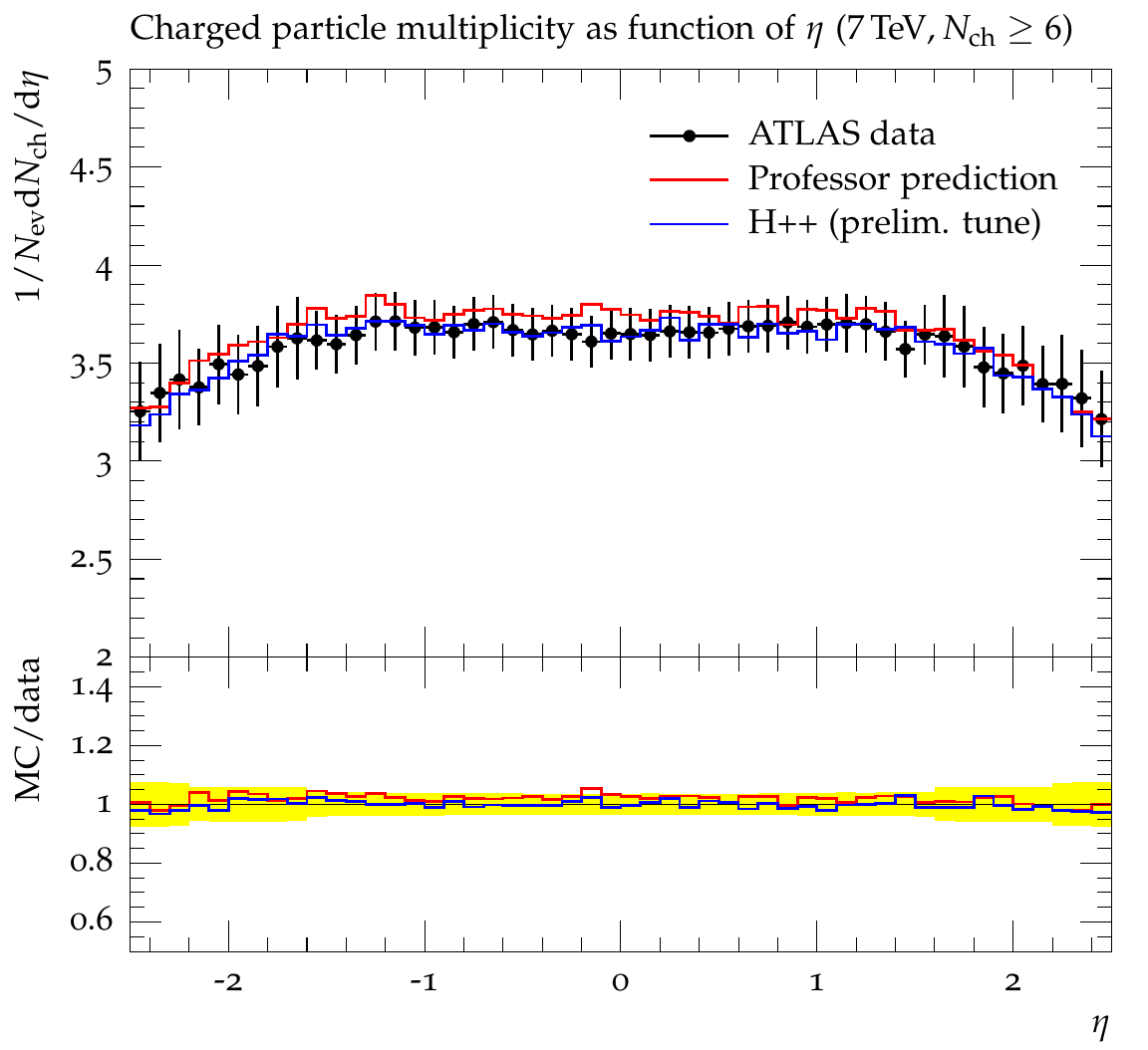}}
    \resizebox{0.45\textwidth}{!}{\includegraphics{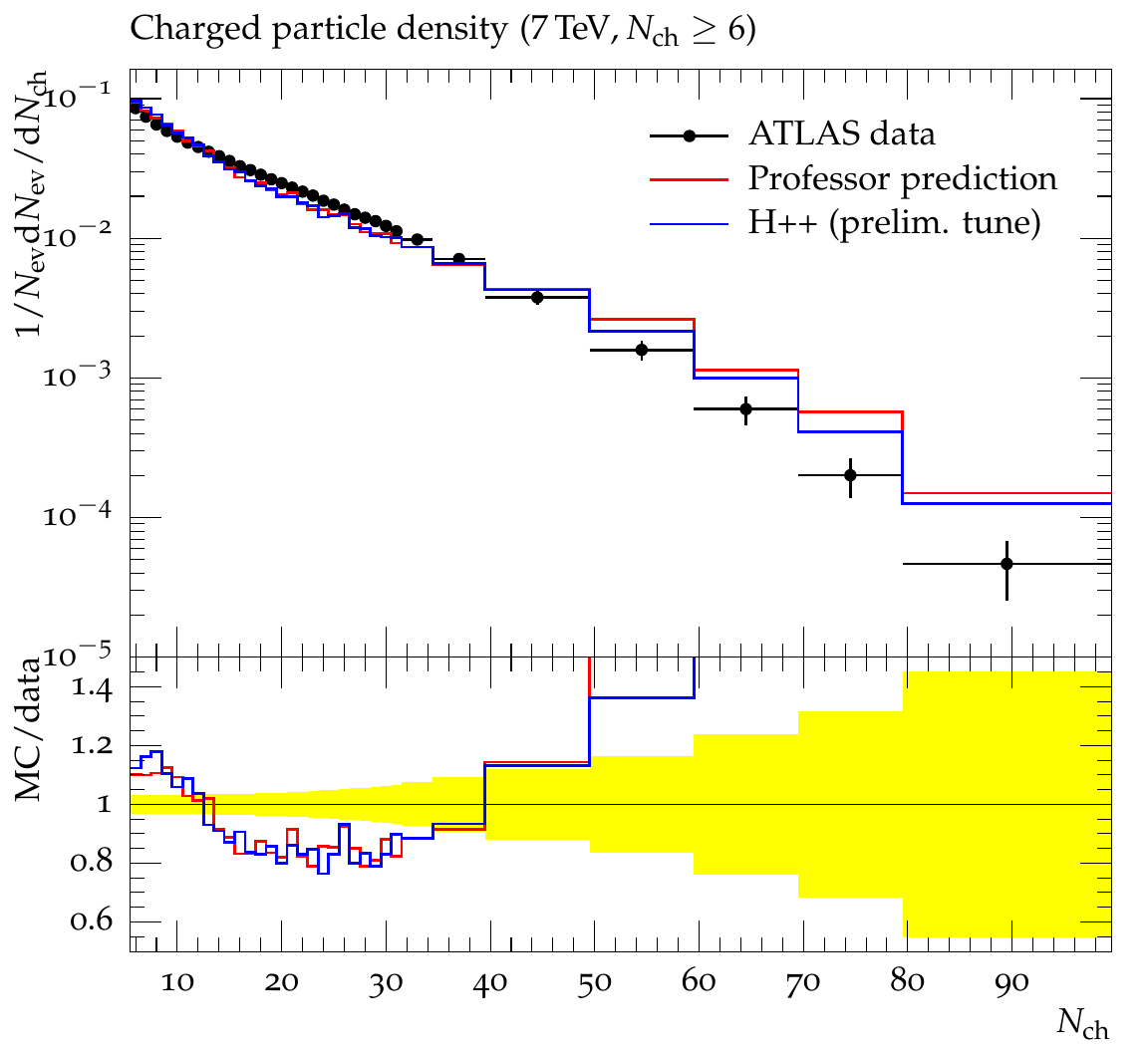}}%
  \end{center}
  \caption{Pseudorapidity and charged particle distribution from the
    ATLAS $N_{\rm ch}\geq 6$ analysis at 7\,TeV compared to
    {\sc Herwig++}. The data points are read off preliminary, but publicly
    available, ATLAS figures. \label{Fig:7tev}}
\end{figure}

Moving on to the LHC, we have used the ATLAS data with a cut $N_{\rm ch}
\geq 6$ in order to remove diffractive events \cite{:2010ir} as we
currently have no diffractive model in {\sc Herwig++}.  We have included the
variation of the colour disruption parameter and the colour reconnection
model and tuned our model to the new data.  We used data from the run at
$\sqrt{s}=900\,$GeV as well as from $\sqrt{s}=7\,$TeV.  The tuned
results are shown in Figs.~\ref{Fig:900gev}~and~\ref{Fig:7tev}.

We find that the overall agreement with the data is very good.  The
expected need for the modification of the colour structure in multiple
partonic interaction events is in fact given. 

\subsubsection{Conclusion}
We have introduced a simple colour reconnection model in {\sc Herwig++} in
order to complete the hadronization of events with multiple partonic
interactions.  We find very good agreement with first data on
non--diffractive Minimum Bias events, measured by ATLAS.  The model is
included in the recent release {\sc Herwig++} 2.5 \cite{Gieseke:2011na}.

\section{Theory of Multi-Parton Scattering}
\label{sec:theory}

The theoretical investigation of MPI has a long history
\cite{Landshoff:1978fq,Takagi:1979wn,Goebel:1979mi,Paver:1982yp,Mekhfi:1985dv}
and has experienced a renewed interest in more recent times
\cite{Snigirev:2003cq,Korotkikh:2004bz,Treleani:2007gi,Calucci:2008jw,Calucci:2009sv,Calucci:2009ea,Gaunt:2009re,Snigirev:2010tk,Calucci:2010wg,Snigirev:2010ds,Diehl:2010aa}, driven by the need to understand the hadronic activity at the LHC. 
The phenomenology of multiple parton interactions relies on a rather
simple and intuitive cross section formula, where multi-parton
distributions are multiplied with the cross sections for each individual
hard scatter, cf. Chapter~\ref{sec:pheno}.  It is natural to ask whether such a factorization formula
can be derived in QCD and to which extent it needs to be modified or
extended. Sections~\ref{diehl} and~\ref{blok}  present two independent efforts to address
   these questions, respectively taking the production of two electroweak
   gauge bosons or of two jet pairs as examples processes for DPS. 

Accurate predictions of DPS cross sections also require good modelling
of double parton distribution functions (dPDFs) used in
phenomenological studies. Typically, they are constructed from
standard single PDFs neglecting possible correlations between the
longitudinal momenta and transverse positions of the two partons
involved. The development of the first set of LO dPDFs in the
framework assuming factorization between the longitudinal and
transverse components is described in Section~\ref{gaunt} of this
chapter. However, as discussed there and in Section~\ref{diehl}, the
validity of the transverse-longitudinal factorization is being contested.

This chapter contains also a discussion, in Section~\ref{treleani}, of
general features of MPI in a probabilistic framework employing a functional approach. The probabilistic picture naturally leads to considering the inclusive and exclusive cross sections, which are linked by the sum rules. They can be used to obtain information on MPI properties, such as two-particle correlations, since inclusive and exclusive cross sections are measured independently. 

The MPI are also very strongly connected to the small-$x$ phenomena of saturation and diffraction. In general, the cut diagrams for MPI differ from multiple parton chain diagrams for saturation and diffraction only by the position of the cut. In high energy $pp$ scattering unitarity requires the presence of MPI as well as a large diffractive cross section. A model to describe diffractive excitations, its relation with the multi-regge formalism and its MC implementation is discussed in Section~\ref{gustafson}.

\allowdisplaybreaks[2]
\graphicspath{{diehl/figs/}}


\renewcommand{\half}{{\textstyle\frac{1}{2}}}

\newcommand{\ms}{\mskip 1.5mu}

\newcommand{\tvec}[1]{\boldsymbol{#1}}

\contribution{Multiple parton interactions: some theoretical considerations}
{Contributing author: M. Diehl}

\label{diehl}


\subsubsection{The basic cross section formula}

As an example for a process to which multiparton interactions contribute
we consider the production of two electroweak gauge bosons ($W$, $Z$ or
$\gamma^*$) with transverse momenta much smaller than their masses or
virtualities.  Since the driving force for studying multiple interactions
is the necessity to describe details of the final state, we keep the cross
section differential in the transverse boson momenta.  For the production
of a single gauge boson there is a powerful theory involving
transverse-momentum dependent parton densities
\cite{Collins:1981uk,Ji:2004wu,Collins:2007ph,Collins:2011}, which can to
a large extent be generalized to the case of multiple hard scattering.

\begin{figure}[b]
\begin{center}
\includegraphics[height=0.28\textwidth]{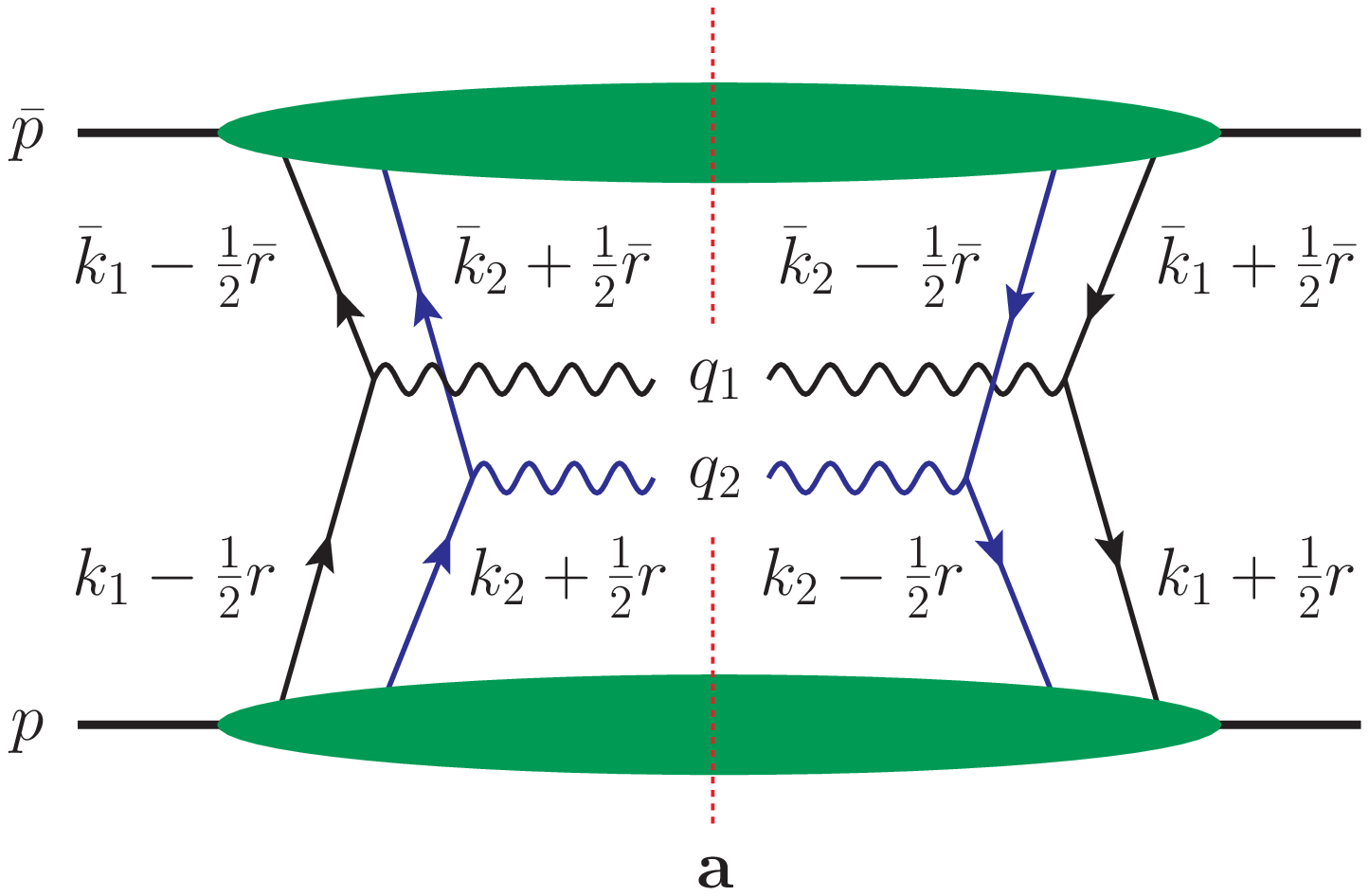}
\hspace{0.5em}
\includegraphics[height=0.28\textwidth]{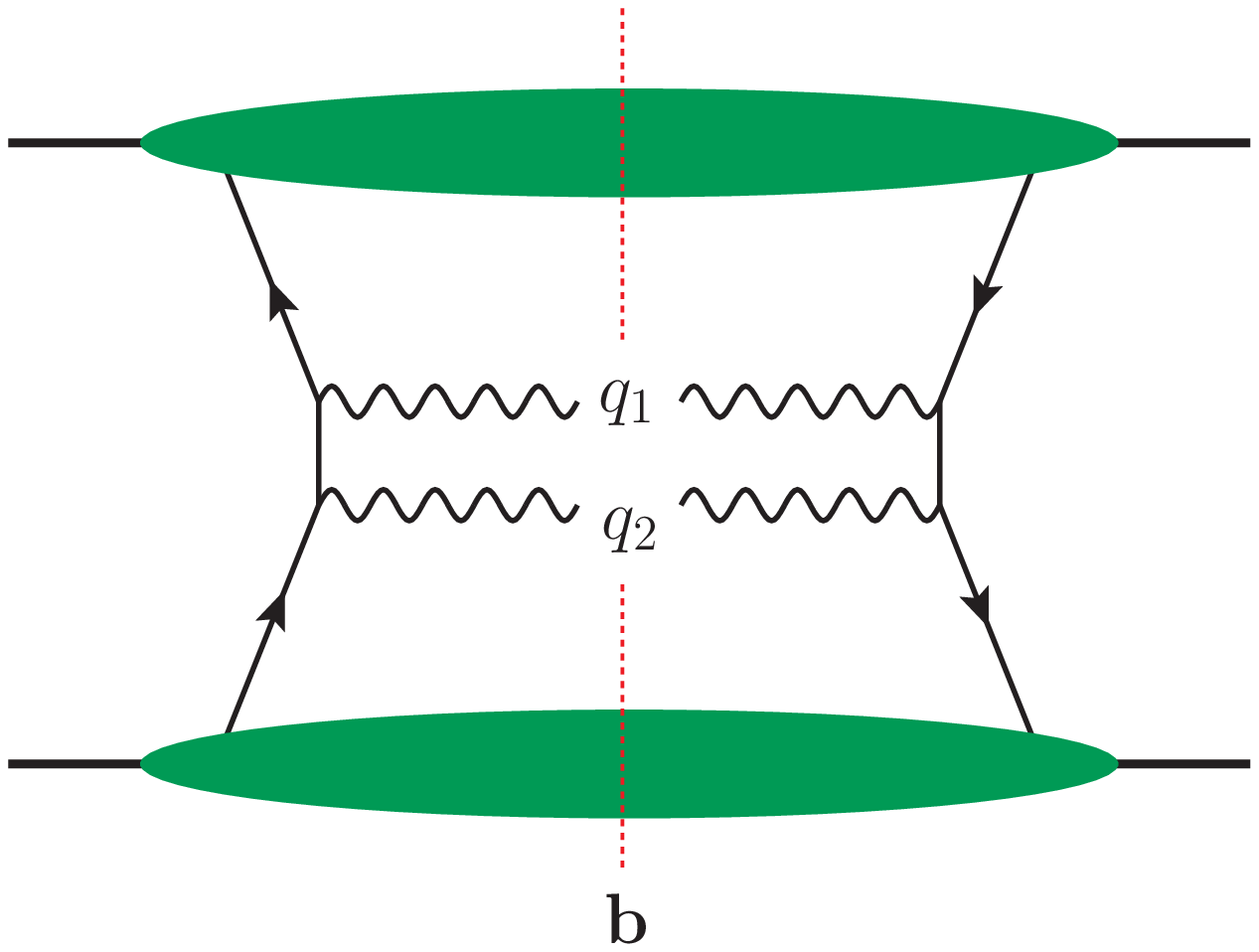}
\end{center}
\caption{\label{fig:scatter} Graphs for the production of two gauge bosons
  by double (a) or single (b) hard scattering.  The dotted line denotes
  the final-state cut.}
\end{figure}

Figure~\ref{fig:scatter}a shows a graph for two-boson production by double
parton scattering.  This graph can be evaluated using the standard
hard-scattering approximations,  neglecting small momentum
components compared with large ones.
Simple kinematic considerations show that the transverse momenta of the
scattering partons are in general \emph{not} equal in the scattering
amplitude and its complex conjugate, as indicated in the figure.  It is
convenient to Fourier transform from the transverse-momentum differences
$\tvec{r}$ and $\bar{\tvec{r}}$ to transverse position variables
$\tvec{y}$ and $\bar{\tvec{y}}$.  The constraint $\tvec{r} +
\bar{\tvec{r}} = \tvec{0}$ from momentum conservation then turns into
$\tvec{y} = \bar{\tvec{y}}$ for the Fourier conjugate positions, and the
cross section reads%
\footnote{In other parts of this chapter, and in
  Chapter~\ref{sec:pheno}, the symbols $D$ or $\Gamma$ are used to
  denote the multiple parton distributions instead of $F$ used
  here. Similarly, transverse coordinates are denoted by $\tvec{b}$, instead of $\tvec{y}$.}
\begin{align}
  \label{X-section}
& \frac{d\sigma}{\prod_{i=1}^2 dx_i\, d\bar{x}_i\, d^2\tvec{q}{}_i}
  \;\bigg|_{\text{Fig.~\protect\ref{fig:scatter}a}}
= \frac{1}{S}
  \sum_{\substack{a_1, a_2 = q, \Delta q, \delta q \\[0.1ex]
          \bar{a}_1, \bar{a}_2 = \bar{q}, \Delta\bar{q}, \delta\bar{q}}}
  \biggl[\, \prod_{i=1}^{2} \,
  \int d^2\tvec{k}_i\, d^2\bar{\tvec{k}}_i\;
    \delta^{(2)}(\tvec{q}{}_i - \tvec{k}_i - \bar{\tvec{k}}_i) \biggr]
\nonumber \\
& \quad \times  
  \hat{\sigma}_{1, a_1 \bar{a}_1}(q_1^2) \;
  \hat{\sigma}_{2, a_2 \bar{a}_2}(q_2^2)
\int d^2\tvec{y}\,
  F_{a_1, a_2}(x_i, \tvec{k}_i, \tvec{y})\;
  F_{\bar{a}_1, \bar{a}_2}(\bar{x}_i, \bar{\tvec{k}}_i, \tvec{y}) \,,
  \phantom{\int}
\end{align}
where $\hat{\sigma}_{i, a_i \bar{a}_i}$ denotes the hard-scattering cross
section for single-boson production.  The statistical factor $S$ is $2$ if
the produced bosons are identical and $1$ if they are not.  Up to power
corrections, the momentum fractions of the colliding partons are fixed by
the measurable momenta as
\begin{align}
x_i^{} &= q_i^+ \big/ p^+ = (k_i^+ \pm \half r^+) \big/ p^+ \,, &
\bar{x}_i^{} &= q_i^- \big/ \bar{p}^{\ms -}
 = (\bar{k}_i^- \pm \half \bar{r}^{\ms -}) \big/ \bar{p}^{\ms -} \,,
\end{align}
where we have introduced light-cone coordinates $v^\pm = (v^0 \pm v^3)
/\sqrt{2}$ for each four-vector $v$.  The definition of the double-parton
distributions in \eqref{X-section} closely resembles the one for a
transverse-momentum dependent single-quark distribution
\cite{Collins:1981uk,Ji:2004wu,Collins:2007ph,Collins:2011}.  For
instance, the distribution of two quarks in the proton with momentum $p$
is given by
\begin{align}
  \label{dist-def}
& F_{a_1,a_2}(x_i, \tvec{k}_i, \tvec{y})
= \biggl[\, \prod_{i=1}^2
       \int \frac{dz_i^- d^2\tvec{z}_i^{}}{(2\pi)^3}\,
       e^{i (x_i^{} z_i^- p^+ - \tvec{z}_i^{} \tvec{k}_i^{})}
    \biggr] \, 2 p^+\!\! \int dy^-
\nonumber \\
 &\qquad \times
    \big\langle p \big|\,
    \bar{q}(- \half z_2)\, \Gamma_{a_2} \, q(\half z_2) \;
    \bar{q}(y - \half z_1)\, \Gamma_{a_1} \, q(y + \half z_1)
    \big| \ms p \big\rangle \Big|_{z_1^+ = z_2^+ = y^+ = 0} \,.
\end{align}
One can identify $\tvec{k}_i$ as the ``average'' transverse momentum of
each quark and $\tvec{y}$ as the ``average'' transverse distance between
the quarks, where the ``average'' refers to the physical scattering
amplitude and its conjugate.  The result \eqref{X-section} thus has an
intuitive interpretation: two bosons with transverse momenta $\tvec{q}_i$
are produced in the collision of partons with average transverse momenta
$\tvec{k}_i^{}$ and $\bar{\tvec{k}}_i$.  The two collisions occur at an
average transverse distance $\tvec{y}$, which is equal to the average
transverse distance between the two partons in each colliding proton.
At a more formal level, $F(x_i, \tvec{k}_i, \tvec{y})$ has the structure
of a Wigner distribution \cite{Hillery:1983ms} in its transverse momentum
and position arguments.

Integrating over the transverse parton momenta, one obtains a collinear
two-parton distribution $F_{a_1,\ms a_2}(x_i, \tvec{y}) = \int
d^2\tvec{k}_1\, d^2\tvec{k}_2\; F_{a_1,\ms a_2}(x_i, \tvec{k}_i,
\tvec{y})$, which can be interpreted as the probability for finding two
quarks with momentum fractions $x_1$ and $x_2$ at a relative transverse
distance $\tvec{y}$ in the proton.  These distributions naturally appear
if one integrates the cross section \eqref{X-section} over the transverse
momenta $\tvec{q}_1$ and $\tvec{q}_2$ of the bosons,
\begin{align}
  \label{X-section-coll}
\frac{d\sigma}{\prod_{i=1}^2 dx_i\, d\bar{x}_i}
  \;\bigg|_{\text{Fig.~\protect\ref{fig:scatter}a}}
&= \frac{1}{S}\, 
  \sum_{\genfrac{}{}{0pt}{1}{a_1,\ms a_2
         = \ms q,\ms \Delta q, \delta q}{%
       \bar{a}_1,\ms \bar{a}_2
         = \ms\bar{q},\ms \Delta\bar{q}, \delta\bar{q}}}
  \hat{\sigma}_{1, a_1 \bar{a}_1}(q_1^2) \;
  \hat{\sigma}_{2, a_2 \bar{a}_2}(q_2^2)
\nonumber \\[0.2em]
&\quad\times
  \int d^2\tvec{y}\;
  F_{a_1,\ms a_2}(x_i, \tvec{y})\, 
  F_{\bar{a}_1,\ms \bar{a}_2}(\bar{x}_i, \tvec{y}) \,.
\end{align}
This is the simple cross section formula mentioned in the introduction.
It was already derived in \cite{Paver:1982yp,Mekhfi:1983az} and underlies
most phenomenological studies of multiparton interactions.  Note that
\eqref{X-section-coll} and its generalization \eqref{X-section} to
measured transverse momenta have an intuitive physical interpretation, but
that they arise from the calculation of lowest-order Feynman graphs as in
figure~\ref{fig:scatter}a, using standard approximations.  One does not
need to appeal to semi-classical arguments to obtain these results.

For each bilinear operator in the matrix element \eqref{dist-def} there
are three relevant Dirac matrices $\Gamma_{a_i}$,
\begin{align}
\Gamma_{q}          &= \half \gamma^+ \, , &
\Gamma_{\Delta q}   &= \half \gamma^+ \gamma_5 \, , &
\Gamma_{\delta q}^j &= \half i \sigma^{j +} \gamma_5
   ~~~\text{with $j=1,2$} \,,
\end{align}
which respectively project on unpolarized, longitudinally polarized and
transversely polarized quarks.  Note that polarized two-parton
distributions exist even in an unpolarized proton, where they describe
spin correlations \emph{between} the two partons.  For small but
comparable $x_1$ and $x_2$ one may well have sizeable spin correlations
between two quarks (which are close in phase space for $x_1 \sim x_2$),
even if there is little correlation between the polarizations of a quark
and the proton (which are far apart in phase space).  The relevance of
such correlations in multiple interactions was pointed out already in
\cite{Mekhfi:1983az} but has to our knowledge not been included in
phenomenology.

If parton spin correlations are sizeable, they can have a strong impact on
observables.  For the production of two gauge bosons one can easily see
that the product $F_{\Delta q, \Delta q}\; F_{\Delta\bar{q},
  \Delta\bar{q}}$ of longitudinal spin correlations enters the cross
section with the same weight as the unpolarized term $F_{q,q}\,
F_{\bar{q},\bar{q}}$.  One also finds that product $F_{\delta q, \delta
  q}\; F_{\delta\bar{q}, \delta\bar{q}}$ of transverse spin correlations
give rise to a $\cos(2\varphi)$ modulation in the angle $\varphi$ between
the decay planes of the two bosons and thus affects the distribution of
final-state particles.


\subsubsection{Power behavior}

It is easy to determine the power behavior of the cross section formula
\eqref{X-section} for double hard scattering.  The parton-level cross
sections $\hat{\sigma}$ behave like $1/Q^2$, where $Q^2 \sim q_1^2 \sim
q_2^2$ denotes the size of the large squared invariant masses of the gauge
bosons.  We find
\begin{align}
  \label{power-double}
\frac{d\sigma}{\prod_{i=1}^2 dx_i\, d\bar{x}_i\; d^2\tvec{q}_i}
  \;\bigg|_{\text{Fig.~\protect\ref{fig:scatter}a}} 
& \sim\, \frac{1}{Q^4 \Lambda^2} \,,
\end{align}
where $\Lambda$ denotes the size of the transverse momenta $\tvec{q}_1
\sim \tvec{q}_2$ or the scale of non-perturbative interactions, whichever
is larger.  To obtain \eqref{power-double}, we have used that the
two-parton distributions scale like $F \sim 1/\Lambda^2$ and that the
typical transverse distance $\tvec{y}$ between the partons is of order
$1/\Lambda$.  The same power behavior as in \eqref{power-double} is
obtained for the case where both bosons are produced in a single hard
scattering, as shown in figure~\ref{fig:scatter}b.  Multiple hard
interactions are therefore \emph{not} power suppressed as long as one
keeps the cross section differential in the transverse momenta of the
particles produced in the hard parton collisions.

The situation changes when one integrates over $\tvec{q}_1$ and
$\tvec{q}_2$.  In the double-scattering mechanism both transverse momenta
are restricted to size $\Lambda$, but for a single hard scattering one has
$|\tvec{q}_1 + \tvec{q}_2| \sim \Lambda$ whereas the individual transverse
momenta can be as large as $Q$.  Because of this phase space effect one
has
\begin{align}
\frac{d\sigma}{\prod_{i=1}^2 dx_i\, d\bar{x}_i}
  \;\bigg|_{\text{Fig.~\protect\ref{fig:scatter}a}} 
& \sim\, \frac{\Lambda^2}{Q^4} \, , & 
\frac{d\sigma}{\prod_{i=1}^2 dx_i\, d\bar{x}_i}
  \;\bigg|_{\text{Fig.~\protect\ref{fig:scatter}b}} 
& \sim\, \frac{1}{Q^2} \,.
\end{align}
In the transverse-momentum integrated cross section, multiple hard
scattering is therefore a power correction.  This is required for the
validity of the usual factorization formulae, which contain only the
single-scattering contribution.


\subsubsection{Additional terms in the cross section}

In the formulae given so far, we have ignored the color structure of the
multiparton distributions.  The quark lines with momentum fraction $x_1$
in figure~\ref{fig:scatter}a can couple to a color singlet (as in
single-parton distributions) but they can also couple to a color octet,
provided that the lines with momentum fraction $x_2$ are coupled to a
color-octet as well.  In the latter case, the color structure of the
operators in \eqref{dist-def} is $(\bar{q}\, \Gamma_{a_2} \lambda^a q)\,
(\bar{q}\, \Gamma_{a_1} \lambda^a q)$.  Such color-octet distributions
contribute to the multiple-scattering cross section, as was already
pointed out in \cite{Mekhfi:1985dv}.  They do not have a probability
interpretation, and little is known about them.  The color structure of
two-gluon distributions is even more involved.

\begin{figure}
\begin{center}
\includegraphics[width=0.55\textwidth]{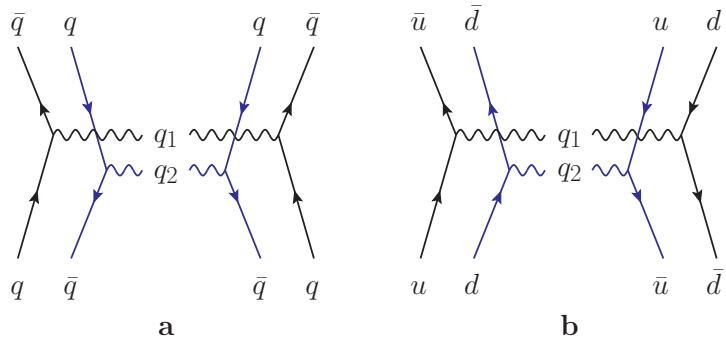}
\end{center}
\caption{\label{fig:interference} Example graphs for interference terms in
  fermion number (a) and in quark flavor (b).  The blobs indicating
  two-parton distributions are not shown.  Labels $q$ and $\bar{q}$
  indicate whether a line is represented by a quark or a conjugate quark
  field in the operator definition.  Momenta are assigned as in
  figure~\protect\ref{fig:scatter}.}
\end{figure}

There are more multiparton distributions that have no probability
interpretation but rather the structure of interference terms.  As shown
in figure~\ref{fig:interference}a, the parton with momentum fraction $x_1$
can be a quark in the scattering amplitude but an antiquark in the
conjugate amplitude, provided that the opposite holds for the parton with
momentum fraction $x_2$. Furthermore, one can have interference in the
quark flavor, as shown in figure~\ref{fig:interference}b.

While it is quite straightforward to include these extra contributions in
the cross section formula, the additional number of two-parton
distributions needed to obtain quantitative results is daunting.  It is
therefore important to have some guidance about the size and behavior of
these functions.


\subsubsection{High transverse momentum and evolution}

The predictive power of the theory is increased in the kinematic region
where the transverse momenta $\tvec{q}_i$ are small compared with $Q$ but
large compared with a typical non-perturbative scale.  At least some of
the transverse parton momenta must then be large as well, and one can
evaluate the corresponding parton distributions in terms of a hard
scattering subprocess at scale $|\tvec{q}_i|$ and parton distributions
that depend on fewer variables.

\begin{figure}
\begin{center}
\includegraphics[width=0.9\textwidth]{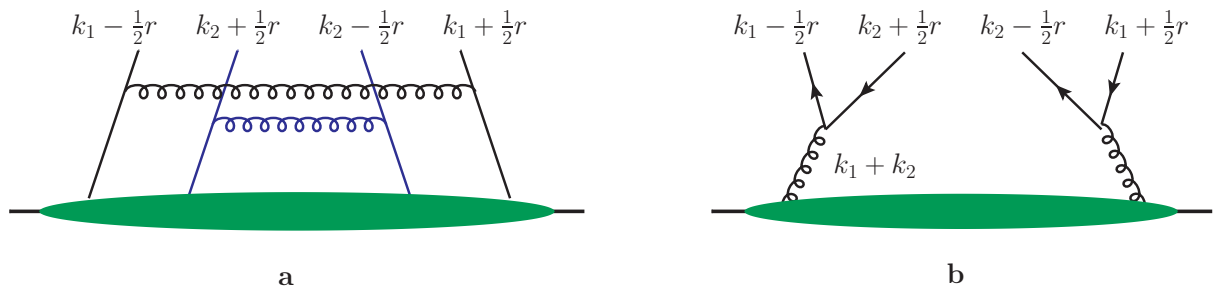}
\end{center}
\caption{\label{fig:high} (a) Ladder graph for a two-parton distribution,
  which generates perturbatively large $\tvec{k}_1$ and $\tvec{k}_2$ at
  small $\tvec{r}$.  (b) Graph for a quark-antiquark distribution, where
  the partons with momentum fractions $x_1$ and $x_2$ originate from the
  splitting of a single gluon.}
\end{figure}

An important example for this are ladder graphs such as the one in
figure~\ref{fig:high}a.  Since the partons with momentum fraction $x_1$
are not connected to those with momentum fraction $x_2$, the momentum
mismatch $\tvec{r}$ is small.  In position space this corresponds to an
inter-parton distance $\tvec{y}$ of hadronic size.  An important feature of
these ladder graphs is that their color factors disfavor the color octet
distributions mentioned in the previous section.  How strong this
suppression is quantitatively remains to be studied.

In the graph of figure~\ref{fig:high}b, the two partons with momentum
fractions $x_1$ and $x_2$ originate from the splitting of a single parton.
Such graphs are relevant for the region of large $\tvec{r}$, i.e.\ of
short distance $\tvec{y}$, and also for large $\tvec{k}_1 - \tvec{k}_2$.
One finds that they generate strong spin correlations between the partons;
the graph in the figure for instance forces the helicities of the quark
and antiquark to add up to zero.

Both graphs in figure~\ref{fig:high} lead to divergent integrals if one
integrates over $\tvec{k}_1$ and $\tvec{k}_2$ at fixed $\tvec{r}$.  Proper
regularization of these divergences gives the evolution equation for
collinear two-parton distributions; in particular splitting graphs as in
figure~\ref{fig:high}b give the famous inhomogeneous term in this equation
\cite{Kirschner:1979im,Shelest:1982dg}.  Remarkably, one finds
however that \emph{no} such inhomogeneous term appears in the evolution of
the distributions $F(x_i, \tvec{y})$ at any finite value of $\tvec{y}$,
provided one uses the definition \eqref{dist-def} with minimal subtraction
of divergences.

\begin{figure}
\begin{center}
\includegraphics[width=0.4\textwidth]{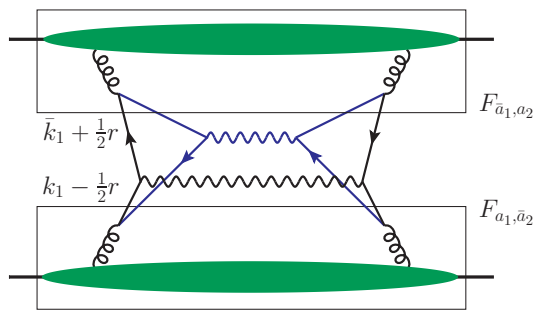}
\end{center}
\caption{\label{fig:high-2-4-X} Graph for the cross section where both
  two-parton distributions $F_{a_1, \bar{a}_2}$ and $F_{\bar{a}_1, a_2}$
  (indicated by boxes) involve the splitting of one into two partons.}
\end{figure}

The splitting mechanism of figure~\ref{fig:high}b is responsible for a
behavior $F(x_i, \tvec{y}) \sim 1/ \tvec{y}^2$ at short distances
$\tvec{y}$, which renders the integral in the cross section
\eqref{X-section-coll} infinite.  The same mechanism already gives
divergent integrals in the differential cross section \eqref{X-section}.
Furthermore, there is a double counting problem associated with this
splitting contribution: the graph in figure~\ref{fig:high-2-4-X} describes
double hard scattering with a $1\to 2$ parton splitting in each proton,
but it also represents a single hard-scattering mechanism, namely
two-gluon fusion $gg\to VV$ into two bosons via a box graph.  Therefore,
either the cross section formulae \eqref{X-section} and
\eqref{X-section-coll}, or the definition \eqref{dist-def} of two-parton
distributions, or both must be modified in a way that removes
singularities at small $\tvec{y}$, and a corresponding prescription for
calculating $gg\to VV$ must be given to avoid double counting.  This
remains a task for future work. %
More
details on the topics and results discussed here can be found in \cite{Diehl:2011tt}.

\graphicspath{{blok/figs/}}

\def\beq{\begin{equation}}
\def\eeq{\end{equation}}
\newcommand \Pomeron {I\!\!P}
\def\beq{\begin{equation}}
  \def\eeq{\end{equation}}
\def\bk{\mathbf{k}}
\def\beq{\begin{equation}}
\def\eeq{\end{equation}}
\def\beeq{\begin{eqnarray}}
\def\eeeq{\end{eqnarray}}
\def\vq{{\bf q}}
\def\vk{{\bf k}}
\def\vq{{\bf q}}
\def\abs#1{\left|#1\right|}

\contribution{The four jet production at LHC and Tevatron in QCD}
{Contributing authors: B. Blok, Yu. Dokshitzer, L. Frankfurt and M. Strikman}
\label{blok}

In spite of an extensive
 theoretical and experimental work,
various aspects  of the high energy hadronic collisions at
Tevatron and LHC are still poorly understood. This is especially
true for the  multijet production which is of a paramount
importance for the understanding of pQCD dynamics at high energy
colliders, and for the search of new particles. 
Here we summarize the first steps of the program to address the topic of MPI starting from the first principles of the perturbative QCD which is necessary for an accurate account of the significant  disbalance of the momenta of the jets (presence of the Sudakov form factors).
Among original results presented here are the derivation of the
formulae in the leading logarithmic  approximation for production
of 4 jets. 
Our key finding is that it is possible to isolate the
kinematics where the leading twist processes $2\to 4$ are not
enhanced. This result will allow to improve the reliability of the
Tevatron studies of the four jet production in the multiparton
kinematics and point out directions for the corresponding analysis
at the LHC.

Another critical issue is the formulation of the
problem in terms of the momentum representation
 double generalized parton distributions and introduction of the mean field approximation for this
  object. This new formulation is very effective for the more detailed studies which are now under way.
   In addition it brings a link with the original formulation in the coordinate space
   \cite{DelFabbro:1999tf,Mekhfi:1983az,Paver:1984ux,Sjostrand:ys,Bahr:2008zr,DelFabbro:2000ds,Accardi:2000ry,Frankfurt:bh,Frankfurt:dq,Frankfurt:2008ve,Rogers:2008qf,Domdey:2009ly,Diehl:2010dr}, and resolves an issue of the value of
   the strength of the double interaction within this approximation. Previously there was a question
   whether a conclusion of ref. \cite{Frankfurt:bh,Frankfurt:dq},  that the observed rate is a factor of two larger
   than the theoretical prediction can be due to
   uncertainties related to many Fourier transforms which were required to convert the HERA data
    to the experimental number.  A new formulation, though mathematically equivalent, has completely
     resolved this issue. This poses serious constraints on the Monte Carlo models of pp scattering at
      collider energies which are  not satisfied
by many of the current models.

These issues are of broad interest for both the theorists and
experimentalists.
\par The standard approach to the multijet production is the QCD improved parton model. It is based on
the assumption  that the cross section of a
hard  hadron--hadron interaction
is calculable in terms of the convolution
of parton distributions within colliding hadrons
with the cross section of a hard two-parton collision.
An application of this approach to the processes
with production of four jets
implies that all jets in the event are  produced  in a
hard collision of {\em two}\/ initial state partons.

\par
The recent data due to CDF and D0 collaborations
\cite{Abe:1997xk,Abazov:2009gc} do not contradict the
dominance of this mechanism in the well defined part of the phase
space. At the same time these data provide the evidence that there
exists a kinematical domain where a more complicated mechanism
becomes important, namely the double hard interaction of two
partons in one hadron with two partons in the second  hadron.

\begin{figure}[t]  
   \centering
   \includegraphics[width=0.4\textwidth]{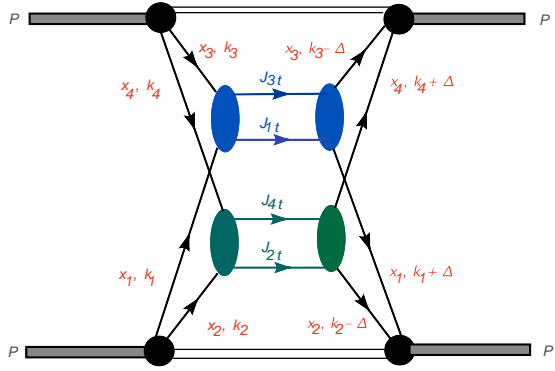}
   \caption{Kinematics of double hard collision - momenta of the colliding partons in in and out states}
   \label{kin}
 \end{figure}

  Within the parton model picture, the four jets produced this way should pair into two groups such
   that the transverse momenta of two jets in each pair compensate each other.
  In what follows we refer to this kinematics as {\em back-to-back dijet production}.
  We consider the dijets for the case
\begin{equation}\label{eq:kinemo}
\delta_{13}^2\equiv (\vec{j}_{1t}+\vec{j}_{3t})^2 \ll \> j_{1t}^2\simeq j_{3t}^2, \>\>
\delta_{24}^2
 \ll \> j_{2t}^2\simeq j_{4t}^2,
\end{equation}
where $\delta$ is the total transverse momentum of the dijet and $j_{it}$ the transverse momentum of
an individual jet (see Fig.~\ref{kin}).
The hardness condition $\delta^2\gg R^{-2}$ is implied, with $R$ the characteristic
 hadron size (non-perturbative scale).
 (The events with disbalances $\delta^2\le R^{-2}$,
 give a small contribution both to
total and differential cross sections, since they are suppressed
by Sudakov form factors. (for a  detailed  detailed discussion of
the issue in  the review \cite{Dokshitzer:1978hw} . Evidently this
nonperturbative contribution is not enhanced in the  Leading
Logarithmic Approximation).

\par
Importantly, in this kinematical region the hard scattering of four partons from the wave
functions of the colliding hadrons remains the dominant source for four-jet production even
 when the pQCD parton multiplication phenomena are taken into account.

The reason for that is the following. When the two partons from each hadron emerge from the
{\em initial state parton cascades}\/ and then engage into double hard scattering, the resulting differential
 distribution of the final state jets lacks the double back-to-back enhancement
 factor $d\sigma\propto \delta_{13}^{-2}\delta_{24}^{-2}$ which is there in the case
  of two independent hard scatterings.
For the two-parton scattering, the characteristic perturbative enhancement
  $d\sigma\propto \delta^{-2}$ results from a coherent enhancement of the
  amplitude due to integration over a large transverse disk, $\rho^2\sim \delta^{-2}\gg j_t^{-2}$.
  The two partons that originate from a perturbative splitting form a relatively compact system in
  the impact parameter space, so that the double hard interaction of such pairs produces only a
  single perturbative enhancement factor, $(\vec{\delta}_{13}+\vec{\delta}_{24})^{-2}$, which
  does not favor the back-to-back dijet kinematics \eqref{eq:kinemo}. The distribution of four
   jets so produced is much more isotropic and can be suppressed by choosing proper kinematical cuts.

\par So, the aim of this section is to consider
the four-jet production in the hard collisions of {\em four}\/
initial state partons. We show that the cross section of
back-to-back dijet production is calculable in terms of new
nonperturbative objects --- the double-parton  Generalized Parton
Distributions 
(DPGPDs).
The properties of the 
DPGPDs can be rigorously studied within QCD. In particular, we
report here the derivation of the geometric picture for multiple
parton collisions in the impact parameter space. Up till now, this
picture was being used based on a semi-intuitive reasoning
 \cite{DelFabbro:1999tf,Mekhfi:1983az,Paver:1984ux,Sjostrand:ys,Bahr:2008zr,DelFabbro:2000ds,Accardi:2000ry,Frankfurt:bh,Frankfurt:dq,Frankfurt:2008ve,Rogers:2008qf,Domdey:2009ly,Diehl:2010dr}.

In  the  kinematical domain \eqref{eq:kinemo} the direct calculation of the light cone Feynman diagrams (momenta of the  partons in the initial  and
final states
are shown in Fig.~\ref{kin})
using the separation of hard and soft scales shows that the four
$\to $ four cross section for the collisions of hadrons "a" and
"b" has the form:
\begin{eqnarray}\label{eq:main_form}
\sigma_4 &=& \int \frac{d^2\overrightarrow{\Delta}}{(2\pi)^2}\int dx_1\int  dx_2\int dx_3\int  dx_4  \nonumber\\[10pt]
&\times &D_a(x_1,x_2,p_1^2,p_2^2, \overrightarrow{\Delta}) D_b(x_3,x_4,p_1^2,p_2^2,-\overrightarrow{\Delta})
\nonumber\\[10pt]
&\times&\displaystyle{\frac{d\sigma^{13}}{d\hat t_1} \frac{d\sigma^{24}}{d\hat t_2}} d\hat t_1d\hat t_2.    
\label{b1}
\end{eqnarray}
Here $D_\alpha(x_1,x_2,p_1^2,p_2^2,  \overrightarrow{\Delta})$
are the new
DPGPDs
 for hadrons "a" and
"b" defined below.  (In the following we will consider the case of
$pp$ collisions and omit the subscript $\alpha$. Summing over
collisions of various types of partons is implied.  In practice
however we will keep hard scattering of  gluons only since it
gives the dominant contribution.). Remember that the light cone
fractions $x_i$ are actually fixed by final jet parameters and
energy momentum constraints.

With account of the radiative pQCD effects, in full analogy with the "DDT formula" for two-body collisions,
the differential distribution \eqref{eq:main_form} acquires Sudakov form factors~\cite{Dokshitzer:1978hw,Catani:1988vd} depending
on the logarithms of the large ratios of scales, $j_t^2/\delta^2$, and the GPDs become scale dependent:
$p_1^2\sim \delta_{13}^2$, $p_2^2\sim\delta_{24}^2$.  It should be mentioned that the structure of the final
formula depends on what one actually measures in the experiment --- whether  energetic single particles with
large transverse momenta in the final state or "jets" --- and on how the jets are precisely defined.
A more detailed account of the pQCD effects will be given in a future publication.

For brevity we will not write explicitly the virtuality scales of
the DPGPDs  and will use the form:
$D(x_1,x_2,\overrightarrow{\Delta} )$.

 \noindent
Note that these distributions depend on the new transverse vector  $\overrightarrow{\Delta}$ that
is equal to the difference of the momenta of partons from the wave function of the colliding hadron
in the amplitude and the amplitude conjugated. Such dependence arises because the difference of parton transverse
momenta within
the parton pair is not conserved. The integration limits in $x_i, \hat t$ are subject to
standard limits determined by kinematic cuts.
\par
Within the parton model approximation the cross section has the form:
 \beq
 \sigma_4= \sigma_1\sigma_2/\pi R^2_{\rm int},
 \label{b2}
 \eeq
where $\sigma_1$ and $\sigma_2$ are the cross sections of two independent hard binary parton interactions. The factor $\pi R^2_{\rm int}$ characterizes the transverse area occupied by the
partons participating in the hard collision. (In  the experimental  \cite{Abe:1997xk,Abazov:2009gc}  and some of the theoretical  papers this factor was denoted  as an effective
cross section. Our Eq.~\ref{b3} below  shows that such wording  is
not satisfactory   since $\pi R^2_{\rm int}$ does not have the  meaning of the
interaction cross section.) The data
\cite{Abe:1997xk,Abazov:2009gc} indicates that
 $\pi R^2_{\rm int}$ is practically constant in the kinematical range studied at the Tevatron.

 Eq.~\ref{b1} leads to  the general model
independent expression for
 \beq
\frac{1}{\pi R^2_{\rm int}}=\int
 \frac{d^2\overrightarrow{\Delta}}{(2\pi)^2}\frac{D(x_1,x_2,-\overrightarrow{\Delta})
 D(x_1,x_2,\overrightarrow{\Delta})}{D(x_1)D(x_2)D(x_3)D(x_4)},
 \label{b3}
 \eeq
 in terms of DPGPDs. Here $D(x_i)$ are the
 corresponding structure functions.

\par The
DPGPDs are expressed through the light cone wave functions of the
colliding  hadrons as follows. Suppose that in a four $\to$ four
process  the two partons in the nucleon in the initial state wave
function have the transverse momenta
$\overrightarrow{k_1},\overrightarrow{k_2}$. Then  in the
conjugated wave function they will have the momenta
$\overrightarrow{k_1}+\overrightarrow{\Delta},\overrightarrow{k_2}-\overrightarrow{\Delta}$.
This is because only sum of parton transverse momenta but not the
difference is  conserved.

The relevant DPGPDs are:
\begin{eqnarray}
 &D&(x_1,x_2,p^2_1,p^2_2,\overrightarrow{\Delta})=\sum_{n=3}^{\infty}\int
\frac{d^2k_1}{(2\pi)^2}\frac{d^2k_2}{(2\pi)^2}\theta
(p_1^2-k_1^2)\nonumber\\[10pt]
&\times& \theta (p_2^2-k_2^2)\int \prod_{i\ne
1,2}\frac{d^2k_i}{(2\pi)^2}\int^1_0\prod_{i\ne
1,2} dx_i\nonumber\\[10pt]
&\times& (\psi_n (x_1,\vec k_1,x_2,\vec k_2,.,\vec k_i,x_i..)
\nonumber\\[10pt]
&\times&\psi_n^+(x_1,\overrightarrow{k_1}+\overrightarrow{\Delta},x_2,\overrightarrow{k_2}
-\overrightarrow{\Delta},x_3, \vec k_3,...) + \, h.c.)
\nonumber\\[10pt]
&\times&  (2\pi)^3\delta( \sum_{i=1}^{i_=n} x_i-1)\delta (\sum_{i=1}^{i=n} \vec
k_i).
\label{b4}
\end{eqnarray}
Note that this distribution is diagonal in the space of all
partons except the two partons involved
 in the collision. Here $\psi$ is the parton
wave function normalized to one in a usual way. An appropriate
summation over color and Lorentz indices is implied. In the case
of kinematics  $1\gg x_1\ge x_2$ we expect only distributions
without the spin flip  to be important. \par Let us stress that it
follows from the above formulae that in the impact parameter space  these GPDs have a
 probabilistic interpretation. In particular  these DPGPD are positively definite  in the impact
   parameter space, cf.\ Eq.~\ref{l1}.
Note that in the same way one can introduce the N-particle GPD, $G_N$,
which can be probed in the production of N pairs of jets.
 In this case  the   first N arguments $k_i$ in Eq.~\ref{b4}  are shifted by
$\overrightarrow{\Delta_i}$ subject to the  constraint $\sum_i \overrightarrow{\Delta_i}=0$. So the cross
section is proportional to
\begin{eqnarray}
\sigma_{2N}&\propto& \int \prod_{i=1}^{i=N}{d\overrightarrow{\Delta}_i\over (2\pi)^2}
D_a(\overrightarrow{\Delta}_1,...\overrightarrow{\Delta}_N)  \nonumber\\[10pt]
 &\times& D_b(\overrightarrow{\Delta}_1,...\overrightarrow{\Delta}_N)
\delta(\sum_{i=1}^{i=N}\overrightarrow{\Delta}_i).
\end{eqnarray}

\par These GPDs can be easily rewritten in the form of the matrix elements  of
the operator product.
For example:
\begin{eqnarray}  D(\Delta )&=& <N\vert \int
d^4x_1d^4x_2d^4x_3  \nonumber\\[10pt]
&\times &G^a_{i+}(x_1)G^b_{j+}(x_2)G^a_{i+}(x_3)G^b_{j+}(x_4)\nonumber\\[10pt]
&\times&\exp(ip_1^+(x_1-x_3)^-+ip_2^+(x_2-x_4)^-  \nonumber\\[10pt]
&+&
 i\vec\Delta_t (\vec{x}_4-\vec{x}_3)_t) \vert N>, 
\label{1b}\end{eqnarray} calculated at the virtualities
$p_1^2,p_2^2$ at fixed $\overrightarrow{\Delta}$. Here we gave an
example for the most relevant case of gluons without a flip in
color and spin spaces. In general a number of distributions can be
written, depending on different contractions of transverse Lorentz
indices and color indices. The classification of the relevant
distributions is the same as the classification of the
quasipartonic operators in ref. \cite{Bukhvostov:1985rn}. Note that the
presence of the transverse external parameter $\vec\Delta$ does
not change the classification, since the corresponding new
structures will be strongly suppressed at high energies. we wrote
the operator expression in the light cone gauge. In arbitrary
gauge we
 shall need Wilson loop W(C) connecting points with contracted color indices
\par In the approximation of
uncorrelated
partons it follows from  Eq.~\ref{b4} that
\begin{equation}D(x_1,x_2,p_1^2, p_2^2, \vec\Delta
)= G(x_1, p_1^2, \vec\Delta)G(x_2, p_2^2, \vec\Delta ),\end{equation}
 where
$G(x,\overrightarrow{\Delta} )$ are conventional one-particle
GPDs. These GPDs can be approximated as  $G_N(x,Q^2, \vec \Delta)=$
$G_N(x,Q^2)$$F_{2g}(\Delta)$, where $F_{2g}(\Delta)$ is the two-gluon
form factor of
 the nucleon extracted from hard exclusive vector meson
 production
  (we suppress here the dependence of $ F_{2g}$ on x)
  \cite{Frankfurt:2002ka} and $G_N(x,Q^2)$ conventional parton distribution of a nucleon.
(Here $Q^2$ is the virtuality due to the radiation, cf.\ discussion
after
 Eq.~\ref{b1}.) Thus :
\beq \frac{1}{\pi R^2_{\rm int}}=\int
\frac{d^2\Delta}{(2\pi)^2}F_{2g}^4(\Delta)=\frac{m^2_g}{28\pi}.\label{b6}\eeq
Here at the last step we used the dipole fit
$F_{2g}(\Delta)=1/(\Delta^2/m^2_g+1)^2$ to the two-gluon form
factor. Using the transverse gluon radius of the nucleon we obtain
\beq R^2_{\rm int}=7/2r^2_g,~~~~~r^2_g/4=d F_{2g}(t)/dt_{t=0}.
\label{b7} \eeq
 This result coincides with the one for the area $\pi R^2_{\rm int}$ obtained earlier in \cite{Frankfurt:bh,Frankfurt:dq}
 using the geometric picture in the impact parameter space.
That derivation
involved
taking  Fourier transform of the two-gluon form factor and calculating  a rather
complicated six-dimensional integral which could potentially lead to large numerical uncertainties.
The form of Eq.~\ref{b7} clearly indicates that numerical
uncertainties are small.
\par It was emphasized in \cite{Frankfurt:bh,Frankfurt:dq} that the experiments on
four-jet production  report a smaller value of $\pi R^2_{\rm int}$
as compared to the one obtained above in the independent particle
approximation (though the issue of how well the contribution of the $2\to 4$ processes was  subtracted
 still remains, cf. discussion in the beginning of this section). 
It is at least a factor of two smaller ---
 that is a four-jet cross section is a factor of two larger ---
than  Eq.~\ref{b7} gives. (The GPDs for sea quarks appear to decrease
with $\Delta$ somewhat faster, resulting in a smaller $1/\pi R^2_{int}$, see discussion in \cite{Strikman:2009bd}.)
\par  It  follows from Eq.~\ref{b3} that the value of
$R^2_{\rm int}$ is determined by the range of integration over
$\Delta$. Hence the characteristic  $\Delta $ in the integral
measures the  effective distance between the parton pairs (which
in principle may differ for different flavor combinations).
 According to the above
evaluation within the independent parton approximation the
integral for $1/R^2_{\rm int}$ is dominated by small
$\Delta^2\sim 0.1 \,m^2_g$.
 The contribution of large $\Delta$ is suppressed
by the two-gluon form factor of a nucleon. This reasoning indicates
the important role of inter-parton correlations.  In other words,
the integral
 over $\Delta$ is effectively cut off  by a scale of the nonperturbative correlations.
  Such correlations
 naturally arise in nonperturbative QCD regime
in a number of nucleon models, such as constituent quark model
(gluon cloud around constituent quark) \cite{Frankfurt:bh,Frankfurt:dq}, or
string model (gluon structure of string).
The detailed analysis of the additional correlations due to the hard-- soft interplay will be reported elsewhere.

 Let us now show that results presented here lead to the intuitive
geometric picture
in the impact parameter space mentioned above
 \cite{DelFabbro:1999tf,Mekhfi:1983az,Paver:1984ux,Sjostrand:ys,Bahr:2008zr,DelFabbro:2000ds,Accardi:2000ry,Frankfurt:bh,Frankfurt:dq,Frankfurt:2008ve,Rogers:2008qf,Domdey:2009ly,Diehl:2010dr}.

 The first step is to make transformation into coordinate space
i.e., to make Fourier transform from variables $k_i$ in
Eq.~\ref{b4} to coordinates $b_i$. Performing integration over
$k_i$ we obtain that transverse coordinates of partons  in the
amplitude and the amplitude conjugated are equal $\rho_i=\rho_f$.
 In the calculation we use the fact  that upper limit of integration over $k^2_t$
is very large compared with the inverse hadron size. Next step is to perform integration over $\Delta$  which produces
$\delta(\vec \rho_1-\vec\rho_2-\vec \rho_3+\vec \rho_4)=\int d^2B \delta (\vec \rho_1-\vec \rho_3-\vec B)
 \delta(\vec \rho_2-\vec \rho_4-\vec B)$.
\par  The delta functions express the fact that within the accuracy $1/p_t$ where $p_t$ is the hard scale,
the interactions of partons
from different nucleons occur at the same point.
$\vec  B$ is
the relative impact parameter of two  nucleons.
\par
\begin{figure}[h]  
   \centering
   \includegraphics[width=0.3\textwidth]{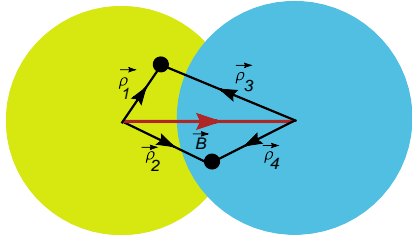}
   \caption{Geometry of two hard collisions in impact parameter picture.}
    \label{bdistr}
 \end{figure}

The expression for the cross  section in the impact parameter  space has the form
 which corresponds to geometry of Fig.\ref{bdistr}
\begin{eqnarray}
\sigma_4&=&\int d^2B
d^2\rho_1d^2\rho_2d^2\rho_3d^2\rho_4D(x_1,x_2,\vec \rho_1,\vec \rho_2)\nonumber\\[10pt]
&\times&D(x_3,x_4,\vec \rho_3,\vec \rho_4)
\delta (\vec b_1+(\vec B-\vec b_3))\delta (\vec b_2+(\vec B-\vec b_4))=\nonumber\\[10pt]
& = &\int d^2B
d^2\rho_1d^2\rho_2 D(x_1,x_2,\vec \rho_1,\vec \rho_2)  \nonumber\\[10pt]
&\times& D(x_3,x_4,-\vec B+\vec \rho_1,-\vec B+\vec \rho_2). 
\label{l1} \end{eqnarray} Here the DPGPD in the impact parameter
space representation  is
 given by
\begin{eqnarray}
&D&(x_1,x_2,\vec \rho_1,\vec \rho_2)= \nonumber\\[10pt]
& = & \sum^{n=\infty}_{n=3}\int \prod_{i\ge 3}^{i=n}\left[dx_i d^2\rho_i\right] \psi_n(x_1,\vec \rho_1,x_2,\vec \rho_2,...x_i,\vec \rho_i,) \nonumber\\[10pt]
&\times &\psi_n^+(x_1,\vec \rho_1,x_2,\vec \rho_2,...,x_i,\vec \rho_i,...)\delta(\sum_{i=1}^{i=n}  x_i\vec\rho_i).
\label{o3}
\end{eqnarray}
where the delta function expresses the center of mass constraint
$\sum_{i=1}^{i=n} x_i\vec\rho_i
=0 $. This is analogous to the case of single parton GPDs, see  \cite{Diehl:2002he,Diehl:2003ny}.
 The functions
$\psi (x_1,\vec \rho_1,x_2,\vec \rho_2,...)$ are just the Fourier
transforms in the impact parameter space of the light cone wave
functions and are given by
\begin{eqnarray}
&\psi_n& (x_1,\vec \rho_1,x_2,\vec \rho_2,...)=\int \prod^{i=n}_{i=1}\frac{d^2k_i}{(2\pi)^2}\exp(i\sum_{i=1}^{i=n}\vec k_i\vec \rho_i)\nonumber\\[10pt]
&\times&\psi_n(x_1,\vec
k_1,x_2,\vec k_2,..)(2\pi)^2
\delta (\sum \vec k_i).\label{o4}
\end{eqnarray}

\par Thus the GPD defined in  Eq.~\ref{b4}  is equivalent to  the representation for cross
 section that indeed corresponds to the
simple geometrical picture, but instead of a triple integral we now  have  an  integral over
one  momentum  $\Delta$. Moreover,  to  determinate the  cross  section
we need to know the
$D(\Delta)$.
The GPD defined in  Eq.~\ref{b4} is useful for calculation of many different processes.
At the same time the knowledge of the full double GPD is necessary for complete description of
 events with a double jet trigger since the pedestal strongly depends on the impact parameter $\vec B$ \cite{Frankfurt:bh,Frankfurt:dq}.

\par Let us stress that this picture is a natural generalization of the
correspondence between momentum representation and geometric
picture for a conventional case of two $\to$ two collisions. Indeed in this case it is easy to
 see that the cross section in the
momentum representation \beq \sigma_2=\int
f(x_1,p^2)f(x_2,p^2)\frac{d\sigma^h}{d\hat t}d\hat t\label{s1}\eeq
has a simple geometric representation
\beq \sigma_2=\int
d^2\rho_1d^2B f(x_1,\vec \rho_1,p^2)f(x_2,\vec
B-\vec{\rho}_1,p^2)\frac{d\sigma^h}{d\hat t}d\hat t,\label{s2}\eeq
 where
$f(x,\vec \rho,p^2)= \psi^+(x,\vec \rho,p^2)\psi(x,\vec \rho,p^2)$
and $\psi(\vec \rho,p^2)$ is the Fourier transform of the light cone
wave function defined above.
\par Let us now summarize our results. We have argued that there exists
the kinematical domain  where the four $\to$ four hard parton collisions
form the dominant mechanism of four-jet production. In this region
we calculated the cross section, see Eqs.~\ref{b1}-\ref{b3}  and
found that it can be expressed through new two particle GPDs (see
Eq.~\ref{b4}), expressed through light cone wave functions. These
GPDs depend on a transverse vector $\vec \Delta$ that measures the
transverse distance  within the parton pairs. (Equivalent
expressions for these GPDs can be easily  given in terms of the operator products.)
In the impact parameter space we derived the
widely used intuitive geometric picture. We argued   that  the
enhancement of a four-jet cross section is  due to
short range correlations in the hadron, as determined by the range
of integral of $\Delta$. The contribution of perturbative
correlations in the appropriate kinematic domain is suppressed.
The detailed study of the interplay of the contribution of hard/soft
correlations will be reported elsewhere.
\par It was argued  recently  \cite{Gaunt:2009re,Gaunt:2010pi}%
, cf.~\ref{gaunt}, %
that the cross
section can be expressed in terms of two parton distribution
functions. Our analysis indicates that  a more detailed treatment
of the QCD evolution effects   is necessary. We found that it is
necessary to introduce the new 2-particle  DPGPDs which depend on
additional parameter $\Delta$.  The parameter $\Delta$ expresses
the fact that the difference in transverse components of the
parton momenta is not conserved and therefore different in $\left|
in \right>$ and $\left<out\right|$ states in the double hard
collisions.

\graphicspath{{gaunt/figs/}}

\newcommand{\bbbar}{b\bar{b}}
\newcommand{\Wp}{W^+}
\newcommand{\Wm}{W^-}
\newcommand{\Zgam}{Z(\gamma^*)}
\newcommand{\Wpm}{W^{\pm}}
\newcommand{\vect}[1]{\boldsymbol{#1}}

\contribution{Double parton distribution functions}
{Contributing author: J. R. Gaunt, W.~J. Stirling}
\label{gaunt}


If we make only the assumption that the hard processes A and B may be factorised, then we may write the cross section for DPS in very general terms as follows: 
\begin{eqnarray} \label{DPSXsec1}
\sigma^D_{(A,B)} = \dfrac{m}{2}\sum_{i,j,k,l}\int\Gamma_{ij}(x_1,x_2,\vect{b};Q_A,Q_B)
\hat{\sigma}^A_{ik}(x_1,x_1')\hat{\sigma}^B_{jl}(x_2,x_2') \\ \nonumber
\times \Gamma_{kl}(x_1',x_2',\vect{b};Q_A,Q_B) 
dx_1dx_2dx_1'dx_2'd^2\vect{b}
\end{eqnarray}

The structure of the DPS cross section formula is similar to that for single parton scattering (SPS), in the sense that it too is expressed in terms of parton level cross sections $\hat{\sigma}$ multiplied by parton distributions $\Gamma$. However, whilst the parton level cross sections used in \eqref{DPSXsec1} are the same as those used for SPS, the parton distributions of \eqref{DPSXsec1} are significantly more complex than the single PDFs. They are the generalised two-parton PDFs (or, using the language of \cite{Blok:2010ge}%
and Section~\ref{blok}, %
the two-parton GPDs). The quantity $\Gamma_{ij}(x_1,x_2,b;Q_A,Q_B)$ can be interpreted as the probability to find a pair of partons in the proton which have flavours $i$ and $j$, longitudinal momenta $x_1$ and $x_2$, and are separated in impact parameter by $\vect{b}$, when the partons are probed at scales $Q_A$ and $Q_B$ respectively. $m$ is a symmetry factor that equals $1$
if $A=B$ and $2$ otherwise. 

Two further assumptions are often applied to \eqref{DPSXsec1}. First, it is assumed that the two-parton GPD may be decomposed into a (typically flavour independent) transverse piece, and a longitudinal piece:
\begin{equation} \label{dPDFlt}
\Gamma_{ij}(x_1,x_2,\vect{b};Q_A,Q_B) \simeq F(\vect{b}) D_p^{ij}(x_1,x_2;Q_A,Q_B)
\end{equation}
In this case the DPS cross section reduces to:
\begin{align} \label{DPSXsec2}
\sigma^{DPS}_{(A,B)} &= \dfrac{m}{2\sigma_{\rm eff}} \sum_{i,j,k,l}
\int dx_1dx_2dx_1'dx_2' \; D^{ij}_p(x_1,x_2;Q_A,Q_B)\;
D^{kl}_p(x_1',x_2';Q_A,Q_B)
\\ \nonumber
&\times \hat{\sigma}^A_{ik}(x_1,x_1')\hat{\sigma}^B_{jl}(x_2,x_2')
\end{align}
with $\sigma_{eff}$ a constant. Second, it is assumed that $D^{ij}_p(x_1,x_2;Q_A,Q_B)$ may be written as a product of single PDFs (sPDFs):
\begin{equation} \label{dPDFfact}
D^{ij}_p(x_1,x_2;Q_A,Q_B) \simeq D^{i}_p(x_1;Q_A)D^{j}_p(x_2;Q_B)
\end{equation}

The analysis of $\gamma + 3$ jet events performed by the CDF collaboration \cite{Abe:1997xk} indicates that \eqref{dPDFfact} approximately holds for sea partons at moderately low $x$. However, it is clear even from elementary considerations that this assumption must be violated on some level - for example, the left hand side of \eqref{dPDFfact} must go to zero at the kinematic bound $x_1+x_2=1$ whilst the product of sPDFs is finite along this line. One might ask as to whether theory can make any predictions on the size of the deviations from \eqref{dPDFfact}.

In 1982, Shelest, Snigirev and Zinovjev derived a `double DGLAP' equation dictating the LO scaling violations of a quantity $D^{ij}_p(x_1,x_2;Q)$ which we shall refer to as the double PDF or dPDF \cite{Shelest:1982dg}. Snigirev identified this quantity with the factorised longitudinal piece of the two-parton GPD for the case in which the two scales are equal ($Q_A=Q_B=Q$) \cite{Snigirev:2003cq}. The equation reads as follows:
\begin{align} \label{dbDGLAP}
Q^2\dfrac{dD^{{j_1}{j_2}}_p(x_1,x_2;Q)}{dQ^2} = \dfrac{\alpha_s(Q^2)}{2\pi} \Biggl[
\sum_{j'_1}\int_{x_1}^{1-x_2}\dfrac{dx'_1}{x'_1}D^{{j'_1}{j_2}}_p(x'_1,x_2;Q)P_{j'_1
\to j_1}\left(\dfrac{x_1}{x'_1}\right)
\nonumber\\
+\sum_{j'_2}\int_{x_2}^{1-x_1}\dfrac{dx'_2}{x'_2}D^{{j_1}{j'_2}}_p(x_1,x'_2;Q)P_{j'_2
\to j_2}\left(\dfrac{x_2}{x'_2}\right)
\nonumber\\
+\sum_{j'}D^{j'}_p(x_1+x_2;Q)\dfrac{1}{x_1+x_2}P_{j' \to j_1
j_2}\left(\dfrac{x_1}{x_1+x_2}\right) \Biggr]
\end{align}

The first two terms on the right hand side of \eqref{dbDGLAP} are associated with changes in the dPDF due to independent branching processes -- processes in which there are a pair of partons, one of which has the appropriate $x$ and flavour, and the other of which splits, either giving rise to the other parton of the appropriate $x$ and flavour, or removing it. The most interesting term is the final inhomogeneous term, which represents the increase in the dPDF due to a single parton with momentum fraction $x_1+x_2$ splitting into a pair with the appropriate $x$ values and flavours. We call this the `sPDF feed term' for obvious reasons. The functions $P_{j\to j_1j_2}(x)$ that appear in this term are known as the $1\to 2$ splitting functions, and may be obtained trivially at LO from the real splitting parts of the usual splitting functions.

 An important prediction of the double DGLAP equation is that even if the dPDF may be taken to be a product of single PDFs at some particular scale $Q_0$, then at any other scale the dPDFs deviate from factorised forms (with more deviation the further away one moves from $Q_0$). 

More recently \cite{Gaunt:2009re}, we demonstrated that certain equalities are preserved by the double DGLAP equation, provided that they hold at some initial scale $Q_0$. By comparing these equations to an equation in conditional probability, we interpreted these equalities as the number and momentum sum rules for the dPDFs, and argued that they should hold at the initial scale $Q_0$. The sum rules are:
\begin{align} \label{mtmsum}
 \text{ Momentum Sum Rule:} &&& \sum_{j_1}\int_0^{1-x_2} dx_1 x_1 D^{{j_1}{j_2}}_p(x_1,x_2;Q) = (1-x_2)D_p^{j_2}(x_2;Q)\\
 \text{ Number Sum Rule:} &&& \int_0^{1-x_2}dx_1D_p^{j_{1v}j_2}(x_1,x_2;Q)= \begin{cases}
N_{j_{1v}} D_p^{j_2}(x_2;Q) & \text{when $j_2 \ne j_1$ or $\overline{j}_1$} \\
(N_{j_{1v}}-1) D_p^{j_2}(x_2;Q) & \text{when $j_2 = j_1$} \\
(N_{j_{1v}}+1) D_p^{j_2}(x_2;Q) & \text{when $j_2 = \overline{j}_1$}
\end{cases}
\label{numsum}
\end{align}

The symbol $j_{1v} \equiv j_1-\overline{j}_1$ ($j_1\ne g$), and $N_{j_{1v}}$ is the number of `valence' $j_1$ quarks in the proton. The first sum rule is a simple statement of the fact that if one observes a parton with momentum fraction $x_2$ in the proton, the momentum fractions of all other partons must add up to $1-x_2$. The second rule states that if one observes a parton with flavour $j$ in the proton, the number of partons of flavour $j$ elsewhere in the proton must be reduced by one (we use the term `number effects' to describe this simple phenomenon). The sum rules give us some information about the deviations of the dPDFs from factorised forms at any scale -- certainly, factorised forms do not obey the relations \eqref{mtmsum} and \eqref{numsum}.

In this contribution, we principally wish to discuss the development of our publicly available set of LO dPDFs -- the GS09 dPDFs -- which have been constructed incorporating pQCD evolution effects and sum rule constraints \cite{Gaunt:2009re}. This discussion may be found in Section \ref{sec:GS09}. However, we should also like to mention some theoretical problems that we have recently uncovered in the aforementioned `dPDF framework' for calculating proton-proton cross sections -- this is covered in Section \ref{sec:dPDFproblems}.

\subsubsection{The GS09 dPDFs} \label{sec:GS09}

The GS09 dPDF package comprises a grid of dPDF values spanning the ranges $10^{-6} < x_1 < 1$, 
$10^{-6} < x_2 < 1$, $1 \text{~GeV}^2 < Q^2 < 10^9 \text{~GeV}^2$, which is available along
with interpolation code from HepForge \cite{HepForgePage}. It has been obtained by constructing
inputs that approximately satisfy the sum rules at $Q_0 = 1 \rm{~GeV}$, and then numerically
evolving these inputs up to higher scales according to the dDGLAP equation. The sPDF set to 
which we have chosen our dPDF set to correspond is (a slightly modified version of) the MSTW2008LO 
set \cite{Martin:2009iq}.

Given the paucity of experimental data regarding the dPDFs, and in accordance with simple arguments
and the CDF results, we base our inputs on factorised products of MSTW2008LO sPDFs. However, we
modify these basic forms in several ways to ensure that the input dPDFs approximately satisfy the 
sum rules. 

First, all of the dPDFs are multiplied by a factor $\rho^{ij}(x_1,x_2)$ which is designed to take account of phase space effects. This factor should ensure the appropriate behaviour of the dPDFs near the kinematic boundary $x_1+x_2=1$ -- namely, a smooth decrease to zero. In previous studies, universal phase factors such as $(1-x_1-x_2)$ and $(1-x_1-x_2)^2$ were used -- however, with the benefit of knowledge of the momentum sum rules, we can see that neither of these options is fully satisfactory. Double PDFs including these factors badly violate the momentum sum rules along the lines $x_1=0$ and $x_2=0$. In these regions, a phase factor of approximately $1$ would be more sensible (to reflect the fact that, ignoring number effects, removing a very low momentum parton does not strongly affect the probability of finding any other parton). Indeed, factorised forms satisfy the sum rules perfectly along the lines $x_1=0$ and $x_2=0$.

It was discovered that the following form for $\rho^{ij}$ gives rise to inputs which satisfy the momentum sum rules (plus appropriate number sum rules) well:
\begin{align}
\rho^{ij}(x_1,x_2)=(1-x_1-x_2)^2(1-x_1)^{-2-\alpha(j)}(1-x_2)^{-2-\alpha(i)} 
\end{align}
$\alpha(i)$ is $0$ if $i$ is a sea parton, and $0.5$ if it is a valence parton. The latter two factors on the right hand side are included to compensate the decrease of the $(1-x_1-x_2)^2$ factor along the lines $x_1=0$ and $x_2=0$.

We recall that there are only a finite number of valence quarks in the proton, as opposed to
an infinite number of sea quarks and gluons. Number effects are therefore
most significant in the context of valence quarks, and on this basis we have chosen to only 
take account of valence number effects in our inputs. This is done by dividing the $u_vu_v$ part of any
dPDF by two, and completely subtracting the $d_vd_v$ part. The reasoning behind
this is that removing one up valence quark essentially halves the probability 
to find another, whilst there is no chance of finding two valence down quarks
in the proton.

Finally, we have added extra terms to input distributions whose flavour indices contain $j\bar{j}$ 
combinations to take account of so-called `$j\bar{j}$ correlations'. These are essentially related 
to sea parton number effects, although they can alternatively be thought of as arising during 
evolution from some lower scale to $Q_0$ via $g \to j\bar{j}$ splittings. It is important to 
include these terms in the equal flavour valence-valence ($j_vj_v$) inputs since the
`$j\bar{j}$ correlation' term is much larger for these than the quasi-factorised piece at low $x$.

With these adjustments, our dPDF inputs satisfy all sum rules to better than $25\%$ accuracy for 
$x \lesssim 0.8$ (in the normal `double human' basis). 

\subsubsection{Is the proton-proton DPS cross section describable in terms of dPDFs?} \label{sec:dPDFproblems}

\begin{figure}
\centering
\includegraphics[scale=0.6]{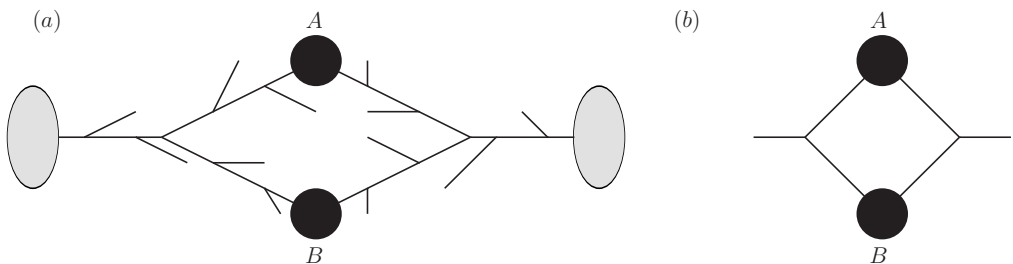}
\caption{\label{fig:DPSloops} {\bf (a):} A diagram that apparently contributes to the leading order DPS cross section according to the `dPDF framework' described in Section \protect\ref{sec:intro}. {\bf (b):} If the dPDF framework of Section \protect\ref{sec:intro} is valid, then it should be possible to derive the $1\to2$ splitting functions by examining the IR singular parts of this parton-level diagram.}
\end{figure}

Let us consider the calculation of the cross section for a particular (equal-scale) DPS process using the framework outlined at the beginning of this section -- that is, folding parton level cross sections together with the dPDFs of \eqref{dbDGLAP} according to \eqref{DPSXsec2}. Due to the presence of the sPDF feed term in the dDGLAP equation, there will be a contribution to the leading order cross section corresponding to diagrams like figure \ref{fig:DPSloops}(a) in this calculation. For figure \ref{fig:DPSloops}(a) to make a contribution to the leading order DPS cross section, it must be the case that there is a $\int dk_{\perp}/k_{\perp}$ singularity associated with every branching, even at the vertices at which the branches on either side of the diagram split into two. Only then will the branches on either side of the diagram contribute a leading $(\alpha_S \log(Q^2))^N$ factor to the cross section.

If the dPDF framework is valid, then it should be possible to extract the $1 \to 2$ splitting functions from a distinct part of figure \ref{fig:DPSloops}(b) that is proportional to $(\alpha_S \log(Q^2))^2/\sigma_{eff}$, just as it is possible to extract (say) the $P_{qq}$ splitting function from a distinct part of the Drell-Yan diagram with one ISR gluon that is proportional to $(\alpha_S \log(Q^2))$ \cite{Ellis:1996ws}. However, in \cite{Gaunt:2011xd} (see also \cite{Gaunt:2011xu}) we demonstrated that for any Standard Model loop with the topology of \ref{fig:DPSloops}(b), there is no natural part of the cross section expression corresponding to the loop that has the required structure. We also showed more generally that there is no natural part of figure \ref{fig:DPSloops}(a) that is proportional to $(\alpha_S\log(Q^2))^N/\sigma_{eff}$.

There is a further, more obvious, problem with the dPDF framework that is actually related to the first problem above. In this framework, the same transverse separation profile is effectively assigned to every part of the dPDF, with the width of the profile being of the same order as the proton radius. This seems sensible for those parts of the dPDF that have arisen as a result of independent branchings. On the other hand, pairs of partons arising as a result of perturbative parton splittings do not `know' about the size of the proton -- how can it be appropriate to assign an effective transverse area of approximately the size of the proton to these pairs?

There is clearly a flaw in the dPDF framework. The root of the problem seems to be the assumption that the two-parton GPD may be decomposed into longitudinal and transverse pieces (as one might have guessed from the previous paragraph). Parton pairs arising from perturbative splittings have a singular transverse separation profile, which differs significantly from the $\vect{b}$ profile of partons produced by independent branching \cite{Diehl:2010aa}. 

The dPDF introduced in Section 1 actually seems to be the integral of the two-parton GPD over $\vect{b}$. As a result it is directly accessed only in DPS processes in which the two particles probing the proton are uncorrelated in transverse space (e.g. the two-nucleon contribution to proton-heavy nucleus DPS \cite{Strikman:2001gz,Cattaruzza:2005nv}). It is nonetheless interesting to ask how `wrong' it is numerically to use \eqref{DPSXsec2} plus GS09 dPDFs to calculate DPS cross sections. At present, we can be reasonably certain that the contribution associated with the sPDF feed parts of both dPDFs being multiplied together should not be included. However, numerically one finds that the accumulated contribution of the sPDF feed term to any dPDF is about $10\%$ at low $x$ \cite{Gaunt:2009re} -- therefore this contribution only represents roughly $10\% \times 10\% = 1\%$ of the cross section. To definitively answer this question of how wrong the dPDF framework is, we must discover the correct framework for calculating proton-proton DPS -- we are currently working on this.

\newcommand{\bbeta}{\mbox{\boldmath$\beta$}}
\newcommand{\esp}[1]{\, e^{\,\,\textstyle {#1}}}
\graphicspath{{treleani/figs/}}

\contribution{MPI: General Features and Consistency Requirements}
{Contributing authors: G. Calucci and D. Treleani}
\label{treleani}

\subsubsection{Incoherence and MPI}

MPI cross sections are expressed by the incoherent sum of the contributions due to different numbers of interactions\cite{Rogers:2008ua,Rogers:2009ke,Corke:2009tk}. A given final state may be however generated by various competing processes, characterized by different numbers of partonic collisions and the cross section is the result of diagonal and off-diagonal contributions. 
Terms with different numbers of interactions, giving rise to the same final state, populate the final state phase space in a different way and, in the kinematical regions where the contributions to the cross section are similar, important interference effects should be expected.

\begin{figure}[h]
\centering
\includegraphics[width=150mm]{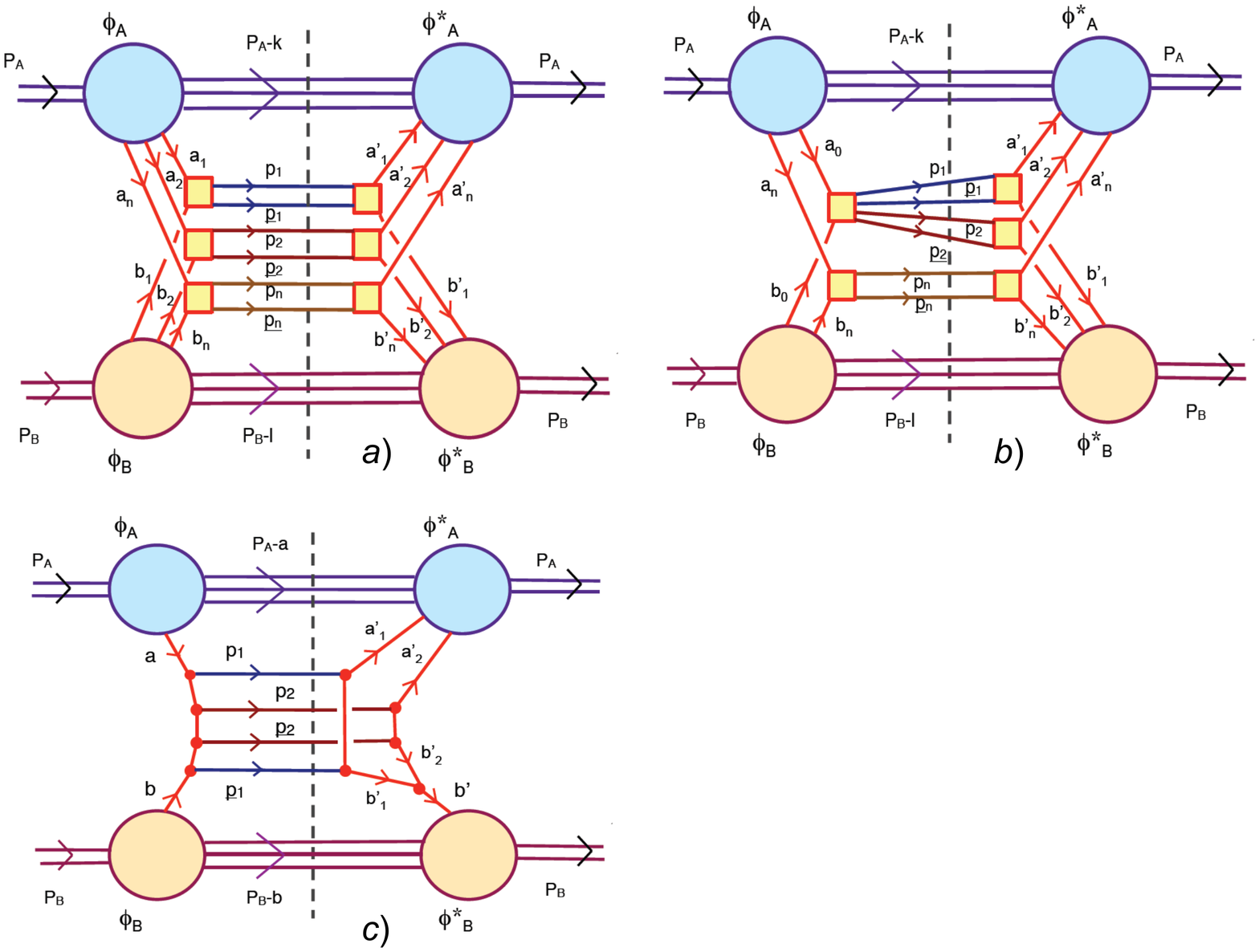}
\caption {{\it a}) A diagonal and {\it b}) a off diagonal contribution to the MPI cross section. {\it c}) an interference and higher order term.}
\end{figure}

A diagonal contribution, corresponding to a term with $n$-partons in the initial state of each interacting hadron, is given by the incoherent super-positions of $n$ disconnected parton interactions, localized in $n$ different points in transverse space, Fig. 1a). On the other hand, the hard component of the interaction, corresponding to the interference between a term with $n$ partons and a term with $n'<n$ partons, is disconnected and localized in no more than $n'$ points in transverse space, Fig. 1b) \cite{Calucci:2009ea}. 
Partons are localized in the hadron by the momenta exchanged in the interaction. When partons are localized inside non overlapping regions, much smaller as compared to the hadron size, they are only connected one with another through soft exchanges and the picture of independent parallel collisions, which characterizes a diagonal contribution to the cross section, is a meaningful one. If, on the contrary, several partons are localized by the interaction inside overlapping regions, much smaller as compared to the hadron size, they are allowed to interact by exchanging momenta of the size of their virtuality. In an interference term more than two initial state partons are localized in the same interaction region\cite{Calucci:2009ea}. Because of the localization in transverse space, the problem of interferences is thus linked to the problem to evaluate the scattering amplitude, at higher orders in the coupling constant and including higher twists in the hadron structure, Fig. 1c).  One may hence argue that interference terms do not represent corrections to the $n$-pairs of partons scattering inclusive cross section. They rather correct the $n'$-pairs of partons, with $n'<n$, scattering inclusive cross section.
 
Given the two very different scales in the process, the hadron size and the large momenta exchanged, the hard component of the interaction may be disconnected, with the different hard parts linked only through soft exchanges and localized in different regions in transverse space. Different MPI terms are hence conveniently understood as the contributions to the final state due to the different disconnected regions, where the hard component of the interaction is localized. In the simplest case, in each different disconnected region the interaction may be evaluated at the lowest order in the coupling constant. In each single partonic collision all transverse momenta balance and a MPI process contributes to the cross section by generating different groups of final state partons where the large transverse momenta are compensated separately.

When MPI are understood in the topological sense described above, different MPI terms, corresponding to different localizations in transverse space, do not interfere and the final cross section is obtained by the simple superposition of the cross sections, due to the contributions of the different topological configurations of the hard component of the interaction.
MPI hence add incoherently. A remarkable consequence is that MPI allow a probabilistic description.

\subsubsection{The Probabilistic Picture of MPI: A Functional Approach, Cancellation of Unitarity Corrections}

As Multiple Parton Interactions add incoherently, the problem may be discussed within a probabilistic framework\cite{Calucci:2008jw}\cite{Calucci:2009sv}. A functional approach is most general. One may start by introducing the probabilities $W_n$, to find the hadron in configurations with $n$-partons
with coordinated $u_1\dots u_n$,  $u_i\equiv(b_i,x_i)$, where $b_i$ are the transverse parton coordinates and $x_i$ the fractional 
momenta, and the multi-parton generating functional, ${\cal Z}$:

\begin{eqnarray}
W_n(u_1\dots u_n),\qquad
{\cal Z}[J]=\sum_n{1\over n!}\int J(u_1)\dots J(u_n)W_n(u_1\dots u_n)
du_1\dots du_n.
\end{eqnarray}

\noindent Probability conservation implies the
normalization condition ${\cal Z}[1]=1$, while the probabilities of the various configurations, $W_n$, are the coefficients of the expansion of ${\cal
Z}[J]$ at $J=0$. The coefficients of the expansion of ${\cal
Z}[J]$ at $J=1$ give the many body densities, i.e. the inclusive
distributions:

\begin{eqnarray}D_1(u)={\delta{\cal Z}\over \delta J(u)}\biggm|_{J=1},\quad
                 \quad\
     D_2(u_1,u_2)={\delta^2{\cal Z}\over \delta J(u_1)\delta J(u_2)}
                  \biggm|_{J=1}\quad\dots\end{eqnarray}

\noindent Correlations, which describe how much the distribution deviates from a
Poissonian, are obtained by the expansion of the logarithm of the generating
functional, ${\cal F}[J]\equiv{\rm ln}{\cal Z}[J]$, at $J=1$:

\begin{eqnarray}{\cal F}[J]=\int D_1(u)[J(u)-1]du+\sum_{n=2}^{\infty}{1\over n!}
\int C_n(u_1&\dots& u_n)\bigl[J(u_1)-1\bigr]\dots\cr
  &\dots&\bigl[J(u_n)-1\bigr]
du_1\dots du_n\end{eqnarray}

\noindent  One has ${\cal F}[1]=0$ and, in the
case $C_n\equiv 0, n\ge 2$, the multi-parton distribution is a Poissonian.

\noindent  Given the multiparton distributions $W_n$, one may express the hard cross section in a functional form:

\begin{eqnarray}\sigma_{hard}=\int d\beta&&\sum_{n,m}{1\over n!}
  {\delta\over \delta J(u_1)}\dots
  {\delta\over \delta J(u_n)}{\cal Z}_A[J]{1\over m!}
  {\delta\over \delta J'(u_1'-\beta)}\dots
  {\delta\over \delta J'(u_m'-\beta)}{\cal Z}_B[J']\cr
&\times&\Bigl\{1-\prod_{i=1}^n\prod_{j=1}^m\bigl[1-\hat{\sigma}_{i,j}(u,u')\bigr]
   \Bigr\}\prod dudu'\Bigm|_{J=J'=0}
\end{eqnarray}

\noindent here $\beta$ is the impact parameter between the two
interacting hadrons $A$ and $B$ and $\hat{\sigma}_{i,j}$ is the
probability for the parton $i$ (of $A$) to have an hard
interaction with the parton $j$ (of $B$).

\noindent The hard cross section is
obtained by summing all contributions due to all different
hadronic configurations (the sums over $n$ and $m$). For each pair
of values $n$ and $m$, one has a contribution to $\sigma_{hard}$
when at least one hard interaction takes place. Given  $n$ and $m$, the probability to have at least one hard interaction
is represented by the term in curly brackets.

\noindent The cross section is analogous to the expression
of the inelastic nucleus-nucleus cross section in the Glauber
model\cite{Bialas:1976ed} and takes into account both disconnected
interactions (which imply $n=m$) and rescatterings (when $n\neq
m$). In the present case one is interested in disconnected interactions. One may hence simplify the expression in Eq.(4) by neglecting
all rescatterings. To that purpose the term in curly
brackets is replaced with the following expression:

\begin{eqnarray}
&&\bigl\{ 1-{\rm exp}\sum_{ij}{\rm ln}(1-\hat\sigma_{ij})\bigr\}
=1-{\rm exp}\biggl[-\sum_{ij}\Bigl(\hat\sigma_{ij}+{1\over
2}\hat\sigma_{ij}\hat\sigma_{ij} +\dots\Bigr)\biggr]\nonumber\\
&&\Rightarrow
  \sum_{ij}\hat\sigma_{ij}-{1\over 2}
  \sum_{ij}\sum_{k\not=i,l\not=j}\hat\sigma_{ij}\hat\sigma_{kl}
  \dots
\end{eqnarray}

\noindent where, in the second line, all repeated indices (which correspond to rescatterings) have been suppressed. In particular only the first term of the expansion of the logarithm in the first line in Eq.(5) needs to be taken into account. The symmetry of the integrand with respect to the indices allows obtaining a compact expression of the hard cross section:

\begin{eqnarray}\sigma_{hard}(\beta)&=&{\rm exp}(\partial)\cdot {\rm exp}(\partial')
  \Bigl[ 1-{\rm exp}\bigl(-\partial\cdot\hat{\sigma}\cdot\partial'\bigr)\Bigr]
  {\cal Z}_A[J]{\cal Z}_B[J']\Bigm|_{J=J'=0}\cr
  &=&\Bigl[ 1-{\rm exp}\bigl(-\partial\cdot\hat{\sigma}\cdot\partial'\bigr)\Bigr]
  {\cal Z}_A[J]{\cal Z}_B[J']\Bigm|_{J=J'=1}\end{eqnarray}

\noindent where all convolutions are understood. 

Eq.(6) includes all MPI, which are identified with the disconnected collisions. The hard cross section $\sigma_{hard}$ is easily expressed as a sum of MPI:

\begin{eqnarray}\sigma_{hard}(\beta)&=&\Bigl[ 1-{\rm exp}\bigl(-\partial\cdot\hat{\sigma}\cdot\partial'\bigr)\Bigr]
  {\cal Z}_A[J]{\cal Z}_B[J']\Bigm|_{J=J'=1}\cr
  &=&\sum_{N=1}^{\infty}{\bigl(\partial\cdot\hat{\sigma}\cdot\partial'\bigr)^N\over{N !} } {\rm e}^{-\partial\cdot\hat{\sigma}\cdot\partial'}
  {\cal Z}_A[J]{\cal Z}_B[J']\Bigm|_{J=J'=1}
\end{eqnarray}

\noindent It's instructive to work out the average number of collisions:

\begin{eqnarray}\!\!\!\!\!\!\!\!\!\!\!\!\!\!\!\!\!\langle N\rangle\sigma_{hard}(\beta)&=&\sum_{N=1}^{\infty}{N\bigl(\partial\cdot\hat{\sigma}\cdot\partial'\bigr)^{N}\over{N !} } {\rm e}^{-\partial\cdot\hat{\sigma}\cdot\partial'}
  {\cal Z}_A[J]{\cal Z}_B[J']\Bigm|_{J=J'=1}\cr
  &=&\partial_{J_1}\cdot\hat{\sigma}\cdot\partial_{J_1'}\sum_{N=0}^{\infty}{\bigl(\partial\cdot\hat{\sigma}\cdot\partial'\bigr)^{N}\over{N !} } {\rm e}^{-\partial\cdot\hat{\sigma}\cdot\partial'}
  {\cal Z}_A[J]{\cal Z}_B[J']\Bigm|_{J=J'=1}\cr&=&\bigl(\partial_{J_1}\cdot\hat{\sigma}\cdot\partial_{J_1'} \bigr){\cal Z}_A[J]{\cal Z}_B[J']\Bigm|_{J=J'=1}\cr
  &=&\int D_A(x_1;b_1)\hat{\sigma}(x_1x_1') D_B(x_1';b_1-\beta)dx_1dx_1'd^2b_1\equiv\sigma_S(\beta)
  \end{eqnarray}

  \noindent where $\hat{\sigma}(x_1x_1')$ in the last line of Eq.(8) is the parton-parton cross section integrated with a cutoff. Given the localization of the interactions in transverse space,  the parton-parton interaction probability has been treated as a $\delta$-function of the transverse coordinates, namely $\hat{\sigma}(u,u')=\hat{\sigma}(x, x')\delta({\bf b}-{\bf b}')$. 
  
  The average $\langle N\rangle\sigma_{hard}$ is hece equal to $\sigma_S$, the single scattering inclusive cross section of the QCD parton model. In an analogous way one obtains

\begin{eqnarray}\!\!\!\!\!\!\!\!\!\!\!\!\!\!\!\!\!\!\!\!{\langle N(N-1)\rangle\over2!}\sigma_{hard}(\beta)&=&{1\over2!}\int D_A(x_1x_2;b_1 b_2)\hat{\sigma}(x_1x_1')\hat{\sigma}(x_2x_2')\nonumber\\&&\quad\times   D_B(x_1' x_2';b_1-\beta, b_2-\beta)dx_1dx_1'd^2b_1 dx_2dx_2'd^2b_2\equiv\sigma_D(\beta)
\end{eqnarray}

\noindent where $\sigma_D$ is the double parton scattering inclusive cross section. The relation is easily extended to any number of MPI. One obtains 

\begin{eqnarray}&&\!\!\!\!\!\!\!\!\!\!\!\!\!\!\!\!\!\!\!\!\!\!{\langle N(N-1)\dots(N-K+1)\rangle\over K!}\sigma_{hard}(\beta)={1\over K!}\int D_A(x_1 \dots x_K;b_1 \dots b_K)\hat{\sigma}(x_1x_1')\dots\hat{\sigma}(x_Kx_K')\nonumber\\&&\quad\qquad\qquad\qquad\qquad\times  D_B(x_1' \dots x_K';b_1-\beta \dots b_K-\beta)\prod_{i=1}^Kdx_idx_i'd^2b_i \equiv\sigma_K(\beta)
\end{eqnarray}

\noindent which proves that, {\it when rescatterings are
neglected}, for any choice of multiparton distributions, the
inclusive cross sections are given by the moments of the
distribution in the number of collisions.

\subsubsection{Inclusive and Exclusive Cross Sections, Sum Rules}

In proton-proton collisions, the inclusive cross sections are thus the moments of the distribution in the number of MPI. 
The most basic information on the distribution in the number of collisions, the average number, is hence given by the single scattering inclusive cross section of the QCD parton model. Analogously the $K$th scattering inclusive cross section gives the $K$th moment of the distribution in the number of collisions and is related directly to the $K$-partons distribution of the hadron structure.

A way alternative to the set of moments, to provide the whole information of the distribution, is represented by the set of the different terms of the probability distribution of multiple collisions. Correspondingly, in addition to the set of the inclusive cross sections, one may consider the set of the "exclusive" cross sections, where one selects the events where only a given number of collisions are present\cite{Calucci:2008jw}. The cross sections called now "exclusive" are in fact partially inclusive cross sections, since one sums over all large $p_t$ partons outside a given phase space interval and over all soft fragments.

Interestingly, in its study of double parton collisions, the CDF experiment\cite{Abe:1997xk} did not measure the double parton scattering inclusive cross section. The events selected where in fact only those containing just double parton collisions, while all events with triple scatterings (about 17\% of the sample of all events with double parton scatterings) where removed. The resulting quantity measured by CDF is hence different with respect to the inclusive cross sections usually discussed in large momentum transfer physics. In fact CDF measured the double parton scattering "exclusive" cross section.

One should emphasize that the "exclusive" cross sections are not given by the usual QDC-parton model expression of large $p_t$ processes. While the inclusive cross sections are in fact linked directly to the multi-parton structure of the hadron, the link of the "exclusive" cross sections with the hadron structure is much more complex. The requirement of having only events with a given number of hard collisions implies in fact that the corresponding cross section (being proportional to the probability of not having any further hard interaction) depends, at least in principle, on the whole series of multiple hard collisions and hence on an infinite non-perturbative input. One has

\begin{eqnarray}
\sigma_{hard}=\sum_{N=1}^{\infty}\tilde\sigma_N,\qquad\quad\sigma_K=\sum_{N=K}^{\infty}\frac{N(N-1)\dots(N-K+1)}{K!}\tilde\sigma_N
\label{mpi:eq51}
\end{eqnarray}

\noindent
where $\tilde\sigma_N$ is the "exclusive" cross section of $N$ partonic collisions, while $\sigma_K$ is the inclusive cross section of $K$ partonic collisions. Notice that Eq.~\ref{mpi:eq51} is a set of sum rules that relate inclusive and "exclusive" MPI cross sections. Notice also that MPI add incoherently in the final cross section, leading to a different probabilistic picture of the process in each phase space interval, in such a way that one may associate a different probability distribution of MPI to each different phase space choice of observing the final state. In a given phase space window only a small number of MPI might give a sizable contribution to $\sigma_{hard}$. If, as an example, the relevant contributions to $\sigma_{hard}$ are at most from triple collisions, one may write

\begin{eqnarray}
\tilde\sigma_1&=&\sigma_S-2\sigma_D+3\sigma_T\nonumber\\
\tilde\sigma_2&=&\sigma_D-3\sigma_T\nonumber\\
\tilde\sigma_3&=&\sigma_T
\end{eqnarray}

\noindent
which shows how to express the "exclusive" cross sections in terms of well defined elements of the non perturbative hadron structure, in the phase space window under consideration.

Any final state phase space window identifies an interval in momentum transfer and in fractional momenta, which represents the domain of definition of the corresponding probability distribution of MPI.
The same interval in momentum transfer and fractional momenta represents also the integration domain of the integrated terms, which appear in the "exclusive" differential cross sections. Integrated terms appear in the "exclusive" differential cross sections because of the normalization of the probability distribution. The limits of the integrated terms are thus fixed unambiguously by normalization and coincide with the kinematical limits adopted to select the final state.

\begin{eqnarray}
d\sigma_S(u,u')&=& D_A(u)d\hat{\sigma}(u,u') D_B(u')\nonumber\\       
d\tilde\sigma_1(u,u')&=& D_A(u)d\hat{\sigma}(u,u') D_B(u')\Bigl[1- \int D_A(u_1)\hat{\sigma}(u_1,u_1') D_B(u_1')du_1du_1'\Bigr]\nonumber\\
&&\qquad-\Bigl[\int D_A(u)d\hat{\sigma}(u,u') C_B(u',u_1')
       \hat{\sigma}(u_1',u_1)D_A(u_1)du_1du_1'
+A\leftrightarrow B\Bigr]\nonumber\\
&&\qquad-\int C_A(u_1,u)d\hat{\sigma}(u,u') C_B(u',u_1')
       \hat{\sigma}(u_1',u_1)du_1du_1'
       \end{eqnarray}

In Eq.(13) the explicit expressions of the single scattering inclusive, $\sigma_S(u,u')$, and "exclusive", $\tilde\sigma_1(u,u')$, differential cross sections are given as a function of the coordinates $u,u'$ of the observed partons. The expression of the inclusive cross section holds at all orders in $\hat\sigma$, while the "exclusive" cross section is at the order $\hat\sigma^2$.

\begin{figure}[h]
\centering
\includegraphics[width=150mm]{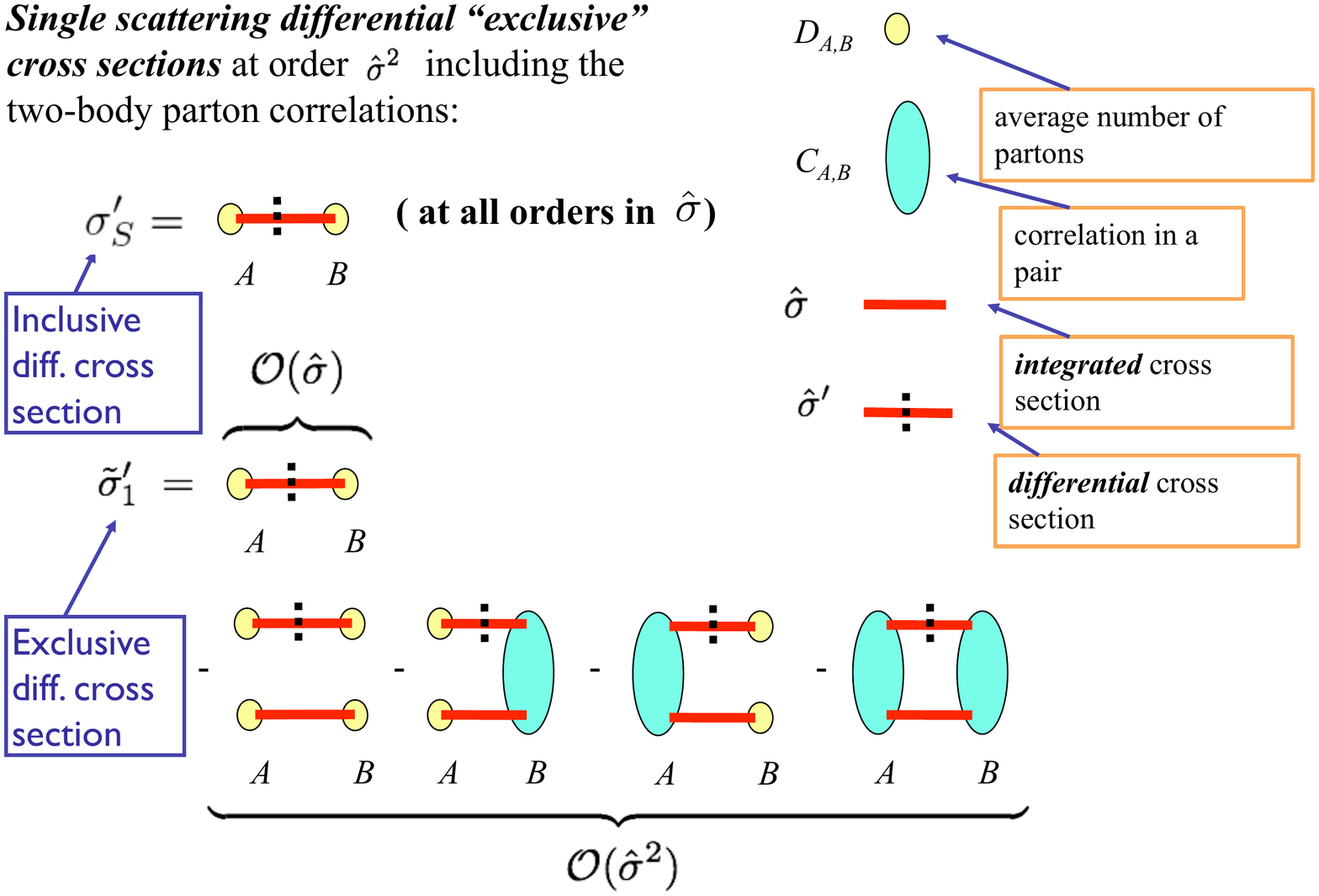}
\caption {Single scattering differential inclusive and "exclusive" cross sections including the 
two-body parton correlations. The "exclusive" cross section is at order $\hat\sigma^2$.}
\end{figure}

By restricting the phase space interval of the observed final state, $\tilde\sigma_1$ is therefore well expressed by the term linear in $\hat\sigma$. In that limit the probability of interaction is well approximated by the average of the distribution, while $\tilde\sigma_2$ is negligibly small and the single parton scattering "exclusive" cross section is well represented by the single scattering expression of the simple QCD parton model $\sigma_S$. When the phase space volume is increased, the single scattering "exclusive" cross section becomes increasingly different from the single scattering of the simple QCD parton model and the difference between $\sigma_S$ and $\tilde\sigma_1$ allows a direct measure of the importance of correlations, Fig.2.

Notice that the components of the hadron structure, namely the terms $D$ and
$C$, are well defined quantities, as they do not mix when the kinematical limits adopted to select the final state are changed. The effect of modifying  the kinematical limits adopted to select the final state is in fact only to change the integration domain of each term, namely to probe the multi-parton structure of the hadron in different domains in $x$ and $Q^2$\cite{Calucci:2008jw}.

Inclusive and "exclusive" cross sections result from independent measurements. By checking the sum rules, comparing the measured integrated inclusive cross section with the sum of the measured integrated "exclusive" cross sections, taken with the proper multiplicity factors, up to a given order in the number of collisions, one has hence a direct indication of the importance of terms with larger numbers of partonic collisions in a given phase space interval. In a phase space interval where, as an example, only single and double collisions give sizable contributions, one may thus obtain information on the effect of two body parton correlations by looking at the difference between the single scattering inclusive and "exclusive" differential cross sections:

\begin{eqnarray}
{d\sigma_S\over dy d{\bf p}_t}-{d\tilde\sigma_1\over dy d{\bf
p}_t}={d\sigma_S\over dy d{\bf p}_t}{\sigma_S\over\sigma_{eff}}
\end{eqnarray}

\noindent
By comparing the behavior, as a function of fractional momenta, of the difference in the left hand side of Eq.(14) with the right hand side one obtains information on the value of the effective cross section and on the dependence of $\sigma_{eff}$ on $y$ and ${\bf p}_t$.

\subsubsection{Concluding Summary}

Given the extended nature of the interacting systems, the hard component of a hadronic interaction is, in general, disconnected. The different terms of a MPI process correspond to the different regions where the hard component of the interaction is localized. In a single partonic collision all transverse momenta balance and a MPI process contributes to the cross section by generating different groups of final state partons where the large transverse momenta are compensated separately. A consequence is that terms with different numbers of hard interactions add incoherently and the MPI cross section is obtained by the superposition of the cross sections, due to the contributions of the different topologies of the hard component of the interaction.

The physical picture of MPI is hence probabilistic, which leads naturally to consider two different sets of cross sections, the inclusive cross sections, given by the moments of the probability distribution of MPI, and the "exclusive" cross sections, given by the different terms of the probability distribution.
The inclusive cross sections depend on all the different terms of the distribution in multiplicity of partonic interactions, each counted with a different multiplicity factor, while each "exclusive" cross section corresponds to a given term of the distribution. A consequence is that inclusive and "exclusive" cross sections, which result from independent measurements, are linked by sum rules, which are implied by the relations expressing the moments of a probability distribution. 

Of course different final state phase space windows lead to different probability distributions of MPI and the number of MPI can be controlled by adjusting the final state phase space interval. The sum rules are thus saturated with a different number of terms in each different final state phase space interval. The number of terms needed to saturate the sum rules in the phase space interval under consideration provides a quantitative measure of the importance of terms with larger numbers of partonic interactions. 
Given the number of terms needed to saturate the sum rules, the non-perturbative input to the "exclusive" cross sections is given explicitly in terms of well defined properties of the hadron structure, which hence allow to evaluate unambiguously also the "exclusive" cross sections in perturbative QCD. 

\graphicspath{{gosta/figs/}}

\contribution{\label{sec:gosta}Multiple interactions, diffraction, and the BFKL pomeron}
{Contributing author: G. Gustafson}
\label{gustafson}


In high energy $pp$ scattering the cross
section for minijet production becomes very large, and unitarity implies that 
multiple interactions, saturation, and diffraction become important. 
Effects of saturation and multiple interactions are most easily described in 
impact parameter space, as parton rescattering is represented by a
convolution in transverse momentum space, which corresponds to a simple
multiplication in transverse coordinate space. 

The proton has an internal substructure, which may be excited in a diffractive
scattering process, and 
diffractive excitation represents large fractions of the cross section in
$pp$ collisions or DIS. In the Good--Walker formalism \cite{Good:1960ba}
diffractive excitation is described by the fluctuations in the scattering 
amplitude. 
In most analyses of $pp$ collisions this mechanism is used only for low mass
excitation, while high mass excitation is
described by a triple-regge formula \cite{Mueller:1970fa,Detar:1971gn},
where regge trajectories and couplings are fitted to experimental data
(for recent analyses see \emph{e.g.} 
Refs.~\cite{Ryskin:2009tj,Kaidalov:2009aw,Gotsman:2008tr}).

The proton substructure is represented by a parton cascade, which at high
energies is described by BFKL evolution. The fluctuations in this 
evolution are known to be very large \cite{Mueller:1996te}. 
An analysis of these fluctuations, within the Lund Dipole Cascade model, 
is able to reproduce the experimental cross
sections for diffractive excitation in $pp$ collisions or DIS 
\cite{Avsar:2007xg,Flensburg:2010kq}. 
This implies that the effective pomeron couplings 
in the multi-pomeron formalism can be estimated without any new free 
parameters.

It should also be noticed that the classification of diffractive events 
varies between different formalisms,
and cannot be uniquely defined. Therefore it is recommended to
study gap events rather than diffractive events. 

\subsubsection{The eikonal approximation and the Good--Walker formalism}

As mentioned in the introduction, diffraction, saturation, and multiple 
interactions are most easily described in impact parameter space.
If the scattering is driven by absorption into inelastic states $i$, 
with weights $2f_i$, the elastic amplitude is given by
\begin{equation}
T = 1-e^{-F},\,\,\,\,\,\mathrm{with}\,\,\,F=\sum f_i.
\end{equation}
For a structureless projectile we find:
\vspace{3mm}
\begin{equation}
\left\{ \begin{array}{l}
d\sigma_{tot} / d^2b =\,\langle 2T \rangle,\,\,\,\,\\
\sigma_{el}/d^2b \,\,= \, \langle T\rangle^2,\,\,\,\,\\
\sigma_{inel}/d^2b =  \,\langle 1-e^{-\sum 2f_i}\rangle \sim \sigma_{tot}-
\sigma_{el}.
\end{array} \right.
\label{eq:sigmat}
\end{equation} 

If the projectile has an internal structure, the mass eigenstates $\Psi_{k}$
can differ from the eigenstates of diffraction $\Phi_n$, which have
eigenvalues $T_n$. With the notation
$\Psi_{k} = \sum_n  c_{kn} \Phi_n$ (with $\Psi_{in} = \Psi_1$)
the elastic amplitude is given by
$\langle \Psi_{1} | T | \Psi_{1} \rangle = \sum c_{1n}^2 T_n
= \langle T \rangle$,
while the amplitude for diffractive transition to mass eigenstate $\Psi_k$ is 
given by
$\langle \Psi_{k} | T | \Psi_{1} \rangle = \sum_n  c_{kn} T_n c_{1n}$.
The corresponding cross sections become
\begin{eqnarray}
d \sigma_{el}/d^2 b &=& (\sum c_{1n}^2 T_n)^2 = \langle T\rangle ^2
\label{eq:sigmae}\\
d\sigma_{diff}/d^2 b &=&\sum_k \langle \Psi_{1} | T | \Psi_{k} \rangle \langle 
\Psi_{k} | T |\Psi_{1} \rangle =\langle T^2 \rangle.
\label{eq:sigmad}
\end{eqnarray}
The diffractive cross section here includes elastic scattering. Subtracting
this gives the cross section for diffractive excitation, which is determined
by the fluctuations in the scattering process:
\begin{equation}
d\sigma_{diff\,ex}/d^2 b  = d\sigma_{diff}- d \sigma_{el} =
\langle T^2 \rangle - \langle T \rangle ^2.
\end{equation}

\subsubsection{Dipole cascade models}


\emph{Mueller's dipole cascade model} 
\cite{Mueller:1993rr,Mueller:1994jq,Mueller:1994gb} is a formulation
of BFKL evolution in transverse coordinate space. 
Gluon radiation from the color charge in a parent quark or gluon is screened 
by the accompanying anticharge 
in the color dipole. This suppresses emissions at large transverse separation,
which corresponds to the suppression of small $k_\perp$ in BFKL.
For a dipole $(\mathbf{x},\mathbf{y})$ the probability per unit rapidity ($Y$) 
for emission of a gluon at transverse position $\mathbf{z}$ is given by
\begin{eqnarray}
\frac{d\mathcal{P}}{dY}=\frac{\bar{\alpha}}{2\pi}d^2\mathbf{z}
\frac{(\mathbf{x}-\mathbf{y})^2}{(\mathbf{x}-\mathbf{z})^2 (\mathbf{z}-\mathbf{y})^2},
\,\,\,\,\,\,\, \mathrm{with}\,\,\, \bar{\alpha} = \frac{3\alpha_s}{\pi}.
\label{eq:dipkernel1}
\end{eqnarray}
The dipole is split into two dipoles, which
(in the large $N_c$ limit) emit new gluons independently. The result is a
cascade, where the number of dipoles grows exponentially with $Y$.

When two cascades collide, a pair of dipoles with coordinates 
$(\mathbf{x}_i,\mathbf{y}_i)$ and $(\mathbf{x}_j,\mathbf{y}_j)$ can interact 
via gluon exchange with the probability $2f_{ij}$, where
\begin{equation}
  f_{ij} = f(\mathbf{x}_i,\mathbf{y}_i|\mathbf{x}_j,\mathbf{y}_j) =
  \frac{\alpha_s^2}{8}\biggl[\log\biggl(\frac{(\mathbf{x}_i-\mathbf{y}_j)^2
    (\mathbf{y}_i-\mathbf{x}_j)^2}
  {(\mathbf{x}_i-\mathbf{x}_j)^2(\mathbf{y}_i-\mathbf{y}_j)^2}\biggr)\biggr]^2.
\label{eq:dipamp}
\end{equation}
Summing over all dipoles in the cascades then reproduces the LL BFKL result.
The elastic scattering amplitude is given by $T=1-\exp(-\sum f_{ij})$, and the
cross sections are given by Eqs.~(\ref{eq:sigmat}, \ref{eq:sigmae},
\ref{eq:sigmad}). 

The \emph{Lund cascade model} \cite{Avsar:2005iz,Avsar:2006jy,Flensburg:2008ag} is a 
generalization of Mueller's model, which includes:

--  NLL BFKL effects

-- Nonlinear effects in the evolution

-- Confinement effects

For an incoming virtual photon splitting in a $q\bar{q}$ pair, the initial 
state wavefunction is determined
by perturbative QCD. For an incoming proton we make an ansatz in form of an
equilateral triangle of dipoles. After evolution the result is rather
insensitive to the exact form of the initial state. The model is also 
implemented in a MC program DIPSY.
The model reproduces successfully the total and (quasi)elastic
cross sections for DIS and $pp$ scattering.

\subsubsection{Fluctuations and diffractive excitation}

The fluctuations in the evolution are large, and the model can also describe
diffractive excitation within the Good--Walker formalism, without new
parameters beyond those adjusted to the total and elastic cross sections
\cite{Avsar:2007xg}. This is similar in spirit to the early analysis by 
Miettinen 
and Pumplin \cite{Miettinen:1978jb}. In DIS saturation effects are not very
important, while in $pp$ collisions saturation effects strongly suppress the
fluctuations, and thus the cross section for diffractive excitation.

\begin{figure}
\begin{center}
\includegraphics[width=0.35\textwidth,angle=270]{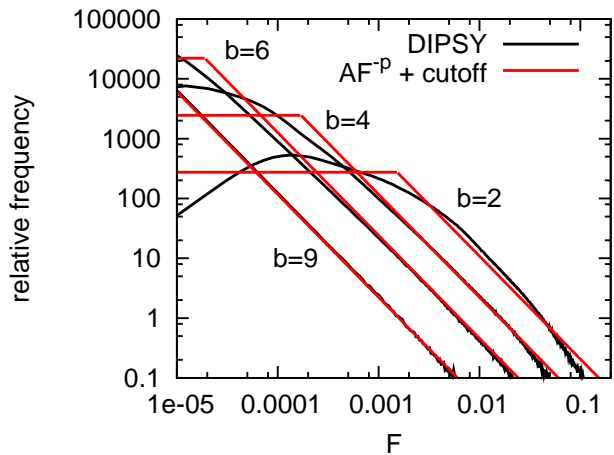}
\end{center}
\caption{Distribution in the one-pomeron
amplitude $F$ in DIS for $Q^2=14$ $\mathrm{GeV}^2$ and $W=220$ $\mathrm{GeV}$.
The photon is here represented by a dipole with size $r=1/Q$. The impact
parameter $b$ is measured in $\mathrm{GeV}^{-2}$.}
\label{fig:f-distDIS}
\end{figure}

\textbf{$\gamma^* p$ collisions}

The distribution in the non-saturated scattering amplitude, $F$, is shown in 
Fig.~\ref{fig:f-distDIS} for different impact parameters. The distribution 
can be approximately described by a power
$\frac{dP}{dF} \approx A\, F^{-p}$ (with a cutoff for small $F$-values), 
which is illustrated by the straight lines in the figure. 
The width of this distribution is rather large, and the approximation gives 
the ratio $d\sigma_{diff.ex.}/  d\sigma_{tot} 
\approx 1-1/2^{2-p}$. In the simulations the power $p$ is rather
independent of the impact
parameter, and therefore this result is also valid for the integrated cross 
sections. This gives $\sigma_{diff}/\sigma_{tot} \sim 0.13$ at 
$Q^2=50 \,\mathrm{GeV}^2$, decreasing for larger $Q^2$, but fairly insensitive
to the energy $W$.

\textbf{$pp$ collisions}

\begin{figure}
\includegraphics[scale=0.5,angle=270]{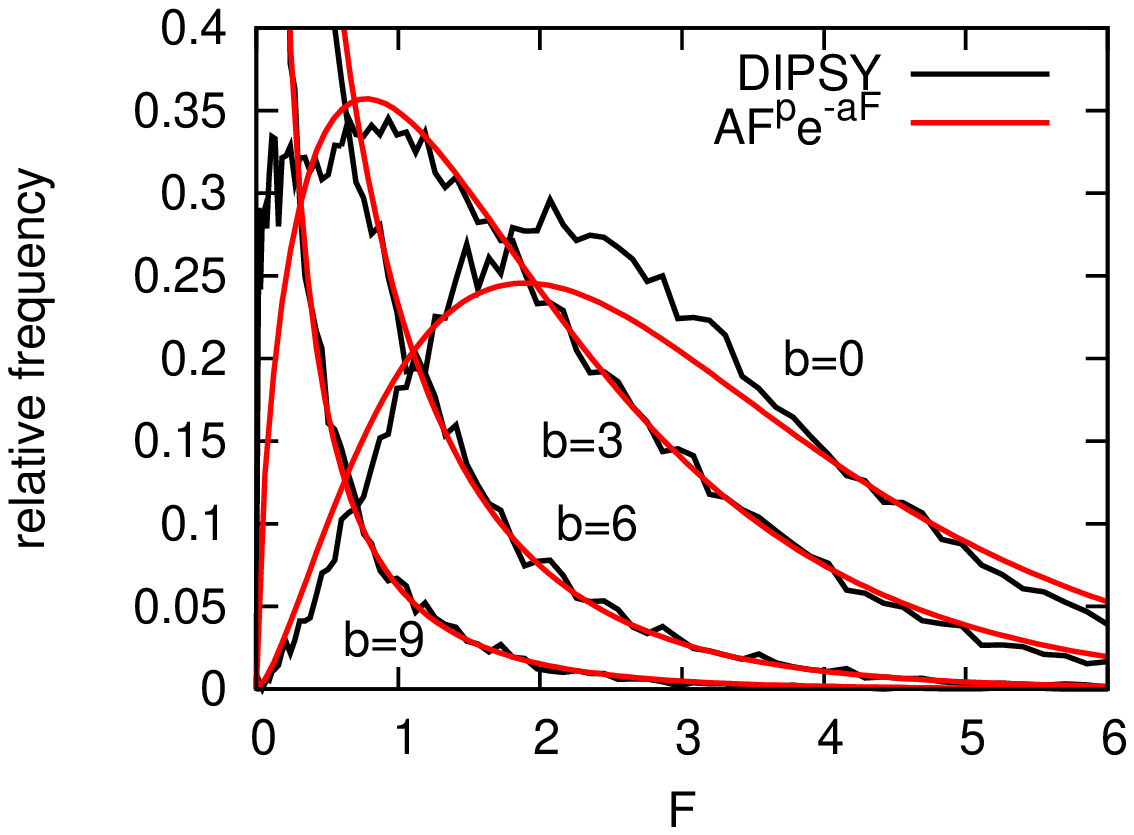}
\includegraphics[scale=0.5,angle=270]{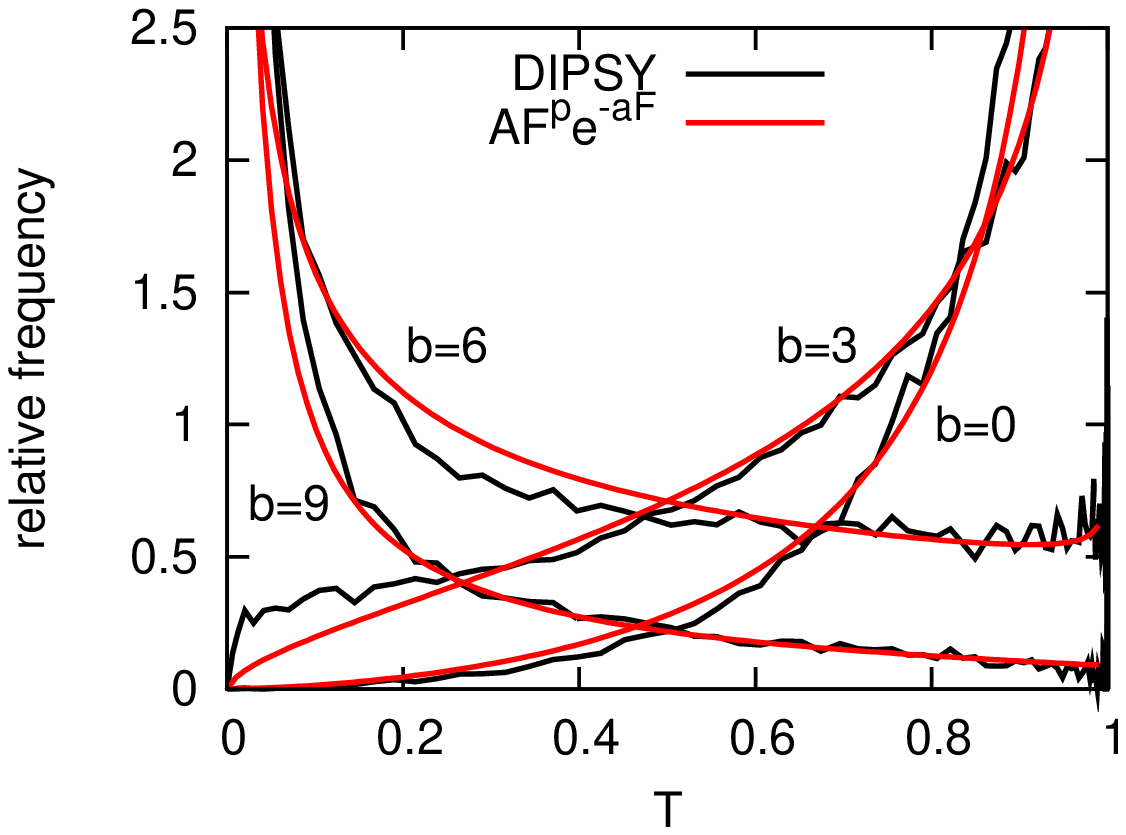}
\caption{Distribution in the one-pomeron
    amplitude ($F$, left) and the uniterized amplitude 
   ($T$, right)
    in $pp$ collisions at 2 TeV. Notation as in Fig.~1.}
\label{fig:ft-distpp}
\end{figure}

In $pp$ scattering the Born amplitude is large, and therefore unitarity
effects are important. Figure~\ref{fig:ft-distpp} shows both the Born amplitude 
and the unitarized amplitude at 2 TeV for different $b$-values. We see that 
the width of the Born amplitude is large, and without unitarization the
fraction of diffractive excitation would be similar to that for  $\gamma^* p$
for lower $Q^2$-values. (The smooth lines are fits of the form $A F^p e^{-aF}$.)

\begin{figure}
\includegraphics[scale=0.5,angle=270]{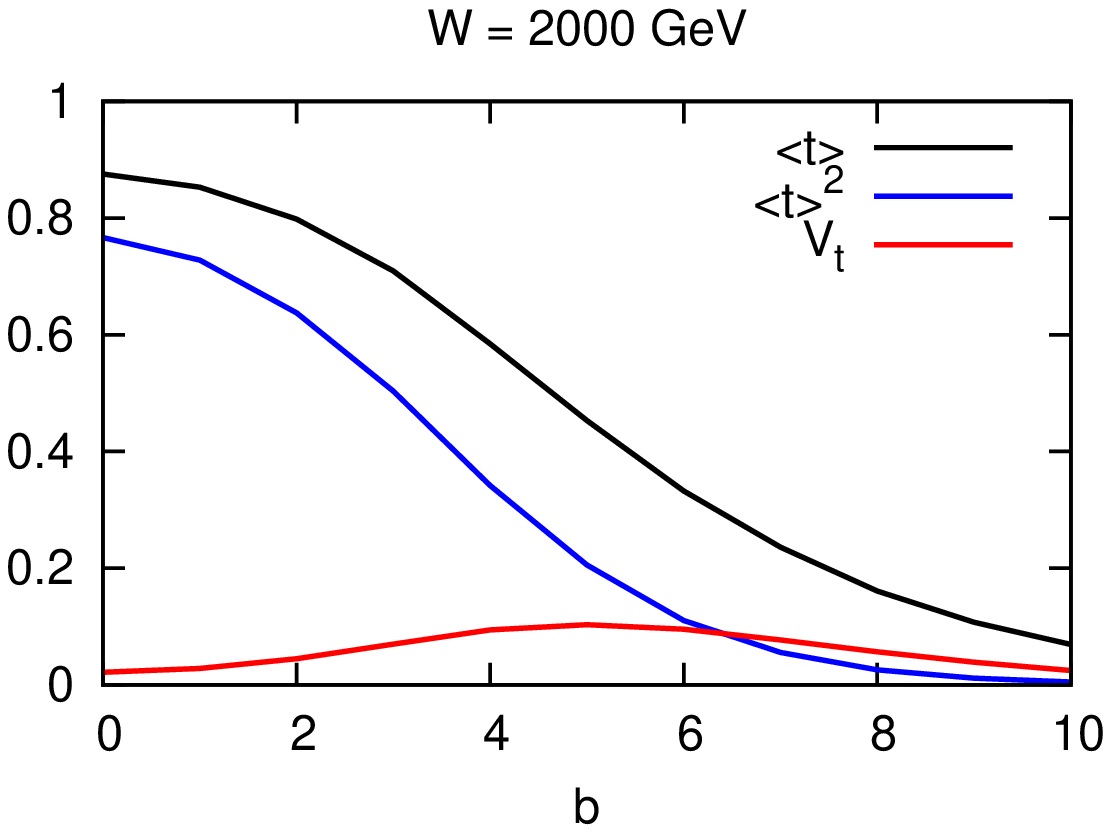}
\includegraphics[scale=0.5,angle=270]{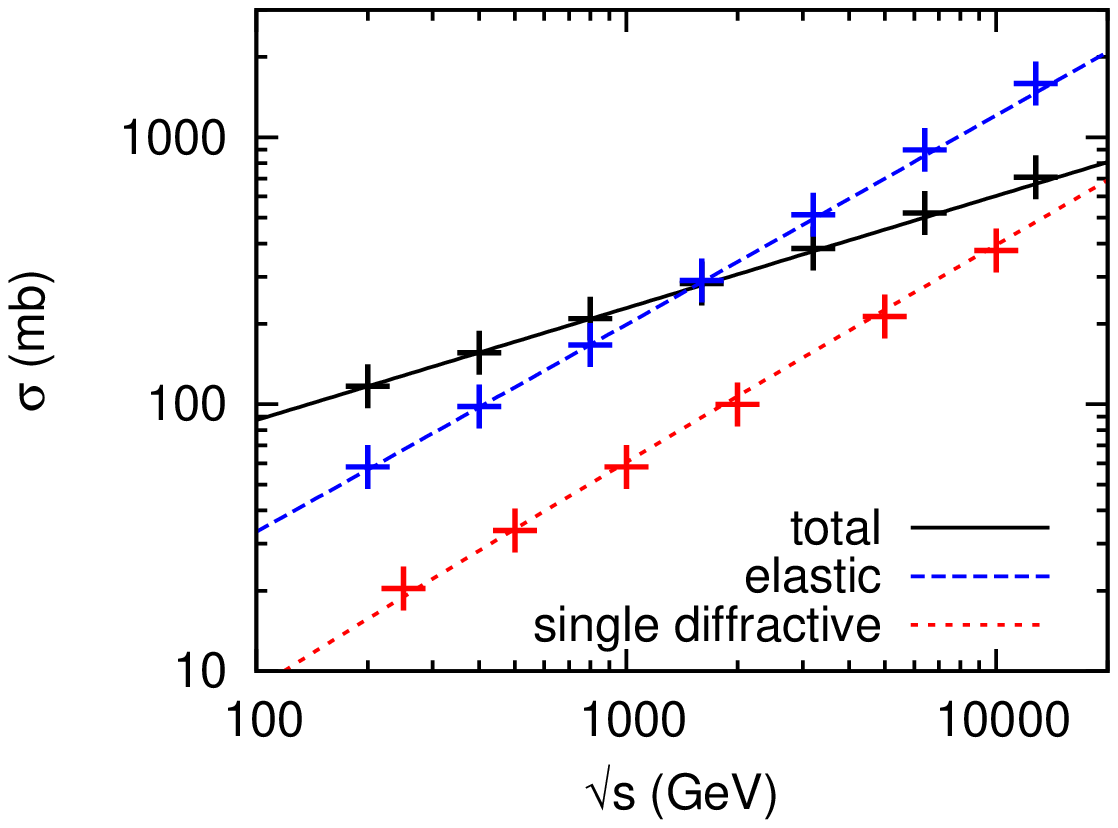}
\caption{\emph{Left}: Impact parameter distributions for 
    $\langle T\rangle=(d\sigma_{\mathrm{tot}}/d^2b)/2$, 
    $\langle T\rangle^2=d\sigma_{\mathrm{el}}/d^2b$, and 
    $V_T=d\sigma_{\mathrm{diff ex}}/d^2b$ in
    $pp$ collisions at $W=2$ TeV. $b$ is in units of GeV$^{-1}$.
    \emph{Right}: The total, elastic, and single diffractive cross
    sections in the one-pomeron approximation. The crosses are 
    model calculations
    and the lines are from a tuned triple-regge parametrisation.}
\label{fig:profile}
\end{figure}

However, the unitarized amplitude is limited
by 1, and the width, and therefore the diffractive excitation, is very much 
reduced. This is in particular the case for central collisions where the
amplitude approaches the black disc limit, as seen in the right panel in 
Fig.~\ref{fig:ft-distpp}. This result corresponds to the  
effect of enhanced diagrams in the conventional triple-regge approach. 
The impact parameter profile is shown in 
Fig.~\ref{fig:profile}. We see that the cross section for diffractive
excitation is largest in a ring with radius $b\sim 1\, \mathrm{fm}$.
This result also implies that factorization is not satisfied when comparing 
diffractive excitation in DIS and $pp$ scattering.

\subsubsection{Comparison with multi-regge analyses}

It is also interesting to compare the results from the Good--Walker analysis
with the multi-regge formalism. To this end we study the contribution from the
\emph{bare pomeron}, meaning the one-pomeron amplitude without contributions 
from saturation, enhanced diagrams or gap survival form factors. 
When $s$, $M_{\mathrm{X}}^2$, and $s/M_{\mathrm{X}}^2$ are all large, pomeron 
exchange should dominate. The results of the MC for the total, elastic, and 
single diffractive cross sections are shown by the crosses in the right panel in
Fig.~\ref{fig:profile}. We note that the results are very well reproduced by
a triple-regge expression with a single pomeron pole, with the parameters
\begin{equation}
\alpha(0)=1.21, \,\,\,\,\alpha' = 0.2\,\mathrm{GeV}^{-2},\,\,\,\,
g_{\mathrm{pP}}(t)=(5.6\,\mathrm{GeV}^{-1})\, e^{1.9 t},\,\, \,\,
g_{3\mathrm{P}}(t) = 0.31\,\mathrm{GeV}^{-1},
\end{equation}
which is shown by the straight lines. Also the $t$-dependence of diffractive
excitation is well reproduced.

These results can also be compared with multi-regge analyses, where \emph{e.g.}
Ryskin \emph{et al.} \cite{Ryskin:2009tj} obtain $\alpha(0)=1.3$, 
$\alpha'\leq 0.05\,\mathrm{GeV}^{-2}$, Kaidalov \emph{et al.} 
\cite{Kaidalov:2009aw} find 
$\alpha(0)=1.12$, $\alpha'=0.22\,\mathrm{GeV}^{-2}$, while Gotsman 
\emph{et al.} \cite{Gotsman:2008tr} find 
$\alpha(0)=1.335$, $\alpha'=0.01\,\mathrm{GeV}^{-2}$. Thus our values are
somewhere in between.
We also note that the Good--Walker results are reproduced by a
single pomeron pole, \emph{i.e.} not by a cut as expected in LL BFKL, or a
series of poles as obtained with a running coupling \cite{Lipatov:1985uk}. 
Also the triple-regge
coupling $g_{3\mathrm{P}}$ is approximately constant, while in LL BFKL it is
proportional to $\sim 1/\sqrt{|t|}$ \cite{Mueller:1994jq,Bartels:2002au}.

\subsubsection{Can diffraction be uniquely defined?}

Multipomeron diagrams are included in the dipole picture,
with fixed multi-pomeron couplings. However, all events with no gap are 
classified as inelastic. This can be compared to the formalism in 
Ref.~\cite{Ryskin:2009tj}, in which a large cross section for overlapping
double diffraction is obtained. 

To compare predictions from different models
with each other and with experiments, we need a unique definition of
diffraction. One attempt might be events with two separate color singlet
systems, containing the original valence quarks. This could be 
obtained by exchange of two (or more) gluons forming a color singlet. 
If such a state is obtained in a calculation in perturbative QCD, a gap could,
however, be filled by final state radiation or nonperturbative strings, or
else a gap could be formed by color reconnection. Thus diffractive events 
cannot be uniquely defined by perturbative QCD. The definition varies between 
different schemes, and for a specified event the diffractive capacity is not an
observable. The solution must be: study observables, meaning events with a gap.

\subsubsection{Summary}

  \begin{itemize}

  \item In high energy $pp$ scattering unitarity implies a high probability for
    multiple interactions, and a large diffractive cross section. In the
    Good--Walker formalism elastic scattering is given by the average of
    the scattering amplitude, while the fluctuations determine the diffractive 
    excitation.

  \item Parton cascades fill the whole rapidity range between projectile
    and target.  The fluctuations in BFKL evolution are large, and within
    the Lund Dipole Cascade model they can 
    describe diffractive excitation within the Good--Walker formalism, 
    to both low and high masses.

  \item  When the interaction approaches the black disc limit in central 
   $pp$ collisions, fluctuations and diffractive excitation is 
   suppressed. This leads to factorization breaking in comparisons of DIS and
   $pp$ scattering.

  \item The result of the model calculations corresponds to a bare pomeron, 
    which is a simple pole, and an almost constant triple-pomeron coupling.

  \item Diffractive excitation is scheme dependent, and cannot be uniquely 
    defined. Study gap events.

\end{itemize}

\section{Phenomenology of Multi-parton Interactions}
\label{sec:pheno}

Due to its high collision energy and luminosity,
the LHC provides a valuable opportunity to observe multiple
parton hard-scatterings, in particular  many DPS
processes. Theoretical investigations of double parton scattering have
a long history, with a large number of studies evaluating the DPS contribution to high energy
processes \cite{DelFabbro:1999tf,Paver:1983hi,DelFabbro:2002pw,Humpert:1983pw,Humpert:1984ay,Ametller:1985tp,Halzen:1986ue,Mangano:1988sq,Godbole:1989ti,Drees:1996rw,Hussein:2006xr,Hussein:2007gj,Domdey:2009bg}.

Compared with the DPS processes already observed, namely final states
involving 4 jets (at the AFS collaboration at the CERN ISR
\cite{Akesson:1986iv}), and $\gamma$ + 3 jets (at the CDF
\cite{Abe:1997xk} and the D0 \cite{Abazov:2009gc} collaborations at
the Fermilab Tevatron), processes at the LHC involve different scales and initial state partons, hence providing complementary
information on DPS. Moreover, large contributions from double parton
scattering could, for example, result in a larger rate for multi-jet production than otherwise predicted, and produce relevant backgrounds in searches for signals of new phenomena.  
It is thus important to know empirically how large the double parton contribution may be and what dependence on relevant kinematic variables it has. Specifically, the differences between final states produced
in SPS and in DPS processes need to be studied in order 
to separate between these processes and gain more detailed experimental
information on DPS.
In addition to its role in general LHC phenomenology, the DPS measurements will have an impact on the development of partonic models of hadrons, since the effective cross section for double parton scattering measures the size in impact parameter space of the incident hadron's partonic hard core.

The DPS phenomenology is based on the general expression for the cross section
$\sigma^{DPS}_{(A,B)}$, cf. Section~\ref{gaunt},
\begin{eqnarray}
\label{eq:sigma_D1}
\sigma^{DPS}_{(A,B)} = \frac{m}{2}\sum_{i,j,k,l}\int
dx_1dx_2dx_1'dx_2'd^2b  \Gamma_{ij}(x_1,x_2,b;t_1,t_2)\Gamma_{kl}(x_1',x_2',b;t_1,t_2)  \hat{\sigma}^A_{ik}(x_1,x_1')\hat{\sigma}^B_{jl}(x_2,x_2') .
\end{eqnarray}
The $\Gamma_{ij}(x_1,x_2,b;t_1,t_2)$ represent double
parton distributions. They may be loosely interpreted as the inclusive
probability distributions to find a parton $i$($j$) with longitudinal
momentum fraction $x_1$($x_2$) at scale $t_1\equiv
\ln(Q_1^2)$($t_2\equiv \ln(Q_2^2)$) in the proton, with the two
partons separated by a transverse distance $b$. The scale $t_1$($t_2$)
is given by the characteristic scale of subprocess $A$($B$).  The
quantity $m$ is a symmetry factor that equals 1 if $A=B$ and 2
otherwise. 
Separating the transverse part,
$\Gamma_{i,j} (x_1,x_2,b; t_1,t_2)= D_{i,j} (x_1,x_2,t_1,t_2)\times F(b)$, Eq.~\ref{eq:sigma_D1} becomes
\begin{eqnarray}
\sigma^{DPS}_{(A,B)} &=&\frac{m}{2 \sigma_{\rm eff}}
\sum_{i,j,k,l} \int dx_1dx_2dx_1'dx_2'  D_{i,j}(x_1,x_2,t_1,t_2)D_{k,l}(x_1',x_2',t_1,t_2) \hat{\sigma}^{A}_{ik}(x_1,x_1',t_1)\hat{\sigma}^{B}_{jl}(x_2,x_2',t_2)
, \nonumber \\ 
\sigma_{\rm eff} &=& \left[\int d^2b(F(b))^2\right]^{-1} .
\label{eq:DPSmaster}
\end{eqnarray}
If one makes the further assumptions that double parton distributions
reduce to the product of
two independent one parton distributions, $D_{i,j} = D_{i}\times D_{j}$,
the DPS cross section can be expressed in the simple form
\begin{equation}
\label{eq:sigma_D3}
\sigma^{DPS}_{(A,B)} = \frac{m}{2} \frac{\sigma_{A}\sigma_{B}}{\sigma_{eff}}.
\end{equation}

In this section the phenomenological studies for the production of two jets in association with a $b\bar{b}$ pair (in section \ref{berger}), same--sign $W$ pair (in section \ref{kom}) and $Z+$ jets (in section~\ref{maina}) are reviewed.

\graphicspath{{berger/}}

\def\be{\begin{equation}}
\def\ee{\end{equation}}
\def\bea{\begin{eqnarray}}
\def\eea{\end{eqnarray}}
\def\met{\slash{\!\!\!\!E}_T}
\def\d0{{D\O}}

\contribution{Dynamical Characteristics of Double Parton Scattering}
{Contributing author: E. Berger}

\label{berger}

\begin{figure}
\begin{center}
\includegraphics[width=0.3\textwidth]{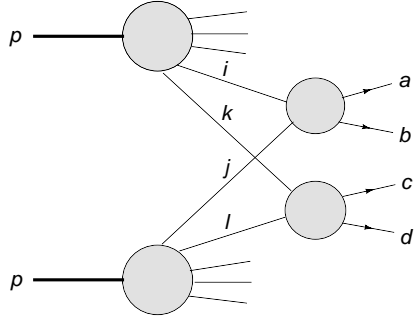}
\end{center}
\caption{Sketch of a double-parton process in which the active partons are 
$i$ and $k$ from one proton and $j$ and $l$ from the second proton.  The 
two hard scattering subprocess are $A(i~j \rightarrow a~b)$ and $B(k~l \rightarrow c~d)$. 
\label{fig:feyn-diag}}
\end{figure}

\begin{figure}
\begin{center}
\includegraphics[width=0.3\textwidth]{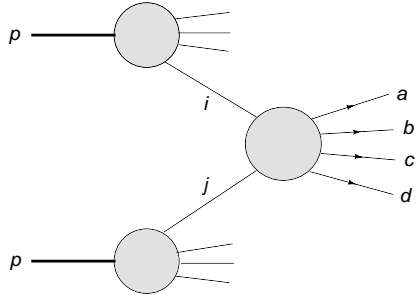}
\end{center}
\caption{Sketch of a single-parton process in which the active partons are 
$i$ from one proton and $j$ from the second proton.  The 
hard scattering subprocess is $A(i~j \rightarrow a~b~c~d)$. 
\label{fig:feyn-diag2}}
\end{figure}

The concept of a DPS process consisting of two short-distance subprocesses occuring in a given hadronic interaction, with two initial partons being active from each of the incident protons, is shown for illustrative purposes in Fig.~\ref{fig:feyn-diag}.  
It may be contrasted with conventional single parton scattering (SPS) in Fig.\ref{fig:feyn-diag2}, in which one short-distance subprocess occurs, with one parton active from each initial hadron.  Both produce the same 4 parton final state. 
Our aims in Ref.~\cite{Berger:2009cm} are to address whether double parton scattering can be shown to exist as a discernible contribution in well defined and accessible final states, and to establish the characteristics features that allow its measurement.  
We show that double parton scattering produces an enhancement of events in regions of phase space in which the ``background'' from single parton scattering is relatively small.  If such enhancements are observed experimentally, with the kinematic dependence we predict, then we will have a direct empirical means to measure the size of the double parton contribution.

From the perspective of sensible rates and experimental tagging, a good process to examine should be the 4 parton final state in which there are $2$ hadronic jets plus a $b$ quark and a $\bar{b}$ antiquark, {\em viz.} $b~\bar{b}~j_1~j_2$.  If the final state arises from double parton scattering, then it is plausible that one subprocess produces the $b~\bar{b}$ system and another subprocess produces the two jets.  There are, of course, many single parton scattering (2 to 4 parton) subprocesses that can result in the $b~\bar{b}~j_1~j_2$ final state, and we identify kinematic distributions that show notable separations of the two contributions.  

The state-of-the-art of calculations of single parton scattering is well developed whereas the phenomenology of double parton scattering is less advanced.  For 
$p p \rightarrow b \bar{b} j_1 j_2 X$, assuming that the two subprocesses 
$A(i~j \rightarrow b~\bar{b})$ and $B(k~l \rightarrow j_1~j_2)$ in Fig.~\ref{fig:feyn-diag} are weakly correlated, and that kinematic and dynamic correlations between the two partons from each hadron may be safely neglected, we employ the common heuristic expression for the DPS differential cross section 
\begin{eqnarray}
\label{bbjj}
d\sigma^{DPS}(p p \rightarrow b \bar{b} j_1 j_2 X) = \frac{d\sigma^{SPS}(p p \rightarrow b \bar{b} X) d\sigma^{SPS} (p p \rightarrow j_1 j_2 X)}{ \sigma_{\rm eff}}.  
\label{eq:dpscross}
\end{eqnarray}
The numerator is a product of single parton scattering cross sections.  
In the denominator, there is a term $ \sigma_{\rm eff}$ with the dimensions of a cross section.  Given that one hard-scatter has taken place, $\sigma_{\rm eff}$ measures 
the size of the partonic core in which the flux of accompanying short-distance partons is confined. Collider data~\cite{Abazov:2009gc} yield values in the range $\sigma_{\rm eff} \sim 12$~mb.  We use this value for the estimates we make, but we emphasize that the goal should be determine its value at LHC energies.  

In Ref.~\cite{Berger:2009cm}, we present the details of our calculation of the double parton and the single parton contributions to $p~p \rightarrow b~\bar{b}~j_1~j_2~X$.  We perform full event simulations at the parton level and apply a series of cuts to emulate experimental analyses.  We also treat  the double parton and the single parton contributions to $4$ jet production, again finding that good separation is possible despite the combinatorial uncertainty in the pairing of jets.   

\subsubsection{Distinguishing variables}
\label{sect:variables}
\begin{figure}[t]
\begin{center}
\includegraphics[width=0.59\textwidth]{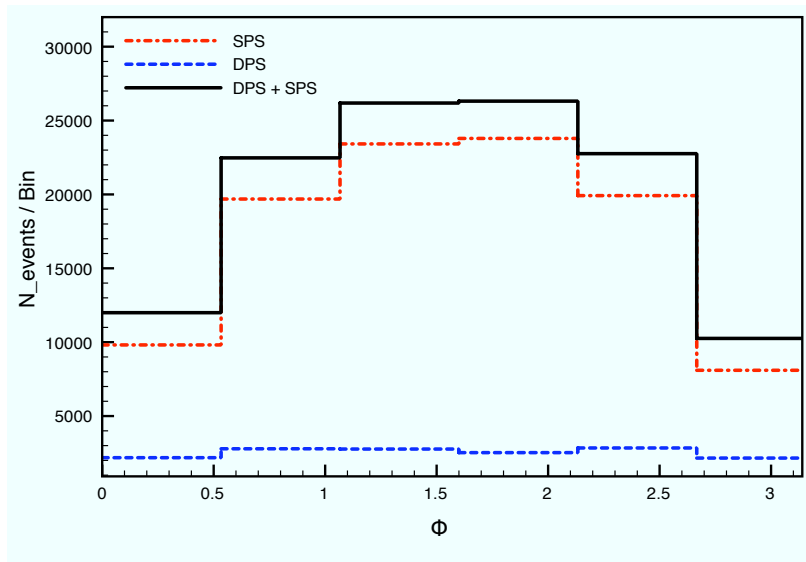}
\end{center}
\caption[]{Event rate as a function of the angle between the two planes
  defined by the $b\bar{b}$ and $jj$ systems.  In SPS events, there is a correlation among the planes which is absent for DPS events.  }
\label{fig:Phi-planes}
\end{figure}

Correlations in the final state are predicted to be quite different between the double parton and the single parton contributions.  For example, we examine the distribution of events as 
function of the angle $\Phi$ between the planes defined by the $b\bar{b}$ system and by the $jj$ system.  If the two scattering processes $i j \rightarrow b \bar{b}$ and $k l \rightarrow j j$ which produce the DPS final state are truly independent, one would expect to see a flat distribution in the angle $\Phi$.  By contrast, many diagrams, including some with non-trivial spin correlations, contribute to the 2 parton to 4 parton final state in SPS $i j \rightarrow b \bar{b} \rm {jet jet} $, and one would expect some correlation between the two planes. 
In Fig.~\ref{fig:Phi-planes}, we display the number of events as a function of the angle 
between the two planes.  There is an evident correlation between the two planes in SPS, 
while the distribution is flat in DPS, indicative that the two planes are uncorrelated.

\begin{figure}[ht]
 \begin{center}
\includegraphics[width=0.59\textwidth]{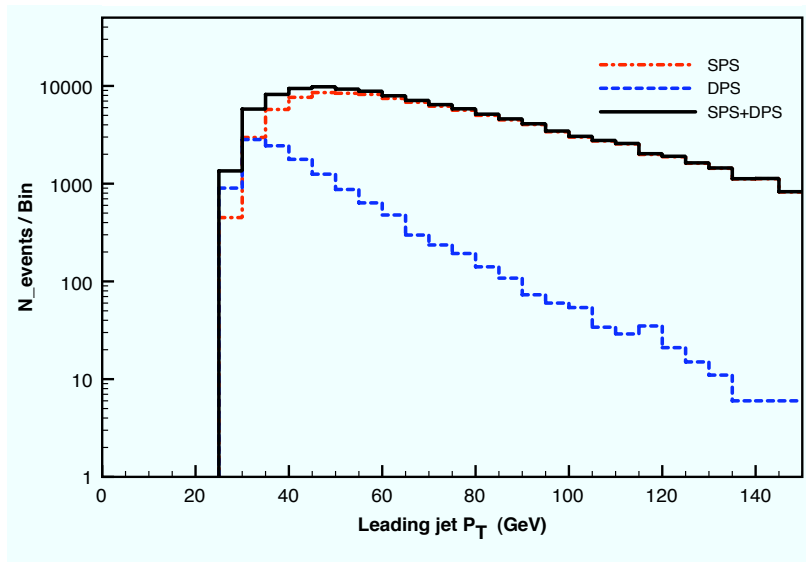}
\end{center}
\caption[]{The transverse momentum $p_T$ distribution of the leading jet in $jjb\bar b$ after minimal cuts. }
\label{fig:ptj_1}
\end{figure}

Another interesting difference between DPS and SPS is the behavior of event rates as a function of transverse momentum.  As an example of this, in Fig.~\ref{fig:ptj_1}, we show the transverse momentum distribution for the leading jet (either a $b$ or light $j$) for both DPS and SPS.  SPS produces a relatively hard spectrum, and for the value of $\sigma_{\rm eff}$ and the cuts that we use, SPS tends to dominate over the full range of transverse momentum considered.  On the other hand, DPS produces a much softer spectrum which (up to issues of normalization in the form of $\sigma_{\rm eff}$) can dominate at small values of transverse momentum.  The cross-over between the two contributions to the total event rate is $\sim 30$ GeV for the acceptance cuts considered.   A smaller (larger) value of $\sigma_{\rm eff}$ would move the cross-over to a larger (smaller) value of the transverse momentum of the leading jet.    

\begin{figure}[htbp]
\begin{center}
\includegraphics[width=0.49\textwidth]{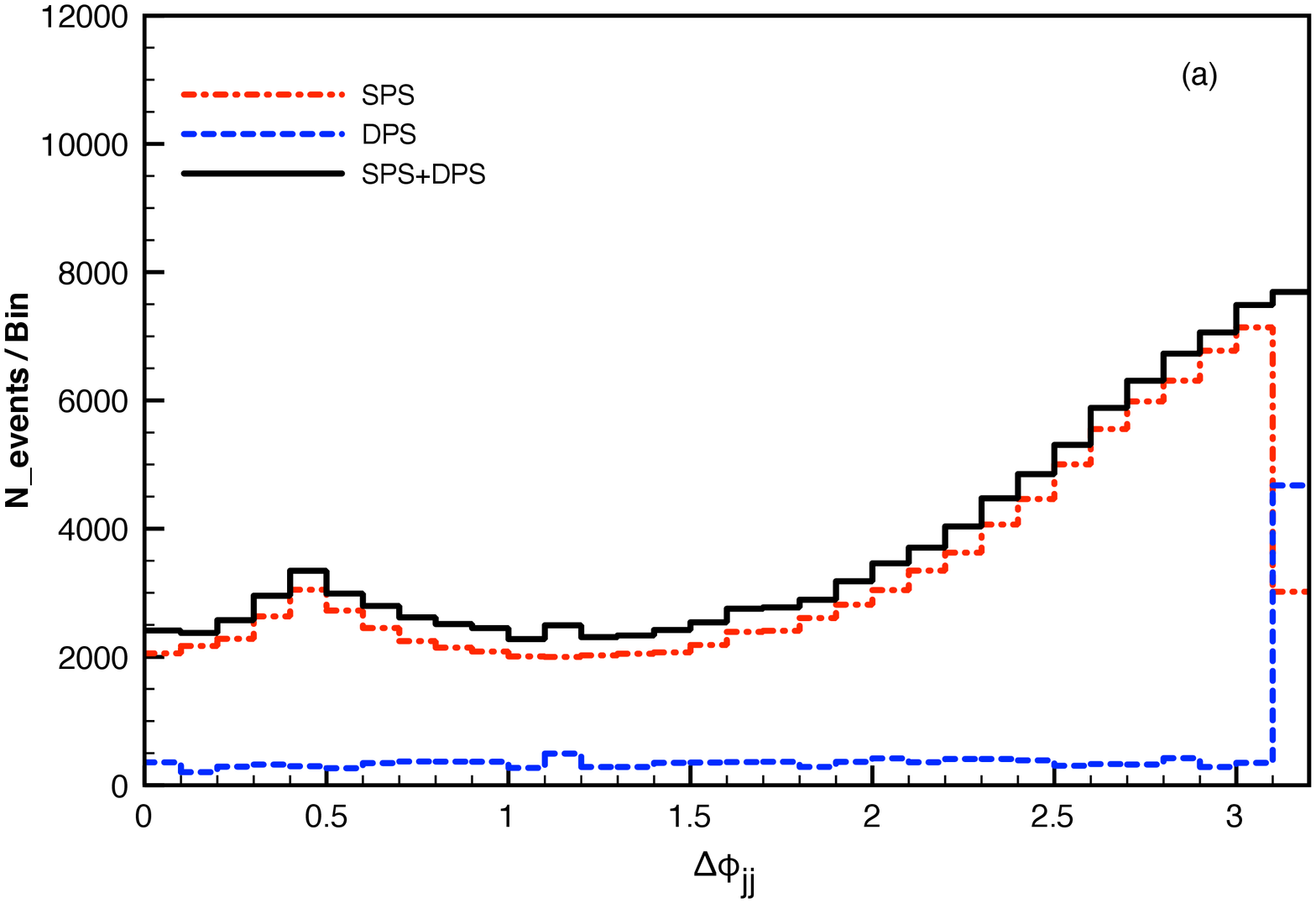}
\includegraphics[width=0.49\textwidth]{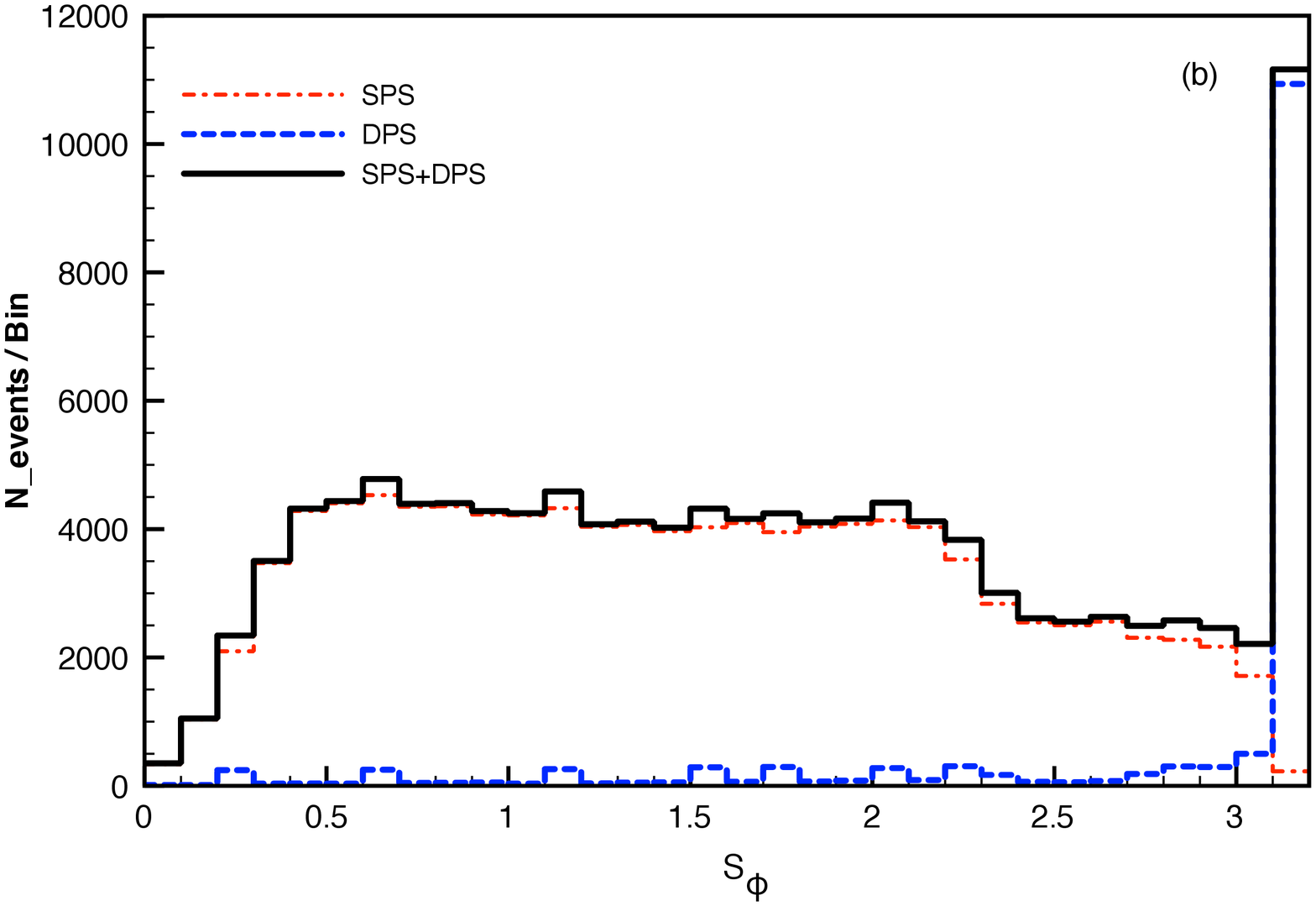}
\caption{(a) The difference $\Delta \phi$ in the azimuthal angles of light jet pairs for DPS and both SPS+DPS events.  The dijet pairs are back-to-back in DPS events.  (b) The variable $S_\phi$ for DPS and SPS+DPS events provides a stronger separation of the underlying DPS events from the total sample when compared to $\Delta\phi$ for any pair.}
\label{fig:delphi}
\end{center}
\end{figure}

We turn next to the search for variables that may allow for a clear separation of the DPS and SPS contributions.  Since the topology of the DPS events includes two $2\to2$ hard scattering events, the two pairs of jet objects are roughly back-to-back.  We expect the azimuthal angle between the pairs of jets corresponding to each hard scattering event to be strongly peaked near $\Delta \phi_{jj} \sim \Delta \phi_{bb} \sim \pi$.  Real radiation of an additional jet, where the extra jet is missed because it fails the threshold or acceptance cuts, allows smaller values of $\Delta \phi_{jj}$.  The relevant distribution is shown for light jets (non $b$-tagged) in Fig.~\ref{fig:delphi}a. There is a clear peak near $\Delta\phi_{jj}=\pi$ for DPS events, while the events are more broadly distributed in SPS events.  The secondary peak near small $\Delta\phi_{jj}$ arises from gluon splitting which typically produces nearly collinear jets.  The suppression at still lower $\Delta\phi_{jj}$ comes from the isolation cut $\Delta R_{jj} > 0.4$.  

The separation of DPS events from SPS events becomes more pronounced if information is used from both the $b\bar{b}$ and $jj$ systems.  As an example, we consider the distribution built from a combination of the azimuthal angle separations of both $jj$ and $b\bar b$ pairs, using a variable adopted from Ref.~\cite{Abazov:2009gc}:
\begin{equation}
S_{\phi}={1\over \sqrt 2} \sqrt{\Delta \phi(b_1,b_2)^2+\Delta \phi(j_1,j_2)^2}.
\end{equation}
In Fig.~\ref{fig:delphi}b, we present a distribution in $S_{\phi}$ for both DPS and SPS+DPS events.  Again, as in the case of the $\Delta \phi$ distribution, the SPS events are broadly distributed across the allowed range of $S_\phi$.  However, the combined information from both the $b\bar{b}$ and $jj$ systems shows that the DPS events produce a sharp and substantial peak near $S_\phi \simeq \pi$ which is well-separated from the total sample.

The narrow peaks near $\Delta\phi_{jj}=\pi$ in Fig.~\ref{fig:delphi}a and near $S_{\phi} = 1$ in 
Fig.~\ref{fig:delphi}b will be smeared somewhat once soft QCD radiation and other higher-order terms are included in the calculation.  

Another possibility for discerning DPS is the use of the total transverse momentum of both the $b\bar{b}$ and $jj$ systems.  At lowest order for a $2 \to 2$ process, the vector sum of the transverse momenta of the final state pair vanishes.  In reality, radiation and momentum mismeasurement smear the expected peak near zero.  Nevertheless, the DPS events are expected to show a reasonably well-balanced distribution in the transverse momenta of the jet pairs.  To encapsulate this expectation for both light jet pairs and $b$-tagged pairs, we use the variable~\cite{Abazov:2009gc}:

\begin{equation}
S_{p_T}^\prime={1\over \sqrt 2} \sqrt{\left({|p_T(b_1,b_2)|\over |p_T(b_1)|+|p_T(b_2)|}\right)^2+\left({|p_T(j_1,j_2)|\over |p_T(j_1)|+|p_T(j_2)|}\right)^2}.
\label{eq:sptprime}
\end{equation}
Here $p_T(b_1,b_2)$ is the vector sum of the transverse momenta of the two final state $b$ jets, and $p_T(j_1,j_2)$ is the vector sum of the transverse momenta of the two (non $b$) jets.  

\begin{figure}[ht]
 \begin{center}
\includegraphics[width=0.49\textwidth]{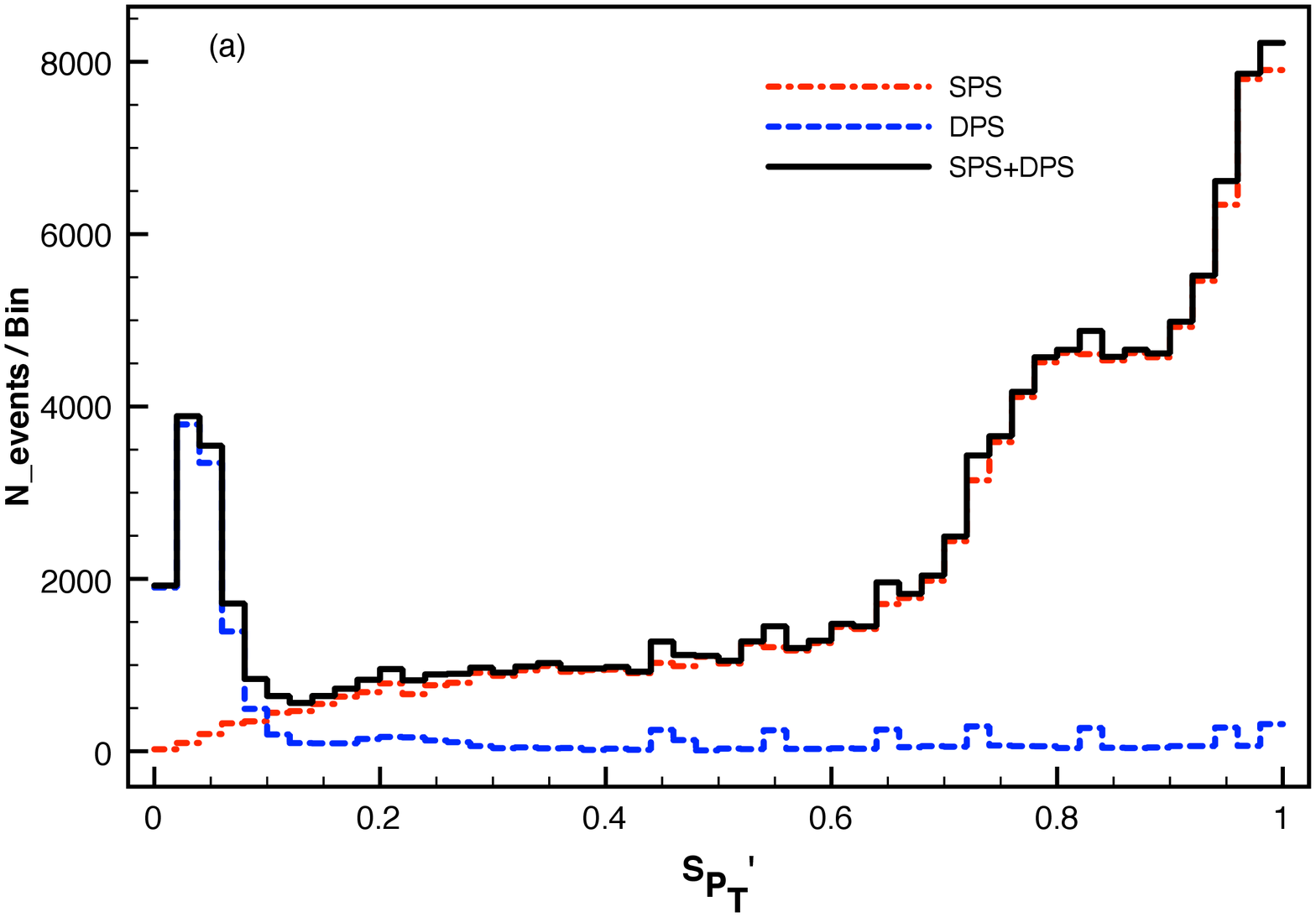}
\includegraphics[width=0.49\textwidth]{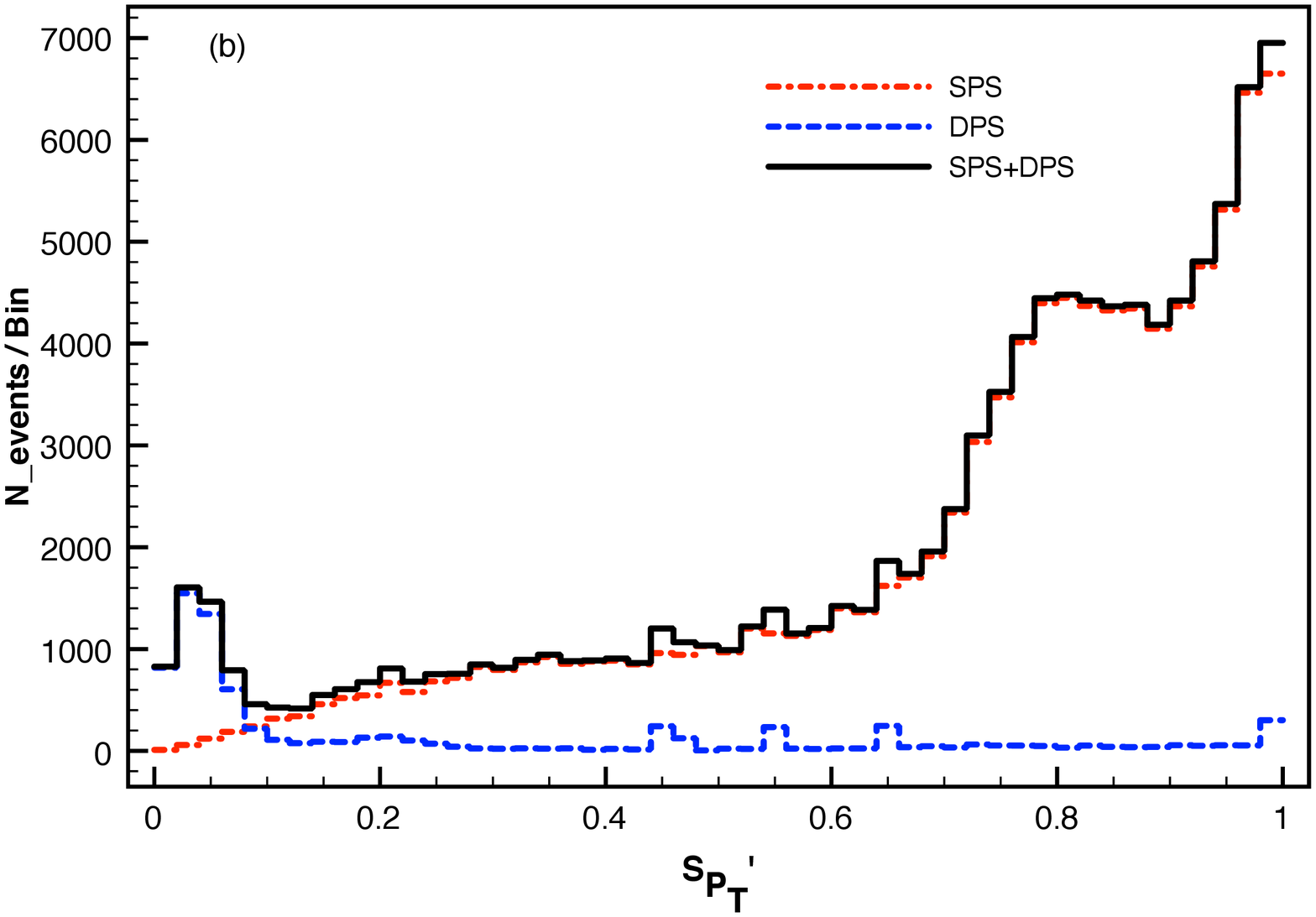}
\end{center}
\caption[]{Distribution of events in $S_{p_T}^\prime$ for the DPS and SPS samples.  Due to the back-to-back nature of the $2\to2$ events in DPS scattering, the transverse momenta of the jet pair and of the 
$b$-tagged jet pair are small, resulting in a small value of $S_{p_T}^\prime$.  In (a) we show the 
$S_{p_T}^\prime$ distribution for our standard cuts, and in (b) we increase the cut on the transverse momentum of the leading jet, $p_T^{j1} > 40$ GeV.  The fraction of DPS  events in the whole sample decreases with increasing $p_T^{j1}$.}
\label{fig:sptprimecut}
\end{figure}

The distribution in $S_{p_T}^\prime$ is shown in Fig.~\ref{fig:sptprimecut}.  As expected, the DPS events are peaked near $S_{p_T}^\prime\sim0$ and are well-separated from the total sample.  The SPS events, on the other hand, tend to be far from a back-to-back configuration and, in fact, are peaked near $S_{p_T}^\prime\sim1$.  This behavior of the SPS events is presumably related to the fact that a large number of the $b\bar{b}$ or $jj$ pairs arise from gluon splitting which yields a large $p_T$ imbalance and, thus, larger values of $S_{p_T}^\prime$.

Our simulations suggest that the variable $S_{p_T}^\prime$ may be a more effective discriminator than 
$S_{\phi}$.   However, given the leading order nature of our calculation and the absence of smearing associated with initial state soft radiation, this picture may change and a variable such as $S_\phi$ (or some other variable) may become a clearer signal of DPS at the LHC.  Realistically, it would be valuable to study both distributions once LHC data are available in order to determine which is more instructive.  

\begin{figure}
 \begin{center}
\includegraphics[width=0.90\textwidth]{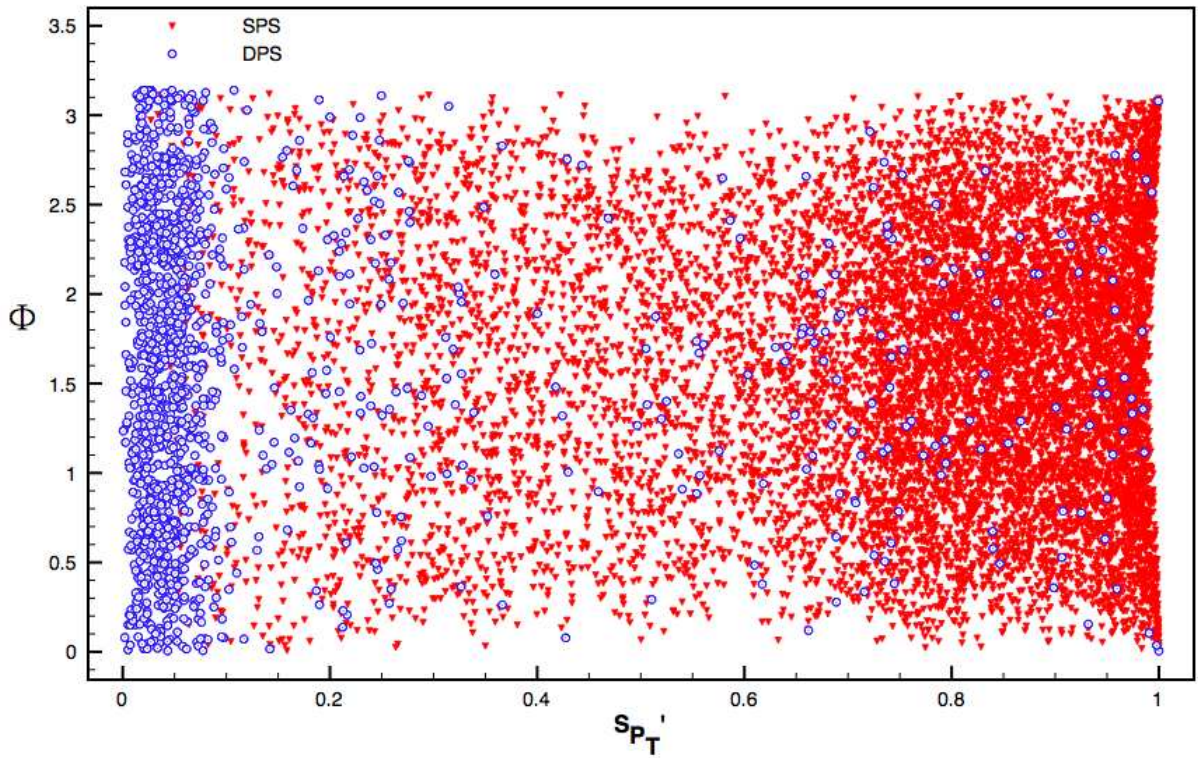}
\end{center}
\caption[]{Two-dimensional distribution of events in the inter-plane angle $\Phi$ and the scaled transverse momentum variable $S_{p_T}^\prime$ for the DPS and SPS samples.}
\label{fig:scatter1}
\end{figure}

The evidence in one-dimensional distributions for distinct regions of DPS dominance prompts the search for greater discrimination in a plane represented by a two dimensional distribution of one variable against another.  One scatter plot with interesting features is displayed in 
Fig.~\ref{fig:scatter1}.   The DPS events are seen to be clustered near $S'_{p_T} = 0$ and are uniformly distributed in $\Phi$.  The SPS events peak toward $S'_{p_T} = 1$ and show a roughly $\sin \Phi$ character.   While already evident in one-dimensional distributions, these two  
features are more apparent in the scatter plot Fig.~\ref{fig:scatter1}.  Moreover, the 
scatter plot shows a valley of relatively low density between $S'_{p_T} \sim 0.1$ and $\sim 0.4$. 
In an experimental one-dimensional $\Phi$ distribution, one would see the sum of the DPS and SPS contributions.  If structure is seen in data similar to that shown in the scatter plot Fig.~\ref{fig:scatter1}, one could make a cut at $S'_{p_T} < 0.1$ or $0.2$ and verify whether the experimental distribution in $\Phi$ is flat as expected for DPS events.  
 
\subsubsection{Strategy and Further Work}
The clear separation of DPS from SPS events in Fig.~\ref{fig:scatter1} suggests a methodology for the study of DPS.  One can begin with a clean process such as $p p \rightarrow b \bar{b} j_1 j_2 X$ and examine the distribution of events in the plane defined by  $S'_{p_T} $ and $\Phi$.  We expect to see a concentration of events near $S'_{p_T} = 0$ that is uniformly distributed in $\Phi$.  These are the DPS events.  Assuming that a valley of low density is observed between $S'_{p_T}  \sim 0.1 $ and $\sim 0.4$, one can make a cut there that produces an enhanced DPS sample.  Relative to the overall sample, this enhanced sample should show a more rapid decrease of the cross section as a function of the transverse momentum of the leading jet, and the enhanced sample can be used to measure 
$\sigma_{\rm eff}$.  A similar examination of other final states, such as 4 jet production, will answer whether the extracted values of $ \sigma_{\rm eff}$ are roughly the same.  Theoretical and experimental studies of other processes can follow, such as $b \bar{b} t \bar{t}$, $W {\rm j j}$, and $H {\rm j j}$.  

On the phenomenological front, next-to-leading order (NLO) expressions should be included for both the SPS and DPS contributions.  The NLO effects are expected to change normalizations and, more importantly, the distributions in phase space.  The sharp peaks near $\Delta\phi_{jj}=\pi$ 
in Fig.~\ref{fig:delphi}a, $S_\phi \simeq \pi$ in Fig.~\ref{fig:delphi}b, and $S_{p_T}^\prime = 0$ in Fig.~\ref{fig:sptprimecut} will be broader and likely displaced somewhat.  

Finally, it would be good to examine the theoretical underpinnings of Eq.~(\ref{eq:dpscross}) and, in the process, gain better insight into the significance of $\sigma_{\rm eff}$.   A firm basis is 
desirable for Eq.~(\ref{eq:dpscross}) starting from the formal expression for the differential cross section in terms of the absolute square of the full matrix element integrated over phase space:
\begin{eqnarray}
d\sigma(p p \rightarrow b \bar{b} j_1 j_2 X) = \frac{1}{2s} | {M (p p \rightarrow b \bar{b} j_1 j_2 X)}|^2 d PS_{b \bar{b} j_1 j_2 X}.  
\end{eqnarray}

The amplitude $M (p p \rightarrow b \bar{b} j_1 j_2 X)$ should include a sum of amplitudes for 
2-parton collisions (one active from each incident hadron, i.e., $2 \rightarrow 4$);  
3-parton collisions (two active from one hadron and one active from the other); and 
4-parton collisions (two active from each hadron or three from one and one from the other), 
and so forth that all yield the same 4 parton final state.  There will be contributions to the final state 
from the squares of individual amplitudes as well as interference terms.  
Specializing to $4 \rightarrow 4$, the DPS case, one would start from a 4-parton $\rightarrow$ 4-parton hard part.   Not evident at this time is how the four-parton matrix element can be reduced to a product 
of two matrix elements for the single parton scatterings, needed for  Eq.~(\ref{eq:dpscross}).  The 
demonstration of clear DPS signals in LHC data would be an important stimulus for further theoretical studies.  

\subsubsection{NLO DPS Study of $p p \rightarrow Wb\bar{b} X \to \ell \nu b\bar{b} X$ }

In a subsequent paper~\cite{Berger:2011ep}, we investigate the possibility to observe double parton scattering at the early LHC in the
process $p p \rightarrow Wb\bar{b} X \to \ell \nu b\bar{b} X$.  Our analysis is done at next-to-leading order in QCD.   It begins with the basic assumption that $Wb\bar{b}$ production consists of two components: the traditional single parton scattering (SPS) process and the double parton scattering (DPS) process where two individual hard scatterings produce the $Wb\bar{b}$ final state.

After identifying the most relevant background processes, we pinpoint a set of observables and cuts which would allow for the best separation between the DPS $W b \bar {b}$ signal and the backgrounds (including the SPS $Wb\bar{b}$ process).  To provide the most precise predictions possible, we generate the DPS $W b \bar {b}$ signal event sample,  the SPS $W b \bar {b}$ sample, and the
dominant  background event samples at next-to-leading order in QCD.  The main obstacles in the extraction of the DPS signal are the backgrounds from
$t\bar{t}$ production and the SPS $Wb\bar{b}$ component.   The most efficient way to suppress the $t\bar{t}$ background is with an upper cut on the missing transverse energy of the event, since top quark decays result in larger values of $\met$.

To separate the DPS component of $Wb\bar{b}$ from the SPS component, we find it useful again to employ observables which take into account information on the full final state rather than observables which involve one or two particles. Examples are the $S_{p_{T}}^{\prime}$ variable and the angle ($\Delta \Theta_{b\bar{b},\ell\nu}$) between the two planes defined by the $b\bar{b}$ and $\ell \nu$ systems, respectively.   By displaying the information from these two observables in two-dimensional distributions, we show
that it is possible to identify distinct regions in phase space where the DPS events reside.  Utilizing cuts on these observables that enhance the DPS $Wb\bar{b}$ sample, we find that the DPS signal can be observed with a statistical significance in the range $S/\sqrt{B} \sim 12 - 15$.

The focus in Ref.~\cite{Berger:2011ep} is to establish double parton scattering as a discernible physics process at LHC energies and measuring the size of its contribution.  Once DPS production of
$Wb\bar{b}$ is observed, it will be interesting to assess its potential significance as a background in searches for other physics, such as Higgs boson production in association with a $W$ boson (where the Higgs boson decays as $H \to b\bar{b}$), and precise studies of single top quark production where new physics could contribute to the $Wtb$ vertex.   A detailed analysis of either of these channels would require a different set of optimized physics cuts.   We limit ourselves in Ref.~\cite{Berger:2011ep} to showing the $b\bar{b}$ invariant mass distribution for the $\ell\nu b\bar{b}$ final state.  We see that the DPS $W b \bar{b}$ component alters the overall shape of the $b \bar{b}$ mass spectrum, enhancing the small mass region.  This feature is consistent with our earlier observation that the $p_T$ spectrum of leading jets is softer in the DPS component.  The DPS component contributes primarily in the region below 120 GeV or so.  At face value, it does not seem to pose a hindrance for searches for Higgs bosons in the $HW$ channel.  However,
$Wb\bar{b}$ DPS could be a significant  background in the search for new particles, with masses in the 50 - 100 GeV range and appearing as resonances in $M_{bb}$, and it should be accounted for in any analysis.

\newcommand{\MZ}{M_Z}
\renewcommand{\bbbar}{b\bar{b}}
\newcommand{\ttbar}{t\bar{t}}

\newcommand{\lag}{\mathcal{L}}
\newcommand{\mM}{\mathcal{M}}
\newcommand{\lam}{\lambda}
\newcommand{\intd}{\int\frac{d^Dq}{(2\pi)^D}}

\newcommand{\MET}{{\not\!\!E_T}}
\renewcommand{\pt}{p_{T}}
\newcommand{\ptlep}{p^l_T}
\newcommand{\ptb}{p^b_T}
\newcommand{\Dphi}{\Delta\phi}

\newcommand{\pdfannotlink}{\pdfstartlink}
\renewcommand{\be}{\begin{equation}}
\renewcommand{\ee}{\end{equation}}
\renewcommand{\bea}{\begin{eqnarray}}
\renewcommand{\eea}{\end{eqnarray}}
\newcommand{\eqa}{\begin{eqnarray}}
\newcommand{\qea}{\end{eqnarray}}
\newcommand{\mc}{\mathcal}
\renewcommand{\gsim}{\gtrsim}
\renewcommand{\lsim}{\lesssim}
\renewcommand{\mb}{\textrm{ mb}}
\newcommand{\mw}{M_{W}}
\renewcommand{\Wp}{W^+}
\renewcommand{\Wm}{W^-}
\renewcommand{\Zgam}{Z(\gamma^*)}
\renewcommand{\Wpm}{W^{\pm}}
\newcommand{\lpm}{l^{\pm}}
\newcommand{\numu}{\nu_{\mu}}
\newcommand{\order}{\mathcal{O}}
\newcommand{\sig}{\sigma}
\newcommand{\sigh}{\hat{\sigma}}
\newcommand{\sigeff}{\sigma_{\textrm{eff}}}
\newcommand{\sigeffN}{\sigma_{N,\textrm{eff}}}
\newcommand{\as}{\alpha_S}
\newcommand{\aw}{\alpha}
\newcommand{\roots}{\sqrt{s}}
\newcommand{\bmix}{B^0\textrm{-}\bar{B}^0}
\renewcommand{\barn}{\textrm{b}}
\newcommand{\BR}{\mathcal{BR}}
\newcommand{\nn}{\langle n\rangle}

\newcommand{\herwig}{\texttt{HERWIG6.510}}
\newcommand{\mcfm}{\texttt{MCFM}}
\newcommand{\mstwlo}{\texttt{MSTW08 LO}}
\newcommand{\vegas}{\texttt{VEGAS}}
\newcommand{\madgraph}{\texttt{MADGRAPH}}
\newcommand{\legacy}{\texttt{LEGACY}}
\def\bib{\bibitem}

\graphicspath{{kom/figs/}}

\contribution{Probing double parton scattering with same-sign W pairs at the LHC}
{Contributing authors: J. R. Gaunt, C.-H. Kom, A. Kulesza, and W. J. Stirling}
\label{kom}

In this section we focus on $\Wpm\Wpm$ production, followed by decays
into same-sign di-leptons (SSDL) plus missing transverse energies
($\MET$).  Our discussion is based on \cite{Gaunt:2010pi} (see
\cite{Kulesza:1999zh} for an early study of this process).  One
advantage in studying DPS $\Wpm\Wpm$ is that single $\Wpm$ production
is well described theoretically, and will be accurately measured at
the LHC.  As a result an approximation with no correlations included
will be well modelled.  In addition, $\Wpm\Wpm$ production through
single parton scattering (SPS) is forbidden at the same order in the
SM, i.e. there is no $q\bar{q}' \to \Wpm\Wpm$ contribution, due to
($U(1)_{\rm EM}$) charge conservation.  The lowest order `background'
process is instead $q\bar{q}' \to \Wpm\Wpm jj$, which is of order
$\order(\aw^4)$ or $\order(\as^2\aw^2)$.  The additional jet activity
can be used to distinguish this background from the DPS signal.

Same-sign $\Wpm\Wpm$ production could then potentially be a benchmark
DPS process.  We will study kinematic properties of this process by
including longitudinal correlation effects, namely constraints from
momentum and valence number conservation.  These constraints are
consistently included in the double parton distribution function
(dPDF) set GS09 \cite{Gaunt:2009re}.  We refer readers to Section~\ref{gaunt} and original literature for details on the construction of
this dPDF set.  We will quantitatively study effects of these
constraints by comparing results using GS09 with other simple dPDF
models that will be defined later.  In the literature, comparison of
DPS and SPS $\Wpm\Wpm$ production has mostly been studied at the level
of production cross sections.  Here we discuss strategies to extract
the signal from the background including important contributions from
heavy flavour and gauge boson pair productions.



\begin{table}[t]
  \centering
  \scalebox{0.93}{
  \begin{tabular}{|c|cccc|}
    \hline
     &$\sigma_{\textrm{GS09}}$&$\sigma_{\textrm{MSTW}_0}$&$\sigma_{\textrm{MSTW}_1}$&$\sigma_{\textrm{MSTW}_2}$\\
    \hline
    $\Wp\Wm$ &0.546 &0.496 &0.409 &0.348\\
    $\Wp\Wp$ &0.321 &0.338 &0.269 &0.223\\
    $\Wm\Wm$ &0.182 &0.182 &0.156 &0.136\\
    \hline \hline
    & &  \hspace{2cm}$R$  & & \\
    \hline
    &0.784&1.00&1.00&1.00\\
    \hline
  \end{tabular}
  }  
  \caption{DPS $WW$ total cross sections (in pb) and ratio $R$ defined
    in Eq.~(\protect\ref{eq:MSTWndef}) for $pp$ collisions at $\sqrt
    s=14$ TeV evaluated using different dPDFs
    sets.}
  \label{tab:xsec_DPS}
\end{table}

\begin{table}[t]
  \centering
  \begin{tabular}{|c|ccc|}
    \hline 
    & &  $\sigma_{\rm GS09}$  & \\
    \hline
      & $\sqrt s=$7 TeV& $\sqrt s=$ 10 TeV &$\sqrt s=$ 14 TeV\\
    \hline
    $\Wp\Wm$ &0.107 &0.250 &0.546\\
    $\Wp\Wp$ &0.0640 &0.148 &0.321\\
    $\Wm\Wm$ &0.0317 &0.0793&0.182\\
\hline \hline
 & &  $R$  & \\
    \hline 
    &0.709&0.751&0.784\\
    \hline
  \end{tabular}
  \caption{DPS $WW$ total cross sections (in pb) and ratio $R$ defined in
    Eq.~(\protect\ref{eq:MSTWndef}) for $pp$ collisions at different CM
    energies $\sqrt s$.}
  \label{tab:xsec_DPSvsCME}
\end{table}

\subsubsection{The $\Wpm\Wpm$ signal}

Our phenomenological investigation for DPS $\Wpm\Wpm$
production is based on the equation~\ref{eq:DPSmaster}, with  $m=1$, $A=B=\Wpm$ and $t_1=t_2=\ln(\mw^2)$. We assume $\sig_{\rm eff}=14.5 {\rm mb}$, the value
obtained by the CDF $\gamma$ + 3$j$~\cite{Abe:1997xk} analysis.  We adopt GS09, which
incorporates momentum and sum rule constraints, as the default
longitudinal dPDF choice.  To see the effect of these constraints, we
compare cross sections using factorised dPDFs of the form:
\begin{eqnarray}
D_{ij}(x_1,x_2) &=&
D_i(x_1)D_j(x_2)\nonumber \\
&&\times\theta(1-x_1-x_2)(1-x_1-x_2)^n\;,\nonumber \\
n&=&0,1,2 \;,
\label{eq:MSTWndef}
\end{eqnarray} 
hereafter referred to as $\textrm{MSTW}_n$ sets.  In the above
expression, the factorization scale is again fixed at $\mu_F=M_W$. The
completely factorised approximation that has been taken in existing
phenomenological studies is obtained by setting $D_{ij}(x_1,x_2) =
D_i(x_1)D_j(x_2)$. The sPDFs $D_i(x)$ are taken from the MSTW
2008 LO set.  In Table \ref{tab:xsec_DPS}, we compare the cross
sections and the values of the ratio

\begin{figure}[!t]
  \begin{center}
    \scalebox{0.7}{ \includegraphics{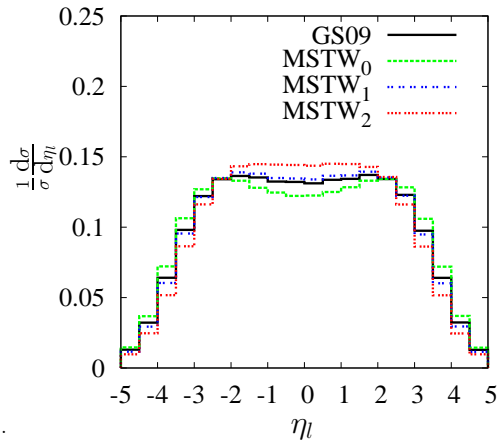} }
    \caption{Normalised $l^+$ pseudorapidity distributions for $pp$
      collisions at $\sqrt s=14$ TeV evaluated using different dPDFs.
      No cuts are applied.  }
    \label{fig:yLtot_vs_dpdf} 
  \end{center}
\end{figure}

\begin{equation}
  R\equiv4\frac{\sigma_{\Wp\Wp}\;\sigma_{\Wm\Wm}}{\sigma^2_{\Wp\Wm}}\;,
\end{equation}
which measures the deviation from the factorisation approach ($R=1$
when factorisation is exact) using different dPDFs.  In
Table~\ref{tab:xsec_DPSvsCME}, the value $R$ as a function of CM
energy using GS09 is evaluated.  We see that using GS09, factorisation
is broken at the 20\% to 30\% level, and the approximation improves at
the higher collider energies as lower $x$ regions are probed.  On the
contrary, the $\textrm{MSTW}_n$ models have $R$ values very close to 1
because effects of the momentum suppression factors $(1-x_1-x_2)^n$ in
the numerator and the denominator of the expression tend to cancel.

Sum rule effects incorporated in GS09 can also be seen in the
distributions of the charged leptons $\lpm$ from $\Wpm$ decays.  To
understand this, note that production of same sign $\Wpm$'s along the
same forward direction is suppressed compared with factorised models
due to both momentum and number sum rule constraints.  This is because
this phase space region is significantly influenced by
$D^{vv}(x_1,x_2)$, where $v=v_u,v_d$ denotes valence up or down
quarks, both of which with relatively high momentum fractions $x_1$
and $x_2$. In fact, $D^{v_dv_d}=0$ identically in GS09 due to the
presence of only one valence down quark in a proton, whereas it is
non-zero for other simple factorised models. The $\lpm$ hence have
more central distributions compared with the $\textrm{MSTW}_n$ models.
The pseudorapidity ($\eta_l$) distributions for $l^+$ are shown in
Fig.~\ref{fig:yLtot_vs_dpdf}.

\begin{figure}[!t]
  \begin{center}
    \scalebox{0.7}{ \includegraphics{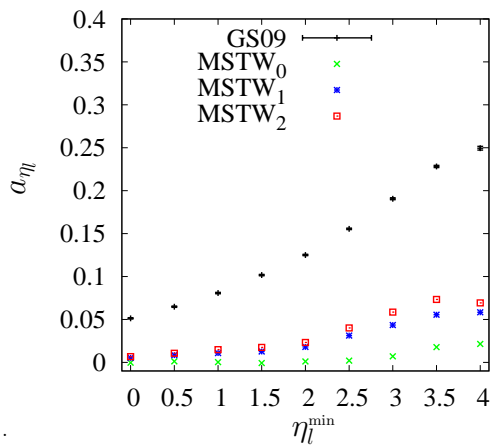} }
    \caption{Pseudorapidity asymmetry $a_{\eta_l}$ for $pp$ collisions at
      $\sqrt s=14$ TeV evaluated using different dPDFs.  No cuts are
      applied.}
    \label{fig:AyLtot_vs_dpdf}
  \end{center}
\end{figure}

Such longitudinal correlations also imply that the SSDL pairs prefer
to lie in opposite hemispheres (i.e. $\eta_{l_1}\times\eta_{l_2} <
0$).  This preference can be quantified by computing pseudorapidity
asymmetry, defined as \bea
a_{\eta_l}=\frac{\sigma(\eta_{l_1}\times\eta_{l_2}<0)-\sigma(\eta_{l_1}\times\eta_{l_2}>0)}{\sigma(\eta_{l_1}
  \times\eta_{l_2}<0)+\sigma(\eta_{l_1}\times\eta_{l_2}>0)}, \eea
where $|\eta_{l_1}|,|\eta_{l_2}|>\eta^{\rm min}_l$.  The value of
$a_{\eta_l}$ for $l^+$ as a function of $\eta^{\rm min}_l$ is
displayed in Fig.~\ref{fig:AyLtot_vs_dpdf}.  It increases with
$\eta^{\rm min}_l$ as the correlations are most important for the
distributions probed at high values of $x$ for both partons in the
same proton, which is reached when the leptons are produced at high
$|\eta_l|$.  For GS09, this effect is even more pronounced for $l^-$,
as only one valence down quark is present in a proton, making
simultaneous extraction of two high $x$ down quarks highly suppressed.


\subsubsection{Single scattering backgrounds}

\begin{figure}[!ht]
  \begin{center}
    \scalebox{0.75}{
      \includegraphics{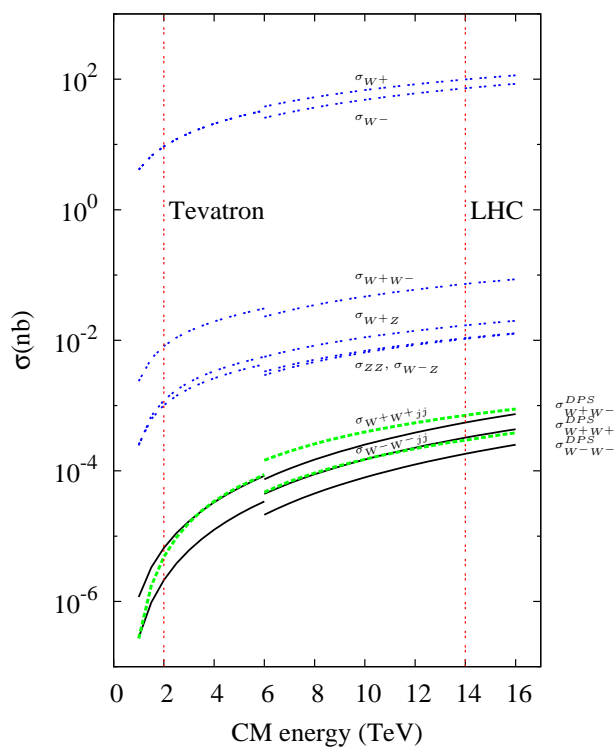}
    }
    \caption{Cross sections of various electroweak processes in
      $p$-$p(\bar{p})$ collisions as a function of $\sqrt{s}$.  The
      dotted curves correspond to single scattering processes, while
      the solid curves correspond to double scattering processes
      computed using GS09 dPDFs.}
    \label{fig:xsecProc} 
  \end{center}
\end{figure}

\begin{figure}[!t]
  \begin{center}
    \scalebox{0.7}{ \includegraphics{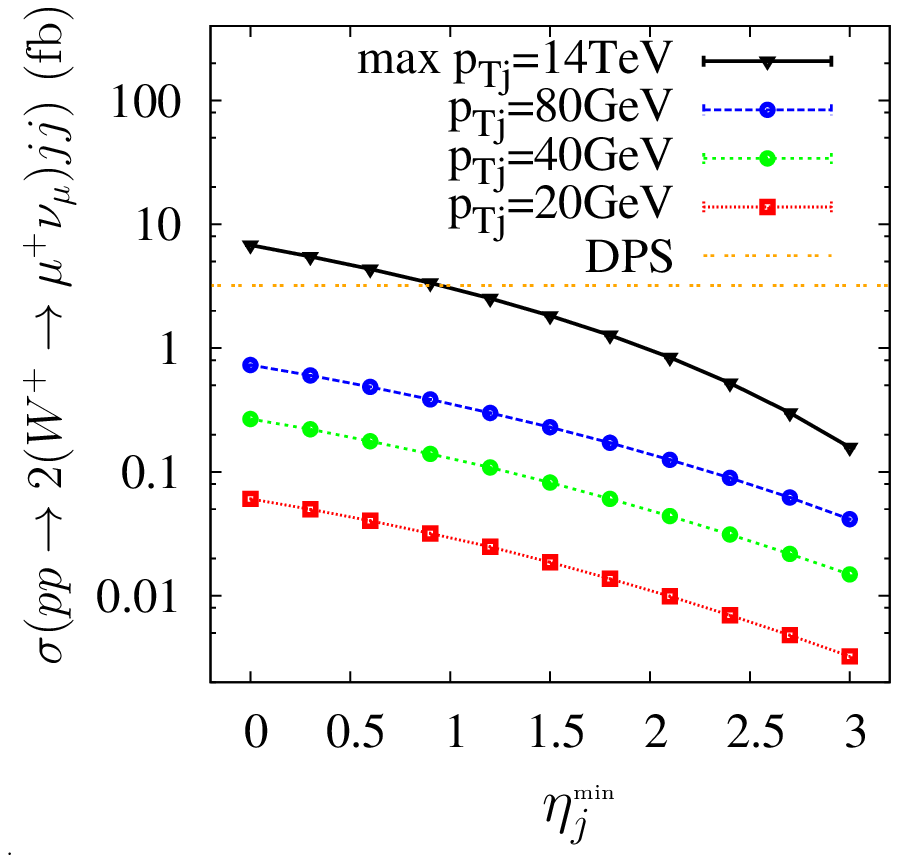} }
    \caption{$\sigma(pp\to\Wp\Wp jj)\cdot[\BR (\Wp\to\mu^+\numu)]^2$~(fb) as
      a function of max $p_{Tj}$ and $\eta^{\textrm{\tiny min}}_{j}$.  No
      other cuts are applied.}
    \label{fig:WWjj_PtjmaxYjmin}
  \end{center}
\end{figure}

We now turn to the single scattering backgrounds.  We show in
Fig.~\ref{fig:xsecProc} the total cross sections of the DPS signal, as
well as electroweak single scattering backgrounds and related
processes.  As discussed before, due to charge conservation the lowest
order `irreducible' SPS background including a pair of same sign $W$'s
is $q\bar{q}' \to \Wpm\Wpm jj$.  At the 14 TeV LHC, the DPS signal
and SPS background total cross sections are of the same order of
magnitude ($\mathcal{O}(0.1 - 1){\rm pb}$ before leptonic branching
ratios).  However the presence of two additional hard jets means that
central hard $\pt$ jet veto can be useful in suppressing this
background.  The SPS cross section as a function of max jet $\pt$ and
min jet $\eta$ at the parton level is displayed in
Fig.~\ref{fig:WWjj_PtjmaxYjmin}.  As we can see, this background can
be suppressed easily, for instance by vetoing events with central jets
with $\pt > 20$ GeV.

Actually other electroweak processes that can lead to the same final
states.  In particular, both $\Wpm\Zgam$ and $\Zgam\Zgam$ gauge boson
production can lead to SSDL + $\MET$ when the `wrong' sign leptons are
not identified, for instance if they lie outside the central tracking
region, or if they are too soft to be reconstructed.  The relevant
processes are \bea q\bar{q}' &\rightarrow& \Wp \Zgam \rightarrow
l^+\nu l^+(l^-), \nonumber \\ q\bar{q}^{\phantom{'}} &\rightarrow&
\Zgam\Zgam \rightarrow l^+(l^-)l^+(l^-), \eea and their
charge-conjugated processes.  In the above expressions, charged
leptons in brackets are not identified.

As can be seen in Fig.~\ref{fig:xsecProc}, the total cross sections
for $\Wpm Z$ and $ZZ$ are about an order of magnitude above the
signal, while the cross sections are significantly larger when
$\gamma^*$ is involved.  However, for these processes the lepton $\pt$
spectra are much harder compared with the signal, hence a max lepton
$\pt$ cut can reduce this background.  A wrong sign lepton-veto in the
central region will also be useful in reducing the $Z$ contribution,
while looking for presence of low invariant mass system of an isolated
charged tracks and a nearby identified lepton \cite{Chanowitz:1994ap}
can help suppressing the $\gamma^*$ contribution.

Another source of background is pair production of heavy flavour
quarks.  Production of $\bbbar$ can lead to SSDL pairs, when a neutral
B-meson is present and undergos $\bmix$ mixing, followed by
semi-leptonic decay for both B-mesons.  The relevant processes are
then: \bea gg &\rightarrow& \bbbar \to B \bar B + ...\; , \nonumber
\\ B &\rightarrow& l^+\nu X, \nonumber \\ \bar{B}^0 &\rightarrow& B^0
\rightarrow l^+\nu \tilde{X}, \eea together with the charge
conjugation processes.

The $\bbbar$ cross section is orders of magnitude larger than that of
the signal.  However, the $\pt$ spectrum of the B-mesons decreases
exponentially.  It is thus very difficult for both charged lepton and
neutrino from the semi-leptonic B decay to acquire large transverse
momenta.  Imposing tight lepton isolation, min lepton $\pt$ and min
$\MET$ cuts will thus help suppressing contribution from this process.

Production of $\ttbar$ pair can also result in SSDL pairs when a top
and the bottom of the other top decay semi-leptonically.  The relevant
processes are
\bea 
t &\rightarrow& \Wp b \rightarrow l^+\nu b, \nonumber \\ 
\bar{t} &\rightarrow& \Wm \bar{b} \rightarrow q\bar q'l^+\nu c.  
\eea
These events have significant jet activities.  Also, the lepton $\pt$
spectrum is much harder compared with the signal.  The leptons,
particularly the ones from B-decays, are usually poorly isolated.
We thus expect tight lepton isolation, central jet veto and a max
lepton $\pt$ cut will be effective in suppressing this background, and
will not consider this process further in our numerical analysis.


\subsubsection{Observing DPS $\Wpm\Wpm$ at the LHC}

\begin{table}[t]
  \centering
  \begin{tabular}{|c|cc|}
    \hline
    & $\sigma_{\mu^+\mu^+}$ (fb) & $\sigma_{\mu^-\mu^-}$ (fb) \\
    \hline
    $\Wpm\Wpm$(DPS) &0.82 &0.46 \\
    \hline
    $\Wpm \Zgam$ & 5.1 &3.6 \\
    $\Zgam \Zgam$ & 0.84 &0.67\\
    $\bbbar\; (\ptb\geq 20 \textrm{~GeV})$ & 0.43& 0.43\\
    \hline
  \end{tabular}
  \caption[]{Cross sections (in fb) of the processes simulated after
    cuts, including branching ratios corresponding to same-sign dimuon
    production.}
  \label{tab:xsec_bcut}
\end{table}

We see in the last section, that despite the backgrounds having much
larger cross sections than the signal, there are handles to suppress
the former processes.  To see to what extent the signal can be
extracted from these backgrounds, we perform a parton level signal +
background simulation\footnote{We refer readers to the original paper
  \cite{Gaunt:2010pi} for a technical account of the simulation.} with
the following criteria:

\begin{itemize}
\item Both leptons in the like sign lepton pair must have
  pseudorapidity $|\eta|<2.5$.
\item Both leptons are required to be isolated:\\ $E_{\textrm{{\tiny ISO}}}^l\le
  E_{\textrm{{\tiny ISO}}}^{\textrm{{\tiny min}}}=10$ GeV, where $E_{\textrm{{\tiny ISO}}}^l$
  is the hadronic transverse energy in a cone of $R=0.4$ surrounding
  each of the like-sign leptons.
\item The transverse momenta of both leptons, $\ptlep$, must satisfy
  $20 \le \ptlep\le 60$ GeV.
\item An event is rejected whenever a third, opposite-signed, lepton
  is identified.  A lepton is assumed to be identified with 100\%
  efficiency when $\ptlep\ge \pt^{\textrm{\tiny id}}$ and
  $|\eta|<\eta^{\textrm{\tiny id}}$, where $\pt^{\textrm{\tiny id}}=10$ GeV and
  $\eta^{\textrm{\tiny id}}=2.5$.
\item The missing transverse energy $\MET$ of an event must satisfy
  $\MET\ge 20$ GeV.
\item Reject an event if a charged (lepton) track with
  $\pt^{\textrm{\tiny id}}\ge\pt\ge 1$ GeV forms an invariant mass $< 1$ GeV
  with one of the same-sign leptons.
\end{itemize}

The cross sections after these cuts are displayed in
Table~\ref{tab:xsec_bcut}.  The largest background comes from the
$\Wpm\Zgam$ processes, which is a factor of a few larger than the
signal.  Unfortunately, we find that many basic kinematic
distributions are fairly similar between this SPS background and the
DPS signal, making further cuts unlikely to be beneficial.

\begin{figure*}[!t]
  \begin{center}
    \scalebox{0.7}{ \includegraphics{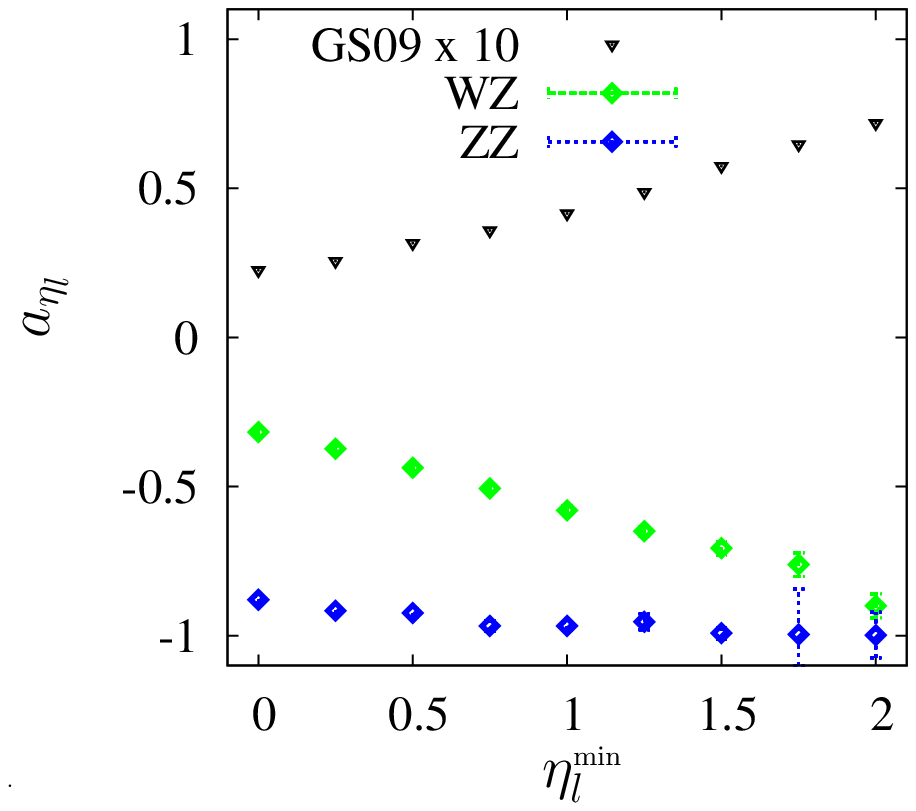} }
    \scalebox{0.7}{ \includegraphics{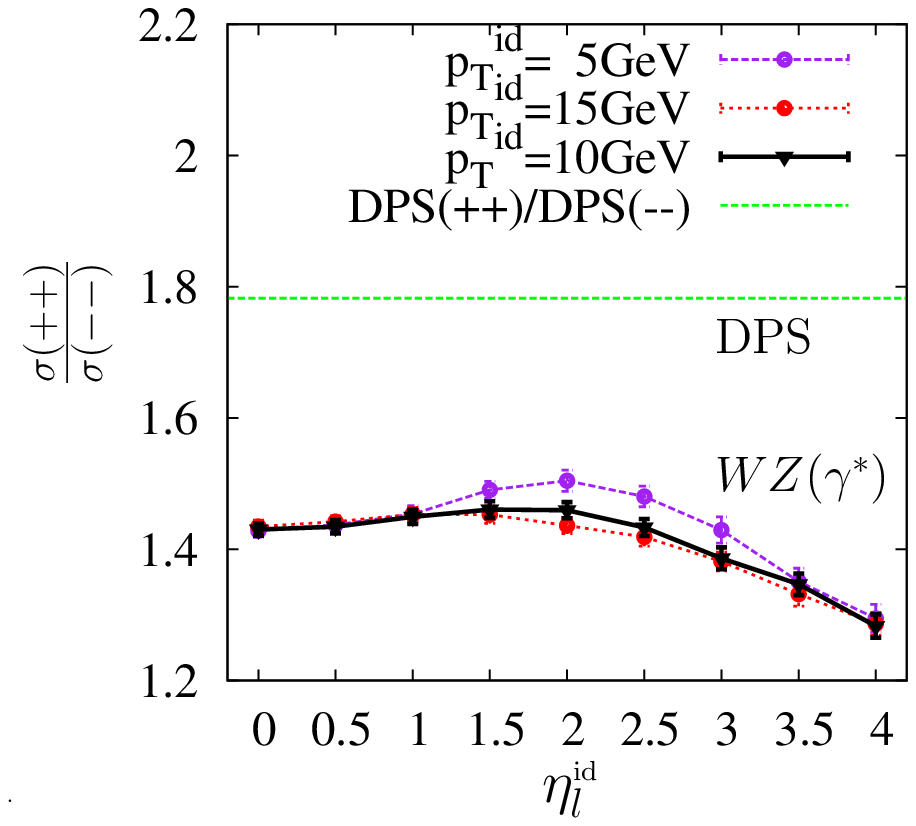} }
    \caption{Left: pseudorapidity asymmetry $a_{\eta_l}$ for the
      positive SSDL+$\MET$ DPS signal and selected SPS background,
      after imposing cuts described in the text.  Right: charge
      asymmetry ratio $(++)/(--)$ as a function of lepton
      identification criteria for different processes.}
    \label{fig:additionalHandles}
  \end{center}
\end{figure*}

On the positive side, it might be advantageous to exploit the fact
that the value of $a_{\eta_l}$ is relatively small, but positive, for
the DPS signal.  On the contrary, $a_{\eta_l}$ tends to be negative
for the background, which reflects the preference for the leptons to
lie close in pseudorapidity space in order to reduce the CM energy of
the system.  This property is illustrated in the diagram on the left
of Fig.~\ref{fig:additionalHandles}, which shows how $a_{\eta_l}$ for
different processes vary as a function of a minimum $\eta_l$ cut.

The ratio of positively charged $(++)$ and negatively charged $(--)$
SSDL events (which we call charge asymmetry ratio) may also be used,
as initial state partons of various flavours and momentum fractions
are involved in different processes, each leading to different charge
asymmetry ratio.  The charge asymmetry ratio for the DPS signal and
SPS $\Wpm\Zgam$ background using different lepton identification
criteria is displayed in the right figure of
Fig.~\ref{fig:additionalHandles}.  An important point that should be
noted is that this ratio is fairly stable when varying the cuts, as
can be inferred from the results in Table~\ref{tab:xsec_bcut}.

To summarise, same-sign $\Wpm\Wpm$ production can potentially be a
benchmark process for studying double parton scattering at the LHC.
We have discussed likely changes to the kinematic properties of the
signal when momentum and sum rule constraints are included in the
description of dPDFs.  We find that, after including important physics
backgrounds previously overlooked in the literature, a small excess of
SSDL DPS events could be observed at the LHC, while further
improvements can be made to enhance the signal.  
Finally, we note that there are other DPS processes with purely leptonic
final states, such as double Drell-Yan and double J/psi production, that
could be interesting.  Recent phenomenological studies of these processes
can be found in Refs \cite{Kom:2011bd,Kom:2011nu,Gaunt:2011rm,Baranov:2011ch,Novoselov:2011ff}.

\newcommand{\ra}{\rightarrow}

\renewcommand{\be}{\begin{equation}}
\renewcommand{\ee}{\end{equation}}

\renewcommand{\bea}{\begin{eqnarray}}
\renewcommand{\eea}{\end{eqnarray}}
\newcommand{\beanon}{\begin{eqnarray*}}
\newcommand{\eeanon}{\end{eqnarray*}}
\newcommand{\ba}{\begin{array}}
\newcommand{\ea}{\end{array}}
\newcommand{\bd}{\begin{description}}
\newcommand{\ed}{\end{description}}
\newcommand{\bi}{\begin{itemize}}
\newcommand{\ei}{\end{itemize}}
\newcommand{\ben}{\begin{enumerate}}
\newcommand{\een}{\end{enumerate}}
\newcommand{\bc}{\begin{center}}
\newcommand{\ec}{\end{center}}
\newcommand{\ul}{\underline}
\newcommand{\ol}{\overline}
\newcommand{\dotp}{\!\cdot\!}

\newcommand{\epem}{\mbox{${\mathrm e}^+{\mathrm e}^-$}\xspace}
\newcommand{\bb}{\mbox{${\mathrm b}{\mathrm b}$}\xspace}
\newcommand{\toptop}{\mbox{${\mathrm t} \bar {\mathrm t}$}\xspace}
\newcommand{\WWL}{\mbox{${\mathrm W_L}{\mathrm W_L}$}\xspace}
\newcommand{\VVL}{\mbox{${\mathrm V_L}{\mathrm V_L}$}\xspace}
\newcommand{\ZZL}{\mbox{${\mathrm Z_L}{\mathrm Z_L}$}\xspace}
\newcommand{\WWT}{\mbox{${\mathrm W_T}{\mathrm W_T}$}\xspace}
\newcommand{\WW}{\mbox{${\mathrm W}{\mathrm W}$}\xspace}
\newcommand{\ZZ}{\mbox{${\mathrm Z}{\mathrm Z}$}\xspace}
\newcommand{\VV}{\mbox{${\mathrm V}{\mathrm V}$}\xspace}
\newcommand{\ZW}{\mbox{${\mathrm Z}{\mathrm W}$}\xspace}
\newcommand{\VW}{\mbox{${\mathrm V}{\mathrm W}$}\xspace}
\renewcommand{\W}{\mbox{${\mathrm W}$}\xspace}
\newcommand{\V}{\mbox{${\mathrm V}$}\xspace}
\newcommand{\VL}{\mbox{${\mathrm V_L}$}\xspace}
\renewcommand{\Z}{\mbox{${\mathrm Z}$}\xspace}
\renewcommand{\pt}{\mbox{${\mathrm p_T}$}\xspace}
\renewcommand{\GeV}{\mbox{${\mathrm GeV}$}\xspace}
\renewcommand{\TeV}{\mbox{${\mathrm TeV}$}\xspace}

\newcommand{\ordEW}{\mathcal{O}(\alpha_{\scriptscriptstyle EM}^6)\xspace}
\newcommand{\ordEWfour}{\mathcal{O}(\alpha_{\scriptscriptstyle EM}^4)\xspace}
\newcommand{\ordQCD}{\mathcal{O}(\alpha_{\scriptscriptstyle EM}^4
  \alpha_{\scriptscriptstyle S}^2)\xspace}
\newcommand{\ordQCDsq}{\mathcal{O}(\alpha_{\scriptscriptstyle EM}^2
  \alpha_{\scriptscriptstyle S}^4)\xspace}

\newcommand{\eqn}[1]{Eq.(\ref{#1})}
\newcommand{\eqns}[2]{Eqs.(\ref{#1}--\ref{#2})}
\newcommand{\eqnsc}[2]{Eqs.(\ref{#1},~\ref{#2})}
\newcommand{\tbn}[1]{Tab.~\ref{#1}}
\newcommand{\tbns}[2]{Tabs.~\ref{#1}--\ref{#2}}
\newcommand{\tbnsc}[2]{Tabs.~\ref{#1},~\ref{#2}}
\newcommand{\fig}[1]{Fig.~\ref{#1}}
\newcommand{\figs}[2]{Figs.~\ref{#1}--\ref{#2}}
\newcommand{\figsc}[2]{Figs.~\ref{#1},~\ref{#2}}
\newcommand{\sect}[1]{Sect.~\ref{#1}}
\newcommand{\subsect}[1]{Sub-Sect.~\ref{#1}}
\newcommand{\subsects}[2]{Sub-Sects.~\ref{#1},~\ref{#2}}
\newcommand{\diags}[2]{diagrams~({#1})--({#2})}

\newcommand{\rf}[1]{Ref.~\cite{#1}}
\newcommand{\rfs}[1]{Refs.~\cite{#1}}  

\renewcommand{\O}{{\mathcal O}}

\def\Ord{\buildrel{\scriptscriptstyle <}\over{\scriptscriptstyle\sim}}
\def\OOrd{\buildrel{\scriptscriptstyle >}\over{\scriptscriptstyle\sim}}

\newcommand{\Phase}{{\sc Phase}\xspace}
\newcommand{\Phact}{{\sc Phact}\xspace}
\newcommand{\Phantom}{{\sc Phantom}\xspace}
\renewcommand{\Pythia}{{\sc Pythia}\xspace}
\newcommand{\MadEvent}{{\sc Madvent}\xspace}
\newcommand{\LHA}{Les Houches Accord\xspace}
\newcommand{\veg}{{\sc Vegas}\xspace}

\graphicspath{{maina/figs/}}

\contribution{Multiple Parton Interactions in $Z+jets$ production at the LHC}
{Contributing author: E. Maina}
\label{maina}

\subsubsection{Multiple Parton Interactions in $Z+jets$ production at the LHC}
\label{sec:calc}

In \rf{Maina:2010vh} the contribution of MPI to $Z+n$--jets production at
the LHC, $n=2,3,4$, where the $Z$ boson is assumed to decay leptonically, has been examined.
These processes have the advantage of a much larger cross section than same--sign
$WW$ production and therefore are more likely to allow detailed studies of MPI
at the low luminosity, about 1 fb$^{-1}$, foreseen for the first two years of
operation at the LHC with $\sqrt{s} = 7$ TeV.

$Z+nj$ production probes initial state parton combinations which are different from those
probed in $W^\pm W^\pm$ processes. The latter, at lowest order,
are always initiated by four--fermion states,
mainly $u\bar{d}u\bar{d}$. The former, on the contrary, typically have at least
two gluons in the initial state since the largest component \cite{Maina:2009vx,Maina:2009sj}
involves a two jet process which is dominated by gluon--gluon scattering.

For comparison we also present the predictions for $\gamma+3j$ production,
the reaction from which the most recent and precise estimates of $\sigma_{eff}$
have been extracted at the Tevatron. The CDF collaboration~\cite{Abe:1997bp,Abe:1997xk}
has measured $\sigma_{eff}=14.5\pm 1.7^{+1.7}_{-2.3}$ mb, a value
confirmed by D0 which quotes
$\sigma_{eff}=15.1\pm 1.9$ mb \cite{Abazov:2009gc}.
In Ref.~\cite{Treleani:2007gi} it is argued, on the basis of the simplest two channel
eikonal model for the proton--proton cross section, that a more appropriate value at
$\sqrt{s}= 1.8$ TeV is 10 mb which translates at the LHC into  
$\sigma_{eff}^{LHC}=12$ mb. Treleani then estimates the effect of the removal by CDF
of triple parton interactions (TPI) events from their sample and concludes that the CDF measurement yields
$\sigma_{eff} \approx 11$ mb at Tevatron energies.
In the following we use  $\sigma_{eff}=12.0$ mb for
all LHC center of mass energies,
with the understanding that this value is affected by an experimental uncertainty
of about 15\% and that it agrees only within 30\% with the predictions of the eikonal model.
Since $\sigma_{eff}$ appears as an overall factor
in our results it is easy to take into account a different value.

It is worth mentioning that at present there is a discrepancy between the value
of $\sigma_{eff}$ extracted by CDF and D0 and the one which is effectively
employed by \Pythia
whose normalization is derived mainly from comparisons with small p$_{T}$ data
which dominate the total cross section.
The description of MPI in \Pythia 8~\cite{Sjostrand:2007gs}
assumes that interactions can occur at different p$_{T}$ values 
independently of each other inside inelastic non--diffractive events.  
The expression for a DPS cross section, here referred to as Double Parton Interactions (DPI) cross section, becomes therefore: 
\begin{equation}
\label{MPI_eq:sigma_3}
    \sigma =  <f_{impact}>\sigma_1 \cdot \sigma_2/\sigma_{ND}/k
\end{equation}
where $\sigma_{ND}$ is the total non--diffractive cross section and $f_{impact}$
is an enhancement/depletion factor chosen event-by-event to account for correlations 
introduced by the centrality of the collision. This quantity is typically averaged during 
an entire run to calculate $<f_{impact}>$ in Eq.~\ref{MPI_eq:sigma_3}.
Typical values at the center of mass energy of 10~TeV are 1.33 for
$<f_{impact}>$ and 51.6~mb for $\sigma_{ND}$.
Comparing Eq.~\ref{MPI_eq:sigma_3} with Eq.~\ref{eq:sigma_D3} tells us that \Pythia 8
predicts an effective $\sigma_{eff}$=$\sigma_{ND}$/$<f_{impact}>$ which
is about a factor three larger than the one actually measured at the Tevatron. I believe
that this issue deserves careful consideration and that new measurements of high p$_{T}$
MPI reactions would be quite welcome.

NLO QCD corrections are or will soon be available for all 
SPS, here referred to as Single Parton Interactions (SPI)
processes leading to an electroweak
vector boson in association with up to four jets
\cite{Campbell:2007ev,Campbell:2002tg,Campbell:2003hd,Ellis:2009zw,KeithEllis:2009bu,Berger:2008sz,Berger:2009ep,Berger:2010vm}.
The Drell-Yan cross section is known at NNLO \cite{Hamberg:1990np}.
Measurements at the Tevatron show good agreement
between NLO calculations and data \cite{Aaltonen:2007ip,Aaltonen:2007cp}. These new developments
open the possibility of validating the predictions using events with large visible
energy, where the MPI
contribution is small, and then using them for a direct measurement of the MPI
cross section at smaller total invariant masses in parallel with more data driven
analysis similar to those of CDF and D0.

In the following we compare the results obtained with the
GS09 dPDF with those obtained with two instances
of fully factorized sPDF: MSTW2008LO \cite{Martin:2009iq} and CTEQ6L1 \cite{Pumplin:2002vw}.
Hence we can estimate, even in the absence of a proper dPDF set based on CTEQ6,
the dependence of MPI predictions on the choice of PDF, a study that to our knowledge
has not been performed before.

The  MPI processes which contribute at leading order to $Z+n$--jets through Double
Parton Interactions are those in which an event producing $k$ jets is superimposed
to an event producing a $Z$--boson and $(n-k)$ jets, $k=2,\dots,n$.

All samples have been generated with the following cuts:

\bea
\label{eq:cuts}
& p_{T_j} \geq 30~{\rm GeV} \, , \; \; |\eta_j| \leq 5.0 \, , 
\nonumber \\
& p_{T_\ell} \geq 20~{\rm GeV} \, ,\; \;
|\eta_{\ell}| \leq 2.5 \, , \\
& p_{T_\gamma} \geq 30~{\rm GeV} \, , \; \; |\eta_\gamma| \leq 2.5 \, , 
\nonumber \\
& \Delta R_{jj} \geq 0.5  \, ,\; \; \Delta R_{jl} \geq 0.1\, ,\; \; \Delta R_{j\gamma} \geq 0.1\nonumber 
\eea

where $j= u,\bar{u},d,\bar{d},s,\bar{s},c,\bar{c},b,\bar{b},g$ and $l=e^+,e^-,\mu^-,\mu^+$.

The $Z+4$--jets sample has been generated with \Phantom
\cite{Ballestrero:2007xq,Ballestrero:1994jn,Ballestrero:1999md}, while all
other samples
have been produced with \MadEvent \cite{Maltoni:2002qb,Alwall:2007st}.
All samples have been generated using CTEQ6L1 \cite{Pumplin:2005rh} 
parton distribution functions.
The QCD scale (both in $\alpha_s$ and in the parton distribution functions)
has been taken as

\be
\label{eq:LargeScale}
Q^2 =\sum_{i=1}^n p_{Ti}^2,
\ee
where $n$ is the number of final state partons, for all reactions with the exception of
$q\bar{q}\rightarrow l^+l^-$ for which the scale has been set at $Q^2 = M_Z^2$.

The results shown in the following under the CTEQ heading have been obtained
combining 
at random one event from each of the reactions which together
produce the  desired final state through MPI.

The results shown under the MSTW and GS09 headings have been obtained through a
reweighting procedure by the appropriate ratio of parton distribution functions
and coupling constants. For instance, an event like $(q_i\bar{q_i} \rightarrow gl^+l^-)
\otimes (gg \rightarrow gg)$, constructed from two events generated separately
with CTEQ6 PDF,
can be transformed in a weighted event with MSTW2008 PDF multiplying its original
weight by  

\be
\label{eq:reweight}
R = \frac{D^{^{MSTW}}_i(t_1)D^{^{MSTW}}_{\bar{i}}(t_1)}{D^{^{CTEQ}}_i(t_1)D^{^{CTEQ}}_{\bar{i}}(t_1)} \times
\frac{\alpha^{^{MSTW}}_s(t_1)}{\alpha^{^{CTEQ}}_s(t_1)} \times
\frac{D^{^{MSTW}}_g(t_2)D^{^{MSTW}}_g(t_2)}{D^{^{CTEQ}}_g(t_2)D^{^{CTEQ}}_g(t_2)}  \times
\frac{\alpha^{^{MSTW}}_s(t_2)^2}{\alpha^{^{CTEQ}}_s(t_2)^2}
\ee


where $t_1,\, t_2$ are the factorization scales for $q_i\bar{q_i} \rightarrow gl^+l^-$
and $gg \rightarrow gg$ respectively. The factorization scales have been read off
from the event files.
The second and fourth factors in \eqn{eq:reweight}
take into account the different values of the strong coupling constants for the two
different sets of PDF: $\alpha_{s,LO}^{^{CTEQ}}(M_Z)=0.130$ while
$\alpha_{s,LO}^{^{MRST}}(M_Z)=\alpha_s^{^{GS09}}(M_Z)=0.139$.
The only difference for the GS09 case would be that
the correlated dPDF $F_{ij}(t_1,t_2)$ would appear instead of the uncorrelated
product $D_i(t_1)D_j(t_2)$ and so on.

All results are obtained with the following values for the electroweak input parameters:
$M_Z$ = 91.188 GeV, $M_W$ = 80.40 GeV, $G_F$ = 0.116639 $\times$ 10$^{-5}$ GeV$^{-2}$. 

The total cross sections for SPI and DPI production for $Z+2$--jets, $Z+3$--jets
and $Z+4$--jets are presented in Table \ref{tab:Znj}.
In our estimates below we have only taken into account the muon
decay of the $Z$ boson. 

\begin{table}[thb]
\begin{center}
\begin{tabular}{|c|c|c|c|c|c|c|c|c|c|}
\hline
 & \multicolumn{3}{|c|}{14 TeV} & \multicolumn{3}{|c|}{10 TeV}  & \multicolumn{3}{|c|}{7 TeV} \\
\hline
  &  CTEQ  & MSTW & GS09   &  CTEQ  & MSTW & GS09   &  CTEQ  & MSTW & GS09 \\
\hline
\multicolumn{10}{|c|}{$Z+2j$}\\ 
\hline
SPI & 52.65 & 60.70 &    & 30.63  & 35.15  &   &  16.56   & 18.88 & \\
\hline
DPI & 11.27  & 14.37 & 15.50  & 4.80  & 6.35  & 6.68 &  1.88 & 2.61 & 2.66\\
\hline
\multicolumn{10}{|c|}{$Z+3j$}\\ 
\hline
SPI & 15.71 & 19.10 &    & 8.46  & 10.23  &   &  4.11   & 4.93 & \\
\hline
DPI & 2.70  & 3.75 & 3.88  & 1.02  & 1.49  & 1.48 &  0.34 & 0.54 & 0.51\\
\hline
\multicolumn{10}{|c|}{$Z+4j$ }\\ 
\hline
SPI & 4.26 & 5.41 &    & 2.00  & 2.53  &   &  0.83   & 1.04 & \\
\hline
DPI & 0.96  & 1.53 & 1.50  & 0.33  & 0.56  & 0.52 &  0.10 & 0.18 & 0.16\\
\hline
\end{tabular}
\end{center}
\caption{$Z+n$--jets, $Z\rightarrow\mu^+\mu^-$ cross sections in pb. Cuts as in \eqn{eq:cuts} with
 $\Delta R_{jj} = 0.5$.
}
\label{tab:Znj}
\end{table}

The total cross sections for SPI and DPI production for $\gamma+3$--jets
are shown in Table \ref{tab:gamma3j}. 
At the LHC
trigger thresholds for single photons are foreseen
to be much higher than those for double leptons \cite{Bayatian:2006zz,Ball:2007zza,Aad:2009wy}.
While pair of leptons are expected to be triggered on for transverse momenta of about
15 GeV, single photons will be detected only when their transverse momenta 
is larger than about 80 GeV at the design energy of 14 TeV.
Since MPI processes are known to decrease sharply with increasing transverse momenta,
we present also
the predictions for $p_{T_\gamma} \geq 80~{\rm GeV}$.

The Single Particle Interaction MSTW results are larger than those obtained with the
CTEQ PDF by an amount which varies between 15\% for $Z+2j$ to 27\% for $ Z+4j$, increasing
as expected with the power of $\alpha_{s}$ in the amplitude.
The Double Particle Interaction MSTW results are larger than those obtained with the
CTEQ PDF by an amount which varies between 30\% and 90\%. The larger shift is due to the
smaller scales for the two individual scatterings compared to a single interaction event
with the same final state particles.
The predictions for the GS09 correlated dPDF are larger than those with
MSTW uncorrelated ones for $\sqrt{s} = 14~{\rm TeV}$ and $\sqrt{s} = 10~{\rm TeV}$
while they are smaller for $\sqrt{s} = 7~{\rm TeV}$. The difference is at most of 15\%.
Taking into account the errors in the measurement of $\sigma_{eff}$ we conclude that
the uncertainties due to the choice of PDF and to correlation effects
are reasonably under control.

These variations should be compared with the uncertainty due to scale variation in
PDF and in the strong coupling constant. 
Changing the scale in \eqn{eq:LargeScale} by a factor of two in either direction
for two limiting cases, namely $Z+2j$ production at $\sqrt{s} = 7~{\rm TeV}$ and
$Z+4j$ production at $\sqrt{s} = 14~{\rm TeV}$ the cross section changes by
+14\%/-13\% in the first case and by +57\%/-29\% in the second.

The effects of higher order corrections are more difficult to estimate since no NLO
calculation for MPI processes is available. QCD one loop calculations are available
for vector boson production with up to four jets and are typically
of order 10\% with the exception of Drell--Yan inclusive production
where they are of the order of 50\%.
NLO corrections for the inclusive jet cross section at the LHC have been presented in
\rf{Campbell:2006wx}. For small transverse momenta, as the ones we are interested in this paper,
they are of the order of 10\%.

The ratio between the MPI and SPI cross sections increases with the collider energy,
that is with decreasing average momentum fractions carried by the incoming partons.
It also increases with the $\Delta R_{jj}$ separation because of the absence of correlations
between the final state partons originating in the independent scatterings which compose
MPI events.
For $Z+nj$ processes and taking $\Delta R_{jj}=0.5$ as an example, the ratio 
is of the order of 10\% for $\sqrt{s} = 7~{\rm TeV}$ and grows to about 25\%
at $\sqrt{s} = 14~{\rm TeV}$.
The results for $\gamma+3$--jets show a similar behaviour with somewhat smaller
fractions of MPI events to SPI ones which however
depend drastically on the $p_{T_\gamma}$
cut. For $p_{T_\gamma} \geq 30~{\rm GeV}$ they range between 5 and 10\% while
for $p_{T_\gamma} \geq 80~{\rm GeV}$ they are at the percent level.

\begin{table}[thb]
\begin{center}
\begin{tabular}{|c|c|c|c|c|c|c|c|c|c|}
\hline
& \multicolumn{3}{|c|}{14 TeV} & \multicolumn{3}{|c|}{10 TeV}  & \multicolumn{3}{|c|}{7 TeV} \\
\hline
 $\gamma+3j$ &  CTEQ  & MSTW & GS09   &  CTEQ  & MSTW & GS09   &  CTEQ  & MSTW & GS09 \\
\hline
\multicolumn{10}{|c|}{$p_{T_\gamma} \geq 30~{\rm GeV}$}\\ 
\hline
SPI &  4516.7 & 5610.2 &    & 2637.2  & 3263.8  &   &  1415.8   & 1744.6 & \\
\hline
DPI & 422.2  & 593.7 & 642.2 & 170.3 & 254.1  & 264.9 &  62.0 & 100.0 & 99.4\\
\hline
\multicolumn{10}{|c|}{$p_{T_\gamma} \geq 80~{\rm GeV}$}\\ 
\hline
SPI &  671.5 & 813.0 &    & 368.30  & 443.38  &   &  177.4   & 212.0 & \\
\hline
DPI & 17.7  & 28.1 & 28.1 & 6.59 & 11.49  & 10.85 &  2.09 & 4.09 & 3.58\\
\hline
\end{tabular}
\end{center}
\caption{$\gamma+3$--jets cross sections in pb. Cuts as in \eqn{eq:cuts}.
}
\label{tab:gamma3j}
\end{table}

\begin{figure}[tb]
\centering
\subfigure{	 
\hspace*{-2.3cm} 
\includegraphics*[width=8.3cm,height=6.2cm]{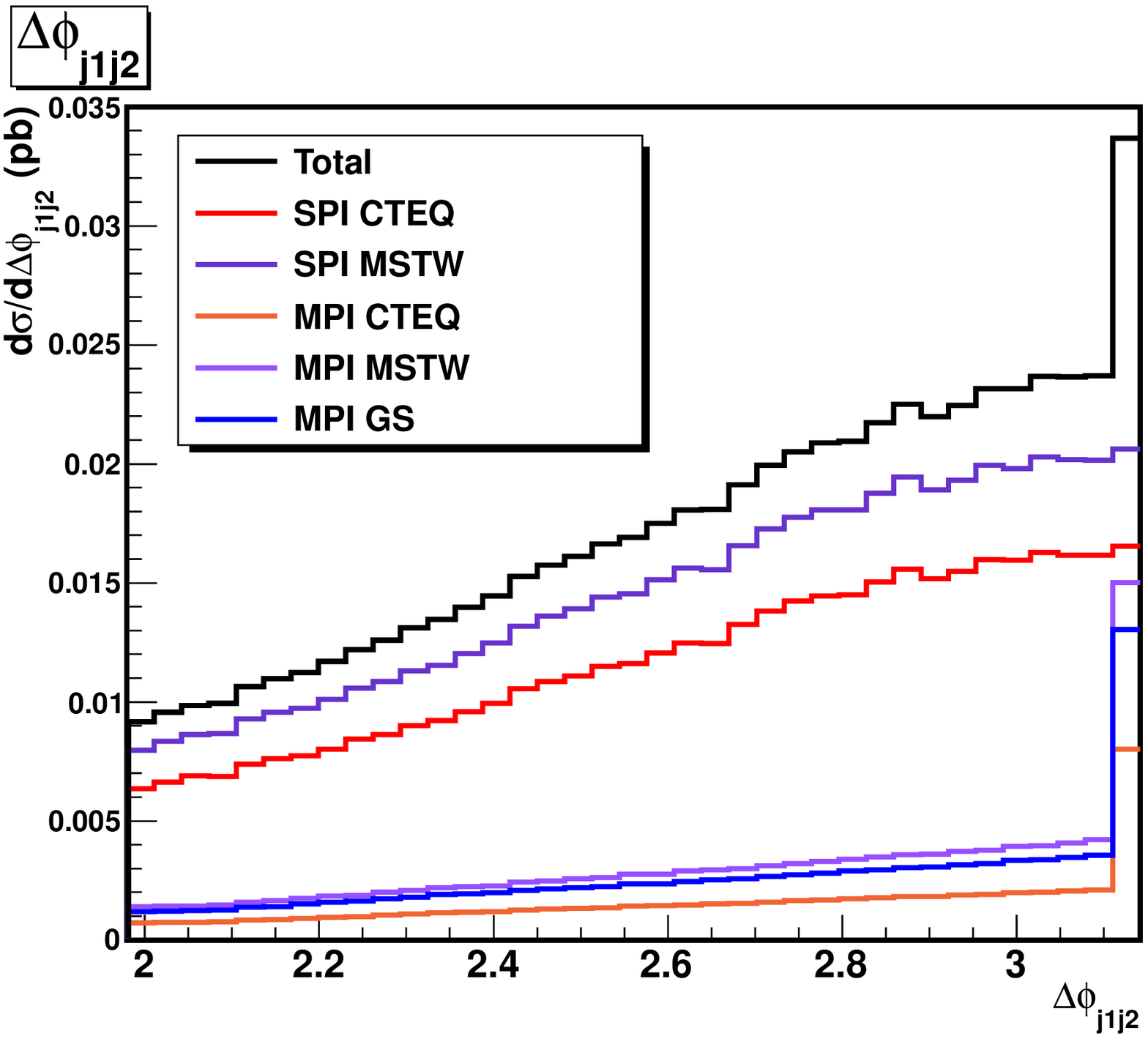}
\includegraphics*[width=8.3cm,height=6.2cm]{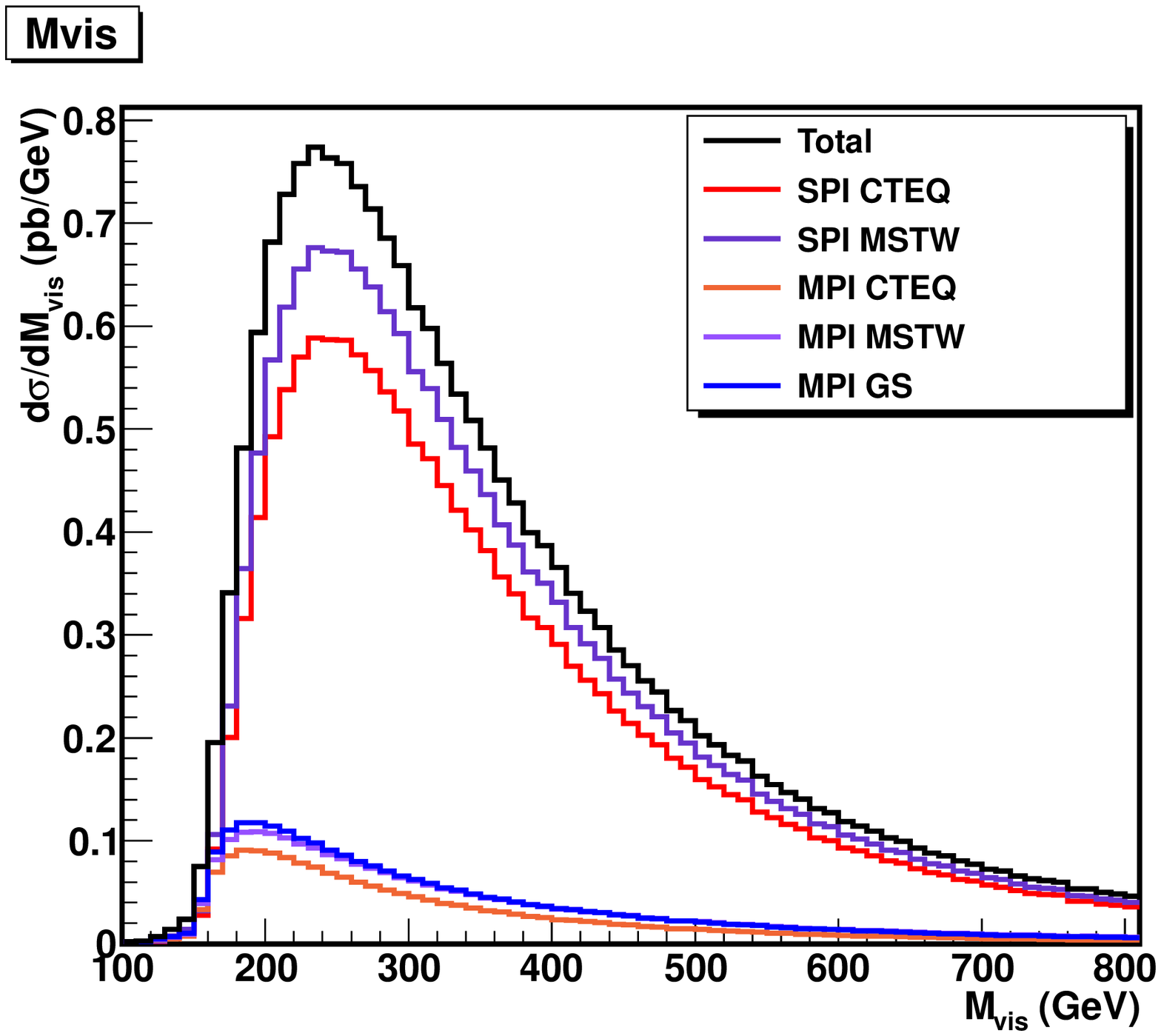}
\hspace*{-3cm}
}
\caption{On the left: distribution of the angular separation 
in the transverse plane between the two highest $p_{T}$ jets  in $Z+4j$ events.
On the right: distribution of the total visible mass,
$(\sum_{i=1}^n p_{i})^2$, in $Z+2j$ events. For both plots $\sqrt{s} = 7~{\rm TeV}$,
$\Delta R_{jj}=0.7$.}
\label{fig:Zjets_dist}
\end{figure}

If we consider the MPI processes as our signal and the SPI ones
as the corresponding background,
we can estimate the prospect of measuring MPI in a given final state from the standard
$S/\sqrt{B}$ significance.
Using for $S$ the result obtained with GS09 PDF and for $B$ the result for the MSTW set
and assuming a luminosity of one inverse femtobarn at 7 TeV, the significancies
extracted from Table \ref{tab:Znj}, in the $Z\rightarrow\mu^+\mu^-$ channel alone,
are 19/7/5 for $Z$+2/3/4 jets .
The corresponding number of expected MPI events are 2600/500/160.
Therefore it appears quite feasible to measure the MPI contribution to  
$Z$+2/3/4 jets already in the first phase of the LHC.

The significance of $\gamma+3$--jets depends on the trigger strategies.
If the threshold for single photon detection can be brought in the $30~{\rm GeV}$
range then the much larger production rate, about ten times that of $Z(\mu\mu)+2j$,
provides the best opportunity for an early measurement of MPI at the LHC.
If, on the contrary, the photon trigger cannot substantially deviate from about
$80~{\rm GeV}$, $Z+2j$ production looks more promising than the $\gamma+3$--jets channel
whose significance becomes similar to that of $Z+3j$.

The contribution to the MPI $Z+n$--jets cross section
due to two jet production in association to $Z+(n-2)$--jets processes
is in all instances the largest one,
therefore, even with more than two jets in the final state, the majority of MPI
events are expected to contain a pair of jets which are back to back in the transverse plane.
This is confirmed by the left hand side of \fig{fig:Zjets_dist}
which displays the distribution of the angular separation $\Delta\phi$
between the two highest $p_{T}$ jets in $Z+4j$ events
at $\sqrt{s} = 7~{\rm TeV}$ and $\Delta R_{jj}=0.7$.


The right hand side of \fig{fig:Zjets_dist} presents the 
total visible mass distribution in $Z+2j$ production with
the same energy and angular separation. It clearly shows that MPI events are produced with
a smaller center of mass energy than SPI ones.

\section{Summary and Outlook}

The understanding  multi-parton interactions (MPI)
in hadronic collisions remains a challenge.  Traditionally, research 
has concentrated on four tasks covered during this workshop:
experimental measurements of underlying events and minimum bias events, models
of soft physics implemented in Monte Carlo (MC) generators, development of the  
theoretical description of MPI, and phenomenological studies.  In particular, a lot of effort, strongly
driven by the experimental measurements, reported here by the ATLAS,
CMS and LHCb collaborations, see Sections~\ref{dkar}, \ref{cms},
\ref{mschmelling}, respectively, focuses on modeling of minimum bias
and underlying event physics in MC generators. This is, of course,
only natural and very much needed given that the MC generators are major tools used in the experimental analyses by the LHC collaborations. In fact, one of the main tasks in LHC phenomenology is the construction and
development of general purpose MC generator fully able to 
describe exclusive states, including contributions from MPI.   This task has enormous
importance for the overall success of the LHC physics
programme.  However, in order to improve the simulation of the MPI in
the MC codes, more elaborate theoretical input, supplemented by information from high-$p_T$ phenomenological studies, is necessary.

The standard MC algorithms are based on the factorization
theorem and the factorization of the QCD amplitudes. This theorem lets 
us to define general algorithms and leads to good predictions for the
LHC. However, the extension of the single interaction picture
to encompass MPI is not a simple task. The main challenge is to
understand the structure and the topology of the MPI based on perturbative QCD, as discussed by D.~Treleani in Section~\ref{treleani}. Two important issues can be identified: the treatment of
the parton distributions and the parton dynamics.

The presence of multiple interacting partons in the initial state enforces the appearance multi parton distribution functions (mPDF).  In the MC programs the mPDF functions are simple scalar functions, resulting from approximating a mPDF function with $n$ partons by a product of single PDF functions. This is a good approximation under the assumption that the correlations in colour, spin and flavour can be neglected, as the most important information about the mPDF in the parton shower context comes from their evolution. Deficiencies in the approximation of the initial-state mPDF reached after backward evolution can be compensated through changes (tuning) in the hadronization.

In the current MC implementations the evolution of the multiple
partons taking part in the same collision relies on well understood
evolution equations for single PDF.   Going beyond this approximation and
moving to full mPDF evolution poses many challenges. The evolution
equation for mPDF scalar functions has been available for some time.  
Recently, J.~R.~Gaunt and W.~J.~Stirling,
cf. Section~\ref{gaunt},  provided a solution and detailed studies
for the double parton case.  However, the question of how double parton distributions should be
    defined and which evolution equation they satisfy has been
    reconsidered since the workshop, see~\cite{Diehl:2011tt,Gaunt:2011xd,Gaunt:2011xu}. Moreover, it is not clear whether the mPDF
functions should be scalar functions or operators in color, spin and
flavor space. As M.~Diehl has pointed out in section~\ref{diehl}, the
non-trivial color, spin and flavor correlations occur in the dPDF case
and they can be as important as the uncorrelated contributions. For
example, non-trivial correlations in colour space can lead to similar
effects as those caused by colour reconnection in the hadronization
model. These issues would have to be resolved in order to achieve
proper description of mPDF in MC generators.

The other important ingredient in the MC models is the treatment of
the parton dynamics.  Initial and final state radiation is well
understood in the standard parton shower models, but the MPI
contributions are more complex. {\sc Herwig++}
(cf. Section~\ref{herwig}) follows a rather minimalistic approach and
considers MPI effects only via the $2\to2$ parton interactions while
{\sc Pythia} studies also effects of rescattering and joint
interaction in the MPI models, see Section~\ref{pythia}. These
contributions are considered in a classical probabilistic
framework. To better understand and improve the existing models it is
important to look beyond the existing approach. In perturbative
parton dynamics there are usually two types of contributions. The
first type is {\em real} radiation, when one or more partons appear in
the final state, and the partons are resolvable. These contributions
can be described by a rather simple splitting function (operator) that
can be obtained from tree level Feynman graphs. The other
contributions are the {\em virtual} and {\em unresolvable}
configurations which are not considered explicitly but are included via
unitary conditions. Usually, the splitting operator of virtual
contributions has some imaginary part that cannot be obtained from
the unitary conditions and thus is not taken into account in the
current MC implementations.  Such imaginary contributions, known in the
literature as the Coulomb gluon terms, can lead to the appearance of
the so-called super leading logarithms in perturbative calculations
for certain quantities. Correspondingly, these effects cannot be
reproduced by the existing parton shower MC codes. Since the imaginary part of the virtual splitting operator can change the colour structure, its effect might be similar in the colour space to the action of colour reconnection. It has been pointed out that colour reconnection, contributing to hadronization, can have a big impact on the predictions, and therefore is important for description of data. Consequently, one should systematically consider all the effects of  perturbative origin, e.g. the colour-changing effects, in order to ensure the universality of the non-perturbative hadronization model.

It is often said that a good MC program has tunable parameters only in the hadronization model. This is more or less is true if we do not have to consider MPI effects, otherwise we need to deal with mPDF which have transverse momentum dependence. Unfortunately, we have very little information on the transverse momentum dependence. The standard practice is to assume the factorization of the longitudinal and transverse dependent parts of the mPDF. In the MC generators, the transverse momentum dependent part is described and parametrized by a simple function for every parton flavour. {\sc Herwig++} makes use of only three tunable parameters while {\sc Pythia} has more complicated parametrization with more parameters. This tunable function describes non-perturbative effects, and it can be shown that the average number of the interactions strongly depends on it. In principle we know very little about goodness of the factorization assumption. Theoretical studies, based on modeling the proton substructure by a dipole cascade model discussed by G. Gustafson in Section~\ref{sec:gosta}, show that the transverse part has a non-trivial dependence on the longitudinal momentum fraction of the incoming parton and the factorization scale. At this point, phenomenological studies of  DPS processes with high $p_T$ final states have an important role to play, as comparisons with experimental results will serve to test the predictions and uncover further information on the transverse momentum dependence of the mPDF.   

The DPS measurements will also provide an important validation
procedure of the MPI models built into MC generators. For this purpose, one needs to identify the processes where the DPS signal is favourable over SPS background. Promising candidate processes such as same-sign $W$ production, $Z$ production in association with jets, four-jet  production or production of a $b\bar{b}$ pair with two jets have been discussed in previous chapters. As pointed out by E.~Maina in Section~\ref{maina}, measuring DPS in different processes, for example in $Z \ + \ jets$ and $W^\pm W^\pm$ production, will deliver complementary information since different initial state parton combinations will be probed.
It is also important to study if one can define quantities more
sensitive to DPS than the ones measured currently. Two examples of such studies for $b\bar{b}$ pair plus two jet production and $p p \rightarrow Wb\bar{b} X \to \ell \nu b\bar{b} X$ were presented by E.~Berger, see Section~\ref{berger}. Alternatively, kinematical regions where the DPS provides the dominant signal can be selected. As reported by B.~Blok, cf. Section~\ref{blok}, in the case of four-jet production such region is constituted by back-to-back dijet production. For gauge boson pair production, the relevant region is where the transverse momentum of each boson is small, as shown in Section~\ref{diehl}. It has to be also checked that the background to the actual final state observed experimentally can be sufficiently suppressed, cf. same-sign lepton final states discussed by C.-H.~Kom in Section~\ref{kom}.

In summary, understanding MPI in hadronic collisions requires further
efforts on both theoretical and experimental fronts. In particular, it
would be helpful to critically reevluate the description of MPI in MC
codes and the tuning strategies. Although comparing results provided
by existing MC codes supplemented with various tunes is certainly
valuable on its own, there is a serious risk that it will not bring
full understanding of the limitations and the systematic errors of 
MC generators. Apart from purely theoretical work and the work
related to development of MC codes, it would be advantageous to, for
example, identify and explore phase-space regions sensitive to MPI
effects, other than those used so far in the experimental
studies. Furthermore, one would also benefit from constructing new
variables probing MPI in particular processes. As already stressed,
such efforts need to be undertaken in common by the experimental and
theoretical communities.\\[2em]

\noindent
{\large \bf Acknowledgements}

The work of RC was supported by the Marie Curie Early Stage Training program
``HEP-EST'' (contract number MEST-CT-2005-019626), the Marie
Curie research training network ``MCnet'' (contract number
MRTN-CT-2006-035606), and the Swedish Research Council (contract numbers
621-2008-4252 and 621-2007-4157). The research presented in Section~\ref{blok} was supported by the
United States Department of Energy and the Binational Science Foundation.
 LF and MS would like to thank the Yukawa International
Program for Quark--Hadron Sciences  for hospitality during
a part of this study.
The work reported in Section~\ref{berger} was done in collaboration with Chris Jackson, Seth Quackenbush, and Gabe Shaughnessy, and it was supported financially by the U.~S.\ Department of Energy under Contract No.\ DE-AC02-06CH11357.

\bibliographystyle{JHEP.bst}
\bibliography{mpi10}


\end{document}